\documentstyle{elsart}
\begin{document}
\input psfig.sty
\hyphenation{MW-PC}
\hyphenation{MW-PCs}
\hyphenation{Pho-to-e-lec-trons}
\hyphenation{pho-to-e-lec-trons}
\hyphenation{Slo-cum}
\font\mittw=cmsy10 at 12 true pt
\def\sT{\hbox{T}}
\def\st{\hbox{t}}
\hoffset=0pt
\voffset=0pt
\begin{frontmatter}
\title{\bf Design, Commissioning and Performance \\
of the PIBETA Detector at PSI}
\author[UVa]{E. Frle\v z\thanksref{author}},
\thanks[author]{Corresponding author. Tel: +1--434--924--6786, 
fax: +1--434--924--4576, e--mail: frlez@virginia. edu (E. Frle\v{z})\hfill}
\author[UVa]{D.~Po\v{c}ani\'c},
\author[UVa]{K.~A.~Assamagan\thanksref{BNL}},
\thanks[BNL]{Current address: Department of Physics, 
Brookhaven National Laboratory, Upton, New York 11973.} 
\author[Tbi]{Yu.~Bagaturia\thanksref{Hei}},
\thanks[Hei]{Current address: Physikalisches Institut der Universit{\"a}t 
Heidelberg, D-69210, Heidelberg, Germany.}
\author[Dub]{V.~A.~Baranov},
\author[PSI]{W.~Bertl},
\author[PSI]{Ch.~Br\"onnimann},
\author[UVa]{M.~Bychkov},
\author[PSI]{J.~F.~Crawford\thanksref{ret}},
\thanks[ret]{Current address: Retired.}
\author[PSI]{M.~Daum},
\author[PSI]{Th.~Fl\"ugel\thanksref{Bas}},
\thanks[Bas]{Current address: Basler Versicherungen, \"Aschengraben 21,
Postfach 2275, CH-4002 Basel, Switzerland.}
\author[PSI]{R.~Frosch\thanksref{ret}},
\author[PSI]{R.~Horisberger},
\author[Dub]{V.~A.~Kalinnikov},
\author[Dub]{V.~V.~Karpukhin},
\author[Dub]{N.~V.~Khomutov},
\author[UVa]{J.~E.~Koglin\thanksref{Col}},
\thanks[Col]{Current address: Columbia Astrophysics Laboratory, 550 West 120 Street, 
New York, NY 10027, USA.}
\author[Dub]{A.~S.~Korenchenko},
\author[Dub]{S.~M.~Korenchenko},
\author[Swi,PSI]{T.~Kozlowski},
\author[PSI]{B.~Krause\thanksref{Desy}},
\thanks[Desy]{Current address: DESY, Deutsches Elektronen Synchrotron, 
22603 Hamburg, Germany.}
\author[Dub]{N.~P.~Kravchuk},
\author[Dub]{N.~A.~Kuchinsky},
\author[UVa]{W.~Li},
\author[ASU]{D.~W.~Lawrence\thanksref{UMas}},
\thanks[UMas]{Current address: Department of Physics, 
University of Massachusetts, Amherst, MA~01003, USA.}
\author[UVa]{R.~C.~Minehart},
\author[Tbi,Dub]{D.~Mzhavia},
\author[PSI]{H.~Obermeier},
\author[PSI]{D.~Renker},
\author[ASU]{B.~G.~Ritchie},
\author[UVa,PSI]{S.~Ritt},
\author[Tbi,PSI]{T.~Sakhelashvili},
\author[PSI]{R.~Schnyder\thanksref{Mikro}},
\thanks[Mikro]{Current address: Zentrum f{\"u}r Mikroelektronik Aargau, 
CH-5210 Windisch, Switzerland}
\author[Dub]{V.~V.~Sidorkin},
\author[UVa]{P.~L.~Slocum\thanksref{Aer}},
\thanks[Aer]{Current address: The Aerospace Corporation, M2/260 PO Box 92957,
Los Angeles, CA 90009-2957, USA.}
\author[UVa]{L.~C.~Smith},
\author[IRB,PSI]{N.~Soi\'c},
\author[UVa]{W.~A.~Stephens},
\author[IRB]{I.~Supek},
\author[Tbi,Dub]{Z.~Tsamalaidze},
\author[UVa]{B.~A.~VanDevender},
\author[UVa]{Y.~Wang},
\author[PSI]{H.~P.~Wirtz\thanksref{phi}},
\thanks[phi]{Current address: Philips Semiconductors AG, Binzstr. 44, CH-8045  Z\"urich, Switzerland.}
\author[UVa]{K.~O.~H.~Ziock\thanksref{ret}}
\address[UVa]{Department of Physics, University of Virginia, 
Charlottesville, VA~22904-4714, USA}
\address[Tbi]{Institute for High Energy Physics, Tbilisi State University, 
GUS-380086 Tbilisi, Georgia}
\address[Dub]{Joint Institute for Nuclear Research, RU-141980 Dubna, Russia}
\address[PSI]{Paul Scherrer Institut, Villigen PSI, CH-5232, Switzerland}
\address[Swi]{Institute for Nuclear Studies, PL-05-400 Swierk, Poland}
\address[ASU]{Department of Physics and Astronomy, Arizona State University, Tempe, 
AZ~85287, USA}
\address[IRB]{Rudjer Bo\v{s}kovi\'c Institute, HR-10000 Zagreb, Croatia}
\vfill\eject

\begin{abstract}
We describe the design, construction and performance of the PIBETA
detector built for the precise measurement of the branching ratio of
pion beta decay, $\pi^+\rightarrow\pi^0e^+\nu_e$, at the Paul Scherrer 
Institute. The central part of the detector is a 240-module
spherical pure CsI calorimeter covering $\sim$3$\pi\,$sr solid angle.
The calorimeter is supplemented with an active collimator/beam degrader 
system, an active segmented plastic target, a pair of low-mass cylindrical 
wire chambers and a 20-element cylindrical plastic scintillator hodoscope. 
The whole detector system is housed inside a temperature-controlled lead brick 
enclosure which in turn is lined with cosmic muon plastic veto counters.
Commissioning and calibration data were taken 
during two three-month beam periods in 1999/2000 with $\pi^+$ stopping rates 
between 1.3$\cdot 10^3\,\pi^+$/s and 1.3$\cdot 10^6\,\pi^+$/s. We examine 
the timing, energy and angular detector resolution for photons, positrons 
and protons in the energy range of 5--150$\,$MeV, as well as the response 
of the detector to cosmic muons. We illustrate the detector signatures for 
the assorted rare pion and muon decays and their associated backgrounds. 
\par\noindent\hbox{\ }\par\noindent
\par\noindent\hbox{\ }\par\noindent
\vfill
\end{abstract}
\end{frontmatter}
\vfill\eject

\section{Introduction}\label{intr}

The PIBETA collaboration~\cite{Poc91} has conducted a program of 
measurements at the Paul Scherrer Institute (PSI) with the ultimate 
goal of determining precisely the branching ratio of pion beta decay ($\pi\beta$),
$R(\pi^+\rightarrow\pi^0e^+\nu_e)$.
$\pi\beta$ decay is one of the most fundamental semileptonic weak interactions.
It is a transition between two spin-zero members of an isospin
triplet, and is therefore analogous to super-allowed pure Fermi transitions
in nuclear beta decay~\cite{Har90}. Due to its relative simplicity and, 
consequently, its accurate theoretical description, Fermi beta decay 
stands today among the prominent successes of the Standard Model of electroweak 
interactions~\cite{Tow98}.

The most recent and most precise determination of the $\pi\beta$ decay rate
$1/\tau_{\pi\beta}$ is due to McFarlane and collaborators~\cite{McF85}, who 
performed a decay-in-flight measurement using the LAMPF $\pi^0$ spectrometer. 
Their result, $0.394\pm 0.015\,{\rm s}^{-1}$, is in good agreement 
with the Standard Model prediction of $0.4027\pm 0.0018\, {\rm s}^{-1}$~\cite{McF93}, 
but with a $\simeq$\,3.8\,\% uncertainty it does not test the full extent 
of the radiative corrections which stand at $\simeq$\,3\,\%~\cite{Mar86,Jau01}. 
When combined with the $\pi^+$ lifetime~\cite{PDG}, the rate measured by McFarlane 
et al. gives the $\pi\beta$ branching ratio relative to all pion decays of
$\Gamma_{\pi\beta}/\Gamma_{\rm total}=(1.026\pm 0.039)\cdot 10^{-8}$.

In the first phase of the PIBETA experiment we have collected $\sim$60,000 
$\pi\beta$ events, and we expect to extract the branching ratio with 
$\simeq$0.4$\,$\% statistical and comparable systematic uncertainty. 
We have cross-calibrated the absolute 
$\pi^+$ beam intensity using the measured rate of the $\pi^+\to e^+\nu$ decays which
is known with the combined statistical and systematic uncertainty 
of $\simeq$\,0.40\,\%~\cite{Cza93,Bri94}. The proposal for a second phase of 
the experiment, intended to reduce the overall uncertainty of both the $\pi\beta$ 
and the $\pi\to e\nu$ measurements to $\sim$0.2\,\% is now under consideration.

The PIBETA apparatus is a large solid angle nonmagnetic detector optimized
for measurements of photons and electrons in the energy range of 
5--150$\,$MeV. The main sensitive components of the apparatus, shown and labelled 
in Fig.~\ref{fig:det1}, are:
\begin{itemize}\setlength{\itemindent}{3ex}
\item[(1)] a passive lead collimator, PC, a thin forward beam counter, BC, two 
cylindrical active collimators, AC$_1$ and AC$_2$, and an active degrader, AD, 
all made of plastic scintillators and used for the beam definition;
\item[(2)] a segmented active plastic scintillator target, AT, used to stop 
the beam particles and sample lateral beam profiles;
\item[(3)] two concentric low-mass cylindrical multi-wire proportional chambers, 
MW\-PC$_1$ and MW\-PC$_2$ surrounding the active target, used for charged particle tracking;
\item[(4)] a segmented fast plastic scintillator hodoscope, PV, surrounding the MWPCs
used for particle identification;
\item[(5)] a high-resolution segmented fast shower CsI calorimeter surrounding
the target region and tracking detectors in a near-spherical geometry;
\item[(6)] a cosmic muon plastic scintillator veto counters, CV, around 
the entire apparatus (not shown).
\end{itemize}

All detector components listed above, together with the
photomultiplier tube (PMT) delay cables, high voltage (HV) supplies, MWPC 
instrumentation and gas system, fast trigger and digitizing electronics, two front-end 
computers, and temperature control system 
are mounted on a single platform and can be moved as a self-contained unit into 
the experimental area (Fig.~\ref{fig:pb_photo1}). When the detector platform is 
precisely positioned and surveyed with respect to the beam line, electrical 
power and Ethernet connections make the detector immediately operational.

The structure of this article is as follows. The secondary beam line elements, the beam tunes 
and the overall layout of the experimental area are presented in Sec.~\ref{beam}. In 
Secs.~\ref{beamc}--\ref{temp}, we describe the design, fabrication and quality control 
of the individual detector components enumerated above. In particular, the CsI calorimeter 
is covered in detail in Sec.~\ref{calo}. Sections~\ref{dsc}--\ref{user} cover
the PMT waveform digitizing system, the fast trigger electronics, and the data acquisition 
system with user analysis software. The calibration procedures and the performance of 
the detector are demonstrated in Secs.~\ref{en_cal}--\ref{track}. There we discuss 
detector timing, energy and angular resolution, particle identification, 
track reconstruction, and signal-to-background ratios for the selected pion and muon 
decay processes. Finally, a short overview of the Monte Carlo (MC) detector simulation
coded using the {\tt GEANT} package~\cite{Bru94} is given in Sec.~\ref{mc}.
At the end, in Sec.~\ref{rad_har}, we review the radiation resistance of 
the active detector elements.  

\bigskip
\section{Beam line and experimental area}\label{beam}
\medskip
The experiment was mounted at the Paul Scherrer Institute (PSI) in Villigen, Switzerland.
The isochronous separated sector cyclotron at PSI accelerates protons to an energy of 590\,MeV.
The $\sim\,$1.5\,mA proton beam is transported along the primary proton 
channel to two target stations where pions and muons are generated and
transported via secondary beam-lines to the experimental areas. 
The accelerator operates at the frequency of 50.63\,MHz producing
a microscopic beam structure of 1\,ns wide proton pulses separated
by 19.75\,ns.

The PIBETA apparatus is set up in the PSI $\pi$E1 experimental area whose 16$\,$m 
long beam line is designed to supply intense low energy pion beams with 
good momentum resolution~\cite{Sin81,Psi94}. Pions are extracted from the second target 
station E at an angle of $8^\circ$ with respect to the incident protons.
During the PIBETA measurements, this graphite target was 60\,mm in length along the axis 
of the proton beam.
Operating in a high-flux optical mode, the $\pi$E1 beam line can deliver a pion beam
with a maximum momentum of 280$\,$MeV/c, a Full-Width-Half-Maximum (FWHM) momentum 
resolution of $<$\,2\% and an accepted production solid angle of 32$\,$msr.
The primary proton current in the ring cyclotron during the PIBETA data 
acquisition periods in years 1999-2001 was 1.6$\,$mA DC on average.

Fig.~\ref{fig:pie1} shows a sketch of the fixed beam line elements between
the pion production target 
and the last pair of focusing magnets (which are embedded in the concrete shielding
of the proton channel). The channel incorporates three 
slit systems that are used to control the beam intensity either by collimating the vertical 
size of the beam (FSH51) and the beam halo (FS51), or by reducing the momentum band 
acceptance (horizontal jaws FSH52), hence improving momentum resolution of 
the transported beam~\cite{For97}. The bending dipole magnet ASZ51 located in 
the front section of the beam line is used to select the beam momentum.

We have developed a 113.4 MeV/c $\pi^+$ beam tune with 
FWHM momentum resolution $\Delta p/p\le 1.3\,$\% and maximum nominal $\pi^+$ beam 
intensity $I_\pi=1.4\cdot 10^6\,$$\pi^+$/s, reached at the full cyclotron 
current of 1.7$\,$mA. Beam tunes with $\pi^+$ fluxes below 200k $\pi^+$/s
have a narrower momentum acceptance of $\Delta p/p<0.28\,$\% (FWHM). The choice of 
a particular beam momentum is governed by the need for good time-of-flight (TOF) separation 
of pions, positrons and muons between the production target E and the first beam defining 
counter BC, as well as between the beam counter BC and the stopping target AT 
(see below). 

To reduce positron contamination even further, a $4\,$mm thick carbon degrader is inserted 
in the middle of the ASY51 dipole magnet. 
The momentum-analyzed pions and positrons have different 
energy losses in the carbon absorber, and are therefore spatially separated in a horizontal
plane.  Unfortunately, this also broadens the beam phase space. We have used TRANSPORT~\cite{Bro73} and 
TURTLE~\cite{Bro74} beam transport codes to develop a nontraditional beam optics 
with foci in both the horizontal and vertical planes at the FSH52 momentum-limiting slit. 
The resulting beam tune reduces the phase space broadening introduced by the carbon degrader. 
A significantly higher luminosity at the PIBETA target position is thus achieved and
the pions are stopped in a laterally smaller region.  

A simplified layout of the PIBETA apparatus
is depicted in Fig.~\ref{fig:hall}. A lead brick collimator PC with a 7$\,$mm pin-hole 
located 3.985$\,$m upstream of the detector center restricts the spatial spread of the incident $\pi^+$ 
beam. The beam 
particles are first registered in a 3$\,$mm thick plastic scintillator BC placed 
immediately upstream of the collimator. The pions are subsequently slowed in 
a 40$\,$mm long active plastic degrader AD and stopped in an active plastic target 
AT positioned at the center of the PIBETA detector. Fig.~\ref{fig:beam_p} shows the TURTLE-calculated 
momentum spectrum of the $\pi^+$ beam incident on the front face of the degrader counter.
The predicted $\pi^+$ FWHM momentum spread at 
this point is 1.2\,MeV/c/113.4\,MeV/c$\simeq$\,1.1\,\%.

The PSI surveying group measured the detector location using a transit theodolite. 
The upstream and downstream detector sides are levelled, with the center 
$1506.0\pm 0.3\,$mm above the floor, corresponding to the nominal beam 
height in the $\pi$E1 channel. The measurement of the detector center is 
reproducible and the alignment of the target axis along the detector 
longitudinal axis is confirmed with an uncertainty of 0.3$\,$mm.

We used the OPTIMA control program~\cite{Roh88} to adjust the currents 
in the dipole and quadrupole beam line magnets that steer and focus the $\pi^+$ beam 
into the target. The goal was to achieve the smallest,
most symmetric beam spot consistent with the high $\pi^+$ intensity. 
The OPTIMA program allows a user to maximize an arbitrary experimental rate 
normalized to the primary cyclotron current by iteratively changing the settings 
of the magnetic elements. We chose to maximize the rate of
four-fold coincidences between the forward beam counter BC, the degrader counter
AD, the active target AT and the accelerator rf signal. These four 
signals are combined in a coincidence unit in such a way that their overlap 
signals a $\pi^+$ particle stopping in the active target. Fig.~\ref{fig:pistop} 
shows the relative timing of the $\pi^+$-stop trigger signals. 
The final optimized settings of beam line magnets are written in a disk file 
as well as saved and updated continuously in the online experiment database.

The beam line shutter that regulates the beam's entry into 
the experimental area is computer operated. This option is very 
useful for restarting automatic data taking in instances when the beam 
goes down temporarily and detector operation has to be restarted from 
a remote location.

\bigskip
\section{Beam-defining detectors}\label{beamc}
\medskip
\subsection{Forward passive beam collimator}\label{beam_col}
\medskip
The lead collimator PC is positioned immediately upstream of the fixed beam 
line's exit window, approximately at the focal plane between the last 
quadrupole doublet (QSL53/54) and the final beam focusing triplet
(QSK51, QSL55, QSK52). Only the beam pions pass through the collimator, while 
positrons (or electrons) stop in the collimator material, since at this point 
they are separated from pions by $\sim$\,40$\,$mm in a horizontal plane by 
the bending magnet ASL51.

This passive collimator consists of two stacked lead brick
blocks with individual dimensions of $250\times250\times50\,$mm$^3$. Both pieces
have a central hole with a diameter of $50\,$mm and a step bore extending the hole 
to $70\,$mm. Removable lead plugs with different apertures can be inserted into
these openings. The upstream collimator plug used in production running
has a central hole diameter of $7\,$mm. The hole in the downstream plug is $10\,$mm
in diameter.
The two collimator pieces are screwed together and then mounted as a unit onto 
a vacuum blind flange which in turn couples to a beam window flange. 
Therefore, the collimator assembly always has a fixed position with respect 
to the beam vacuum tube.
The complete collimator setup can be moved continuously in both the horizontal 
and vertical directions with a quasi-rack and pinion mechanism in order to adjust 
its pin hole position relative to the pion beam axis.

The pion beam transmittance 
is checked periodically by taking Polaroid photographs of the beam spot at 
the vertical plane in front of the collimator's central hole.
Fig.~\ref{fig:beam_sp} shows a photograph from our experimental logbook.
A one minute beam exposure reveals the resolved $\pi^+$ and $e^+$ beam spots, 
separated at this point by $\sim$\,40mm. The circle centered on a pair of 
cross-hairs represents the position and size of the collimator opening.

\bigskip
\subsection{Forward beam counter}\label{b0}
\medskip
The forward beam counter BC is the first active detector counter. It is placed immediately
downstream of the lead collimator and just upstream of the vertical bending magnet SSL51.
This counter tags beam particles that pass through the collimator. 

The central part of the beam counter is a rectangular piece of BICRON 
BC-400 plastic scintillator with dimensions $25\times25\times2\,$mm$^3$. 
The scintillator is optically coupled on all four edges to
tapered acrylic light guides. One light guide is glued to the scintillator
edge surface with BICRON BC-600 optical cement. The other three light guides 
are coupled to the scintillator via air gaps.
Both the scintillator and the light guides are mounted inside a light-tight 
enclosure which consists of an aluminum frame with outer dimensions 
$150\times150\times50\,$mm$^3$, covered by two thin ($30\,\mu$m) aluminum windows.
The frame is attached to the lead collimator and fixes the
counter position. Mechanical feedthroughs at the four sides of the box hold the light 
guides and the magnetic shield cylinders of the PMTs, as seen in Fig.~\ref{fig:fbeam}.

Each of the four light guides is air-gap coupled to a Hamamatsu R7400U mini-PMT. 
The PMT voltage dividers were custom designed at the University of Virginia 
to operate at typical counter rates $>$\,2\,MHz. 
The four analog PMT signals are electronically summed in a NIM LeCroy Model 428F 
linear fan-in unit. The two identical outputs representing summed analog BC signals 
are distributed to the fast trigger electronics and the digitizing electronics 
branch (see Sec.~\ref{dsc}). Fig.~\ref{fig:b0_adc} shows a pulse charge spectrum
of BC signals digitized
in a 10-bit CAMAC ADC LeCroy Model 2249 with a 25\,ns integration gate. 
Accounting for the pion energy straggling in the thin counter,
the 12.7\,\% rms resolution of the ADC spectrum corresponds to  
a scintillator light output of $\simeq$\,103\,photoelectrons/MeV. 

\bigskip
\subsection{Active beam collimator pair}\label{col}
\medskip
Beam-related particles can reach the PIBETA calorimeter accidentally in coincidence 
with a $\pi$-stop signal. The accidental coincidences arise from:
\begin{itemize}\setlength{\itemindent}{3ex}
\item[(1)] beam halo particles entering the detector off axis;
\item[(2)] beam particles scattering off detector support structures;
\item[(3)] beam pions undergoing hadronic interactions in air or in
passive and active detector elements;
\item[(4)] particles from pion and muon decays in flight occurring outside 
the active target volume.
\end{itemize}

Two active beam collimators AC$_1$ and AC$_2$ suppress backgrounds
caused by detector hits that are not associated with a $\pi$-stop event.
Both collimators are cylindrical rings made of 25.4$\,$mm thick 
BICRON BC-400 plastic scintillator. The first (upstream) counter has 
an outer diameter of 120$\,$mm and a central hole diameter of 50$\,$mm. 
The second (downstream) active collimator has an outer 
diameter of 172$\,$mm and a central hole diameter of 90$\,$mm. 
The collimator AC$_1$ is placed around the beam vacuum tube, about 550$\,$mm upstream 
from the center of the detector. The collimator AC$_2$ is attached to the aluminum 
ring mounting of the active degrader and covers its struts. It is located
310$\,$mm upstream of the center of the detector.
The collimator dimensions 
were chosen so as to ensure that beam tube elements and mechanical support structures
are shielded without intersecting the calculated envelope of the beam.

The collimator rings are flattened on the outer envelope at four points by 2$\,$mm 
deep cuts, every 90 degrees.  Light is collected by Hamamatsu R7400U mini-PMTs glued
at each of these flat surfaces. The analog pulses from the four PMTs 
are summed in a NIM LeCroy Model 428F linear fan-in unit located in the electronics hut.  This
provides one common analog signal for each active collimator counter.      

Typical collimator counting rates at $\simeq$\,1\,MHz stopping $\pi^+$ flux 
are 10\,kHz and 20\,kHz for AC$_1$ and AC$_2$, respectively. These comparatively 
low hit rates allow us to use the collimator signals for veto cuts in 
the offline analysis without introducing an additional dead time inefficiency.

\bigskip
\subsection{Active degrader}\label{deg}
\medskip
In order to maximize pion beam transmission along the length of
the beam line, pions reaching the central detector area have a relatively high 
momentum of 113.4\,MeV/c. They are subsequently moderated by the active degrader AD. 
The degrader counter reduces the average pion kinetic energy from 40.3\,MeV to 27.6\,MeV. 
We continually monitor the ratio of counting rates $t=N({\rm BC\cdot AD})/\-N({\rm BC})$
to keep the beam transmittance stable. The theoretical value calculated for perfect 
beam transport with pions decaying in flight in a 3.82\,m long region is 46.0\,\%. 
The measured ratio is in the range $0.45<t<0.48$, reflecting an efficient beam
transport with minimal losses.

The active degrader counter is made of BICRON BC-400 plastic scintillator
and has the shape of a truncated cone. Fig.~\ref{fig:act_deg} shows the cross section of
the degrader counter. This geometry ensures that the degrader's downstream
projection covers the whole target area (40\,mm diameter) while particles entering
parallel to the beam axis always traverse the same 30\,mm scintillator thickness. 
Furthermore, the degrader's radial dimensions are compatible with the
inner MWPC cylinder, which has a 90\,mm minimal inner diameter. 
The outer surfaces of the degrader counter are therefore slanted and connect to four 
acrylic light guides, each equipped with a Hamamatsu R7400U PMT. 
This geometry is required so that the PMTs are located outside of the inner region, 
while the degrader projection matches the target area. 
The slant angle is small, optimizing light collection. 
The fish-tail shaped light guides are bent into quarter circles. 
The light guides couple to the whole outer area of the scintillator. 
Each light guide ends up in a square cross section $6 \times 6$\,mm$^2$,
which matches the round window of the PMT. The photomultiplier tubes are 
coupled with an air gap.

The degrader counter is supported by a plastic stand which also holds 
the PMTs and voltage divider bases. It is mounted at the ends of four steel 
threaded rods protruding from the outer support structure of the detector into 
its center from the upstream side, leaving a 5\,mm gap to the active target face. 
The inner active collimator is mounted to the same mechanical support structure 
and covers the rods in the direction of the beam. 

Fig.~\ref{fig:deg_adc} shows the pulse-height spectrum of the active degrader for
the $\pi^+$-in-beam trigger (see Sec.~\ref{bpt}) taken with a short, 25\,ns gate
in a 10-bit CAMAC ADC unit. The spectrum has an excellent rms energy 
resolution of 7.8\,\%, corresponding to photoelectron statistics of 
$\simeq$\,270\,pho\-to\-ele\-ctrons\-/\-MeV. It easily identifies the events 
with two or more beam particles, from either a single beam pulse or two adjacent 
beam pulses.

\bigskip
\subsection{Active PIBETA stopping target}\label{tag1}
\medskip
The active target is a cylindrical plastic scintillator counter 
with a 50\,mm length and a 40$\,$mm diameter. The counter is segmented into 
9 elements, as shown in Fig.~\ref{fig:tgt9a}. These 9 pieces are 
optically isolated by wrapping them individually in aluminized Mylar foil.
The segments are pressed together and the whole assembly is wrapped 
with black plastic tape. Each target element is coupled to a miniature 
(8$\,$mm photo-cathode) Hamamatsu R7400U PMT
via a 60$\,$mm long, tapered acrylic light guide. Fig.~\ref{fig:tgt9b} is a
photograph of the partially assembled target counter.

The analog signal from each of the target segments is divided into two branches
by a custom-made passive splitter. One side is discriminated in a Phillips Scientific
PS 7106 16-channel time-over-threshold 
discriminator packaged in a single width CAMAC module. 
This output is digitized in a FASTBUS time-to-digital converter (TDC) LeCroy Model 1877, 
as well as counted with a scaler unit (Phillips PS 7132H) that is read out every 10$\,$s. 
The second branch is connected to an 12-bit FASTBUS analog-to-digital converter (ADC)
LeCroy Model 1882F, gated with a 100$\,$ns event trigger gate.

Fig.~\ref{fig:tgt_adc} shows the FASTBUS ADC spectrum in a single target detector
for $\pi^+$-in-beam trigger events. The $\simeq$\,10\,\% rms width of the 
$\pi^+$ stopping peak at 27.6\,MeV is consistent with $\simeq$\,8\,photoelectrons/MeV.
PMT gains of the target counters are equalized using 
a 4.1$\,$MeV $\pi\to \mu \nu$ stopping muon line as a reference.

\bigskip
\subsection{Beam stopping distribution and beam composition}\label{beam_dis}
\medskip

We have extracted the counting rates of $\pi$-stop events and beam positrons in 
the individual target segments from the gain-matched ADC spectra.
Fig.~\ref{fig:extr_2d} shows the 2-dimensional shape of the $\pi^+$ stopping 
distribution using the counting rates of individual segments superimposed on an 
outline of the target. The $\pi^+$ beam spot measured by this method agrees with 
the independent back-tracking tomography analysis that uses the $\pi^+\to e^+\nu_e$ 
positron tracks recorded by the MWPCs. The beam spot is almost symmetric and is centered 
on the target, with horizontal and vertical profile rms widths of 
7.6\,mm and 8.4\,mm, respectively.
The 2-dimensional $x$-$y$ beam profile is represented in our analysis by
two separated, rotated, and then modulated Gaussian distributions. 
The details of our algorithm are described in a separate publication~\cite{Frl01c}.

The longitudinal distribution of stopping pions in the active target scintillator is
calculated in a {\tt GEANT} simulation and agrees very well with the back-tracking
tomography reconstruction of the $z$ coordinate beam profile. Fig.~\ref{fig:pistop_z}
shows the Monte Carlo histogram with the $\sigma_z$ width of 1.7\,mm.

The temporal stability of the beam stopping distribution, namely its position inside 
the target and its spread, are monitored continuously with the back-tracking 
tomography algorithm. The lateral and the longitudinal centroids of the $\pi^+$ 
stopping spot varied with rms widths of $\simeq$\,0.05\,mm and 0.2\,mm, respectively,
during the three month calibration period.

The distances from the forward beam counter BC and the pion production E target 
to the center of the stopping target are 3.87\,m and 16.83\,m,
respectively. The $e^+$ and $\mu^+$ times
relative to the $\pi^+$ arrival time, calculated under the assumption of no 
decay-in-flight contamination, are then $-$7.3\,ns and $-$2.8\,ns for 
the BC--AT path and $-$12.6\,ns and $-$12.1\,ns for the E--AT path. 
Fig.~\ref{fig:contam} shows the beam composition revealed by this method. 
The expected $e^+/\mu^+$ TOF values quoted above are indicated by the cross 
markers and agree reasonably well with the measured relative timings.
The mismatch is due to timing walk in the beam counter discriminator.   
The extracted $e^+$ and $\mu^+$ beam contaminations in the $\pi$-in-beam 
trigger measured in TOF spectra are small, 0.4$\,$\% and 0.2$\,$\%, respectively.

Calibration runs were performed using a relaxed trigger
configuration to study the pion beam contamination.  We retain 
a remarkably clean $\pi^+$ beam spot with less than 1.0\,\% nonpionic contamination
even after omitting the rf signal 
in the $\pi^+$ beam coincidence and reducing the degrader discriminator threshold 
by a factor of three, well below the muon and positron energy depositions.
In addition, the measured $e^+$ beam fraction value was confirmed by using an extended target
with a 50\,mm long passive front section that stopped $\pi^+$ and $\mu^+$ components.  
Monte Carlo simulations with $\pi^+$s generated at the forward beam counter 
position predict an $e^+$ contamination of 0.5\,\% at the target
position, arising mostly from $\mu^+$ decays in flight.

\bigskip
\section{Cylindrical MWPCs}\label{mwpc}
\medskip

The primary need for a tracking detector in the experiment is due to
the $\pi e2$ decay trigger.  Individual positrons from $\mu\to e\nu\nu$ (Michel) 
decays can be distinguished from the $\pi e2$ decay positrons by their lower energy.
However, in a high-rate measurement, accidentally coincident Michel events will give 
rise to a substantial background under the $\pi e2$ positron energy peak.

The solid angle resolution of the calorimeter is $\approx$\,0.004 of 4$\pi$\,sr.
This is insufficient for the suppression of accidental Michel coincidences occurring 
at high pion stopping rates ($>1\cdot10^6$\,/s). In addition to the double-track 
resolution requirement, the following constraints
were used in choosing the design of the tracking detector:
\begin{itemize}\setlength{\itemindent}{3ex}
\item[(1)] low mass, in order to minimize the $\gamma$'s converting into 
$e^+e^-$ pairs;
\item[(2)] high efficiency -- better than 99.9\,\%;
\item[(3)] high rate capability -- up to 10$^7$ minimum-ionizing 
particles (MIP) per second;
\item[(4)] stable operation and good radiation hardness;
\item[(5)] cylindrical geometry.
\end{itemize}

After considering different design options we concluded that MWPCs provide 
the best solution. Moreover, wire-chamber technology is mature and reliable.
We selected a design with a pair of cylindrical chambers, each 
having one anode wire plane along the $z$ direction, and two cathode strip planes 
in a stereoscopic geometry. The MWPCs were manufactured at the Joint Institute 
for Nuclear Research, Dubna. Table~\ref{tab1} lists their basic parameters.
Ref.~\cite{Kra94} gives a general description of the design and operation of
the ``DUBNA''-type cylindrical chambers.
Refs.~\cite{Kar98,Kar99} provide a detailed description of the PIBETA wire-chambers.
In this section we elaborate on the angular 
resolution of the chambers measured with cosmic muons and the chamber 
detection efficiencies measured with minimum ionizing particles.

The angular resolution of the MWPCs is an important parameter of
the chamber response. Cosmic muons are ideally suited for the 
calibration and resolution studies of the MWPCs. A clean sample of 
cosmic muon events is collected throughout the experiment during weekly 
beam-off cyclotron maintenance periods. The trigger logic requires 
the coincidence of the cosmic muon veto counter (see Sec.~\ref{cosm}) 
and the ``low threshold'' (LT$\simeq$\,5\,MeV) CsI calorimeter signal 
(see Sec.~\ref{elect}).

In off-line data analysis, two additional, stringent cuts 
are imposed: CsI calorimeter energy deposition more than 200\,MeV, 
and exactly two reconstructed hits in each MWPC. These constraints 
effectively remove any extraneous noncosmic background.
The cosmic muon track is reconstructed from a pair of hits in one MWPC 
and the intersections of that track with the other chamber is calculated.
The difference between the calculated
and measured intersection coordinates represents 
the spatial resolution of the chamber. Fig.~\ref{fig:mwpc_r} shows the azimuthal angle
$\Delta\phi$ and longitudinal coordinate $\Delta z$ resolutions of MWPC$_1$,
properly weighted for the nonuniform angular distribution of cosmic muons.
These measurements fix the chambers' angular rms resolution at $\le 0.25^\circ$. 

The detection efficiency of the inner chamber for 
minimum ionizing particles $\epsilon_{\rm MWPC_1}$ can be measured using copious 
$\mu^+\to e^+\nu_e\bar{\nu}_\mu$ positrons emanating from the target:
\begin{equation}
\epsilon_{\rm MWPC_1}={ {N {(\rm AT{\circ} MWPC_1{\circ} MWPC_2{\circ} PV{\circ} CsI)}}\over
{N {(\rm AT{\circ} MWPC_2{\circ} PV{\circ} CsI)}} },
\end{equation}
where the $N$'s represent the number of Michel events for which all the detectors
in the parentheses register coincident hits above the discriminator threshold.
The CsI calorimeter signal is discriminated with the LT level, while the
window cut on the PV pulse-height spectrum selects the MIP events. An equivalent 
expression, with indices 1 and 2 exchanged, holds for the outer chamber MWPC$_2$.
The average MWPC detection efficiencies at the $\simeq$\,1\,MHz $\pi^+$ beam 
stopping rate are $>$\,94\,\% and $>$\,97\,\% for the inner and outer
chambers, respectively. The combined detection efficiency is therefore $>$\,99.8\,\%.
Fig.~\ref{fig:mwpc_phi} demonstrates that the experimental charged particle detection
efficiency is uniform in the azimuthal angle for both chambers.

\bigskip
\section{Plastic veto hodoscope}\label{pv}
\medskip
The  plastic veto hodoscope PV is located in the interior of 
the calorimeter surrounding the two concentric wire chambers.
The hodoscope array consists of 20 independent plastic scintillator staves
arranged to form a complete cylinder 598$\,$mm long with a $132\,$mm radius.
The long axis coincides with the beam line and the target
axis (see Fig.~\ref{fig:pv_geom}(i)). The PV hodoscope covers the entire 
geometrical solid angle subtended by the CsI calorimeter as seen from 
the target center.
A single PV detector consists of four main components:
\begin{itemize}\setlength{\itemindent}{3ex}
\item[(1)] plastic scintillator stave;
\item[(2)] two light guides;
\item[(3)] two attached photomultiplier tubes;
\item[(4)] aluminized Mylar wrapping.
\end{itemize}

The BC-400 scintillator staves were procured from the Bicron Corp. (now part
of Saint-Goban/Norton Industrial Ceramics Corp.). 
The maximum light emission of the scintillator occurs at 
a wavelength of $423\,$nm, while the light output falls to 20\% of maximum at 
$410\,$nm and $480\,$nm. The rise and decay times of the fast scintillator 
pulses are $0.9\,$ns and $2.4\,$ns, respectively.  The bulk medium attenuation length 
is $160\,$cm. The light response of the scintillator has negligible temperature 
dependence from $-60^\circ$C to $+20^\circ$C~\cite{Bic89}.

The dimensions of the individual plastic staves are 
$3.175\times41.895\times598\,$mm$^3$ (see Fig.~\ref{fig:pv_geom}(ii)). They
are rigid enough to maintain their shape with support only at the two ends. 
The design geometry ensures that each charged particle path emanating 
from the target volume towards the calorimeter sphere intersects 
the PV array. An 18$^\circ$ angle on one side of each stave
minimizes the passive PV volumes along the likely particle trajectories
(see Fig.~\ref{fig:pv_geom}(iii)).

The light guides flare out to allow more space between 
the PV PMTs and to provide better access to the central part of the detector, facilitating
servicing operations. Each light guide is made of two acrylic pieces that were first glued 
together and then firmly attached to the ends of the PV stave using Bicron two-component
epoxy glue transparent to the scintillator light. Care was taken to eliminate
air bubbles in the glue material. After an overnight drying period, the glued joints 
were polished to maintain a precise $3.175\,$mm thickness of the stave. The joint areas were
polished to a smooth reflective surface by first using a fine grain sandpaper with 
$\sim$\,600 grit mixed with water and subsequently applying a buffing wheel and a polishing gel 
equivalent to 10,000-grit sandpaper. The scintillator crazing that can result from 
the pressure caused by the weight of the light guides was alleviated by annealing all staves 
in a warm water bath for a period of several hours before installation.
Each plastic scintillator along with its attached light guides is wrapped in a 0.25$\,\mu$m
thick aluminized Mylar foil to separate it optically from the adjacent staves and to provide 
the reflective surface which maximizes the amount of light reflected towards the phototubes.

The assembled hodoscope array is supported by the outside surface of
a $530\,$mm long carbon fiber cylinder with a total thickness of 1\,mm, 
equivalent to $5.3\cdot 10^{-3}$ radiation lengths. The hodoscope modules 
are held tightly around the support cylinder by a helically wound, thin plastic 
string, tensioned at the extreme ends of the detector stand.

Before the veto hodoscope was assembled inside the PIBETA calorimeter, 
the light response of each plastic counter was calibrated in a specially designed 
cosmic muon tomography apparatus~\cite{Frl00}. We found the average attenuation length of 
the scintillator light to be 396$\pm$13$\,$mm. The number of pho\-to\-e\-lec\-trons 
per MeV of deposited energy varies between 21$\,$pho\-to\-e\-lec\-trons/MeV and 
63\,$\,$pho\-to\-e\-lec\-trons\-/\-MeV, with the average being 
38.3\,photoelectrons/MeV. Fig.~\ref{fig:pv_adc} shows the measured energy spectrum of the
positrons and protons in the PV detector, corrected for the angle of incidence.
The energy resolution measured for minimum ionizing particles is $\sigma_E/\bar{E}$=33.2$\,$\%.

The scintillator light is viewed at both detector ends by two Burle Industries S83062E 
photomultiplier tubes. These tubes are 28$\,$mm head-on fast PMTs with 10 dynode stages.
The photocathodes are sensitive to wavelengths between 300$\,$nm and 660$\,$nm.
The phototubes are operated at an average voltage of 1400$\,$V.

The individual PMT analog signals from the PV counters are split into two branches: one 
is connected to the trigger electronics and the other is fed into a
1-to-2 splitter whose appropriately delayed output is wired to the FASTBUS ADCs 
and TDCs. The signals at the trigger side are summed using standard NIM
logic fan-in modules. The discriminated PV OR signal is used routinely as a trigger
in MWPC tests. 

The PV array is an essential part of the charged particle tracking system. It supplements
the MWPC tracking and CsI calorimetry by providing:
\begin{itemize}\setlength{\itemindent}{3ex}
\item[(1)] efficient charged particle detection, particularly when combined 
with the MWPC data;
\item[(2)] reliable discrimination between minimum ionizing particles
(cosmic muons, positrons/electrons) and protons;
\item[(3)] crude measurements of charged particle azimuthal angle ($\pm 9^\circ$);
\item[(4)] precise charged particle timing information ($\pm 0.8\,$ns).
\end{itemize}

Particle identification is accomplished by cuts on the two-dimensional 
$E_{\rm PV}$--$(E_{\rm PV}+E_{\rm CsI})$ histogram.  This is especially useful for
discerning between protons and minimum ionizing positrons. Fig.~\ref{fig:part_id} 
shows the measured two-dimensional energy spectra of positrons and 
protons in the PV detector and CsI calorimeter, corrected for the angle of incidence 
using MWPC data. The calibrated energy boundaries expressed in MeV units are 
conveniently parameterized by the functions
\begin{eqnarray}
E^{\pi^+}_{\rm PV} \ge & 0.20\cdot \exp {\left[-0.007\cdot 
                             \left(E_{\rm PV}+E_{\rm CsI}\right)\right]} < E^p_{\rm PV},
\label{eq:id1}\\
E^p_{\rm PV} \ge & 2.30\cdot \exp {\left[-0.007\cdot 
                             \left(E_{\rm PV}+E_{\rm CsI}\right)\right]}.\hfill
\label{eq:id2}
\end{eqnarray}
Single positron events fall in a band between the two curves while single proton events 
lie above the second curve. The $e^+/p$ identification efficiency $\epsilon_{\rm MP}$ 
was studied in low intensity runs and was determined to be $99.8\pm 0.2$\,\%. 

The PV charged particle detection efficiency is evaluated separately for
mi\-ni\-mum-ion\-izing positrons with total energies above 5$\,$MeV and 
nonrelativistic protons with kinetic energies in the range 10-150$\,$MeV.
A charged particle track is defined by coincident hits in
the active target AT, the wire chambers MWPC$_1$ and MWPC$_2$, and the CsI
calorimeter CsI.
The PV detection efficiency $\epsilon_{\rm PV}$ is defined as the following ratio:
\begin{equation}
\epsilon_{\rm PV}={ {N {(\rm AT{\circ} MWPC_1{\circ} MWPC_2{\circ} PV{\circ} CsI)}}\over
{N {(\rm AT{\circ} MWPC_1{\circ} MWPC_2{\circ} CsI)}} },
\end{equation}
where the $N$'s represent the number of events satisfying the condition in the
parentheses. The average positron detection efficiency $\epsilon_{\rm PV}$ measured 
during the detector commissioning period was $\ge$\,99.2\,\% 
(Figs.~\ref{fig:pv_eff}~and~\ref{fig:pv_eff_zoom}). 
At the nominal 1\,MHz $\pi^+$ stopping rate, the hit rate in an individual plastic 
scintillator phototube is $\simeq$\,130\,kHz while the hit rate in the whole PV hodoscope 
system is $\simeq$\,0.88\,MHz. Under these conditions, the charged particle tracking 
system, combining the information from the PV hodoscope and MWPC pair has a MIP 
tracking inefficiency in the range $(1.0\pm 0.2)\cdot 10^{-5}$.

\bigskip
\section{Modular pure CsI calorimeter}\label{calo}
\medskip
\subsection{Calorimeter geometry}\label{geom}
\medskip
The heart of the PIBETA detector is the shower calorimeter.
Its active volume is made of pure Cesium Iodide~\cite{Kob87,Kub88a,Kub88b}.
The optical and nuclear properties of 
pure CsI are summarized in Table~\ref{tab2}.
The experimental signature of a {$\pi^{+}\to\pi^{0}e^{+}\nu$\/ event is 
the prompt decay $\pi^0 \to \gamma\gamma$. The calorimeter must be 
able to handle high event rates and cover a large solid angle with high efficiency for 
$\pi^0 \to \gamma\gamma$ detection.  Efficient suppression of background events
requires good energy and time resolution.  Furthermore, the system must operate 
with low systematic errors and be subject to accurate calibration. The geometry 
of the shower calorimeter is therefore chosen subject to the following conditions:
\begin{itemize}\setlength{\itemindent}{3ex}
\item[(1)] high rate-handling ability---low pile-up and low dead time;
\item[(2)] good energy and time resolution for background suppression;
\item[(3)] optimum light collection uniformity for individual modules;
\item[(4)] uniform solid angle coverage by each module for uniform distribution
of counting rates;
\item[(5)] maximum economy of different module shapes compatible with the above
requirements.
\end{itemize}

These requirements are best met in the nearly-spherical geometry obtained
by the ten-frequency class II geodesic triangulation of an icosahedron~\cite{Ken76}.
In this respect,
our design bears similarities to the SLAC Crystal Ball~\cite{Ore80}. However,
in order to enhance the light collection efficiency, we opted for truncated 
pentagonal, hexagonal, and trapezial pyramids (see Fig.~\ref{fig:csi_shapes})
rather than the triangular pyramids of the Crystal Ball.

The PIBETA calorimeter consists of 240 pure CsI crystals
(Fig.~\ref{fig:balls}). Our geodesic division 
results in 220 truncated hexagonal and pentagonal pyramids 
covering the total solid angle of 0.77$\times$$4\pi$\/ sr. An additional 20 
crystals cover two detector openings for the beam entry and exit and 
act as electromagnetic shower leakage vetoes. The inner radius of the calorimeter 
is 26$\,$cm, and the module axial length is 22$\,$cm, corresponding to 12 CsI 
radiation lengths ($X_0$=1.85$\,$cm). There are nine different module 
shapes: four irregular hexagonal truncated pyramids (we label them 
HEX--A, HEX--B, HEX--C, and HEX--D), one regular pentagon (PENT) and 
two irregular half-hexagonal truncated pyramids (HEX--D1 and HEX--D2), 
plus two trapezohedrons which function as calorimeter vetoes (VET--1 and VET--2).
The volumes of the CsI modules vary from 797$\,$cm$^3$ 
(HEX--D1/2) to 1718$\,$cm$^3$ (HEX--C). Fig.~\ref{fig:csi_shapes} shows
all the crystal shapes in nine successive panels.

The first 25 crystals acquired were manufactured at the Bicron Corporation 
facility in Newbury, Ohio. The remaining 215 were grown in 
the Institute for Single Crystals in Harkov (AMCRYS), Ukraine. Preliminary
quality control, including the measurement of optical and mechanical properties,
was done at the factory sites.

The crystal sphere was designed to be self-supporting.  Thus,
tight mechanical tolerances were required to avoid construction difficulties. 
Manufacturing tolerances were specified for 
the linear dimensions ($+$150 $\mu$m/$-$50 $\mu$m) and the angular 
deviations ($+$0.040$^\circ$/$-$0.013$^\circ$). The
machined crystals were measured upon delivery at PSI 
using a computer-controlled distance-measuring device ({\sl WENZEL
Precision}). The machine was programmed to probe automatically the 
surfaces of the crystal to be tested with a predefined shape. Each crystal surface 
plane was scanned with a touch head at six points which were then mathematically fitted 
with a plane. Body vertices, measured with 
an absolute precision of 20 $\mu$m and reproducible to within 2 $\mu$m, 
were then compared with the expected values. Crystals 
that failed the stringent geometrical tolerances were returned to 
the manufacturer for reuse as raw crystal growing material.
\medskip
\subsection{CsI detector surface treatment}
\medskip
After completing the measurements, all crystal surfaces were hand-polished 
with a mixture of 0.2 $\mu$m aluminum oxide powder and ethylenglycol.
The surfaces of each CsI crystal were then painted with a special organo-silicon 
mixture developed by the Harkov Single Crystals Research Institute~\cite{xxx}.
The lacquer was a wavelength-shifting ladder organosilicon copolymer 
with chemical composition PPO+POPOP+COUM.1, where PPO is 
2.5--Di\-phe\-nyl\-ox\-a\-zole, POPOP represents 
1.4-Di-2-(5-Phe\-nyl\-ox\-a\-zolile-Benzene) 
and COUM.1 is 7-Diethylamine-4-Methylcoumarin~\cite{hand71}.
It provides an optical treatment of crystal surfaces superior to more common
matting or wrapping treatments. 

The total light output and timing resolution of the painted crystal 
modules were unchanged, but the energy resolution was improved due to the more 
favorable wavelength modulation of axial light collection probability.
The protective surface coating also guards the scintillators 
against potential surface deterioration and minimizes changes of 
the detector response over the duration of the experiment.

The contribution of the fast decay component to the total light
output, the photoelectron statistics and the temperature dependence of
the light output were measured for each CsI crystal using a
radioactive source in a specially designed scanning apparatus.  These
measurements are covered in depth in a separate article~\cite{Frl01a};
here we only summarize the optical properties of CsI crystals in
Table~\ref{tab2}.

Electron Tubes Inc.\ (formerly EMI) custom 9822QKB fast 10-stage
PMTs~\cite{EMI} with 78$\,$mm diameter end windows were glued to the
back faces of the hexagonal and pentagonal CsI crystals using a 300
$\mu$m layer of silicone Sylgard 184 elastomer (Dow Corning RTV
silicon rubber plus catalyst). The resulting crystal--PMT couplings
are strong and permanent, but can be broken by application of a
substantial tangential force. The smaller half-hexagonal and
trapezial detector modules were equipped with 46$\,$mm 10-stage EMI
9211QKB PMTs~\cite{EMI}. Both photocathodes have quartz windows
letting through UV light down to 175$\,$nm.  The window transparency
peaks at $\sim\,$380$\,$nm~\cite{EMI} and is thus approximately
matched to the spectral excitation of the pure CsI fast scintillation
light component which peaks at $\sim\,$310 nm at room
temperature~\cite{Woo90}.

The PMT high voltage dividers are model UVA 131-a and UVA 131-b,
designed and built at the University of Virginia.  The dividers were
designed to meet several concurrent design goals: (a) high pulse
linearity over a signal dynamic range exceeding $100:1$, (b) high gain
stability over a wide range of counting rates, up to several hundred kHz,
and (c) low thermal dissipation in order to minimize the heat load on
the detector temperature stabilization system.  In this design the
voltages of the last four dynode stages were regulated using 600\,V
MOSFET transistors, while the cathode-to-first-dynode voltage was
fixed by the use of Zener diodes. In this way the dividers reduce the
so-called ``super-linearity'' exhibited by many PMTs well below the
onset of saturation ($>50\,\mu$A average anode current).  The choice
of SbCs instead of the standard BeCu dynode material in the custom
9822QKB tubes further reduces the PMT gain shift that occurs at small
anode currents (1--10$\,\mu$A).  The maximum PMT nonlinearity measured
in a test with a pair of light-emitting diodes was less than 2$\,$\%
over the full dynamic range encountered in the $\pi\beta$\/ decay rate
measurement~\cite{Col95}. More details concerning the design,
construction and performance of the UVA voltage dividers in
conjunction with the Electron Tubes PMT's will be given in a
forthcoming publication.

Two LeCroy 1440 high voltage mainframes provide the high voltage for
the PMTs. This 340-channel HV system has computer control capability
and thus allows for frequent, remote changes in the HV supplied to the
PMTs. The demand HV values can be set in 1\,V increments with
reproducibility of $\simeq$\,1\,V, which corresponds to a gain change
of $<$\,0.5\% for our 10 stage PMTs operating in the range of 1500\,V
to 2200\,V.

\medskip
\subsection{Assembly of the CsI calorimeter sphere}
\medskip

The CsI calorimeter is mounted inside a forged
spherical steel shell which is itself supported by a steel frame
(the calorimeter mass is approximately 1,500\,kg).
The shell has two large axial openings, 550\,mm in diameter, 
for beam entrance and exit, and 220 smaller holes distributed over its surface. 
These smaller holes are aligned with the axes of the individual crystals and are intended 
to allow access to the PMTs (the 20 veto crystals are accessed from a neighboring
crystal's hole). Inside the shell beam openings are placed cast-iron plugs 
(one on each end) in the shape of pentagonal truncated pyramids.
The plugs press the stacked crystals together and hold them in place. The plugs have 
a cylindrical borehole (diameter 270\,mm) for the beam which also allows access 
to the interior of the CsI sphere, e.g., the wire chambers, the target counters, etc.).

Assembly of the calorimeter proceeded as follows: 

The empty spherical shell was equipped with a single
plug insert. The beam hole of this plug was used to hold a micrometer gauge
with a touch ball in place. This device allowed a precise measurement of
the distance from a crystal's front surface to the center of the sphere.
The whole frame was rotated by 90$^\circ$ such that the plug was at the bottom. Through 
the other beam hole, now at the top of the sphere and not yet closed with its plug, 
all crystals, already glued to their PMTs, were moved into the interior 
of the spherical shell. First came 10 pieces of VET-1 and VET-2 type crystals 
which were positioned on the flat surfaces of the bottom
plug. The correct distance to the center of the sphere was adjusted using
threaded bolts mounted on the plug, supporting each crystal's rear face. 

The next crystal layer
was stacked on top of the veto layer. Threaded tubes with boreholes large enough
to house a PMT and $\mu$-metal shielding were turned through the smaller holes in the shell.
The front of these tubes
pressed against the rear surfaces of the crystals and thereby allowed for a
precise adjustment of their positions along a radius of the sphere.
The high voltage divider base was placed in the tube and
connected to the PMT. Bases for the veto crystals had to be mounted somewhat 
differently as there are no holes available for them on the shell due to the close
proximity of the axial plugs. The closest hole, later employed for a crystal on
the next layer, was used to move each veto crystal's base into the sphere, then 
sideways until connection to the veto's PMT was possible. Small supporting structures
were mounted to lightly press the base against the PMT to prevent loosening
due to gravity for vertically mounted crystals. 

When the sphere was half filled with crystals, the procedure had to be
changed somewhat.  In order to prevent the crystals in the upper hemisphere from sliding
down towards the sphere's
center, the measurement device at the center was replaced by a pin-holder with a
pin for each remaining crystal, adjusted in length to give the correct distance to the center.

After the last layer of 10 veto crystals was complete, the second
pentagonally shaped plug was mounted on top and pressed against the crystal ball underneath. 
Calculations made prior to the assembly indicated that the plug would have to be pushed
10\,mm against the crystal ball in order to pack all crystals tightly. 
This prediction was fulfilled exactly, giving us
confidence in the correct positioning of all modules. The completed sphere was rotated
several times while being closely monitored for stability before the pin-holder at the center 
was disassembled through the beam entry/exit holes.

\medskip
\subsection{Calorimeter clustering and nomenclature}
\medskip
The one-arm calorimeter energy trigger is a basic element of the trigger logic.
A preliminary simulation study~\cite{Ass95} of the calorimeter response to
photons from $\pi\beta$ decay at rest and 70\,MeV positrons
from $\pi e2$ decays indicated that:
\begin{itemize}\setlength{\itemindent}{3ex}
\item[(1)] electromagnetic shower profiles of the mean deposited energies are 
similar for photons and positrons, in particular for $\theta_c\le 12^\circ$,
with $\theta_c$ being a half-angle of a conical bin concentric with the direction 
of an incident particle;
\item[(2)] the average deposited energy and the corresponding energy resolution
of the calorimeter both reach saturation within a cone of 12$^\circ$ half-angle;
\item[(3)] a centrally hit CsI module receives on average 90\,\% of the deposited energy;
at most three modules contain a significant part of shower energy; and
a group of 9 detectors (a CsI ``cluster'') constitutes an excellent summed
energy trigger as it registers on average $\ge$\,98\,\% of the incident particle energy.   
\end{itemize}

Therefore, the building blocks of our physics triggers are overlapping
clusters of nine CsI detectors. Excluding the CsI shower vetoes from the 
scheme we define 60 such detector groupings (see Table~\ref{tab3}). 
Every CsI cluster has a symmetric partner
in the antipodal calorimeter hemisphere. In addition, each CsI module belongs 
to no more than three clusters. This limitation helps to minimize the degradation 
of analog pulses due to excessive signal splittings.
In trigger design studies that looked at the energy captured by a single cluster
as a figure of merit, it was found that this clustering scheme, in conjunction with 
a 50\,MeV discrimination threshold, gives 99.3\,\% and 98.6\,\% triggering 
efficiency for 70\,MeV photons and positrons, respectively.  

The overall calorimeter topography and the CsI module labels are shown in a Mercator
projection in Fig.~\ref{fig:calo_map}. One type of CsI cluster is centered around 
a PENT module (10 pentagons total), and five different types of clusters are 
centered around each of the HEX-B modules (50 hexagon Bs total). Six adjacent 
CsI clusters are in turn grouped into a CsI ``supercluster'': there are ten
such superclusters in the calorimeter each containing 24 individual CsI modules.
In trigger parlance, a supercluster fires if at least one of its constituent
clusters fires. A cluster fires if the summed energy of its modules is greater 
than the preset discriminator threshold.

Ten supercluster labels and crystal types are the basis of the CsI modules' nomenclature. 
The last digit of the crystal number designates the supercluster in which a crystal 
is located. We start by labeling all of them according to their central pentagon 
with numbers 0--9. By separating the calorimeter crystals into two hemispheres 
the beam-upstream crystal labels end in digits 0--4 while the names of downstream 
crystals terminate with digits 5--9. Antipodal crystals carry labels differing 
by 5 units. The crystal map shown in Fig.~\ref{fig:calo_map} contains module labels 
as well as the manufacturer's serial numbers (S) of all crystals. In this map, 
the module shape types (P, A, B, C, D, H, and V) precede the crystal numbers.  

The serial numbers assigned to the CsI modules run from S001 to S240: the pentagons
carry the serial numbers S001--S010, the hexagon As S011--S060, etc. The CsI vetoes receive 
the last twenty designations S221--S240. The labels associated with the clusters and
superclusters relate to the actual positions of the modules in the calorimeter, 
while the serial numbers link the module to the online database of crystal properties,
such as the shape, machining accuracy, fast-to-total light output ratio,
photoelectron statistics, light collection uniformity etc. 
The same calorimeter database is also used in the {\tt GEANT} simulation of the complete
detector response, described below in Sec.~\ref{mc}. 

\medskip
\subsection{Calorimeter electronics}\label{calo_ele}
\medskip
Each calorimeter PMT is shielded by a metal cylinder that covers both the phototube 
housing and the voltage divider attachment.  The voltage divider has two anode signal
outputs and a high voltage input connector. One analog output is delayed 
in a 75$\,$m (385$\,$ns) long Russian-made PK-50-2-16 coaxial cable. These delay cables 
are smaller in diameter than the more commonly used (and more expensive) RG-58 cables 
(namely 3.5$\,$mm vs 5.0$\,$mm). However, they have nearly identical signal propagation 
speeds, rise times, and peak attenuations when tested with a CsI pulse. For a 
385\,ns delay, they also have $\sim$\,60$\,$\% smaller signal attenuation when 
compared with RG-58s at 50\,MHz and 100$\,$MHz. 

The delay cables are bundled in groups of sixteen, spooled on six 
vertical posts mounted on the detector platform and connected to the digitizing 
electronics branch. The points of connection to the PIBETA electronics are UVA 139 
ADC signal splitters that were custom-made at the University of Virginia. Each unit is 
a 16-channel 50\,Ohm balanced passive splitter with chip resistors. All 16 signal 
inputs accepting LEMO connectors are provided on the back side of the unit, while the 
front splitter plate features two sets of outputs: a column of individual LEMO outputs 
and a single-header 2$\times$17 pin connector to couple to a twisted ribbon cable. 
Five UVA 139 units carrying 80 split channels each are fitted into a single housing 
strip with a standard 483\,mm electronics rack width.

The LEMO outputs of UVA 139 are routed to CAMAC Phillips PS 7106 discriminators with
two sets of outputs. One set of discriminated signals is taken to a CAMAC Phillips PS 7132H scaler 
while the other 16-signal twisted ribbon output is connected to a FASTBUS LeCroy Model 1877 multi-hit 
TDC unit. The 2$\times$17 simple-header outputs of UVA 139 are split again in the ratio 1:3 by
a custom made plug-in splitter. The split branch with a third of the charge
is connected to FASTBUS LeCroy Model 1882F ADC, while the parallel line is routed to 
the domino-sampling chip digitizing system (Sec.\ref{dsc}). 

The second set of PMT outputs (identical to the first) are connected
to the fast trigger electronics section via standard 64$\,$ns long RG-58 coaxial cables
which couple to a feedthrough plate where shorter (0.5--10\,ns) cables can be inserted to
equalize the time delays of all CsI signals relative to the time of
the slowest one. The simultaneous signals are then split in the custom-made 
UVA 126 3-way resistive splitter. This unit has LEMO-type inputs and three
outputs per input. Each splitter unit contains 60 channels and fits into a standard 
483\,mm wide electronics rack. Up to three copies of CsI analog signals available at 
the outputs are distributed into dedicated linear summer/discriminator modules according 
to the cluster logic detailed in the previous section. 

The custom-made single-width NIM UVA 125 summer/discriminator is the most important
trigger unit, performing fast and efficient calorimeter energy summing and discrimination.
A single unit has 4 identical sections, each with a 9-input summing amplifier followed 
by a bi-level discriminator. The amplifier output is fed simultaneously to two voltage 
comparators after passing through a protective diode voltage limiting circuit.  Each
comparator has its own reference (threshold) voltage, one low and
the other high. The best timing is achieved with the low threshold
comparator where time slewing versus pulse height is less pronounced than it is for the upper
threshold comparator. The upper level comparator
triggers a monostable multivibrator which then arms a second monostable
multivibrator for a period during which the latter can be triggered by the
delayed signal from the low level comparator. Thus, the low level (LT) and high
level (HT) discriminator outputs are essentially synchronous.

\bigskip 
\section{Cosmic lead house and active cosmic vetoes}\label{cosm}
\medskip
The PIBETA detector is shielded from background radiation
in the experimental hall as well as from the cosmic ray background. 
The active parts of the detector are doubly protected. A lead house 
enclosure provides the inner passive shielding. It is in turn lined 
on the outside with active cosmic veto counters. 

The lead house structure is supported by a steel skeleton frame mounted 
on the detector platform. The skeleton consists of steel beams with 
100\,$\times$\,100\,mm$^2$ square profiles separated by a 400\,mm gap. 
The upstream side of the support structure 
is closed by a separate lead brick wall fixed to the area floor. This 
wall has a central hole for the beam vacuum pipe.
Successive lead brick rows are shifted by half of a brick length starting from 
the rail pedestal plane. Every $\sim$\,300\,mm the bricks are attached to 
the support frame. The lead house platform is coupled to the detector platform. 
Both have a common rail system on which the cosmic house can be driven slowly, 
powered by an electric motor, until it touches the front wall and completes 
the shielding box which is then open only at floor level.
A 3\,mm thick steel plate is placed on the top of the whole structure. 

The active cosmic muon veto consists of 5 extensive scintillator planes shielding
all four sides and the top of the lead house. Each of these planes is made 
of plastic scintillator sandwiched between two green-emitting wavelength shifting 
bars, air-gap coupled to its sides (see Fig.~\ref{fig:cv_princ}). 
The plastic scintillators are assembled from 300\,mm wide, 25\,mm thick panels 
with lengths of up to 2.29\,m. They were manufactured at the Scientific Research 
Concern of the Institute for Single Crystals, Harkov, Ukraine. 
A few scintillator pieces are made smaller in order to fit modular planes within 
the full detector width. The wavelength shifter material is BICRON BC-482\,A 
extruded into rectangular rods with a cross section of $19\times29$\,mm$^2$.
The cosmic veto counters lining the top and side walls are extended 
10\,cm beyond the walls of the lead house. This geometry compensates for the passive
volume of the frame structure. All atmospheric cosmic muons which hit 
the PIBETA calorimeter must also intersect at least one of the active cosmic veto detectors.   

The optical properties of the plastic counters were tested beforehand and found to be
comparable with those of BICRON BC-400 plastic scintillator. 
The response of the cosmic veto counter to cosmic muons is illustrated in
the ADC vs. TDC scatter-plot in Fig.~\ref{fig:cosm_e_t}. The linearized TDC values
are plotted along the horizontal axis. The $t=0$ point
corresponds to the CsI calorimeter hit. The ADC readings in noncalibrated
channel numbers are shown on the vertical axis. Due to the large
area of the veto planes and the opportunity for the scintillator light to reflect
many times in the thin and extensive counters, the time resolution is limited to 15\,ns
FWHM. No timing correction is made to account for the varying position of 
the cosmic muon intersection with the veto planes. The energy resolution is 
consistent with statistics of $\simeq$\,5\,photoelectrons per MeV.
The mean detection efficiency is defined as the fraction of cosmic
events with reconstructed tracks in the calorimeter and the MWPC-PV system
which also register a TDC hit within a $\pm$15\,ns window in at least
one cosmic veto plane.  The mean detection efficiency was measured to be $>$95\,\%.

A cosmic event detected in our apparatus often produces energetic back-to-back 
showers in the calorimeter. If such a cosmic event is in random coincidence with 
a $\pi$-stop trigger and the charged particle detection system is inefficient,
it might be misinterpreted as a $\pi\beta$ decay event. Fig.~\ref{fig:cosm_en} 
(top panel) shows the calorimeter energy spectra of cosmic muons taken
with a cosmic trigger. The trigger is defined as a single CsI 
cluster firing above the low discriminator threshold {\sc LT}$\simeq$\,5 MeV.
While the most probable energy deposition is $\simeq$\,170\,MeV,
calorimeter cosmic energies span the range from the {\sc LT} threshold
for tracks that just graze the CsI calorimeter, all the way up to almost
1000\,MeV for the extended cosmic showers.  

The calorimeter $\pi\beta$ trigger requires two coincident superclusters to be
located in the opposite calorimeter hemispheres firing above the high discriminator 
threshold {\sc HT}$\simeq$\,52\,MeV. The measured sum of the 
two cluster energies for $16,916$ cosmic triggers (a single 1.3 hour long
cosmic run) shown in the bottom panel of Fig.~\ref{fig:cosm_en} has a low energy 
cutoff at $\sim$\,200\,MeV. We can use this value as an upper level cut in 
the offline $\pi\beta$ event analysis and thus achieve a minimum suppression 
factor of $2.3\cdot 10^{-6}$. The raw counting rate of {\sc HT} cosmic triggers is 
$\sim$\,5\,Hz while the nominal $\pi^+$ stopping rate $\le$\,1\,MHz. With logic signals 
about 10\,ns wide, the random cosmic-$\pi$-stop overlap rate is 
5$\cdot$10$^{-2}$\,Hz. 

Combining the calorimeter energy cut defined above and the $1\cdot 10^{-5}$ tracking 
system inefficiency discussed previously we find that the cosmic background 
contributes to the $\pi\beta$ event sample at a level of $\le$10$^{-4}$. Additionally, 
information from the cosmic muon veto counters, which enclose the detector,
suppresses this contamination by another order of magnitude, making it 
negligible. 

\bigskip

\section{Temperature stabilization control}\label{temp}
\medskip
The design goals for the temperature stabilization system were:
\begin{itemize}\setlength{\itemindent}{3ex}
\item[(1)] constant CsI calorimeter temperature of 22$^\circ\,$C stable to $\pm 0.2^\circ\,$C;
\item[(2)] relative humidity inside the detector thermal housing $\le$\,50\%. 
\end{itemize}
If the air is cooled in the heat exchanger to just above the freezing point,
the humidity condenses and the water is removed from the thermal volume 
through a drain pipe. A temperature of 0.2$^\circ$\,C gives a vapor pressure 
of 460\,Pa, corresponding to a relative humidity of 25\,\% at 22$^\circ$. With
this method, dry air is produced, and the water sensitive CsI crystal
surfaces are protected.

We can calculate the cooling power requirement as follows:
the surface area of the thermal enclosure is 28\,m$^2$. 
The Styrodur 4000S panels used for the thermal enclosure have a heat conductivity 
of 0.028\,W/m$\cdot$K. Assuming an extreme operating 
temperature of 10$^\circ\,$C with an outside temperature of 35$^\circ\,$C and insulator
thickness of 0.04 m we find a maximum heat load of
\begin{equation}
{\rm P = {28 m^2\cdot 25 K\cdot 0.028\,W \over 0.04 m\cdot m\cdot K} = 490\,W.}
\end{equation}
This power dissipation has to be increased by another 500-600\,W to account for 
the detector PMT bases, and 60\,W for the air circulation blowers.
A HX-200 Recirculating Chiller from NESLAB Instruments~\cite{nes}
suffices and allows some margin of error.  It is able to dissipate $\sim$2000 W at
0$^\circ$C. This unit is
designed to maintain the temperature of the chilled liquid (usually water) to
within $\pm 0.1^\circ\,$C. The chiller unit can also be regulated from an external
temperature sensor. It requires an analog input signal and produces an
analog output signal.
The cooling fluid is kept at a constant temperature of 0.2$^\circ$C using
the regulation of the HX-200 unit while a second feedback loop is used to regulate
the temperature of the air in the enclosure with a heater in the air stream.

Since the temperature is above its freezing point, water (with some antifreeze for safety)
is an acceptable cooling fluid. The high-flow centrifugal pump (such as the PD2 Model from
Sta-Rite Corp.~\cite{starite}) can handle 11.3 $\ell$/min 
which provides a temperature differential of 1.5$^\circ\,$C.  
To avoid potential erosion problems, two heat exchangers in parallel, both 
running at about 10 $\ell$/min are used giving a temperature difference of 0.85$^\circ$C 
in the cooling water. This configuration also has the advantage of permitting 
cold air to be blown in from two sides.

Two heaters, each
with a heating power of 2.1\,kW, are used in the air pipes just after the
cooling heat exchanger to keep the temperature stable under varying conditions.
A regulator turns 
the heaters on and off using a Triac circuit with a period of about one second. 
Switching happens at the zero crossing of the line power, in order to reduce 
the influence on the trigger and data acquisition electronics. The duty cycle is 
determined by an external voltage in the range 0--5 V, which can be either 
generated manually with a potentiometer or controlled by a computer. 

The air temperature is measured just after both heaters, and the CsI 
temperatures are measured with eight sensors glued directly on the front 
and back faces of CsI crystals at different positions. The slow control computer 
implements two proportional integral-differential (PID) loops in software. 
The outer loop tries to keep the CsI temperature constant by means of varying 
the temperature of the heated air. The time constant of this loop was experimentally
determined to be about one hour due to the large mass of the CsI
crystals and the metal frame. The inner loop receives the demand value for
the air temperature from the outer loop, and controls the heating power in
order to keep the air temperature at the demand value. The time constant for 
this loop is determined by the mass of the heating coils and was set at
about 30 seconds.

The combination of the two PID loops keeps the
temperature of the CsI crystals extremely stable.
The stability is demonstrated in a four-day snapshot of the CsI 
temperature, averaged over 6 sensors (Fig.~\ref{fig:temp}). The sensor values 
are sampled several times each second and sent to the history subsystem of the data
acquisition system whenever they change significantly. There they are written
to a hard disk. The temperature oscillations have an rms value of 0.02$^\circ$\,C. 
They are due to day-night changes of the outside temperature and do not cause detectable 
CsI light output variations. During the same time period the relative humidity 
inside the detector enclosure was regulated within a $\sim\,\pm 3\,$\% range.

\bigskip
\section{500 MHz Domino Sampling Chip}\label{dsc}
\medskip
The Domino Sampling Chip (DSC) is a significant addition to our electronics
arsenal.  It is used to digitize waveforms from every phototube in the 
detector~\cite{Bro99}. Waveform digitization helps achieve
higher background suppression and thus, improved calorimeter energy resolution at higher 
event rates. Just as important is the improvement of time resolution.
Our current analysis indicates that 200\,ps rms 
accuracy in the relative time of the leading edge of digitized CsI 
pulses is attainable. The ultimate accuracy of the method depends on the
noise level in the actual signal pulse shapes.

The DSC was originally considered for a more advanced stage of the
experiment.  However, successful initial tests made the chip available
sooner. A dedicated PIBETA circuit board with zero suppression
and appropriate readout features was developed at PSI. The device
is now fully integrated into the PIBETA apparatus and routinely operational.
In addition to its other benefits, the DSC is also fast (up to 1.2\,GHz) and inexpensive. 

\bigskip
\subsection{DSC scale calibration}\label{dsc_sca}
\medskip
The DSC was first used with success, albeit with only 5 beamline detectors 
(BC, ACs, AD, and AT), during detector commissioning in 1999.
Full implementation, with digitization of all detector channels, was attained in the
summer of 2000.  Each chip operates at $\simeq$500\,MHz producing waveforms 256$\,$ns 
wide in steps of about 2$\,$ns. The analog waveforms are pre-analyzed 
in the DAQ front-end computer. Zero suppression is applied in software to reduce 
the data rate by a factor of ten, while retaining the interesting sections 
of the waveforms for offline analysis.

The domino speed and thus the sampling frequency varies from channel to channel. Though 
all channels have been set to $\simeq$\,500\,MHz by manually adjusting the voltage levels which 
control the domino speed, a precise calibration is accomplished in software, taking care 
of the differences in the preset voltages as well as temporal time scale drifts.
The random trigger induces a 100\,ns calibration signal in order to determine the exact
sampling speed. The signal, capacitively coupled to each DSC input, consists of two 
voltage spikes set exactly 100\,ns apart.
The exact positions of the spikes are determined by fitting a second-order 
polynomial through three peak points, resulting in sub-bin time resolution. 
DSC speed in a specific run is determined by analyzing and averaging each channel
over all random triggers. Since the random trigger events come at $\sim$\,1\,Hz frequency 
interlaced with our physics triggers, the calibration process runs permanently 
in the analyzer and updates the DSC calibration constants about once an hour.
The examination of DSC calibration runs taken one week apart testifies to the stability of the
domino speed. The maximum run-to-run difference is 0.3\,\% which corresponds to 
0.4\,ns for an average peak in the center of the time scale.

\bigskip
\subsection{DSC waveform fitting}\label{dsc_fit}
\medskip
A digitized PMT waveform can be decomposed into a collection of (``ADC'',\-``TDC'') 
pairs corresponding to the individual signals present. The waveform analysis 
thus reduces to determination of the exact shape of an average waveform (a ``system function'' $g$) 
for each particle type (pion, positron, photon or proton) in each PMT channel, followed by 
fitting of all waveforms with these functions. If there is more than one hit in a given 
256\,ns wide window, the analyzer uses a sum of $N$ system functions, each with a different 
multiplier (related to the ADC values) and adjustable time offsets (corresponding to 
the TDC values): 
\begin{equation}
f(t)=\sum_{i=1}^N {\rm DADC}_i \cdot g(t-{\rm DTDC}_i),
\end{equation}
where $g(t)$ is a system function for a given PMT channel and $f(t)$ is the function  
being fitted to a waveform by varying ${\rm DADC}_i$ and ${\rm DTDC}_i$ values in order 
to minimize the $\chi^2$ of the fit.

Fig.~\ref{fig:dsc_fun} shows a typical system function for a $\pi^+$ in the active degrader 
counter, derived by averaging waveforms of $\simeq$\,100,000 $\pi^+$-beam events which contained 
only one hit in that particular channel. 
Ripples such as the one around $t=175\,$ns are due to either reflections on imperfect
50\,Ohm terminations in the signal path, or to the delayed arrival of positive ions
inside the PMT. In either case, their amplitude is proportional to the main signal amplitude.
These ripples are present in this channel for every event and are correctly 
taken care of by the system function. Other techniques, such as using multi-hit TDCs, would falsely 
identify this peak as a real hit. The system functions for all DSC channels are stored in 
a waveform database file which resides in the analyzer directory and must be present to 
run the code successfully.

The single event display in the Fig.~\ref{fig:dsc_fun} shows a representative
$\pi^+$ waveform from the active degrader channel (dark line) with the fitted function
superimposed (lighter line). The flat waveform sections arise from
the front-end zero suppression, which stores only the data around real hits. The sections 
with small amplitude and small curvature are not taken into account in fitting. Open circles 
indicate the extracted DTDC values. The time axis on the presented plot is not 
calibrated, but one horizontal unit corresponds to $\simeq$\,2\,ns.

Comparison can be made between the data derived from the waveform digitizer 
and traditional FASTBUS and CAMAC ADC/TDC values since we acquire both data streams 
in parallel for all PMT signals. Four panels of Fig.~\ref{fig:dsc_comp} show correlation 
plots between selected variables. The top left panel shows the DTDC values plotted against 
the FASTBUS TDC values. The gap around zero is caused by the trigger veto, which suppresses 
prompt events. 
To make a meaningful comparison with the ADC data, further cuts are helpful. 
The second panel (top right) shows raw DSC data on the vertical scale vs. CAMAC ADC data plotted 
on the horizontal scale. Since the width of the ADC gate is only 25\,ns, analog signals that 
are not in time with the calorimeter trigger are not completely digitized during a CAMAC ADC 
gate, and therefore show smaller ADC values. This limitation is not present in the DSC data, 
which covers the full range of 256\,ns. One can see on the same panel a distribution 
which, when projected onto the vertical axis, has a peak near channel 270 corresponding 
to a valid pion signal. The projection onto the horizontal axis produces only a broad distribution 
extending all the way to zero time, corresponding to events where the incoming pion comes 
$>$\,25\,ns before the trigger and is therefore absent in the CAMAC ADC data. Only the band along 
the main diagonal shows a clean correlation between DSC and ADC data. Selecting prompt trigger 
events, as in the third panel, only this band is present. The correlation arises from the fact that
prompt events signals always come at $t=0$\,ns and are therefore fully contained inside 
the ADC gate. The last panel of Fig.~\ref{fig:dsc_comp} shows DSC ADC vs. DSC TDC data. 
There is no visible dependency, as expected, unlike for the CAMAC ADC data.

A closer look at the time resolution reveals the 0.5\,ns resolution of the LRS 1877 TDC; there 
is also a timing jitter between the DSC and TDC variables. The differences between DTDCs and FASTBUS 
TDCs have a distribution with 0.4\,ns rms which is dominated by the limited LRS 1877 TDC resolution. 
A more detailed presentation of the PIBETA waveform digitizer system will be published in
a separate paper.

\bigskip
\section{PIBETA detector fast trigger logic}\label{elect}
\medskip
Selective, bias-free triggers capable of handling high event rates are 
an essential requirement of the detector system. 
Fast analog triggers were optimized to accept simultaneously:
\begin{itemize}\setlength{\itemindent}{3ex}
\item[(1)] beam particle events;
\item[(2)] prompt interaction events;
\item[(3)] delayed pion and muon decay processes;
\item[(4)] cosmic muon background;
\item[(5)] true random calibration events. 
\end{itemize}
A complete list of the 12 triggers used is given in Table~\ref{tab4}. The table
summarizes trigger names, trigger logic definitions, prescaling 
factors (if any), and typical raw event rates recorded at the production $\pi^+$
beam flux of $\le$\,1MHz. Here we discuss the predefined trigger classes in order 
of increasing their complexity.

\medskip
\subsection{Random (ADC pedestal) trigger}
\medskip
A 190$\times$20$\times$8$\,$mm$^3$ plastic scintillator counter made of BICRON BC-400 
material is placed above the electronics racks, parallel to the area floor, 
about 3$\,$m away from the main PIBETA detector. By virtue of its position, 
the counter is shielded from the experimental radiation area by a 50$\,$mm thick 
lead brick wall as well as a 500$\,$mm thick concrete wall. Operating with 
a high discriminator threshold, it counts only cosmic muons and random background
events at about 1-2$\,$sec$^{-1}$, and has a stable counting rate independent of
the beam. Its discriminated signals define our true random trigger.

The random trigger $\rm T_R$ is connected to the ``trigger mixer'' logic unit
and during production runs is always included among the enabled triggers.
A single production run is limited to 200,000 events, which includes
$\sim$10$^4$ random trigger events.
The online analysis program updates the ADC pedestal spectra 
of all ADC channels for every random event. The pedestal peaks are
then analyzed and realigned automatically by an end-of-run routine.
A detailed description of the ADC pedestal correction algorithm is
given in an earlier paper~\cite{Frl01b}.

We note parenthetically that in off-line data analysis, the ADC values for all 
detectors in random events are written to separate files. These random event
energy depositions are used subsequently in a {\tt GEANT} simulation~\cite{Bru94} 
of the PIBETA detector, enabling us to account rigorously for the residual 
ADC noise and accidental event pile-ups in the energy spectra. Fig.~\ref{fig:rcsi}
shows the calibrated calorimeter energy spectrum for the random trigger events.
The average deposited energy per single CsI module
associated with the random events at our production $\pi^+$ stop rate is 0.15\,MeV. 
The random trigger comes most often in accidental coincidence with a Michel 
decay event. Therefore, the random calorimeter spectrum in the figure has a steep 
cut-off corresponding to the $\simeq$\,53\,MeV Michel end-point energy minus
$\sim\,$4\,MeV energy losses in the target and PV counters.

\medskip
\subsection{Beam particle triggers}\label{bpt}
\medskip
The particle-in-beam trigger \sT$_{\rm B}$ is defined by a four-fold coincidence between 
the forward beam counter BC, the active degrader AD, the active target AT, 
and the rf accelerator signal:
\begin{equation}
\rm \sT_B=BC{\circ} AD{\circ} AT{\circ} rf.
\label{eq:no6}
\end{equation}
Minimum-ionizing positrons in the BC, AD, and AT counters deposit
0.6$\,$MeV, 7.2$\,$MeV, and 9.0$\,$MeV, respectively. The corresponding energy depositions
for 114.0$\,$MeV/c pions (with kinetic energy 40.6$\,$MeV) are 0.7$\,$MeV, 12.7$\,$MeV, 
and 28.0$\,$MeV. By appropriately adjusting the discriminator thresholds and the relative
timing of inputs into the quadruple coincidence~(\ref{eq:no6}), the $\pi$-in-beam 
and $e$-in-beam triggers \sT$_{\pi}$ and \sT$_e$
are set up. Fig.~\ref{fig:pistop} shows a digital oscilloscope snap-shot of the four
input signals constituting a $\pi$-in-beam coincidence.

Moreover, we define a less restrictive $\pi$-beam signal $\rm {B}_\pi$ that tags 
a $\pi^\pm$ beam particle inside the detector via an overlap between the forward beam 
counter pulse BC and the active degrader pulse AD:
\begin{equation}
\rm {B}_\pi=BC{\circ} AD.
\end{equation}
Discriminator thresholds and relative times in the beam particle coincidence
\sT$_{\rm B}$ are adjusted to preferentially select through-going pions.

Each pion stopping in the target initiates a pion gate {\small $\pi$G}, 
a 180 ns long window, whose delay is adjusted to start 50\,ns ahead of
the pion stop time $t_0$. A 10\,ns wide gate {\small $\pi$S}
coincident with a pion stop pulse is also generated.
$\rm \sT_B$, {\small $\pi$G}, and {\small $\rm\pi$S} are used as 
the building blocks of more complex triggers described below. Higher order 
triggers are generated on the basis of a coincidence (or anti-coincidence) between 
one of these gates and shower signal(s) in the calorimeter.

\medskip
\subsection{One-arm calorimeter trigger}
\medskip
The 15 UVA 125 discriminator/summers define 60 discriminated CsI cluster 
pulses and provide both low and high threshold discriminated outputs.
Fig.~\ref{fig:trig_ele} shows a schematic 
layout of the fast trigger logic beginning at the discriminator outputs.
Two CAMAC LeCroy Model 4564 OR logic units are used to combine these 60 clusters 
into 10 CsI supercluster pulses. Again, both high (HT) and low (LT) supercluster 
variants of the output signals are implemented. The simplest calorimeter trigger 
is the one-arm CsI trigger $\rm C_{S}$.  It requires the firing of at least one 
CsI supercluster:
\begin{equation}
\rm C_{S}= 0+1+2+3+4+5+6+7+8+9,
\label{eq:no7}
\end{equation}
where the numbers 0-9 represent ten CsI superclusters. The OR logic of Eq.~(\ref{eq:no7})
is implemented in a single PSI LB 500 logic box, described in the next section.

$\pi e2$ decays, for example, are identified by means of a prescaled trigger 
that requires detection of a single localized calorimeter shower with deposited energy 
exceeding the HT level during a prescaled pion gate {\small $\rm\pi G_{PS}$:
\begin{equation}
\rm T_S^H=C_S^{H}{\circ} \pi G_{PS}{\circ} \bar{B}_\pi,
\end{equation}
where $\rm \bar{B}_\pi$ represents a 10\,ns wide prompt veto signal overlapping 
the $\pi$ stop time. This coincidence logic is illustrated in Fig.~\ref{fig:trig_log} 
which shows a digital oscilloscope snap-shot of the trigger inputs.

\medskip
\subsection{Two-arm calorimeter trigger}
\medskip
The default $\pi\beta$ event calorimeter logic requires a coincidence between 
any CsI supercluster and a logic OR of five superclusters that are not its neighbors. 
The bottom panel of Fig.~\ref{fig:balls} shows the calorimeter hemisphere
complementary to a single CsI supercluster.
This relaxed requirement simultaneously samples
a broad unbiased non-$\pi\beta$ background. The logic requirement could be
made more stringent for higher running rates.  However, the experiment also has
the goal of measuring radiative pion and muon decays with broader 
angular correlation coverage and this requires the logic as implemented.

In terms of the CsI supercluster indices, the two-arm calorimeter trigger 
$\rm C_{S\bar{S}}$ is defined as:
\begin{eqnarray}
\rm C_{S\bar{S}} = &&
0{\circ} (2 + 3 + 5 + 6 + 9) +
1{\circ} (3 + 4 + 5 + 6 + 7) +
2{\circ} (0 + 4 + 6 + 7 + 8) + \nonumber \\
&& 3{\circ} (0 + 1 + 7 + 8 + 9) +
4{\circ} (1 + 2 + 5 + 8 + 9) + \nonumber \\
&& 5{\circ} (0 + 1 + 4 + 7 + 8) +
6{\circ} (0 + 1 + 2 + 8 + 9) +
7{\circ} (1 + 2 + 3 + 5 + 9) + \nonumber \\
&& 8{\circ} (2 + 3 + 4 + 5 + 6) +
9{\circ} (0 + 3 + 4 + 6 + 7),
\label{eq:csi_pbl}
\end{eqnarray}
where the first two lines refer to the ``northern'' calorimeter hemisphere,
and the last two lines refer to the ``southern'' hemisphere.
The trigger OR involves the logic sum of 10 conjugate ORs that are 
implemented in the first programmable LB 500 logic box.
The actual logic function implemented omits the repetition of two-fold
AND terms. 

The two-arm trigger has three versions, one with high discriminator
thresholds on both superclusters  $\rm C_{S\bar{S}}^{H}$, another with 
two low discriminator thresholds $\rm C_{S\bar{S}}^{L}$, and a third
version with one high and one low threshold $\rm C_{S\bar{S}}^{LH}$.  The latter
versions allow sampling of the radiative $\pi$ and $\mu$ decays.

$\pi\beta$ decays as well as radiative $\pi\to e\nu\gamma$ events
are identified by means of a dedicated trigger requiring two antipodal 
CsI showers during the {\small $\rm\pi G$} with energy deposited in
the calorimeter exceeding HT:
\begin{equation}
\rm T_{S\bar{S}}^H=C_{S\bar{S}}^{H}{\circ} \pi G{\circ} \bar{B}_\pi.
\end{equation}
The radiative muon decays $\mu\to e\nu\nu\gamma$ are captured with
a prescaled two-arm low threshold calorimeter trigger:
\begin{equation}
\rm \sT_{S\bar{S}}^L=C_{S\bar{S}}^{L}{\circ} \pi G{\circ} \bar{B}_\pi.
\end{equation}
Finally, the third version of the two-arm trigger samples dominantly
the radiative $\pi\to e\nu\gamma$ decays with a low energy positron or
photon:
\begin{equation}
\rm \sT_{S\bar{S}}^{LH}=C_{S\bar{S}}^{LH}{\circ} \pi G{\circ} \bar{B}_\pi.
\end{equation}

\medskip
\subsection{Three-arm calorimeter trigger}
\medskip
The three-arm calorimeter trigger
$\rm C_{3S}$ is defined as:
\begin{eqnarray}
\rm C_{3S} = &&
0{\circ} 2{\circ} 9 +
0{\circ} 3{\circ} 7 +
0{\circ} 7{\circ} 9 +
1{\circ} 3{\circ} 5 +
1{\circ} 4{\circ} 8 + \nonumber \\
&& 1{\circ} 5{\circ} 8 +
2{\circ} 4{\circ} 6 +
2{\circ} 6{\circ} 9 +
3{\circ} 5{\circ} 7 +
4{\circ} 6{\circ} 8.
\label{eq:csi_3}
\end{eqnarray}
The ten terms in the logic sum represent all possible combinations of 
three nonadjacent low-threshold supercluster overlaps. Like the others,
the three-arm trigger is implemented in the LB 500 unit.
We use this trigger to collect events with three coincident particles 
in the final state following the $\pi$ stop signal:
\begin{equation}
\rm \sT_{3S}=C_{3S}{\circ} \pi G{\circ} \bar{B}_\pi.
\end{equation}
This trigger is dominated by the random coincidences of three
Michel positrons but it also includes the Dalitz decays $\pi^0\to e^+e^-\gamma$,
and rare $\mu^+\to e^+\nu_e\bar{\nu}_\mu e^+e^-$ decays.  

\bigskip
\subsection{Prompt event trigger}\label{prompt_tr}
\medskip

The prompt event trigger $\rm T_P$ uses the one-arm CsI logic $\rm C_S$
and an adequately prescaled version of the pion stop $\rm\pi S_{PS^\prime}$:
\begin{equation}
\rm T_P= C_S^{H}\circ \pi S_{PS^\prime}.
\end{equation}
During production running we select a high threshold version of
the $\rm C_S$ input, whereas the energy spectrum of prompt protons 
in the calorimeter shown in Fig.~\ref{fig:prot_en} is taken in a calibration
run using the low-threshold version in order to record the entire energy spectrum
of prompt events.

\bigskip
\subsection{Cosmic muon event trigger}\label{cosm_tr}
\medskip

There are two different versions of the cosmic muon trigger.
In routine production running with high intensity $\pi^+$ beam,
a cosmic event is defined as a coincidence between a prescaled signal 
from the cosmic muon veto counter $\rm CV_{PS}$ and a high threshold
one-arm CsI signal $\rm C_S^H$, vetoed by the beam:
\begin{equation}
\rm T_c^H=   CV_{PS} \circ C_S^H\circ  \bar{B}_\pi.
\end{equation}
The low-threshold one-arm CsI version of the cosmic trigger $\rm T_c^L$
defined as
\begin{equation}
\rm T_c^L=   CV \circ C_S^L\circ  \bar{B}_\pi,
\end{equation}
uses a nonprescaled cosmic muon counter signal $\rm CV$.
It is included in the trigger mix only during periods where there is no
$\pi^+$ beam in order to take cosmic muon data for calibration purposes. This
usually happened during planned accelerator maintenance -- normally one day each week.

\bigskip
\subsection{The PSI LB 500 programmable logic unit}\label{LB500}
\medskip

The several triggers described in the previous sections were originally
implemented in traditional CAMAC coincidence logic units with specially made twisted
pair cables. This solution proved to be unreliable and inflexible.
Therefore, a new logic unit denoted LB 500 was developed at PSI. This
single width CAMAC unit contains ECL-TTL converters for 64 input/output
lines and a Lattice Complex Programmable Logic Device (CPLD) 
ispLSI 2128E~\cite{Lat}. This chip has 6000 programmable gates which can be 
combined in virtually any way and a pin-to-pin propagation delay of 5\,ns. 
Unlike field programmable gate arrays (FPGAs), the CPLD does not contain static 
RAM cells for logic functions.  Therefore, it can function exactly as a conventional 
coincidence unit in overlap mode, preserving the times of the input signals 
at the output. The LB 500 unit can be reprogrammed via CAMAC in a few seconds. 
This allows remote modification of the trigger in a reliable way. 
The 64 ECL lines at the front panel of the LB 500 unit can be configured 
in groups of four as either inputs or outputs by changing internal resistors.

One LB 500 unit is used for defining the two-arm and three-arm triggers and another
unit is used for prescaling selected triggers. This solution is more
powerful than that using conventional prescalers with clocks and coincidence
units, since the prescaling factor can be remotely adjusted over a wide range.
Internally, the unit uses 15-bit synchronous counters, which have
propagation delays independent of the prescaling factor. Reshaping of the
output signal preserves the original input pulse width and time.

A third LB 500 unit is used for the trigger mix. The signal timing inside the
LB 500 unit can only be adjusted in steps of $\simeq$2.5\,ns  by 
means of inverter chains. An external programmable CAMAC delay LeCroy Model 4418 
is used to adjust the timing of the calorimeter triggers and the various beam 
gates with an accuracy of 1\,ns. In normal production mode, all predefined triggers 
are enabled except the low threshold cosmic trigger.

Fig.~\ref{fig:trig_ele} referenced above is a flow chart of the signal processing 
described above. The design details for all three LB 500 units are 
available on the World-Wide-Web at {\tt www.\-pibeta.\-psi} or 
{\tt www.\-pibeta.\-phys.\-virginia.\-edu}. 

\bigskip
\section{The MIDAS data acquisition system}\label{daq}
\medskip

A new, general purpose data acquisition system called MIDAS (Maximum Integrated Data 
Acquisition System)~\cite{midas} was developed for the PIBETA experiment.
It is now used in a number of experiments at several
laboratories. The system is suitable for experiments ranging from small test set-ups
with only one personal computer (PC) connected to CAMAC, through medium scale experiments 
with several front-end and backend computers. The system runs under Linux, Windows NT,
and VxWorks.

An integrated slow control system includes a fast online database and a
history system. Drivers exist for CAMAC, VME, FASTBUS, high voltage systems,
GPIB and several PC plug-in DAQ boards. A framework is supplied which can be
extended by user code for a front-end readout on the one side, and data analysis
on the other. The online data can be presented in real time via the CERN Physics
Analysis Workstation library ({\tt PAW}) in
histogram and N-tuple format~\cite{Bru93}. The run control is done via a Web
interface, which allows remote monitoring through the Internet.

MIDAS software consists of a set of library functions and applications. The library
is designed in two layers. The lower layer covers all calls dependent on the
operating system, mainly shared  memory access and semaphore operations. The upper
layer is entirely independent of the operating system. Since the lower layer contains
only about 7\,\% of the code, porting the library 
to a different operating system is quite easy.

The MIDAS library contains routines for buffer management, a message system, a history 
system and an online database (ODB). MIDAS buffers are FIFOs which support multiple producers 
and multiple consumers. Consumers may request a certain subset of event types from a buffer. 
They can also specify whether they want to receive all events of a given type or only as 
many as they can process without blocking the producers. A watchdog scheme has been designed 
which removes crashed clients from a buffer to prevent them from blocking the whole system.
The transfer speed between a producer and a consumer is on the order of $\sim\,$20\,MB/s if both
run on the same computer and $\sim\,$10 MB/s over 100BaseT Ethernet. The history system is used
to store data on a hard disk and produce value-versus-time plots. It is capable of storing
several thousand events per second and of changing the event definition on-the-fly during 
an experiment.

The online database provides central data storage and contains all relevant experiment
variables like logging channel information, event definitions, slow control variables, front-end 
parameters and histogram definitions. It is kept entirely in shared memory for fast 
access of up to 50,000 read/writes per second locally and 500 read/writes per second remotely. 
The database is hierarchically structured, similar to a file system, with directories 
and sub-directories. The data are stored in key/value pairs, where a key corresponds to 
a file name and the value to the file contents. Keys can reside in directories which
can be subdirectories themselves.
ODB keys can be created, edited and deleted dynamically during run-time. 
Each key can contain either a single value of any type (an integer, real, string, etc.) or 
an array of values of the same type.

Entries in the ODB can be changed in three ways. The first possibility is direct 
editing using ODBEdit, a general purpose editor developed to view and change values 
in the ODB. This method is used mostly when running online. Parameters can be 
changed and the effect of the change can be inspected immediately by looking
at the online N-tuples and histograms.
Access to the online database is also possible through a Web
interface. Fig.~\ref{fig:control} shows an example of the main PIBETA experiment
control window. This makes it possible to control an experiment from any computer 
running a Web browser.
The final way to change ODB variables is by 
loading configuration files. Subtrees of the ODB can be saved and loaded in a simple 
ASCII format. 

A client can register a "hot-link" of a local C-structure 
to a database sub-tree. Whenever a value in the sub-tree is changed by 
someone, the client C-structure automatically receives an update of the modified parameter. 
This scheme makes it very easy to control dynamically the behavior of front-end programs 
and analyzers. The slow control system is based on the same principle. Whenever a demand value 
is changed in the online database, the slow control front-end receives an update and can 
propagate it to the proper hardware via a device driver. An image of the whole database 
contents can be written to the logging channels to reflect the current status of 
the experiment.

Several applications have been written using the MIDAS library. A general-purpose logger 
supports multiple logging channels to disks, magnetic tapes and ftp sites. Different logging 
channels can receive different types of events in different formats. Three data formats 
are currently supported: ASCII, an optimized MIDAS format and the YBOS~\cite{ybos} format. 

A general-purpose analyzer framework was designed which incorporates
CERN {\tt HBOOK} library routines~\cite{hbook}. Raw and derived data values are booked 
automatically as N-tuples and can be histogrammed and presented with {\tt PAW} commands 
online. The identical analyzer can be used for an offline replay analysis. 

The MIDAS implementation used by the PIBETA experiment has
two front-end PC computers connected to CAMAC, FASTBUS, VME and a LeCroy LRS 1440 
high voltage system. 
The event response of the first front-end computer running under Windows NT
using a polling scheme is 8\,$\mu$s. The second front-end computer measures
temperature, humidity and gas flow values of the MWPC system and controls
the high voltage and detector temperature as described in Sec.~\ref{temp}.

The backend computer running under Linux performs a full online data analysis
and stores data on disk and magnetic tape.
Data files are written to a dual digital linear tape drive (DLT TTi 2200) with
20 GB capacity and  1.5 MB/s transfer rate. In addition, the data
files are written to a disk, from which they are migrated automatically to the
PSI file archive. The archive system uses a HP/CONVEX SPP1000 mainframe with
several DLT tape robots and UniTREE software. Data can be transferred
using the FTP protocol with a dedicated Ethernet connection from the PIBETA
experiment. Logically, a user can think of the backup archive simply as another
disk directory tree. From a user perspective, archive files appear as if they were on disk.
All 3.4\,TB of data recorded by the PIBETA experiment could be replayed directly
from the archive.

The MIDAS source code for Unix (Digital Unix/Ultrix/Linux) and 
the executables for Windows NT (Intel) and MS-DOS are available at the MIDAS http server 
{\tt http://midas.psi.ch/} or {\tt http://midas.triumf.ca}. The Concurrent Version System 
(CVS) tree that lets developers access the latest code is available via 
the cvsweb interface.
\bigskip

\section{User DAQ code}\label{user}
\medskip

\subsection{PIBETA analyzer}\label{ana}
The PIBETA analyzer program is written in the C language. 
It consists of two parts: a system 
part which is responsible for reading and writing events in various formats and 
a user part which actually does the experiment-specific data analysis.  The analyzer format 
is modular and user-friendly. It is easy to install, use and extend.

In order to make the data analysis more flexible, a multi-stage concept is 
chosen as the backbone of the analyzer structure. A raw event is passed through several 
stages in the analyzer, where each stage has a specific task. The different stages read 
an event, analyze it and append the results of the analysis back to 
the event. Therefore, each stage in the chain has access to all results from previous 
tasks. The first stages in a chain typically deal with data calibration, while the last 
stages contain the ``physics'' analysis code, which calculates kinematical and
physical variables. 
The multi-stage concept allows the collaboration members to use standard modules for 
the calibration stages ensuring that all experimentalists deal with identically 
calibrated data, while the last stages can be modified by individuals to look 
at different aspects of the data.

This method is different from the usage of  data summary tapes (DSTs) 
in other experiments. Instead of producing some intermediate data which get
distributed, analysis is always performed on the original raw data. 
The advantage of this method is that one still has access to the rudimentary data
(such as raw TDC and ADC values) even in the most advanced stages of the analysis. 
If one is in doubt about the calibration, one can always go back and test 
a different calibration method, which is not possible if one works on precalibrated 
DST data.

A bank system used for event storage reflects the multi-stage concept in 
the data structures. A bank contains a sub-structure of an event. It can be of 
variable length (like the sparsified ADC data) or of fixed length. A fixed size 
bank can contain different data types like integers and floating point values, 
while a variable length bank can only contain a single data type.
The front-end program produces banks related to different parts of the detector
electronics, like a raw ADC bank, a raw TDC bank, a scaler bank and so on. 
The first analyzer stages use these banks to produce calibrated data 
such as energy deposition expressed in MeV, residing in a calibrated ADC bank, 
a charged particle track intersection with the horizonatal plane in millimeters 
in a MWPC bank and so on. Different users can add private banks which contain 
variables they are interested in.

Several analyzer modules use banks from the online system to produce new, calculated 
banks. At the end of an analysis process, all events are written to disk. Each bank has 
a database flag telling the system if this bank should be included in the output file. 
By suppressing some extensive online banks, the amount of data in the output file 
can be reduced significantly.

Since the contents of banks are defined in the online database, the system part of 
the analyzer knows how to interperet the contents of an event. Thus, 
N-tuples can be booked automatically by the system. When running the analyzer offline, 
column-wise N-tuples (CWNTs) are used. Each event bank is booked as a ``block'' in 
a CWN-tuple. The CWNT identifier is identical to the event number.

At present the PIBETA analysis at each analyzer stage is performed by 16 independent modules. 
Each module is written in a different source file and exports routines which are 
called for each event at the beginning and end of a run. Currently, the modules have 
to be written in the C language, though MIDAS does permit FORTRAN analyzers.
Each module uses a set of parameters, which are 
stored in the online database. The advantage of this scheme is that 
the parameters can be changed without recompilation of the analyzer.

Modules can be placed into either standard modules or private (user) modules. 
Standard modules are generally accepted by all collaboration members and used 
in every analysis pass, e.g., modules for energy and time calibration,
ensuring uniform calibration for all users. User modules are written by individuals 
to look into specific aspects of the data. Over time, user modules which are 
accepted by the collaboration can be promoted to the standard modules. If a specific method 
is established during offline analysis, it can go straight to the online analysis of 
the next beam time, so one gets online the same results that previously came only after
offline analysis.

The online database stores all variables which concern a specific experiment. 
The same database is used both online and offline. When running the analyzer offline, 
user configuration files can be loaded, overwriting the ODB parameters stored 
in the raw data file. Thus, different configuration parameters can be loaded 
for different individual runs or different series of runs.

The analyzer task can read data from hard-disk files, tapes or the online 
DAQ system. It can write events or specific event banks to the output files.
The program currently supports three different output file formats: MIDAS 
binary, pure ASCII files, and {\tt HBOOK} RZ files with both Column-Wise and Row-Wise N-tuples. 
While the RZ file format can only be used for the analyzer
output, binary and text files are readable by the analyzer. Moreover, these types of 
files can be written and read directly in GNU-zipped format: the data stream is 
compressed and decompressed on-the-fly. While this method reduces disk space requirements
by 50\,\%, it takes about 20\,\% more CPU time.

The following online run types are predefined for the PIBETA experiment:
\begin{itemize}\setlength{\itemindent}{3ex}
\item[(1)] physics data acquisition run;
\item[(2)] pedestal run;
\item[(3)] timing run;
\item[(4)] cosmic run.
\end{itemize}
The pedestal run is performed when bringing the detector online after long 
operational interruptions. Its routines update the ADC pedestal values for all detector 
channels. A run in the timing mode should follow by checking the trigger times and alerts 
the user if any adjustments in the time delays of individual sub-detectors are necessary. 
During production running, the physics DAQ mode is chosen for stable beam-on periods 
while the cosmic running mode is used during longer beam-off or low beam intervals.
Depending on the run mode, the analyzer books different
histograms optimised for that run mode. Since the analyzer has access to the
whole hardware through the ODB, it reconfigures the trigger electronics and
operates the beam blocker (for cosmic runs) automatically. At the end of a
calibration run histograms are automatically evaluated to obtain
pedestal values, software gains for ADC and timing offsets for TDC data.

\bigskip
\subsection{Single event display}\label{sed}
\medskip
The PIBETA single event display (SED) was written in the Java 
programming language. It has proven to be a valuable debugging tool for 
verifying the correct operation of the user analyzer code. It has also
proven to be indispensable in the development of the DSC signal fitting algorithms. 
The display program can run online by reading from the data stream as well 
as offline by taking the input from raw MIDAS files.
The graphics interface menu offers the following display choices:
\begin{itemize}\setlength{\itemindent}{3ex}
\item[(1)] a segmented CsI calorimeter map in the Mercator projection;
\item[(2)] the front or side cross-sections of the detector with
the active target, MWPCs, plastic veto and the CsI calorimeter;
\item[(3)] a digitized waveform from a single DSC channel.
\end{itemize} 
The calibrated ADC/TDC values, i.e., energies and event times, can 
be shown associated with the individual detector shapes, like PV segments 
and CsI modules. The cross-sectional views show the reconstructed
charged particle tracks, while different particle types (positrons,
protons, photons and cosmics) are identified by differently colored line 
types.

The DSC option shows both the digitized waveform of the selected channel
and the sum of fitted system functions in a 125\,ns time window. The number 
of hits found is indicated by circles located at the DSC TDC timings.

Program options include the selection of a particular 
trigger or an arbitrary trigger combination. The scanned events can 
be displayed sequentially, or, alternatively, triggers can be stepped through 
in a loop. A single selected event can also be specified by its serial number.

The top panel of Fig.~\ref{fig:3e} shows the SED of a $\mu^+\to e^+\nu_e\bar{\nu}_\mu
e^+e^-$ event in a lateral ($x$--$y$) cross section through the detector. 
The bottom panel illustrates a digitized waveform of 
one active target segment (ch.~283) displayed in a SED window. In this example 
the fitting routine finds two hits, one associated with a stopping pion signal 
(first, larger peak in channel 24, i.e., at $\cdot$48\,ns), followed by a stopping 
muon pulse from the $\pi^+\to \mu^+\bar{\nu}_\mu$ decay 
(second pulse, in channel 42, i.e., at $\cdot$84\,ns). 

\bigskip
\subsection{Experiment WWW control page}\label{wwwpb}
\medskip
Detector operation is monitored remotely at all times by experiment 
collaborators who could be residing in different world time zones.
Monitoring is facilitated by the implementation of the WWW interface 
to the online experiment. At least two shift takers were usually present
locally at PSI during production running. Off-site workers followed the
experiment via DSL or modem lines.

At a glance, the WWW experiment control page provides an overview of 
status of the experiment, as well as the conditions of the detector 
subsystems. The data taking can be started, stopped, paused or resumed 
from the control page. The experiment URL is linked to the ODB and
CAMAC CNAF command page, a window with the latest automatically generated
messages from the experiment tasks, the electronic logbook system,
alarm system page, the menu of all experiment tasks, the history
page and the online help link.  

A separate menu line leads an experimentalist to the trigger settings, detector 
rates and ratios of rates, the PSI accelerator status display, the WebPAW 
interface, and the online PIBETA detector handbook. A user can also 
examine trigger and scaler variables, check all demand and measured high 
voltages, as well as the wire chambers and the area beam line settings.
Also highlighted, due to their importance, are the activities of the 
logger channels, the sizes of data files for the current run, the available 
free disk space and the progress in writing the data to the magnetic 
tape and the remote ftp archive.
About half of all detector malfunctions can be successfully
corrected by user intervention from a remote location.

\bigskip
\subsection{Electronic logbook}\label{elog}
\medskip
A traditional paper logbook was not a practical option for a number of reasons.
To adress this,
the MIDAS system contains an electronic logbook. Shift workers can 
enter logbook messages with optional attachments through the same Web interface
used for experiment control. Online histograms can be attached in addition to
the variables defined in the online database. Messages are classified into certain 
groups, which makes it simple to search for a specific entry. In addition, the analyzer
has access to the electronic logbook and can submit regular entries concerning, for
example, the automated gain calibration.

Use of the electronic logbook greatly enhanced access to the experiment and
mutual communication among collaborators both on-site and remotely.
The e-logbook files were copied on a daily basis between PSI and the University of
Virginia for local access and backup. A stand-alone version of the
electronic logbook is available for offline browsing of the logbook 
entries~\cite{elog}.

\bigskip
\subsection{Alarm system}\label{alarm}
\medskip
To optimize the efficiency of data taking, malfunctions of the detector 
or the beam line must be quickly detected and promptly remedied. 
MIDAS contains an alarm system for that purpose. This 
system regularly checks the operation of all experiment computers and programs. 
In addition, any parameter value defined by the slow control system and/or 
known to the analyzer can be verified. We use this facility to check:
\begin{itemize}\setlength{\itemindent}{3ex}
\item[(1)] temperatures in the detector enclosure and the electronics hut;
\item[(2)] gas pressure in the MWPC system;
\item[(3)] detection efficiency of the MWPCs;
\item[(4)] operation of the tape drives and uninterrupted archive backup;
\item[(5)] rates of the pion beam stops and triggers;
\item[(6)] scaler rates for all CsI crystals and plastic veto counters.
\end{itemize}
These computerized checks ensure good stability of the experiment over long
measurement periods. The detector rates are especially sensitive to
beam line problems and electronic unit failures. In case of an
alarm, the shift persons, both local and remote, are notified by telephone calls through 
a modem connected to the back-end computer. This arrangement makes it possible to run 
the detector unattended over certain periods while retaining the ability to react to problems in
less than 15 minutes at any time.

\bigskip
\section{Energy calibration of the PIBETA calorimeter}\label{en_cal}
\medskip
Energy calibration of the PIBETA calorimeter involved two
correlated processes: equalizing discriminator thresholds 
(expressed in MeV) for 220 CsI detector signals that define 
the calorimeter trigger and calibrating signal gains of 240 
CsI detectors at the opposite analog ADC/DSC branch. We accomplished 
first the trigger branch threshold equalization by adjusting the 
high voltages of the individual CsI PMTs. Once all CsI discriminator 
thresholds, expressed in calibrated MeV units, were matched (usually within
$\pm$1\,V, or 0.5\,\% in signal gain), the CsI analog signals 
measured in the FASTBUS ADC branch were matched by introducing software 
gain factors for individual detectors. 
These two adjustments can be done both manually by the experiment operators 
on duty, as well as automatically by a computer program.

\bigskip
\subsection{Calorimeter PMT gain adjustment}\label{g_thr}
\medskip
Using a digital oscilloscope we confirmed that the individual 
daisy-chained UVA 125 cluster discriminator/summers all have identical 
demand threshold levels (within 1\,\%).
The difference (if any) between a signal gain of an individual CsI detector 
at the trigger branch on one side, and at the input to a digitizing
ADC/DSC system on the other side, could arise from different 
signal attenuations in the delay cables of two branches, differences in 
the resistor values in UVA 126 and UVA 139 passive signal splitters 
and/or small mis-matches in time offsets of the two branches.
We measured ratios of the fast trigger pulses to delayed pulse charges
using an oscilloscope in the single shot mode and found the normalized
ratios to vary up to 10\,\% for the same CsI detector. The conclusion
is that once  CsI signals are perfectly gain-matched at the trigger inputs,
thus giving the best overall trigger energy and timing resolution, the signals 
in the ADC branch have to be simultaneously scaled by 0.9--1.1 to attain
the best possible calorimeter energy resolution.

For a detector equipped with an $n$-stage PMT, a normalized real gain change 
$g$, relating two different configurations 1 and 2, depends on the ratio of 
the software gains $s_i$ and the corresponding high voltages ${\rm HV}_i$:
\begin{equation}
g = {{s_1} \over {s_2}}\cdot 
\left({{\rm HV_1} \over {\rm HV_2}}\right) ^n.
\end{equation}
The above equation relates the net gain factor of a scintillator detector
in configuration 2 to the gain in configuration 1. 

To facilitate and automate the detector energy calibration,
an analyzer routine was developed that could optionally:
\begin{itemize}\setlength{\itemindent}{3ex}
\item[(1)] apply an overall gain factor by varying all CsI PMT high voltages or
software gain multipliers;
\item[(2)] adjust the CsI ADC gains at the
trigger discriminators via HV changes by comparing the counting
rates in the $\pi\to e\nu$ peak and the high energy edge of the 
$\mu\to e\nu\nu$ spectrum; 
\item[(3)] match the CsI ADC gains using the fitted $\pi\to e\nu$ peak histograms;
\item[(4)] match the CsI ADC gains using the fitted $\mu\to e\nu\nu$ energy spectrum;
\item[(4)] adjust the PV detector geometric-mean ADC gains using the
fitted peaks of minimum ionizing $e^\pm$s.
\end{itemize}
At detector start-up time, the gain-match routine is run 
manually by an experienced operator. Several iterations of the CsI PMT high 
voltage adjustments are necessary before the $\mu^+\to e^+\nu_e\bar{\nu}_\mu$ to 
$\pi^+\to e^+\nu_e$ counting rates for the high threshold trigger are fixed to 
the default 3:1 ratio. The software 
gains of individual CsI detectors are then set by matching the positions 
of the 69.8 MeV $\pi^+\to e^+\nu_e$ peaks. Consequently, all CsI high voltages 
are adjusted automatically by the computer program when the preset cumulative 
statistics in the reference histograms are attained. This dynamic gain monitoring 
occurs daily and the gain-matching routine automatically submits the appropriate
e-log entries.

\bigskip
\subsection{Calorimeter energy resolution}\label{eres}
\medskip
We calibrated the 5--150\,MeV dynamic range of the calorimeter
by employing 
\begin{itemize}\setlength{\itemindent}{3ex}
\item[(1)] monoenergetic 69.8\,MeV $\pi^+\to e^+\nu_e$ positrons;
\item[(2)] the energy line shape of reconstructed $\pi^0$s following 
the $\pi\beta$ decay at rest;
\item[(3)] the energy line shape of the SCX $\pi^0$s
produced on the $\rm CH_2$ target;
\item[(4)] the $\mu^+\to e^+\nu_e\bar{\nu}_\mu$ energy
spectrum;
\item[(5)] the cosmic muon energy spectrum.
\end{itemize}

The most important calibration point comes from the $\pi^+\to e^+\nu_e$
line: the experimental spectrum shown in Fig.~\ref{fig:pienu_en}
is used to fix the single conversion factor to the absolute energy scale 
as well as to fine-tune the parameters (mostly the photoelectron statistics)
in the detector Monte Carlo simulation.
Once established, these parameters are ``frozen'' in the offline
data analysis and MC simulation. The energy distributions for all other
reaction channels and probes are then required to agree independently with the same
{\tt GEANT} MC simulation.

The $\pi^+\to e^+\nu_e$ calorimeter peak position is determined by 
energy losses in the active target, plastic veto scintillator, and the
insensitive layers in front of the CsI crystals, positron 
annihilation losses, photoelectron statistics of individual
CsI modules, and axial and transverse coefficients parameterizing
the nonuniformities of CsI light collection. At the measured peak position 
of $62.55\pm 0.03$\,MeV, the FWHM fractional resolution $\Delta E/E$ is 
$12.8\pm 0.1$\,\%.

The demonstration that the energy response of the calorimeter is simulated
correctly also at the lower energy scale between 5\,MeV and 50\,MeV is given 
by the Michel $\mu^+\to e^+\nu_e\bar{\nu}_\mu$ energy spectrum. The experimental
histogram accumulated for 10 runs as well as the expected Monte Carlo spectrum are
shown in Fig.~\ref{fig:michel_en}. The $\chi^2$ per degree of freedom 
is 1.3.

The precise trigger-defining absolute high energy and low energy 
discriminator thresholds can also be determined from the Michel energy spectra.
We find that the average values in the production running were
$E_{\rm HT}=51.8\pm 0.1$\,MeV and $E_{\rm LT}=4.5\pm 0.1$. The corresponding 
threshold widths, important in the correct simulation of the detector 
response were measured to be $\sigma_{\rm HT}=2.4\pm 0.1$\,MeV, and
$\sigma_{\rm LT}=1.2\pm 0.1$\,MeV, respectively.

Fig.~\ref{fig:pb_en} shows the experimental $\pi^0$ spectra.
The superimposed {\tt GEANT} simulation of the $\pi^0$ line shapes agrees 
very well with the data. The calorimeter fractional FWHM resolution at the
energy of $128.70\pm 0.06$\,MeV is $\Delta E/E=10.0\pm 0.1$\,\%.

\bigskip

\section{Timing response of PIBETA calorimeter}\label{tim}
\medskip
The calorimeter time resolution depends on the intrinsic time
resolution of the individual CsI modules, the spread in the arrival 
of analog PMT signals at the trigger point where the analog CsI summing 
is done, and the uncertainties of the software time offsets. 
Before assembling the calorimeter we measured the intrinsic time 
resolutions of all component CsI modules using cosmic muons as a probe. 
CsI times are determined relative to a small plastic scintillator 
counter. The average CsI detector rms TDC resolution specified in such 
a way is 0.68$\,$ns. The details of these measurements are provided 
in Ref.~\cite{Frl01a}.

\bigskip
\subsection{Multi-hit TDC data}\label{tdc_tcal}
\medskip
The discriminated signals of all PMTs are timed in 
96-channel FASTBUS LeCroy Model 1877 multi-hit TDC units. The instruments 
were operated in common stop mode with 0.5\,ns least significant 
count, measuring both leading and trailing edge time information 
over an interval of 512\,ns.
Up to 10 sequential leading and trailing edge hits per event per TDC channel are stored 
in the analyzer bank and, optionally, in the corresponding column-wise N-tuple. 
However, the physics analysis dictates a more convenient way to present these data.
We chose a scheme which saves information on up to three hits in any channel
(leading edge time plus width) along with the information on the number of
hits in the channel. The scheme is flexible and can easily be modified
to record more information.

\bigskip
\subsection{Trigger timing}\label{tdc_trig}
\medskip

The timing spread of the 220 trigger-defining CsI analog signals is checked
periodically in timing calibration runs with the prompt trigger. The idea is to find 
the time difference between a single reference detector, in our case the active 
degrader, and each CsI counter. This type of timing histogram, associated with a given CsI 
detector, is incremented only if a charged particle track is identified as a fast proton 
($E_p\ge 60$\,MeV) in the plastic veto hodoscope and 80\,\% of the shower energy 
is contained in that module. The total energy contained by the calorimeter is used 
to calculate the time-of-flight correction, a term that is as large as 1.0\,ns for
100\,MeV protons. We use the proton events because $\sim$\,1\,\% statistics in TDC spectra
is acquired within one hour of data taking. The peaks of the timing histograms
are fitted at the end of the run and the peak positions are ordered
relative to the slowest CsI detector. The resulting information is used
to add trigger cable delays, available in 0.5\,ns increments,
to the faster CsI lines, at the patch panel preceding CsI discriminator/summer units.
Three iterations of this procedure resulted in a 0.86\,ns relative trigger rms timing spread
(Fig.~\ref{fig:csi_tr_tim}).
 
\bigskip
\subsection{TDC calibration: zero offsets and slewing}\label{tdc_cal}
\medskip
TDC calibration is accomplished via two independent corrections, both applied in software.
The primary TDC correction compensates 
for different cable delays of the digitizing branch. The zero time is defined as the center of 
gravity of the self-timing peak for each detector channel.
The self-timing peaks are evaluated in the end-of-run routine of a special analyzer 
module. If the appropriate analyzer flag in the ODB is enabled, all raw TDC histograms are fitted 
with Gaussian functions. The results, namely timing peak offsets and the rms values of the peaks, 
are stored in arrays in the ODB database. Since the algorithm needs a clean self-timing 
peak, this procedure is performed only on the prompt event histograms.
All raw TDC values are then modified in the analyzer by subtracting the zero point 
offset. Negative TDC values clearly correspond to signals which come before the trigger, 
positive time values are associated with the delayed calorimeter events. 

The secondary TDC correction linearizes the slewing of 
TDC time caused by the differing amplitudes of ADC signals. A smaller amplitude signal
takes more time to rise to the fixed discriminator threshold than a larger signal.  The
result is an artificial energy dependence of TDC values with lower energy signals
registering later times.
The data for each individual channel are fitted to determine the TDC versus ADC dependence.
The secondary TDC correction is implemented 
in offline analysis by subtracting the fitted function from each TDC value.
We have implemented this solution using the basic functional form: 
\begin{equation}
{\rm CTDC}={\rm TDC_0}+a\cdot ({\rm ADC}-b)^{c}
\end{equation}
where TDC$_0$, $a$, $b$, and $c$ are free parameters of the fit, ADC is the
calibrated ADC value proportional to the deposited energy, and the result, CTDC 
is the corrected TDC value.
Fig.~\ref{fig:slew} shows the energy dependence of one representative
CsI TDC and the reduction in the time slewing after applying the correction. 
This procedure is automated in the analyzer and includes all PMT 
signals. The experimental data are fitted automatically and the parameters
from the above equation are stored in the ODB database.

We use the coincident photon pairs following the decay of $\pi^0$s
copiously produced by single charge exchange (SCX) inside the
degrader and target material to quantify the improvement in 
the time resolution due to the algorithm described above. 
Fig.~\ref{fig:cal_t} shows the relative timing 
between two $\pi^0$ photons before and after the correction. Both 
distributions are fitted with the sum of a constant term and 
a Gaussian function. The algorithm reduces the rms width from 1.30\,ns to 
0.84\,ns, or 0.59\,ns per single photon shower. The signal-to-background ratio, 
defined as the ratio of the height of the Gaussian peak P2 to the constant 
term P1, increases more than fiftyfold, from 24 to 1214.

Fig.~\ref{fig:pienug_coin} shows the dependence of the online timing resolution
on the $\pi^+$ stopping rate. The top panel is the radiative $\pi^+$ decay
signal recorded with one-arm calorimeter trigger at production rate of
$\sim 7.3\cdot 10^5\,\pi^+$/s. The same $e^+$-$\gamma$ timing difference shown 
on the bottom panel corresponds to the data for the lower intensity beam of 
$\sim 1.4\cdot 10^5\,\pi^+$/s. The low-beam rms calorimeter timing resolution
is again 0.6\,ns per shower.

We have continuously monitored the absolute $t=0$ point using the precise 
accelerator primary beam structure. The precise PSI accelerator frequency is 50.63280(4)\,MHz, 
and the phase stability of the primary quartz oscillator is better than 0.01$^\circ$.
This stability translates to the time interval between pulses of 
19.750\,\,ns with $\Delta T_{\rm rf}=0.0028$\,\%. The proton beam bunch rms width is
$\simeq$\,0.03\,ns.

We have studied the timing offset accuracy that we can achieve relying on SCX $\pi^0$ 
event timings. The relevant variable defining the timing of SCX $\pi^0$'s relative to 
a $\pi^+$ stop signal is:
\begin{equation}
[(t_{\gamma_1}+t_{\gamma_2})/2]-t_{\rm \pi G}, 
\end{equation}
where $t_{\gamma_1}$ and $t_{\gamma_2}$ are the TDC values
for two neutral CsI calorimeter showers that reconstruct to $\pi^0$ and 
$\rm \pi G$ is a TDC datum of the $\pi^+$ gate signal. The gate width was set
to $\simeq$\,190\,ns and thus captures 10 beam pulses. The beam pulse
closest to $t=0$ is suppressed in the trigger by the beam veto; therefore 
we do not use the pion stop peak in the analysis. 

As an example of our approach we show the SCX $\pi^0$ timing peaks for
a set of 236 runs in Fig.~\ref{fig:beam_time}. We fit the TDC lineshapes
with Gaussian functions and extract the peak positions and widths. 
The average non-linearized timing peak width is $\sigma=1.493\pm 0.014$\,ns.
A least-squares {\tt MINUIT} fit of the extracted peak times of
the 9 beam bursts with a period $T_{\rm rf}$ left as 
a free parameter gives:
\begin{equation}
t_0=-4.003\pm 0.026,\ {\rm and}\ T_{\rm rf}=19.751\pm 0.010.
\end{equation}

\bigskip
\section{Angular resolution of PIBETA calorimeter}\label{pres}
\medskip
The calorimeter granularity is such that if a module is hit 
centrally, it will contain about 90$\,$\% of the shower energy. Three 
CsI modules at most should receive significant portions of a single particle's 
shower energy. The median number of calorimeter modules with energy 
depositions above 0.25$\,$MeV recorded in low beam flux runs (with no 
accidental pile-up) with a one-arm high threshold trigger is 25.  

The angular direction $(\theta_c,\phi_c)$ of a particle initiating 
an electromagnetic shower in the PIBETA calorimeter is found 
as an energy-weighted mean:
\begin{equation}
\theta_c = { {\sum\limits_{i=0}^{N} \omega_{mi}(E_i)\theta_i}\over
{\sum\limits_{i=0}^{N} \omega_{mi}(E_i)} },
\end{equation}
where $\theta_i$ is the polar angle of
an individual CsI module, and $N$ is the number of 
nearest neighbors of the crystal with maximum energy deposition 
$E_i^{max}$. The sum is over the central crystal ``0'' and all of its nearest
neighbors and thus contains 6 to 8 terms, 
depending on the shape of the centrally hit module. The formula for the azimuthal
angle $\phi_c$ is identical to that for the polar angle with the replacement
$\theta_i \to \phi_i$.

Three different weighting functions $\omega_{mi}(E_i)$ are examined 
(see Refs.~\cite{Ako77,Dav77,Bin81,Bin83,Bug86,Bar91,Awe92,Bit92,Cri97}):
\begin{equation}
\omega_{mi}(E_i)=
\cases{
\omega_{1i}=E_i & {\rm linear\ weighting}, \cr 
\omega_{2i}=E_i^\alpha & {\rm power\ weighting}, \cr 
\omega_{3i}=\mbox{max}\big[ 0, a_0+\ln (E_i)-\ln (E_{tot})\big] &
{\rm logarithmic\ weighting}, \cr
}
\end{equation}
where $\alpha$ and $a_0$ are empirical parameters and $E_{tot}$ 
is the total energy deposited in the $N+1$ modules that define 
the shower cluster.
The linear weighting formula predicts an rms angular resolution
of 2.2$^\circ$ when coded into a {\tt GEANT} Monte Carlo that generates
70$\,$MeV positrons emanating from the target center. A slightly
better resolution of 2.0$^\circ$ is achieved by
power weighting with $\alpha = 0.7$, though both algorithms overemphasize the
calorimeter granularity. 
The logarithmic weighting method of shower localization~\cite{Awe92} is
motivated by the exponential fall-off of the shower's transverse
energy deposition~\cite{Bug86}. The log-weighting algorithm gives 
the best directional resolution and it was employed in the offline analyzer code.

The optimum values for the power-weighting exponent $\alpha$ and 
the logarithmic algorithm parameter $a_0$ are found by two different
methods: {\tt GEANT} Monte Carlo simulations of the detector uniformly
illuminated by positrons and photons in the energy range 30--80$\,$MeV
and comparison of the particle directions reconstructed from
the calorimeter response to 70$\,$MeV $\pi e2$ positrons and 
the positron MWPC tracks.
The {\tt GEANT} optimization gives $\alpha=0.5$ and $a_0=5.5$ for 
the best overall angular resolution. 
Fig.~\ref{fig:pos_resol} shows the distribution of differences between the
charged particle direction deduced from MWPC hits and the direction found
from power-weighted calorimeter energies.
The rms value of 2.0$^\circ$ between the MWPC
tracks and calorimeter shower centroids agrees well with the Monte
Carlo prediction quoted above.

\bigskip
\section{Energy clustering and tracking algorithm}\label{track}
\medskip
The cluster finding algorithm operates in the online analyzer program
and makes the first attempt at identifying calorimeter shower clumps due
to the interaction of a single particle. The initial inspiration for our algorithm
was the Crystal Ball idea of ``a clump discriminator function''~\cite{Ore80}.
The algorithm first constructs a list of up to 6 nonadjacent calorimeter
modules in order of decreasing deposited energy (``clump centers''). The minimum 
calibrated energy allowed for a clump center in online running is 1.0$\,$MeV,
which is lower than the low energy discriminator threshold of $\simeq$4.5\,MeV.

The clump energies $E_{C_i}$ are calculated next by adding the calibrated energies 
of neighboring modules to the energy recorded in the clump center, provided that 
the TDC hits $t_{C_i}$ of neighbors fall within a specified time window. The default 
value of the window width is $\pm5\,$ns with respect to a clump center time. 
Energies of crystals outside that time window 
are not included in the energy sum, under the assumption that they are related to 
accidental coincidences. The number of nearest neighbors varies from 5 for
a centrally hit pentagon, to 7 neighbors for the outlying HEX-C modules.  
When acquiring data at lower beam intensities, with correspondingly lower 
accidental pile-up rates, a provision is made for two rings of neighboring modules 
to be included in the clump sum. The time associated with a clump is
calculated as the energy weighted average of all clump members.
The clumping algorithm finally saves energies, times and the relative angles
between all clump pairs in an analyzer bank. A similar bank is also
created for up to 6 plastic veto ``clumps'' which contain 
energies and times of single hit PV detectors. 

The tracking algorithm begins by finding the ``best''
pairs of hits in the inner and outer MWPC chambers, using the criterion
that the most probable tracks point to the stopping target in the center of 
the detector. The charged tracks identified in this way are then
associated with the CsI and PV clumps defined previously. Finally, all
CsI clumps that do not pair with the charged MWPC tracks are defined
as neutral tracks. 
The single track information sets the stage for particle identification.  
The target, PV and CsI clump energies (Eqs.~\ref{eq:id1}~and~\ref{eq:id2}) and 
corresponding times, and MWPC directions of the identified charged tracks (if any)
are used to associate a positron, proton, photon, neutron or cosmic muon label 
with each track. This final piece of information is appended to the track analyzer bank.

\bigskip
\section{Detector Monte Carlo description}\label{mc}
\medskip
A complete {\tt GEANT} Monte Carlo description of the PIBETA 
detector that includes all major sensitive as well as passive detector
components was developed~\cite{Bru94}. 
The user code is written in modular form in standard {\tt FORTRAN 77} 
and organized into over 300 subroutines and data files~\cite{Frl97}. A graphical
user interface to the code based on {\tt Tcl} routines supplements the program. 
  
Standard {\tt GEANT 3.21} routines are used to describe the electromagnetic shower processes in CsI
(leak-through, lateral spreading, and backsplash) as well as the electromagnetic
interactions in the active target and 
tracking detectors (bremsstrahlung, Bhabha scattering, and in-flight annihilation for the positrons, 
and pair production, Compton scattering, etc. for photons).

The individual PIBETA detector components, both passive and active, can be repositioned or 
disabled without recompiling the code. The following detectors are defined in 
the Monte Carlo geometry:
\begin{enumerate}
\item a beam counter, active plastic degrader and several versions of segmented stopping targets;
\item two concentric cylindrical multiwire proportional chambers used for charged particle
tracking;
\item a 20-piece cylindrical plastic hodoscope used for charged particle discrimination;
\item a 240-module pure CsI calorimeter sphere;
\item a 5-plate cosmic muon veto scintillator system;
\item a passive calorimeter stand and individual detector support systems; 
\item detector phototubes and HV divider bases, and a lead brick shielding.
\end{enumerate}

User input to the simulation code requires the detector version, e.g., 
the run year, beam properties, selected reaction/decay final states, and the 
version of ADC and TDC simulation codes. A user can select one particular final state 
or any combination of different final states with the relative probabilities defined 
by the reaction cross sections and  decay branching ratios. The simulation of ADC values
and TDC hits accounts for individual detector photoelectron statistics, axial 
and transverse light collection nonuniformities, ADC pedestal variations, 
electronics noise, and event pile-up effects. Photoelectron statistics and
individual detector light collection nonuniformity coefficients are initialized
from the RASTA database file~(\cite{Frl01a}). The calculation of accidental coincidences 
up to fourth order is optional.

Selectable options include the effects of:
\begin{enumerate}
\item[(1)] lateral and axial extent and the divergences of the stopping pion beam;
\item[(2)] positron and muon beam contamination; 
\item[(3)] photonuclear and electron knockout reactions in CsI material~\cite{photo};
\item[(4)] aluminized Mylar wrapping of CsI modules and plastic veto staves;
\item[(5)] the gaps between the CsI detector modules;
\item[(6)] the temperature-dependent light output coefficients of 
the individual calorimeter modules;
\item[(7)] gain instability and drifts of the calorimeter detector modules; 
\item[(8)] electronic discriminator thresholds and ADC gate widths. 
\end{enumerate}

Particular attention was paid to the correct accounting of individual CsI module 
software gains. 
The set of gain constants depends on the optical properties of individual
CsI crystals as defined in our {\tt GEANT} database, but also upon the  
incident particle chosen for the Monte Carlo calibration runs and its 
energy because of differences in shower developments of photons and positrons.
In the simulation calculation the values of the detector 
software gains allow the same user control as the detector high voltages in
the operation of the physical detector. The software gains were determined in an iterative
procedure constrained by the Monte Carlo $\pi^+\to e^+\nu_e$ positron ADC spectrum in
each CsI detector. Similarly, in the real experiment, the high voltages of the
individual CsI detectors are also set by matching the positions of the 69.8 MeV 
positron peaks. 

The output of the calculation saves the selected experimental layout and the
final states chosen in the calculation, as well as the production cross sections
and branching ratios used. A number of physical variables that could be of interest
are optionally filled in 1- and 2-dimensional histograms. Row-wise and Column-wise 
{\tt PAW} N-tuples~\cite{Bru93} are used to digitize individual events. Simulated 
energy depositions and the ADC and TDC values associated with the sensitive detectors 
are saved on an event-by-event basis. The single event display operating on 
the output can be used to examine a particular event in detail.

\bigskip
\section{Radiation hardness and aging of active detector elements}\label{rad_har}
\medskip
During production running in 1999 and 2000 the detector 
operated in the pion beam for a total of 297 days accumulating radiation doses of
up to $2\cdot 10^6$\,rads. During that period the active target registered a total of 
$1.4\cdot 10^{13}$ $\pi^+$ stops. Table~\ref{tab5} lists the radiation doses 
received by individual active detector components. The more detailed analysis of 
the radiation resistance of the PIBETA detector is presented in a separate
publication~\cite{radia}.

The gain of the forward beam counter decreased approximately linearly over 
a period of one calendar year due to exposure to 2\,Mrad of radiation. 
The decrease in gain was $\simeq$\,20\,\%/Mrad. The energy resolution 
for through-going pions was also slightly degraded, changing from 13.1\,\% 
to 13.9\,\%.

In the course of data taking the active degrader counter was irradiated 
with a 1.4\,Mrad dose, producing an average gain decrease of
15\,\%/Mrad. The temporal degradation in 
the counter energy resolution is much smaller but is still noticeable.  

The five central target segments AT$_0$--AT$_4$ received the bulk of the beam exposure,
and consequently show the largest temporal variation in the gain factor and energy resolution 
over the 15 month period. Table~\ref{tab5} details the responses of three different segments.
The cumulative absorbed radiation dose is calculated to
be 0.5\,Mrad per segment. The reduction in the gain of the central target scintillator 
is considerable ($\simeq$\,52\,\%/Mrad). The energy resolution 
was degraded by a full 80\,\%. This level of radiation damage forced us 
to manufacture two additional copies of the active target detector.
The target assembly was replaced on an annual basis. 

The temporal dependence of the light output was similar for 
all 20 PV hodoscope staves. Table~\ref{tab5} lists the data for
two individual PV detectors as well as the average numbers for the
complete PV system. The gain factor dropped by 0.25\,\%/krad, giving 
a cumulative degradation over the production period of 10\,\%. 
The energy resolution on the other hand shows a barely measurable change.

The overall CsI calorimeter gain and energy resolution were
determined from calibrated energy spectra of 240 CsI detectors.
The results are again summarized in Tables~\ref{tab5}.
The PIBETA calorimeter gain decreased 17\,\% and the overall online rms energy 
resolution for the 70\,MeV $e^+$s was lowered from 5.5\,\% to 6.0\,\%. For 
an illustration we present the initial and final gain and resolution 
values for 6 individual CsI detectors.

In conclusion, we find measurable decreases in energy gains and 
degradations in energy resolutions of the PIBETA detector subsystems, 
in particular for the beam detectors that were exposed to high radiation 
doses as well as for the more radiation-sensitive pure CsI calorimeter modules. 
These changes in detector responses are monitored online throughout 
the production running and documented in the replay data analysis.
The changes affect the energy calibration of the PIBETA detector elements 
and have to be taken into account properly when defining the
energy cuts in the physics analysis of rare pion and muon decays. 

\bigskip
\section{Results and conclusions}\label{res_con}
\medskip
We designed, built and commissioned the PIBETA detector---a nonmagnetic 
large solid angle detector with an active stopping target. It was
used for studies of rare pion and muon decays at PSI. In this paper we have 
provided a comprehensive description of the design of the individual detector 
subsystems and their calibration and performance in the intense $\pi^+$ beam.

The major performance parameters of the detector's CsI calorimeter are:
\begin{itemize}\setlength{\itemindent}{3ex}
\item[(1)] the rms energy resolution for 70\,MeV positrons $\Delta E/E$ of $\simeq$\,5.0\%;
\item[(2)] the rms energy resolution for decay-at-rest $\pi^0$s 
$\Delta E/E$ of $\simeq$\,4.0\%;
\item[(3)] the two-arm relative time resolution for two $\pi^0$-decay-at-rest photons 
$\simeq$\,1.3\,ns online and $\simeq$\,0.8\,ns offline; per-shower timing resolution
$\sim\,$0.6\,ns;
\item[(4)] the angular resolution of 70\,MeV positron tracks $\simeq$\,2.0\,$^\circ$.  
\end{itemize}

The detector's charged particle tracking system incorporating a pair of cylindrical
MWPCs and a plastic veto hodoscope has a stable charged particle detection 
inefficiency of $\le\,2\cdot 10^{-5}$. Positron identification, when used in 
conjunction with data from CsI calorimeter, is better than 99.8\% accurate.  

The pion beam geometry and beam transmission, temperature and humidity inside 
the detector housing, relative trigger rates, detector time offsets 
and PMT gain factors are proven to be very stable during the three calendar 
years of the data acquisition. Detector operation is nearly 
100$\,$\% automated, requiring only a few experimenters to be physically present 
and on call at the PSI site during the routine data acquisition, with other collaborators 
working remotely.

The apparatus can easily handle pion stopping rates of 1$\,$MHz and lower. 
It can detect virtually background free rare $\pi\beta$ decay events, as well as radiative pion 
and muon decays. Minimal backgrounds at the highest stopping rate arise from accidental 
coincidences with ordinary muon decay events and cosmic muons. So far we have collected 
physics and calibration data based on $2.3\cdot 10^{13}$ stopping pions. The analysis 
of the $\pi\beta$ data sample and radiative pion and muon decay events is underway and
will soon be ready for publication.

\bigskip
\section{Acknowledgements}
\medskip
We thank K.~J.~Keeter who helped with some of the early preparatory work
preceding the PIBETA experiment proposal.
Our engineer Z.~Hochman provided valuable technical assistance 
throughout the detector assembly and commissioning phases. The support from 
the Hallendienst and many other PSI staff members is gratefully acknowledged.
D.~Mzhavia and Z.~Tsamalaidze express gratitude to Profs. Nodar Amaglobeli and
Albert Tavkhelidze for showing great interest and providing help in
this project.

The PIBETA experiment has been supported by the National Science Foundation,
the Paul Scherrer and the Russian Foundation for Basic Research.
This material is based upon work supported by the National Science
Foundation under Grant No.\ 0098758.

\clearpage


\vspace*{\stretch{1}}
\begin{figure}[!tpb] 
\caption{A schematic cross section of the PIBETA
apparatus showing the main components: forward beam counter (BC), two active
collimators (AC1, AC2), active degrader (AD), active target (AT), two MWPCs 
and its support, plastic veto detectors (PV) and PMTs, pure CsI calorimeter 
and PMTs.}
\label{fig:det1}
\end{figure}

\begin{figure}[!tpb] 
\caption{A photograph of the PIBETA assembly located in the $\pi E1$
experimental area: the air-conditioned enclosure housing the PIBETA calorimeter
is in the center, focusing quadrupole magnets are on the left, and the
lead-wall cosmic house is opened at right. The analog delay cables, high
voltage supplies and air-conditioning system are partially blocked from
view by the cement shielding at front, while the fast electronics house is
visible at the back of the photo.}
\label{fig:pb_photo1}
\end{figure}

\begin{figure}[!tpb] 
\caption{A layout of the $\pi E1$ secondary beam line in the main experimental hall
of the Paul Scherrer Institute's ring accelerator.  Focusing quadrupole magnets
are labeled with ``Q'', bending dipoles are labeled with acronyms beginning
with ``A'', and passive collimators are labeled with acronyms beginning
with ``F''.}
\label{fig:pie1}
\end{figure}

\begin{figure}[!tpb] 
\caption{A layout of the PSI $\pi E1$ experimental area for the PIBETA
experiment (top view). The PIBETA detector is located on the platform
which also carries the fast electronics, air conditioning system, high 
voltage supplies, and analog cable delays. The focusing quadrupole magnets
in the beam line channel (Qs), the forward beam counter (BC) and
the passive collimator (PC) are also indicated.}
\label{fig:hall}
\end{figure}

\begin{figure}[!tpb] 
\caption{The momentum spectrum of the positive pions at the face of 
the active degrader as calculated by the {\tt TURTLE} beam transport program.}
\label{fig:beam_p}
\end{figure}

\begin{figure}[!tpb] 
\caption{The pion signal (``$\pi$-stop'') is defined as a four-fold coincidence
between a forward beam counter (channel~1), an active degrader (2), an active
target (3), and a 19.75$\,$ns rf cyclotron signal (4). In the above digital
oscilloscope snapshot the channel~4 scale shows three rf periods.}
\label{fig:pistop}
\end{figure}

\begin{figure}[!tpb] 
\caption{A Polaroid photograph of the beam spot taken in front
of the forward passive collimator PC. The brighter ellipsoid right of the center
is the $e^+$ beam stopped in the collimator material. 
The fainter spot in the center is the momentum-analyzed $\pi^+$ beam. 
The pions and positrons at that point are separated by 40\,mm in the horizontal 
plane.}
\label{fig:beam_sp}
\end{figure}

\begin{figure}[!tpb] 
\caption{A schematic drawing of the forward beam counter inside 
the light-tight housing. A feedthrough and a single photomultiplier tube 
are drawn only on the right side. The beam direction is perpendicular 
to the plane of paper.}
\label{fig:fbeam}
\end{figure}

\begin{figure}[!tpb] 
\caption{The uncalibrated pulse-height spectrum of the 40.6\,MeV incident $\pi^+$ 
beam as detected in the thin forward beam counter. Gaussian fit parameters for 
the lower energy part of the $\pi^+$ peak are showed in the statistics window.}
\label{fig:b0_adc}
\end{figure}
\vfill
\vspace*{\stretch{2}}
\clearpage
\vspace*{\stretch{1}}

\begin{figure}[!tpb] 
\caption{A cross sectional view (left) and a frontal view (right) of the active 
degrader counter. The degrader counter has the shape of a beaker vessel. Beam pions 
incident parallel to the central detector axis are moderated in 30\,mm of scintillator 
material before stopping in the center of active target. 
The four light guides which end up in a cross section of $6 \times 6$\,mm$^2$ are 
also shown.}
\label{fig:act_deg}
\end{figure}

\begin{figure}[!tpb] 
\caption{The uncalibrated ADC spectrum of moderated $\pi^+$s in the active degrader 
scintillator. A 25\,ns ADC gate is used with a 10-bit CAMAC unit. The Gaussian fit is 
superimposed on the data points.}
\label{fig:deg_adc}
\end{figure}

\begin{figure}[!tpb] 
\caption{A sketch of the regular PIBETA active target,
composed of 9 detectors.  The segment sizes are chosen to balance 
the scaler counting rate.}
\label{fig:tgt9a}
\end{figure}

\begin{figure}[!tpb] 
\caption{A photograph of the regular PIBETA active target. Acrylic tapered
light guides are glued to 9 optically isolated target segments.}
\label{fig:tgt9b}
\end{figure}    

\begin{figure}[!tpb] 
\caption{The pulse-height spectrum of pions stopping in the central active 
target segment. The peak-to-tail area ratio depends on the beam divergence.}
\label{fig:tgt_adc}
\end{figure}

\begin{figure}[!tpb] 
\caption{Reconstructed 2-dimensional shape of the $\pi^+$ beam
superimposed on an outline of the segmented active target. Data for 
target counting rates are collected during one 48 hour long series of runs.}
\label{fig:extr_2d}
\end{figure}

\begin{figure}[!tpb] 
\caption{The longitudinal ($z$) component of the $\pi^+$ stopping distribution in 
the active target calculated in a {\tt GEANT} simulation using the momentum 
distribution given in Fig.~\ref{fig:beam_p}.}
\label{fig:pistop_z}
\end{figure}

\begin{figure}[!tpb] 
\caption{Beam particle identification using the time-of-flight (TOF) method.
The TOF difference between the forward beam counter BC and the active
target AT, which are separated by 3.87$\,$m is plotted along the vertical axis. 
The horizontal axis shows the particle TOF between the production target E 
tagged by the 50$\,$MHz accelerator rf signal and the active target AT.
The beam path length between these two points is 16.83$\,$m.}
\label{fig:contam}
\end{figure}

\begin{figure}[!tpb] 
\caption{MWPC$_1$ angular resolutions in the azimuthal angle $\phi_{\rm CH}$ (top panel), 
and axial coordinate $z_{\rm CH}$ (bottom panel) extracted from a sample of cosmic muon events.
The data points for $\Delta\phi$ and $\Delta z$ resolutions are fitted with a sum of two Gaussians, 
the broader one accounting for the cosmic muon scattering in the apparatus..}
\label{fig:mwpc_r}
\end{figure}

\begin{figure}[!tpb] 
\caption{The MWPC$_1$ detection efficiency for minimum ionizing charged
particles as a function of the azimuthal angle (top panel). The bottom panel 
shows the equivalent histogram for the outer chamber MWPC$_2$.}
\label{fig:mwpc_phi}
\end{figure}
\vfill
\vspace*{\stretch{2}}
\clearpage
\vspace*{\stretch{1}}

\begin{figure}[!tpb] 
\caption{A {\tt GEANT} rendition of the plastic veto hodoscope array
surrounding the active target in the center of the PIBETA detector (top panel).
The dimensioned side and end views of one plastic veto scintillator stave with
the light guides glued at the both ends are shown in the middle and bottom
panels, respectively.}
\label{fig:pv_geom}
\end{figure}

\begin{figure}[!tpb] 
\caption{The calibrated energy spectrum of minimum ionizing positrons (left peak) and
5--150\,MeV protons (right peak) in the PIBETA plastic veto hodoscope.}
\label{fig:pv_adc}
\end{figure}

\begin{figure}[!tpb] 
\caption{The $\Delta E$-$E$ particle discrimination between protons and minimum 
ionizing positrons using the calibrated ADC values of plastic veto and CsI calorimeter.}
\label{fig:part_id}
\end{figure}

\begin{figure}[!tpb] 
\caption{The detection efficiency of the PIBETA plastic veto hodoscope for 
the minimum ionizing charged particles plotted as a function of the azimuthal angle.}
\label{fig:pv_eff}
\end{figure}

\begin{figure}[!tpb] 
\caption{An expanded view of the detection efficiency of the plastic veto hodoscope 
as a function of the azimuthal angle: the inefficient regions correspond
to the $\sim$0.08$\,$mm gaps between the 20 individual PV staves.}
\label{fig:pv_eff_zoom}
\end{figure}

\begin{figure}[!tpb] 
\caption{Nine different CsI scintillator shapes that make up the PIBETA 
calorimeter: (i) PENTA, (ii) HEX-A, (iii) HEX-B, (iv) HEX-C, (v) HEX-D, 
(vi) HEXD1, (vii) HEXD2, (viii) VETO1, and (ix) VETO2. The calorimeter proper
is built from 220 CsI crystals while 20 CsI veto crystals surround the beam entry and 
exit regions.}
\label{fig:csi_shapes}
\end{figure}

\begin{figure}[!tpb] 
\caption{The top panel shows the geometry of the pure CsI
shower calorimeter. The sphere is made up of 240 elements, truncated
hexagonal, pentagonal, and trapezoidal pyramids; it covers about 77\% of
4$\pi$ in solid angle. The bottom panel separates one supercluster and its 
complement. The PIBETA calorimeter is made out of 10 
nonoverlapping superclusters.}
\label{fig:balls}
\end{figure}

\begin{figure}[!tpb] 
\caption{The breakdown of the PIBETA calorimeter in the Mercator projection
with the numbering scheme adopted for the detector clusters and superclusters.
The top indices refer to the module shapes, the bottom labels represent the crystal
serial numbers. The beam orientation as well as the polar and azimuthal coordinate axes 
are indicated.}
\label{fig:calo_map}
\end{figure}

\begin{figure}[!tpb] 
\caption{Two drawings of the cosmic veto counters: the top panel shows
the make-up of a single scintillator plane. The bottom panel
shows the details of the light readout geometry via wavelength-shifting bars.}
\label{fig:cv_princ}
\end{figure}

\begin{figure}[!tpb] 
\caption{The energy and time resolution of the cosmic muon veto
detector. The CV timing with respect to a calorimeter
TDC hit is plotted against the uncalibrated CV ADC values.}
\label{fig:cosm_e_t}
\end{figure}

\vfill
\vspace*{\stretch{2}}
\vspace*{\stretch{1}}

\begin{figure}[!tpb] 
\caption{(i) Cosmic muon low-threshold trigger: the energy spectrum 
of the most energetic clump ($E_{\rm 1}>5\,$MeV) in the CsI calorimeter. 
(ii) Energy sum of the two high threshold CsI clumps ($E_{\rm C1,C2}>52\,$MeV) 
with a relative opening angle greater than 150$^\circ$ (log scale).
The Monte Carlo predictions are shown with full line functions.}
\label{fig:cosm_en}
\end{figure}

\begin{figure}[!tpb] 
\caption{The temperature stability of the pure CsI calorimeter: temperature
sensor readings in 10$\,$seconds intervals spanning 100 hours are shown. The
maximum temperature variation of $\le$0.1$^\circ\,$C corresponds to a variation 
of $\le$0.15\% in the CsI light output.}
\label{fig:temp}
\end{figure}

\begin{figure}[!tpb] 
\caption{``The system function'' describing a default $\pi^+$ waveform in
the active degrader counter. The functions of that type are derived by averaging 
100,000 single hit waveforms in the 250\,ns DSC gate.}
\label{fig:dsc_fun}
\end{figure}
\vfill

\begin{figure}[!tpb] 
\caption{Comparison between the DSC-derived data and the conventional
ADC and TDC measurements. Detailed interpretation is given in the text.}
\label{fig:dsc_comp}
\end{figure}

\begin{figure}[!tpb] 
\caption{The calibrated ADC spectrum of a single CsI detector measured using 
a random event trigger at the beam intensity of $8\cdot 10^5\,\pi^+$/s.
The mean accidental energy deposited in a single CsI detector averaged over
all calorimeter modules at that beam flux is 0.15\,MeV.}
\label{fig:rcsi}
\end{figure}

\begin{figure}[!tpb] 
\caption{A block diagram of the fast trigger electronics. The PIBETA trigger logic 
is implemented in three PSI LB500 logic units that use the ``on-board'' 
reprogrammable logic device ispLSI 2128E-LT165 from Lattice Semiconductor~\cite{Lat}.}
\label{fig:trig_ele}
\end{figure}

\begin{figure}[!tpb] 
\caption{A digital oscilloscope snap-shot of a $\pi$-stop signal (channel~1), 
a pion gate (2), a beam signal (3), and a ``CsI high threshold'' signal (4).  
The ``$\pi\to e\nu$ high'' calorimeter trigger is defined as a 3-fold coincidence
$2\cdot \bar{3}\cdot 4$.}
\label{fig:trig_log}
\end{figure}

\begin{figure}[!tpb] 
\caption{The calibrated pulse-height spectrum of the most energetic CsI ``clump''
for the events identified in the plastic veto hodoscope as prompt protons (top).
The equivalent prompt photon and positron energy spectra in the calorimeter are shown 
in the middle and bottom panels. The low CsI threshold of the trigger discriminators 
is set at about 4.5$\,$MeV.}
\label{fig:prot_en}
\end{figure}

\begin{figure}[!tpb] 
\caption{The Netscape experiment control window: different users can remotely monitor 
the state of the detector, the status of the data acquisition system, and the current
values of the database variables. The collaborators with the password-protected
access can submit electronic logbook entries, stop or pause an ongoing run or start 
a new run.}
\label{fig:control}
\end{figure}

\begin{figure}[!tpb] 
\caption{The radial-azimuthal projection of a $\mu^+\to e^+\nu_e\bar{\nu}_\mu e^+e^-$ 
event in the PIBETA single event display (SED, top). The digitized and
fitted waveform of the incident $\pi^+$ stopping in the active target,
followed by the $\pi^+\to \mu^+\nu_\mu$ decay as seen in the SED (bottom).}
\label{fig:3e}
\end{figure}

\begin{figure}[!tpb] 
\caption{The background-subtracted $\pi^+\to e^+\nu_e$ energy spectrum in 
the CsI calorimeter. Software gains of all CsI detectors are automatically 
adjusted by a fitting program. The {\tt GEANT}-simulated detector response 
is represented by the full line histogram.}
\label{fig:pienu_en}
\end{figure}
\vfill
\vspace*{\stretch{2}}
\clearpage
\vspace*{\stretch{1}}

\begin{figure}[!tpb] 
\caption{The agreement between the measured $\mu^+\to e^+\nu_e\bar{\nu}_\mu$ 
energy spectrum in the CsI calorimeter and the Monte Carlo simulation. The data
are collected over 10 runs. The $\chi^2$ per degree of freedom is 1.3.}
\label{fig:michel_en}
\end{figure}

\begin{figure}[!tpb] 
\caption{The calibrated energy spectrum (sum of energies of coincident photon
pairs) for the final $\pi\beta$ event sample (top). Two-photon energy spectrum 
produced by $\pi^-$ single charge exchange (SCX) on the degrader and active target 
counters (bottom). The rms energy resolution of 4.0\,\% is consistent with 
the calorimeter response obtained for clean $\pi\beta$ events.}
\label{fig:pb_en}
\end{figure}

\begin{figure}[!tpb] 
\caption{The timing spread of 220 trigger-defining CsI analog
signals at the inputs of the UVA 125 discriminator/summer units. 
The mean time difference between the fast proton TDC hit in
a CsI module corrected for a time-of-flight delay and
the TDC value in the reference active degrader counter has
rms width of 0.86\,ns.}
\label{fig:csi_tr_tim}
\end{figure}

\begin{figure}[!tpb] 
\caption{Uncorrected (top panel) and linearized (bottom panel) 
ADC vs TDC scatter plots for one representative CsI channel.
Displayed data points are accumulated over 10 consecutive runs.}
\label{fig:slew}
\end{figure}

\begin{figure}[!tpb] 
\caption{Measured time difference in ns between two coincident photons
from $\pi^0$ produced via SCX in the active target. The top histogram is
formed using raw TDC values. The bottom histogram demonstrates the 
effectiveness of the TDC linearization.}
\label{fig:cal_t}
\end{figure}

\begin{figure}[!tpb] 
\caption{Signal-to-background ratio for the events with one minimum ionizing and one 
neutral particle inside the $\pi$-stop gate. The cuts imposed on the high threshold
one-arm calorimeter trigger are listed. The top panel corresponds to the $\pi$-stop rate
of $\sim 7.3\cdot 10^5\,\pi^+$/s while the bottom panel shows the data for
the low intensity beam of $\sim 1.4\cdot 10^5\,\pi^+$/s.}
\label{fig:pienug_coin}
\end{figure}

\begin{figure}[!tpb] 
\caption{The online time distribution of the $\pi^0$ events recorded with
the $\pi\beta$ trigger with respect to the $\pi$-stop signal. 
The average value of the two photon TDCs whose energies and directions 
reconstruct to the $\pi^0$ is shown. The narrow-peak pattern corresponds 
to the 50.63$\,$MHz radio-frequency cyclotron beam structure.}
\label{fig:beam_time}
\end{figure}

\begin{figure}[!tpb] 
\caption{The angular resolution of the PIBETA calorimeter: differences between 
the direction of a minimum ionizing positron calculated using the power algorithm with
energy depositions in the nearest-neighbor CsI clusters and a reconstructed track 
direction from the MWPC hits.}
\label{fig:pos_resol}
\end{figure}
\vfill
\vspace*{\stretch{2}}
\clearpage


\vspace*{\stretch{1}}
\bigskip
\begin{table}[ph]
\caption{Geometrical specifications of the two cylindrical chambers MWPC$_1$ and MWPC$_2$ 
used for the charged particle tracking.}
\label{tab1}
\medskip
\begin{tabular}{lcc}
\hline
\multicolumn{1}{c}{Chamber Parameter} 
&\multicolumn{1}{c}{Inner MWPC$_1$} 
&\multicolumn{1}{c}{Outer MWPC$_2$}\\
\hline\hline
Active length [mm]                   &  350     &   540     \\
All length [mm]                      &  580     &   730     \\
Min. diameter [mm]                   &   90     &   206     \\
Max. diameter [mm]                   &  152     &   255     \\
Diameter anode [mm]                  &  120.3   &   240.2   \\
Diameter inner cathode [mm]          &  115.2   &   235.2   \\
Diameter outer cathode [mm]          &  125.0   &   245.2   \\
Inner cathode--anode gap [mm]        &  2.45    &     2.5     \\
Volume of chamber [cm$^3$]           &  660     &  2035     \\
Total chamber thickness [mg/cm$^2$]  &   53.9   &    74.8   \\
Total chamber thickness [rad. length]&  $1.4\cdot 10^{-3}$&  $2.0\cdot 10^{-3}$\\

Anode wires                                                  \\
\hline

Number of anode wires                &  192     &   384      \\
Anode wire spacing [mm]              &    1.96  &     1.96   \\
Anode wire tension (min-max) [N]     & 0.44--0.48&  0.44--0.51 \\
Resistance of anode wire [Ohm]       &  110     &   155      \\
Resistance between anode-anode [Ohm] &  560k    &   560k     \\
Capacity between anode-ampl. [pF]    & 1000     &  1000      \\

Cathode strips                                               \\  
\hline

Number of inner cathode strips       &  2$\cdot$64  &   192      \\
Angle of slope inner cathode [deg]   &   36.960     &    44.602  \\
Number of outer cathode strips       &  2$\cdot$64  &   192      \\ 
Angle of slope outer cathode [deg]   &  $-33.650$   &  $-42.339$  \\
Width of cathode strips [mm]         &    3.0       &     2.4    \\
Gap between strips [mm]              &    0.4       &     0.3    \\
Resistance of strips (6u Al) [Ohm]   & $\sim$1.5   &$\sim$2.0    \\
\hline
\end{tabular}
\end{table}
\vspace*{\stretch{2}}
\clearpage


\vspace*{\stretch{1}}
\bigskip
\begin{table}[ph]
\caption{Optical properties of the pure CsI scintillators used for
the PIBETA calorimeter (Manufacturers: Bicron Corporation and Harkov
Institute for Single Crystals).}
\label{tab2}
\medskip
\begin{tabular}{lll}
\hline
\multicolumn{1}{l}{Quantity} &\multicolumn{1}{l}{Average Value} & Reference \\
\hline\hline
Density                              & 4.53$\,$g/cm$^3$      & \protect\cite{PDG} \\
Radiation Length                     & 1.85$\,$cm            & \protect\cite{PDG} \\
Refractive Index at 500\,nm          & 1.80                  & \protect\cite{PDG} \\
Refractive Index at 315\,nm          & 1.95                  & \protect\cite{Bic89} \\
Nuclear Interaction Length           & 167$\,$g/cm$^2$       & \protect\cite{PDG} \\
$dE/dx\vert _{\rm min}$              & 1.243$\,$MeV/g/cm$^2$ & \protect\cite{PDG} \\
Photonuclear Absorption ($\pi^0$ photons)& 0.675$\,$\%       & \protect\cite{Frl95} \\
Light Attenuation Length             & 103$\,$cm             & \protect\cite{Frl89} \\
Fast Component Wavelength            & 305$\,$nm             & \protect\cite{Kub88a,Kub88b} \\
Fast Component Decay Time            & 7$\,$ns               & \protect\cite{Kub88a,Kub88b} \\
Slow Component Wavelength            & 450$\,$nm             & \protect\cite{Kub88a,Kub88b} \\
Slow Component Decay Time            & 35$\,$ns              & \protect\cite{Kub88a,Kub88b} \\
Lower Wavelength Cutoff              & 260$\,$nm             & \protect\cite{Bic89} \\
Fast-to-Total Light Output Ratio     & 0.76                  & \protect\cite{Frl00,Frl01a} \\
Light Output                         & 66.2$\,$Photoel./MeV  & \protect\cite{Frl00,Frl01a} \\
Light Output Nonuniformity           & $+0.28\,$\%/cm        & \protect\cite{Frl00,Frl01a} \\
Light Output Temperature Coefficient & $-1.56\,$\%/$^\circ$C & \protect\cite{Frl00} \\
Time Resolution (wrt PV tag)       & 0.68$\,$ns            & \protect\cite{Frl01a} \\
Radiation Resistance                 & See Sec.~\ref{rad_har}& \protect\cite{radia} \\
Stability                            & Slightly Hygroscopic  & \protect\cite{Kub88a,Kub88b} \\ 
\hline
\end{tabular}
\end{table}
\vspace*{\stretch{2}}
\clearpage



\vspace*{\stretch{1}}
\hbox{\ }
\vglue -2cm
\begin{table}[ph]
\tiny 
\caption{CsI cluster definitions for the PIBETA calorimeter. Sixty overlapping
clusters CL00--CL59 each contain nine CsI modules. The first six clusters CL00--CL50 
make the first CsI supercluster SC00, and so on.}
\label{tab3}
\medskip
\begin{tabular}{rrrrrrrrrrrrrrrrrrrr}
\hline
 \multicolumn{1}{r}{Cl/CsI}
&\multicolumn{1}{r}{1}
&\multicolumn{1}{r}{2}
&\multicolumn{1}{r}{3}
&\multicolumn{1}{r}{4}
&\multicolumn{1}{r}{5}
&\multicolumn{1}{r}{6}
&\multicolumn{1}{r}{7}
&\multicolumn{1}{r}{8}
&\multicolumn{1}{r}{9}
&\multicolumn{1}{r}{Cl/CsI}
&\multicolumn{1}{r}{1}
&\multicolumn{1}{r}{2}
&\multicolumn{1}{r}{3}
&\multicolumn{1}{r}{4}
&\multicolumn{1}{r}{5}
&\multicolumn{1}{r}{6}
&\multicolumn{1}{r}{7}
&\multicolumn{1}{r}{8}
&\multicolumn{1}{r}{9}\\
\hline\hline
CL00& 60& 200& 160& 120&  20&  10& 110&   0& 100& CL05& 65& 215& 165& 125&  25&  15& 115&   5& 105\\
CL10& 70& 120& 170& 180& 130&  30&  20& 154&  80& CL15& 75& 125& 175& 185& 135&  35&  25& 159&  85\\ 
CL20& 80&  30& 130& 197& 188& 140&  40& 147& 178& CL25& 85&  35& 135& 192& 183& 145&  45& 142& 173\\
CL30& 90&  50&  40& 140& 190& 171& 150& 138& 181& CP35& 95&  55&  45& 145& 195& 176& 155& 133& 186\\
CL40&100& 110&  10&  50& 150& 161& 210& 121&   0& CL45&105& 115&  15&  55& 155& 166& 205& 126&   5\\
CL50&  0&  10&  20&  30&  40&  50&  70&  80&  90& CL55&  5&  15&  25&  35&  45&  55&  75&  85&  95\\
CL01& 61& 201& 161& 121&  21&  11& 111&   1& 101& CL06& 66& 216& 166& 126&  26&  16& 116&   6& 106\\
CL11& 71& 121& 171& 181& 131&  31&  21& 150&  81& CL16& 76& 126& 176& 186& 136&  36&  26& 155&  86\\
CL21& 81&  31& 131& 198& 189& 141&  41& 148& 179& CL26& 86&  36& 136& 193& 184& 146&  46& 143& 174\\
CL31& 91&  51&  41& 141& 191& 172& 151& 139& 182& CL36& 96&  56&  46& 146& 196& 177& 156& 134& 187\\
CL41&101& 111&  11&  51& 151& 162& 211& 122&   1& CL46&106& 116&  16&  56& 156& 167& 206& 127&   6\\
CL51&  1&  11&  21&  31&  41&  51&  71&  81&  91& CL56&  6&  16&  26&  36&  46&  56&  76&  86&  96\\
CL02& 62& 202& 162& 122&  22&  12& 112&   2& 102& CL07& 67& 217& 167& 127&  27&  17& 117&   7& 107\\
CL12& 72& 122& 172& 182& 132&  32&  22& 151&  82& CL17& 77& 127& 177& 187& 137&  37&  27& 156&  87\\
CL22& 82&  32& 132& 199& 185& 142&  42& 149& 175& CL27& 87&  37& 137& 194& 180& 147&  47& 144& 170\\
CL32& 92&  52&  42& 142& 192& 173& 152& 135& 183& CL37& 97&  57&  47& 147& 197& 178& 157& 130& 188\\
CL42&102& 112&  12&  52& 152& 163& 212& 123&   2& CL47&107& 117&  17&  57& 157& 168& 207& 128&   7\\
CL52&  2&  12&  22&  32&  42&  52&  72&  82&  92& CL57&  7&  17&  27&  37&  47&  57&  77&  87&  97\\
CL03& 63& 203& 163& 123&  23&  13& 113&   3& 103& CL08& 68& 218& 168& 128&  28&  18& 118&   8& 108\\
CL13& 73& 123& 173& 183& 133&  33&  23& 152&  83& CL18& 78& 128& 178& 188& 138&  38&  28& 157&  88\\
CL23& 83&  33& 133& 195& 186& 143&  43& 145& 176& CL28& 88&  38& 138& 190& 181& 148&  48& 140& 171\\
CL33& 93&  53&  43& 143& 193& 174& 153& 136& 184& CL38& 98&  58&  48& 148& 198& 179& 158& 131& 189\\
CL43&103& 113&  13&  53& 153& 164& 213& 124&   3& CL48&108& 118&  18&  58& 158& 169& 208& 129&   8\\
CL53&  3&  13&  23&  33&  43&  53&  73&  83&  93& CL58&  8&  18&  28&  38&  48&  58&  78&  88&  98\\
CL04& 64& 204& 164& 124&  24&  14& 114&   4& 104& CL09& 69& 219& 169& 129&  29&  19& 119&   9& 109\\
CL14& 74& 124& 174& 184& 134&  34&  24& 153&  84& CL19& 79& 129& 179& 189& 139&  39&  29& 158&  89\\
CL24& 84&  34& 134& 196& 187& 144&  44& 146& 177& CL29& 89&  39& 139& 191& 182& 149&  49& 141& 172\\
CL34& 94&  54&  44& 144& 194& 170& 154& 137& 180& CL39& 99&  59&  49& 149& 199& 175& 159& 132& 185\\
CL44&104& 114&  14&  54& 154& 160& 214& 120&   4& CL49&109& 119&  19&  59& 159& 165& 209& 125&   9\\
CL54&  4&  14&  24&  34&  44&  54&  74&  84&  94& CL59&  9&  19&  29&  39&  49&  59&  79&  89&  99\\
\hline
\end{tabular}
\end{table}
\vspace*{\stretch{2}}
\clearpage


\vspace*{\stretch{1}}
\bigskip
\begin{table}[ph]
\caption{Physics triggers built into the PIBETA electronics logic.
``PS'' labels the prescaled trigger events, ``D'' represents dedicated 
triggers, ``C'' stands for the CsI calorimeter, ``B'' is the $\pi^+$ beam.
``LT'' and ``HT'' denote a low discriminator threshold ($E_{\rm LT}=4.5
\,$MeV) and high discriminator threshold  ($E_{\rm HT}=52.0\,$MeV),
respectively.
}
\label{tab4}
\medskip
\begin{tabular}{lllcc}
\hline
\multicolumn{1}{l}{Number/}
&\multicolumn{1}{l}{Trigger}
&\multicolumn{1}{l}{Trigger}
&\multicolumn{1}{l}{Prescaling}
&\multicolumn{1}{l}{Typical}\\
\multicolumn{1}{l}{Label}
&\multicolumn{1}{l}{Name}
&\multicolumn{1}{l}{Logic}
&\multicolumn{1}{l}{Factor at 1$\,$MHz}
&\multicolumn{1}{l}{Rate (Hz)}\\
\hline\hline
1/$\rm T_{S\bar{S}}^H$ & Two-Arm HT D        & $\rm C_{S\bar{S}}^{H}{\circ} \pi G{\circ} \bar{B}_\pi$ & 1.0      &  8  \\
2/$\rm T_{S\bar{S}}^L$ & Two-Arm LT PS       & $\rm C_{S\bar{S}}^{L}{\circ} \pi G_{PS}{\circ} \bar{B}_\pi$ & 512 &  6  \\
3/$\rm T_S^L$          & One-Arm HT PS       & $\rm C_S^{H}{\circ} \pi G_{PS}{\circ} \bar{B}_\pi$ & 16               & 40  \\
4/$\rm T_S^H$          & One-Arm LT PS       & $\rm C_S^{L}{\circ} \pi G_{PS}{\circ} \bar{B}_\pi$ & 32768            &  4  \\
5/$\rm T_P$            & Prompt HT PS        & $\rm C_S^{H}\circ \pi S_{PS^\prime}$               & 1024             &  2  \\
6/$\rm T_C^L$          & Cosmics LT D        & $\rm CV\circ C_S^L\circ  \bar{B}_\pi$              & 1                & 46  \\
7/$\rm T_{\pi}$        & $\pi$-in-Beam HT PS & timed $\rm BC{\circ} AD{\circ} AT{\circ} rf$       & $1.6\cdot 10^6$  & 0.5 \\
8/$\rm T_C^H$          & Cosmics LT PS       & $\rm CV_{PS}\circ C_S^H\circ  \bar{B}_\pi$         & 8                & 0.5 \\
9/$\rm T_e$            & $e$-in-Beam HT PS   & timed $\rm BC{\circ} AD{\circ} AT{\circ} rf$       & 32               & 0.2 \\
10/$\rm T_R$           & Random Trigger PS   & random                                                         & 1 &  2  \\
11/$\rm T_{3S}$        & Three-Arm LT PS     & $\rm C_{3S}^{L}{\circ} \pi G_{PS}{\circ} \bar{B}_\pi$ & 64     &  8  \\
12/$\rm T_{S\bar{S}}^{LH}$ & Two-Arm LT-HT PS& $\rm C_{S\bar{S}}^{HL}{\circ} \pi G{\circ} \bar{B}_\pi$ & 4    &  5  \\
\hline
\end{tabular}
\end{table}
\vspace*{\stretch{2}}
\clearpage


\vspace*{\stretch{1}}
\bigskip
\begin{table}[ph]
\caption{Summary of changes in gain factors and energy
resolutions of active elements due to radiation exposure.}
\label{tab5}
\medskip
\begin{tabular}{lccccc}
\hline
\multicolumn{1}{l}{Detector} &
\multicolumn{1}{l}{Radiation} & 
\multicolumn{1}{l}{Starting} &
\multicolumn{1}{l}{Final} & 
\multicolumn{1}{l}{Starting} &
\multicolumn{1}{l}{Final} \\
\multicolumn{1}{l}{\ } &
\multicolumn{1}{l}{Dose (rads)} & 
\multicolumn{1}{l}{Gain} &
\multicolumn{1}{l}{Gain} & 
\multicolumn{1}{l}{Res. (\%)} &
\multicolumn{1}{l}{Res. (\%)} \\
\hline\hline
BC    & $2.0\cdot 10^6$& 1.00& 0.66& 13.1& 13.9  \\
AD    & $1.4\cdot 10^6$& 1.00& 0.75&  7.8& 8.5  \\
AT$_0$& $5.6\cdot 10^5$& 1.00& 0.76&  7.2& 10.8  \\
AT$_1$& $6.3\cdot 10^5$& 1.00& 0.89&  8.4& 8.6  \\
AT$_5$& $1.5\cdot 10^5$& 1.00& 0.74&  7.7& 8.9  \\
PV$_0$& $4.4\cdot 10^4$& 1.00& 0.99& 31.5& 32.1  \\
PV$_1$& $4.0\cdot 10^4$& 1.00& 0.99& 29.9& 31.1  \\
PVeto & $4.0\cdot 10^4$& 1.00& 0.95& 26.2& 27.9  \\
CsI$_0$& 45.9&           1.00& 0.89&  5.1&  5.5 \\
CsI$_2$& 44.5&           1.00& 0.92&  4.9&  5.2\\
CsI$_{11}$& 116.5&       1.00& 1.02&  6.0&  6.1 \\
CsI$_{19}$&  93.0&       1.00& 0.65&  5.0&  5.3 \\
CsI$_{102}$& 151.7&      1.00& 0.86&  6.0&  6.5 \\
CsI$_{165}$& 125.5&      1.00& 0.74&  5.4&  5.8\\
Calo&        118.5&      1.00& 0.83&  5.5&  6.0 \\
\hline
\end{tabular}
\end{table}
\vspace*{\stretch{2}}
\clearpage



\vspace*{\stretch{1}}
\centerline{\psfig{figure=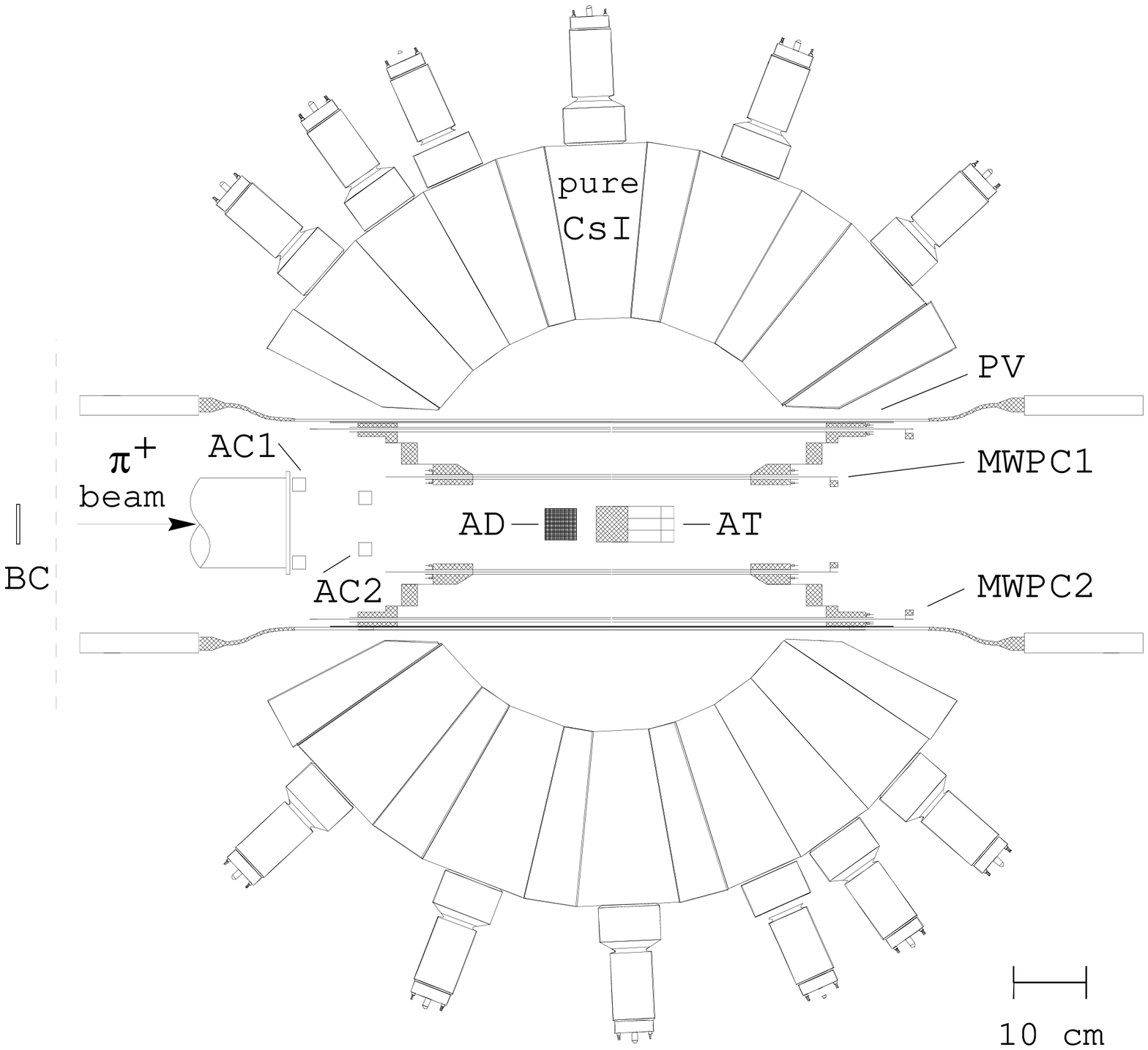,width=15cm}}
\vglue 1.5cm
\centerline{FIGURE~\ref{fig:det1}}
\vspace*{\stretch{2}}
\clearpage

\vspace*{\stretch{1}}
\centerline{\psfig{figure=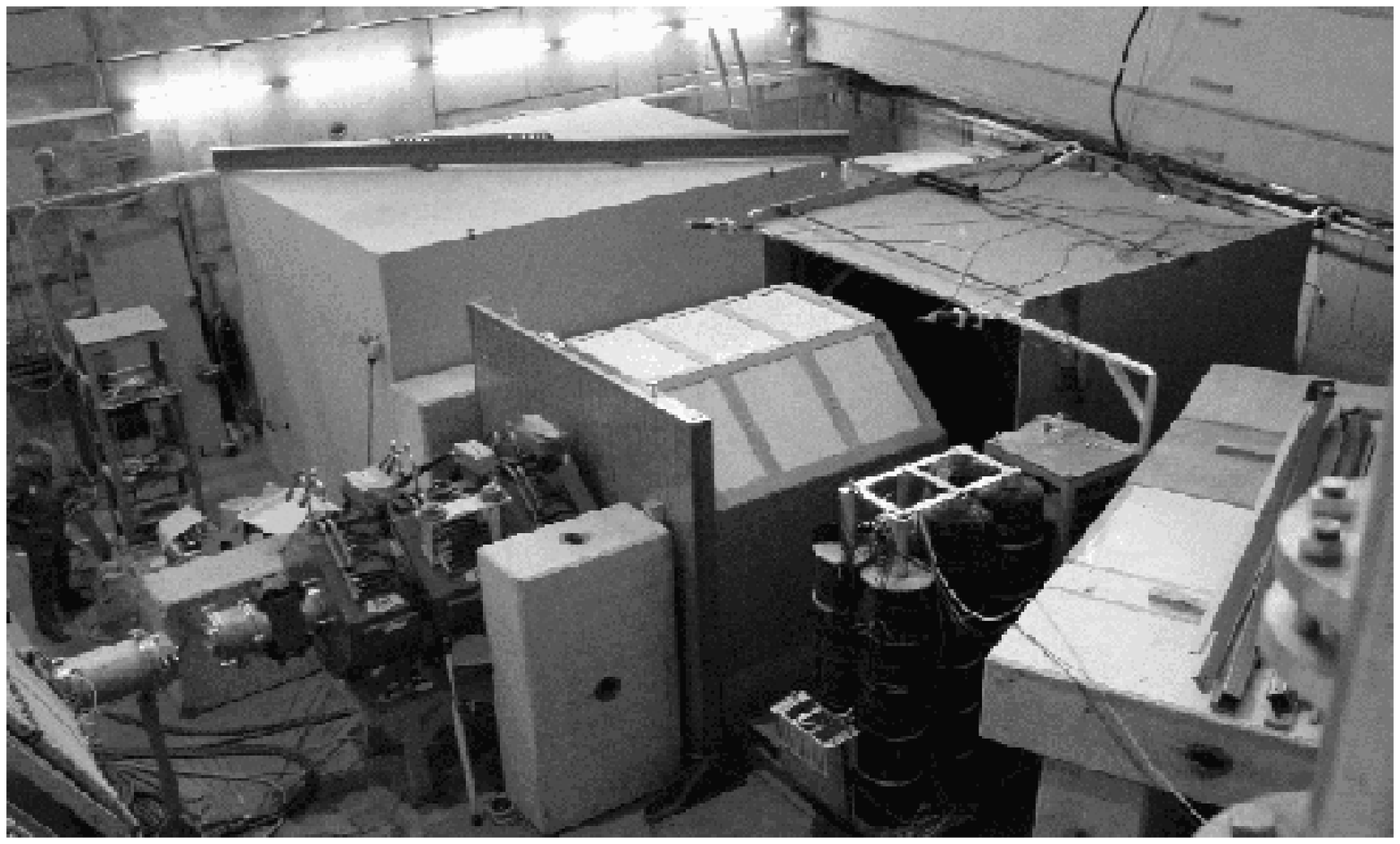,width=16cm}}
\vglue 1.5cm
\centerline{FIGURE~\ref{fig:pb_photo1}}
\vspace*{\stretch{2}}
\clearpage

\vspace*{\stretch{1}}
\hbox{\ }\vglue -3cm
\centerline{\psfig{figure=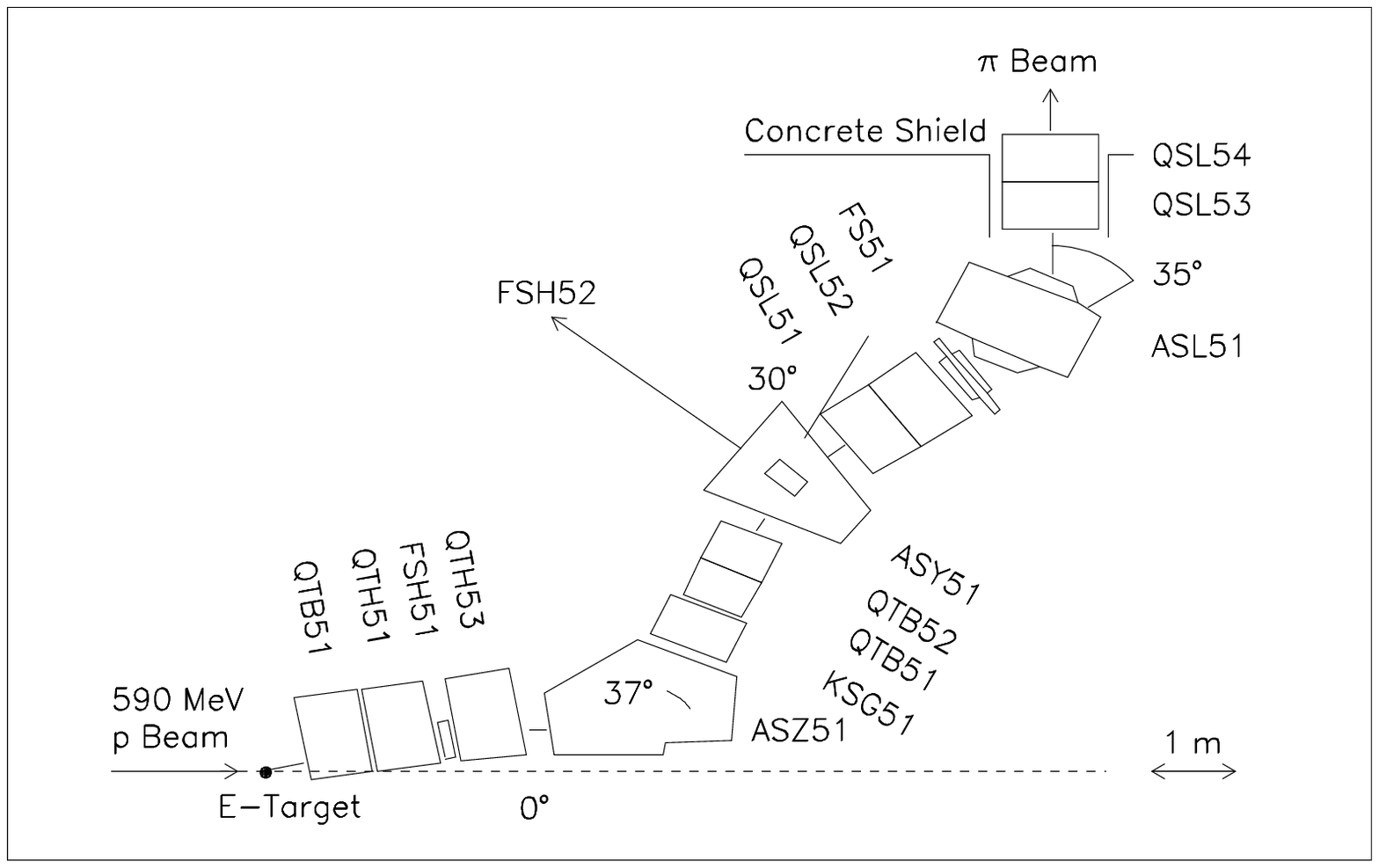,height=24cm}}
\bigskip
\vglue -12.0cm
\centerline{FIGURE~\ref{fig:pie1}}
\vspace*{\stretch{2}}
\clearpage

\vspace*{\stretch{1}}
\centerline{\psfig{figure=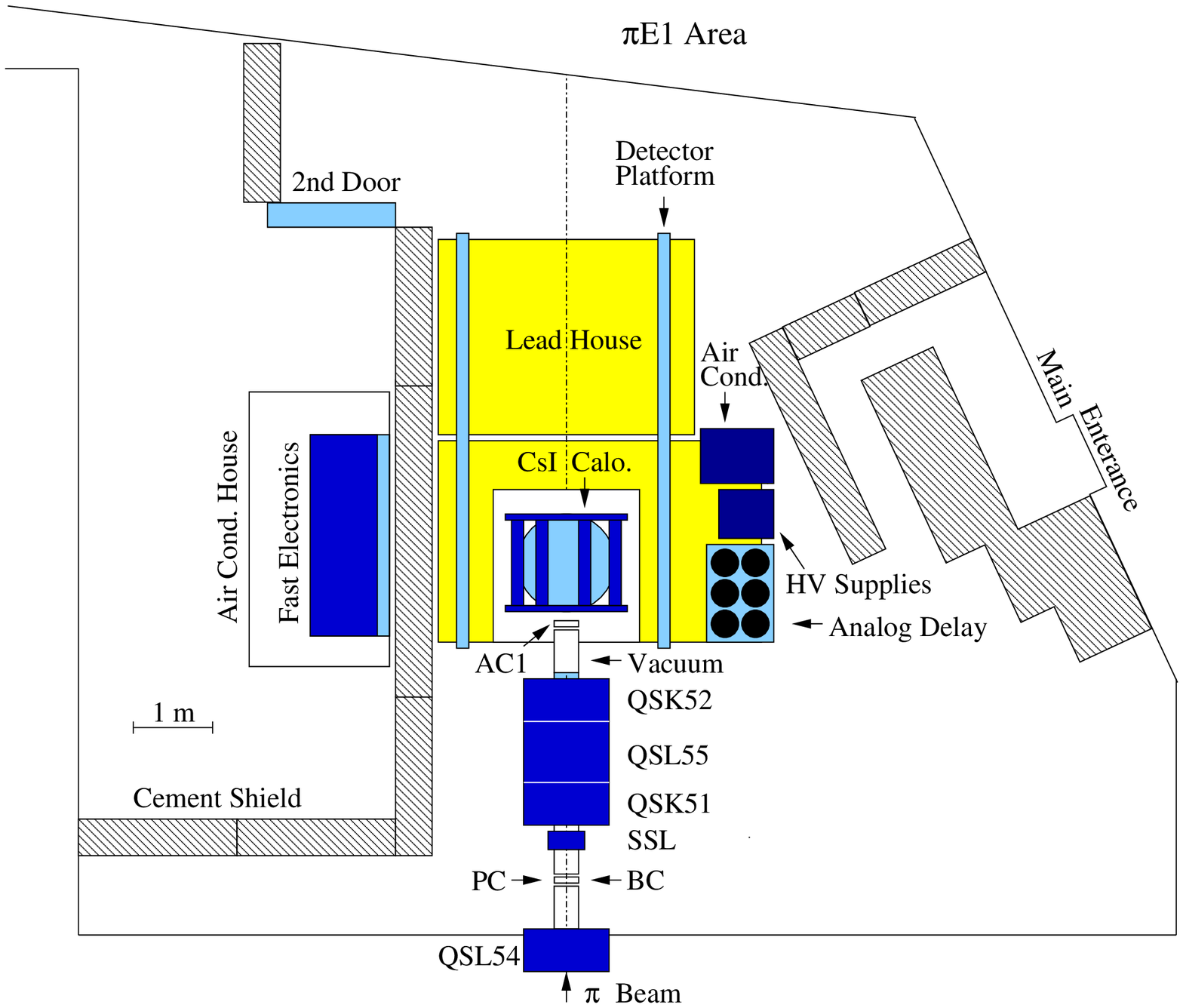,width=17cm}}
\vglue 1.5cm
\centerline{FIGURE~\ref{fig:hall}}
\vspace*{\stretch{2}}
\clearpage

\vspace*{\stretch{1}}
\centerline{\psfig{figure=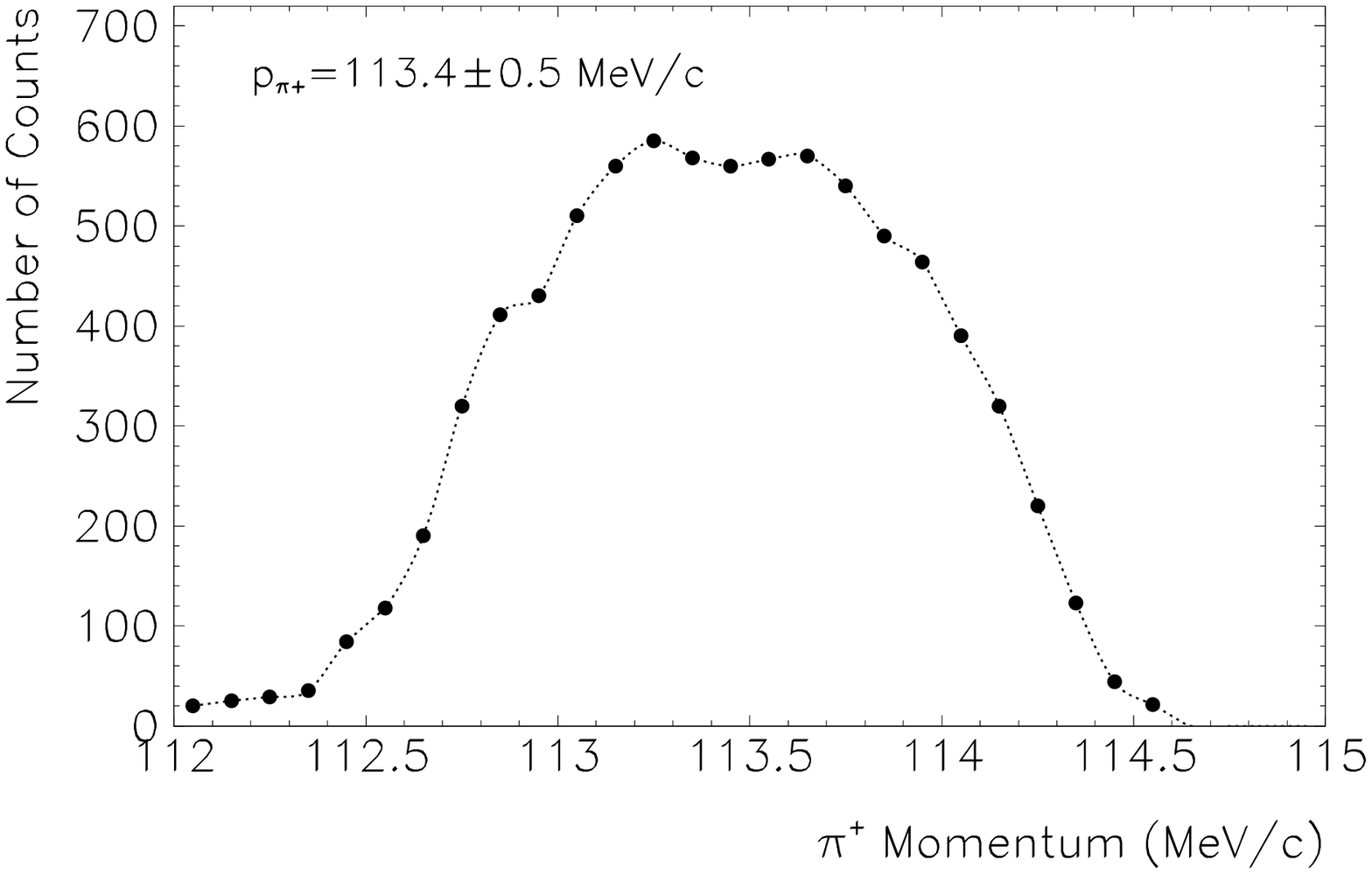,height=22cm}}
\vglue -9.5cm
\centerline{FIGURE~\ref{fig:beam_p}}
\vspace*{\stretch{2}}
\clearpage

\vspace*{\stretch{1}}
\centerline{\psfig{figure=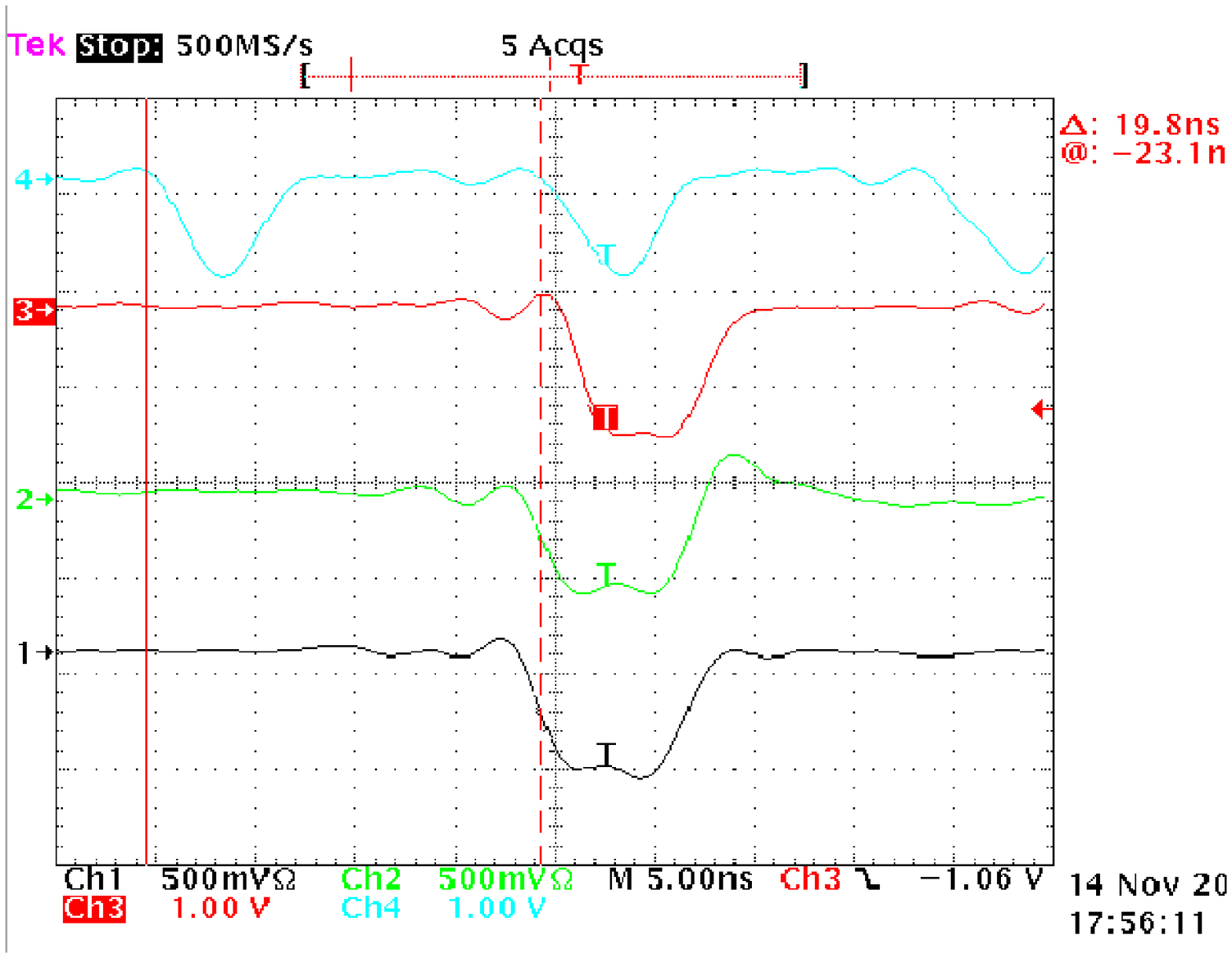,height=12cm}}
\bigskip\bigskip
\centerline{FIGURE~\ref{fig:pistop}}
\vspace*{\stretch{2}}
\clearpage

\vspace*{\stretch{1}}
\centerline{\psfig{figure=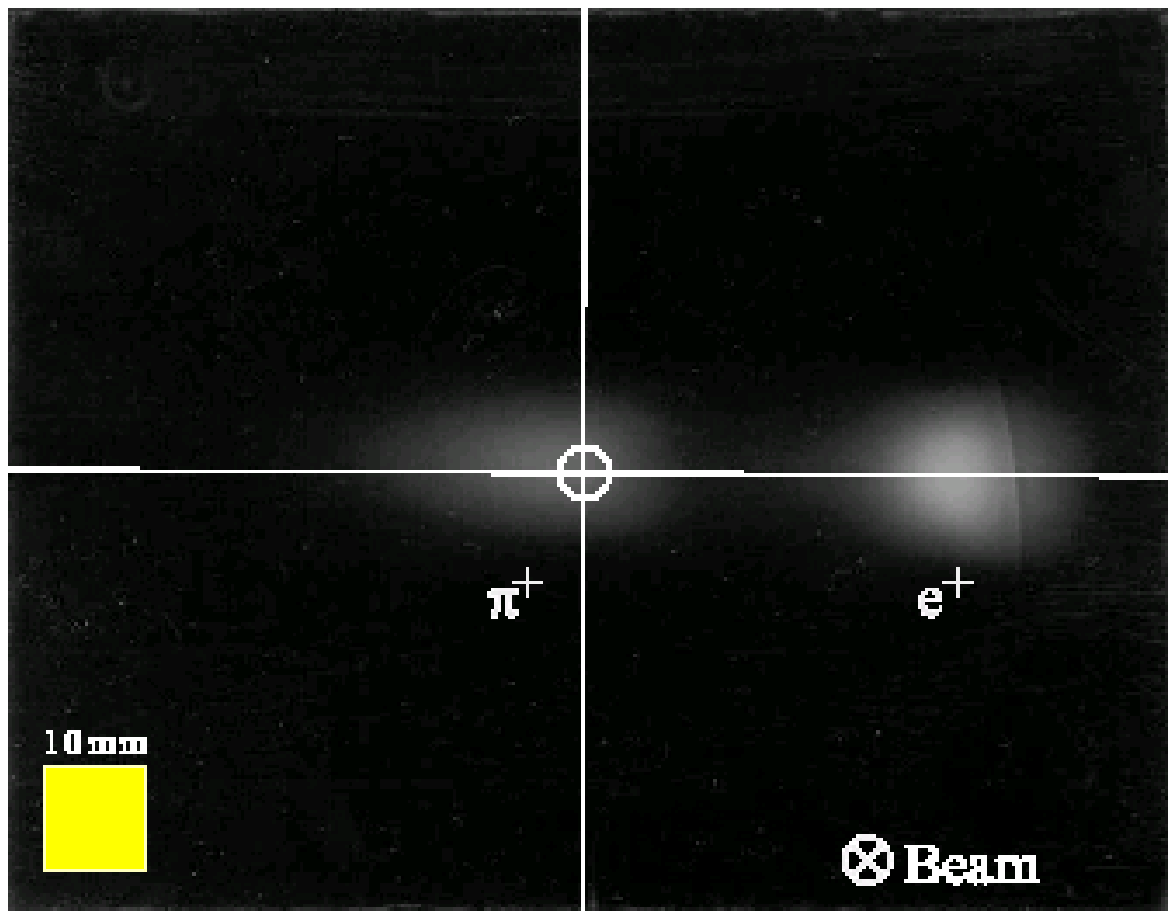,width=14cm}}
\vglue 2.0cm
\centerline{FIGURE~\ref{fig:beam_sp}}
\vspace*{\stretch{2}}
\clearpage

\vspace*{\stretch{1}}
\centerline{\psfig{figure=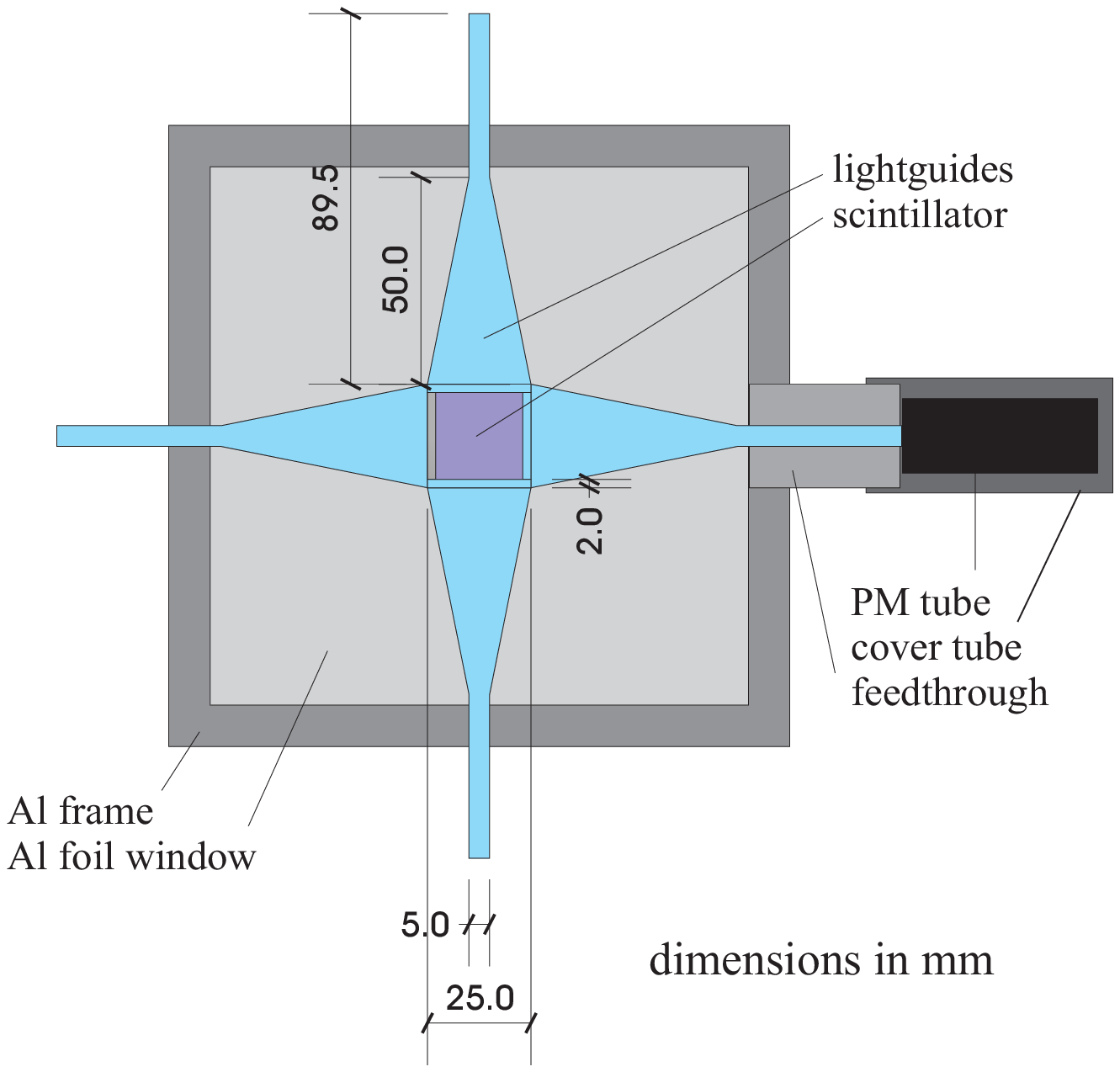,width=12cm}}
\vglue 1.5cm
\centerline{FIGURE~\ref{fig:fbeam}}
\vspace*{\stretch{2}}
\clearpage

\vspace*{\stretch{1}}
\centerline{\psfig{figure=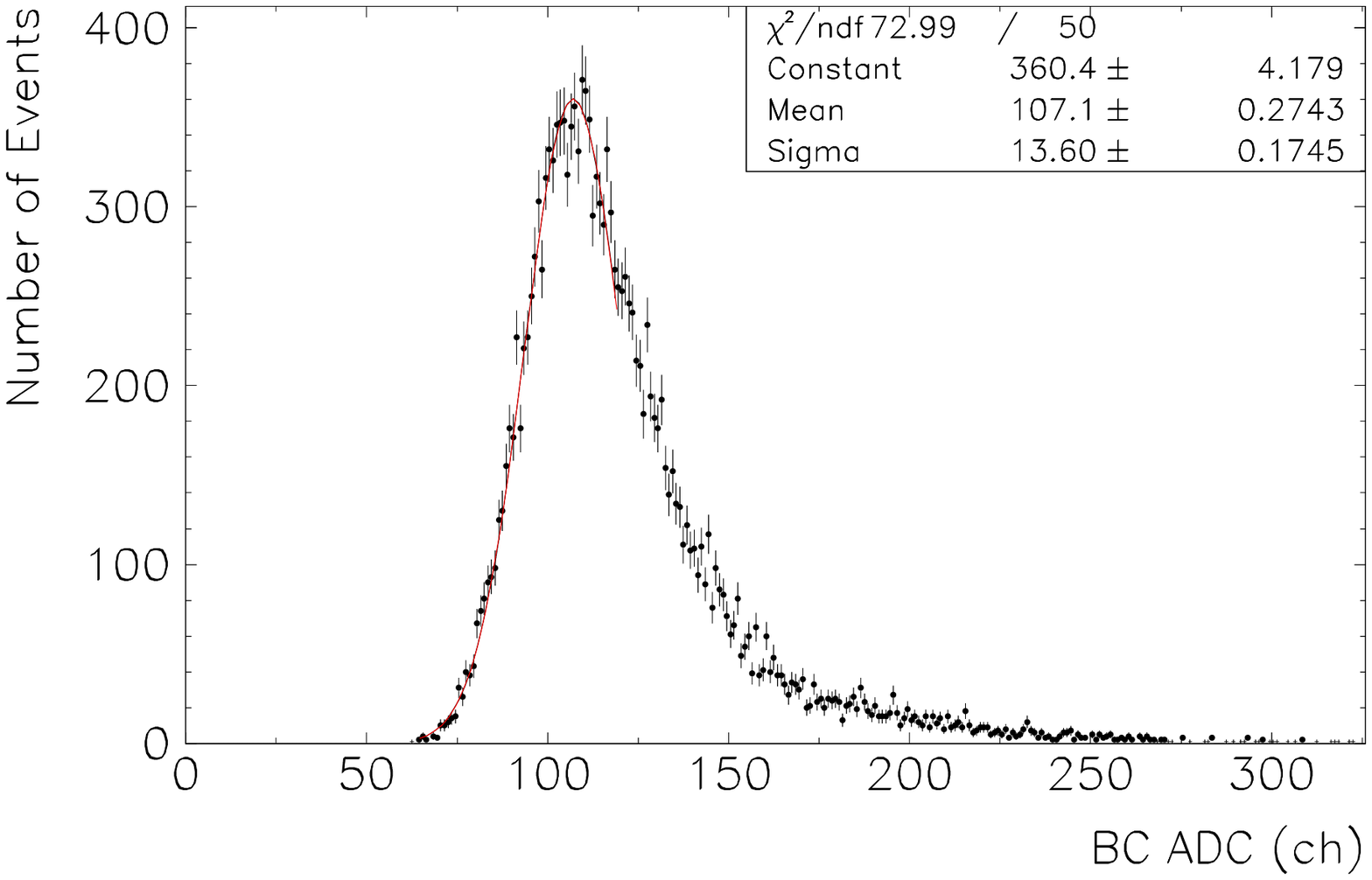,height=11cm}}
\vglue 1.0cm
\centerline{FIGURE~\ref{fig:b0_adc}}
\vspace*{\stretch{2}}
\clearpage

\vspace*{\stretch{1}}
\centerline{\psfig{figure=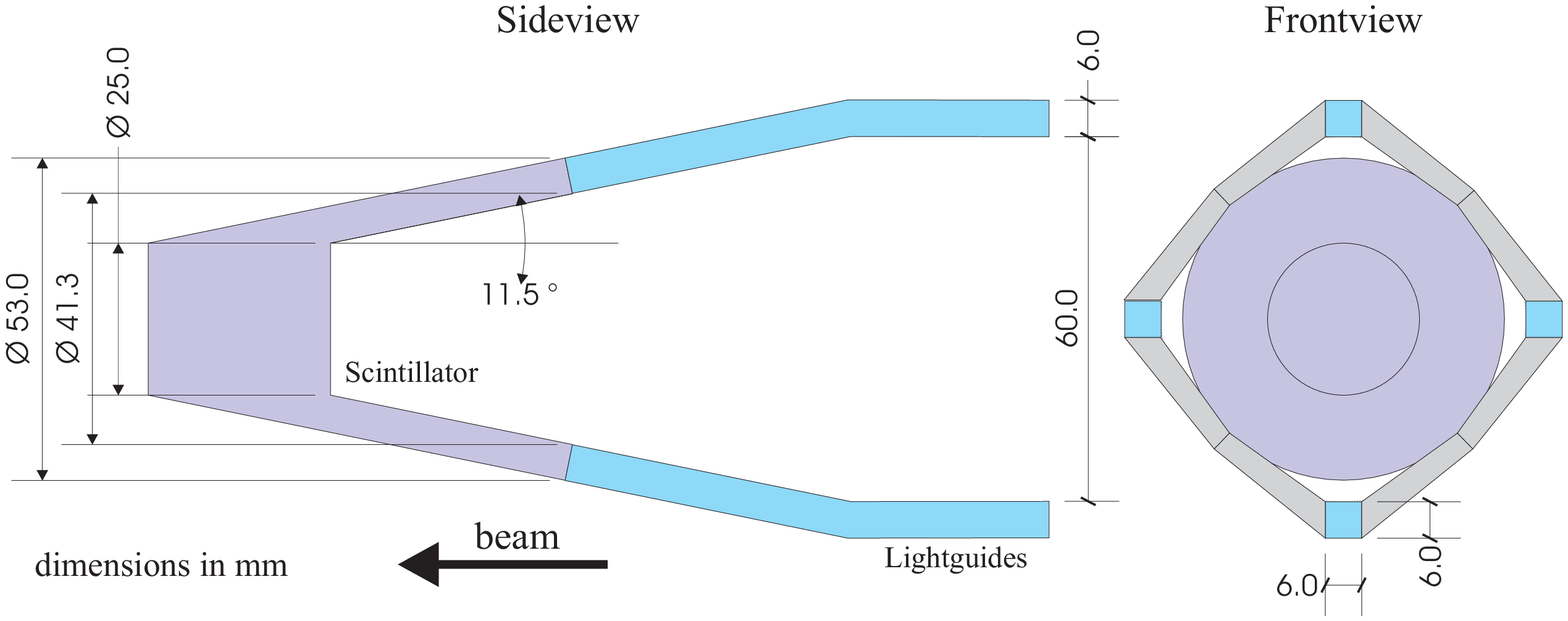,width=15cm}}
\vglue 1.5cm
\centerline{FIGURE~\ref{fig:act_deg}}
\vspace*{\stretch{2}}
\clearpage

\vspace*{\stretch{1}}
\centerline{\psfig{figure=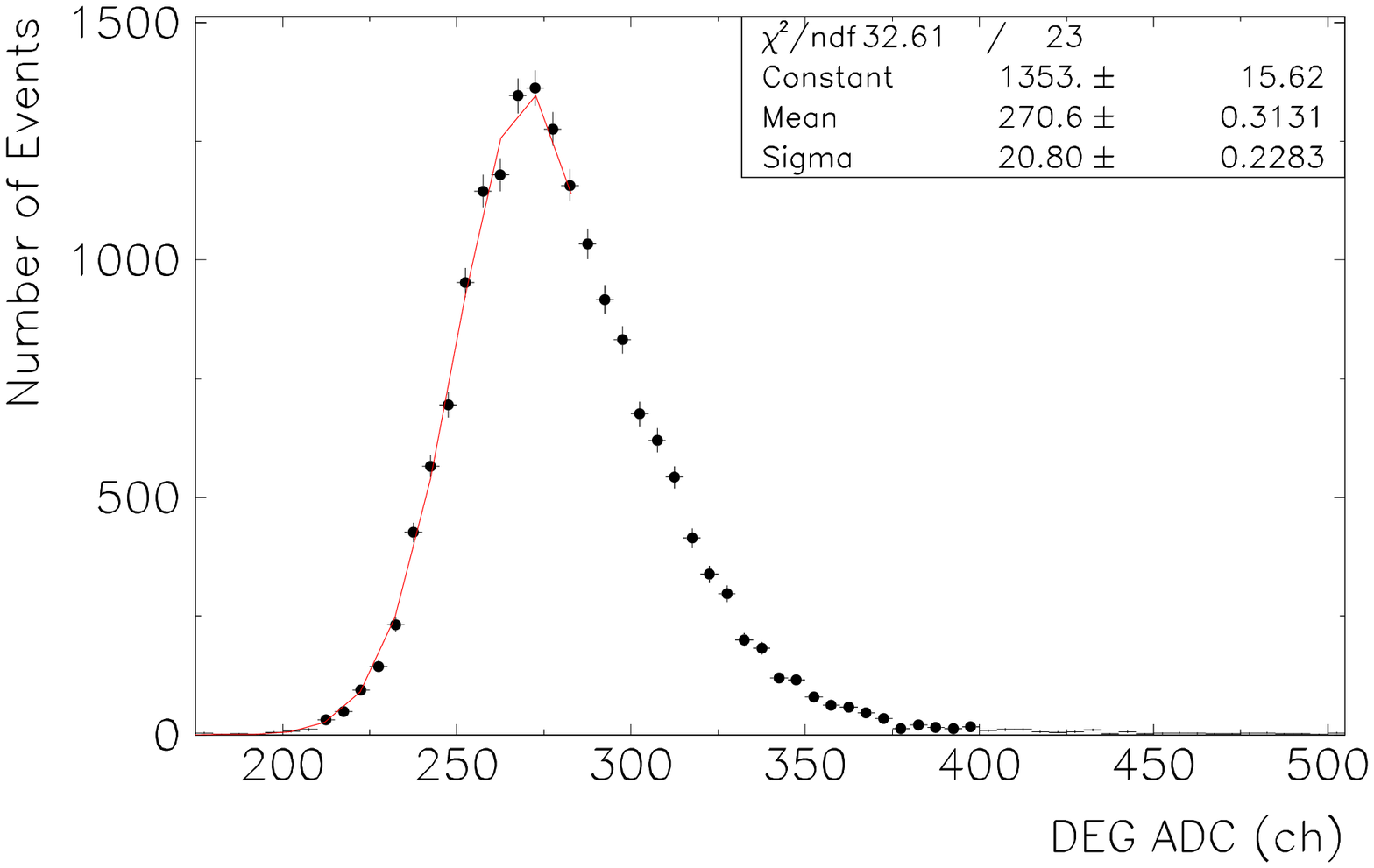,height=11cm}}
\vglue 1cm
\centerline{FIGURE~\ref{fig:deg_adc}}
\vspace*{\stretch{2}}
\clearpage

\vspace*{\stretch{1}}
\centerline{\psfig{figure=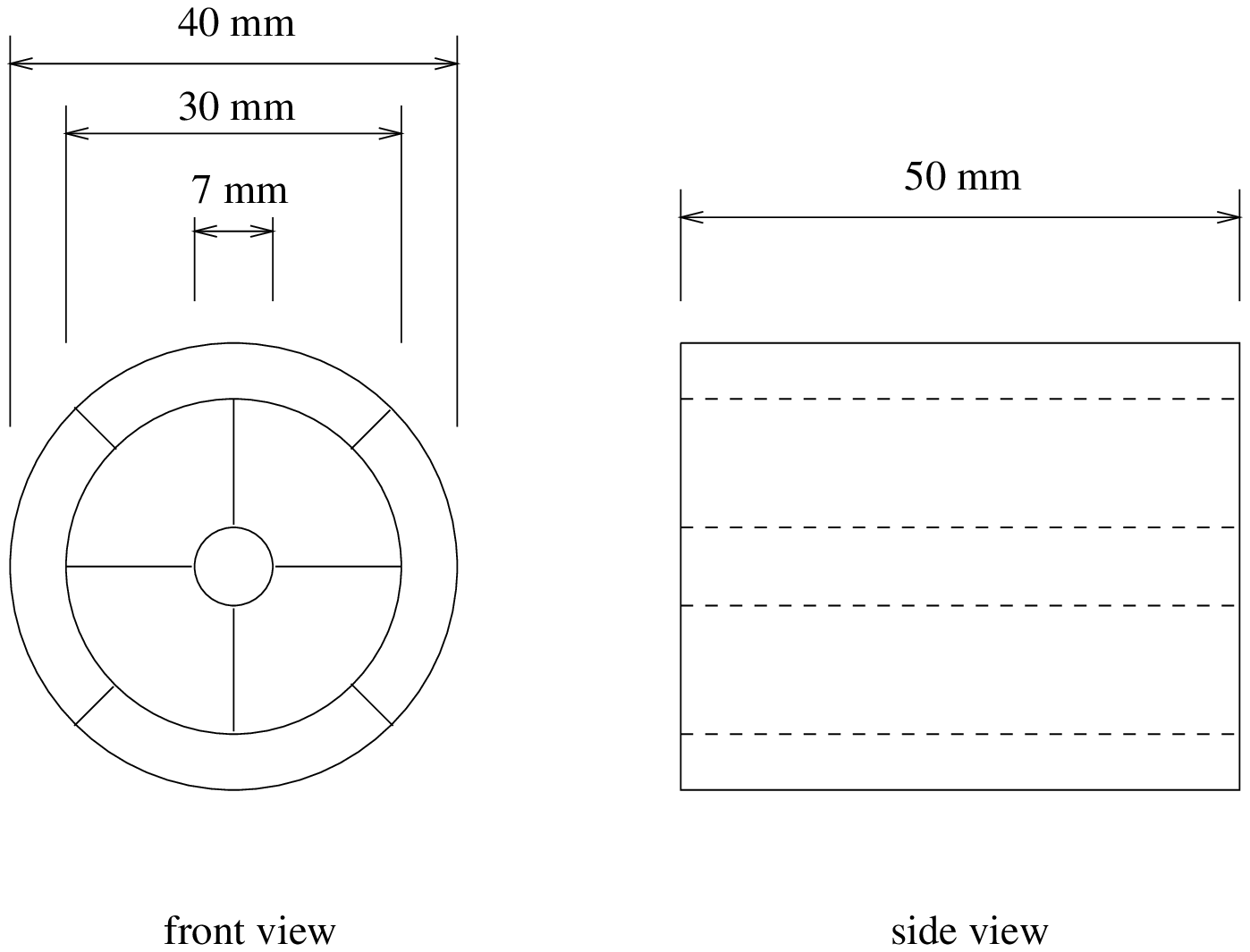,height=8cm}}
\vglue 2.0cm
\centerline{FIGURE~\ref{fig:tgt9a}}
\vspace*{\stretch{2}}
\clearpage

\vspace*{\stretch{1}}
\centerline{\psfig{figure=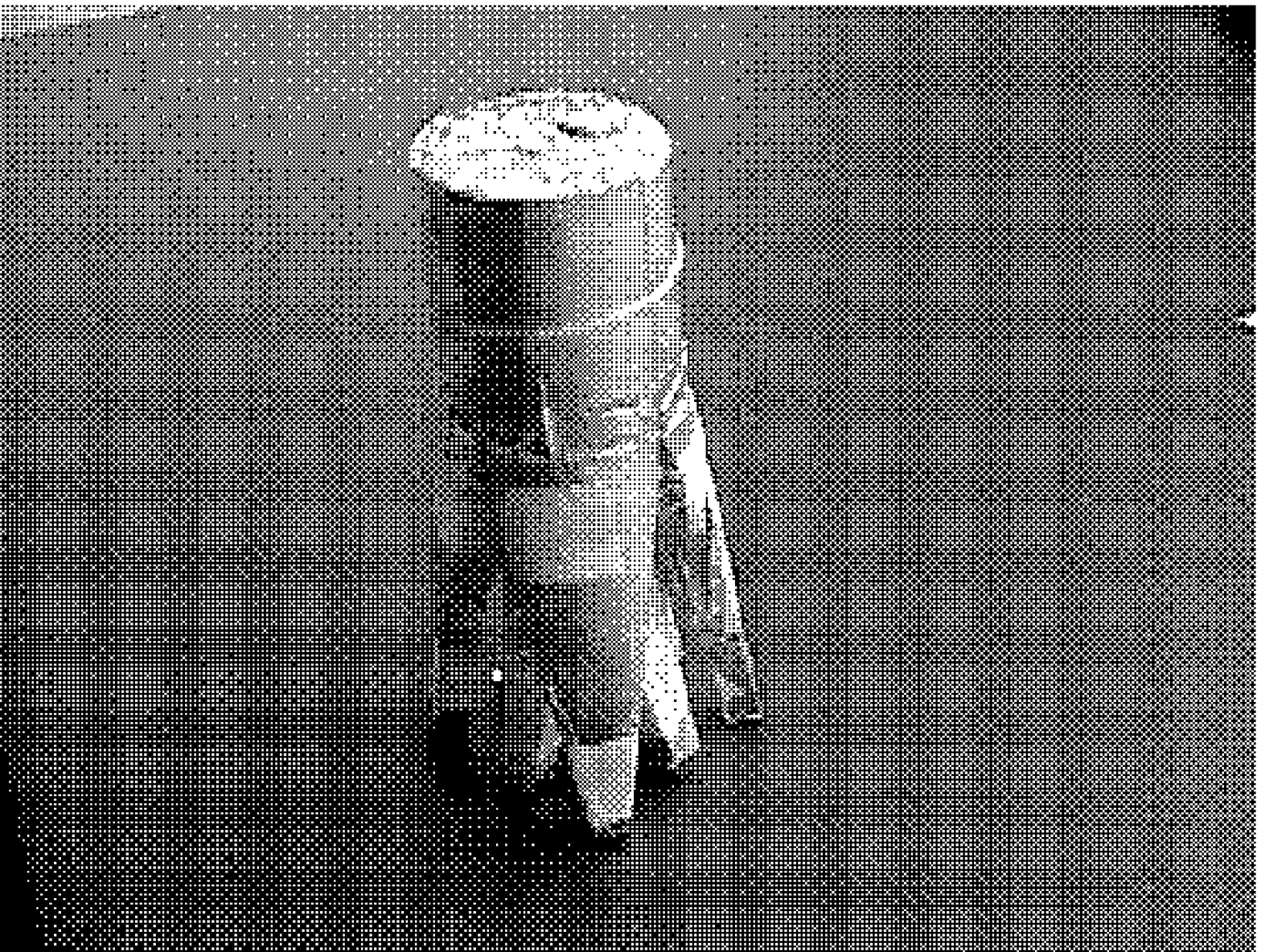,height=10cm}}
\bigskip
\bigskip
\bigskip
\centerline{FIGURE~\ref{fig:tgt9b}}
\vspace*{\stretch{2}}
\clearpage      

\vspace*{\stretch{1}}
\centerline{\psfig{figure=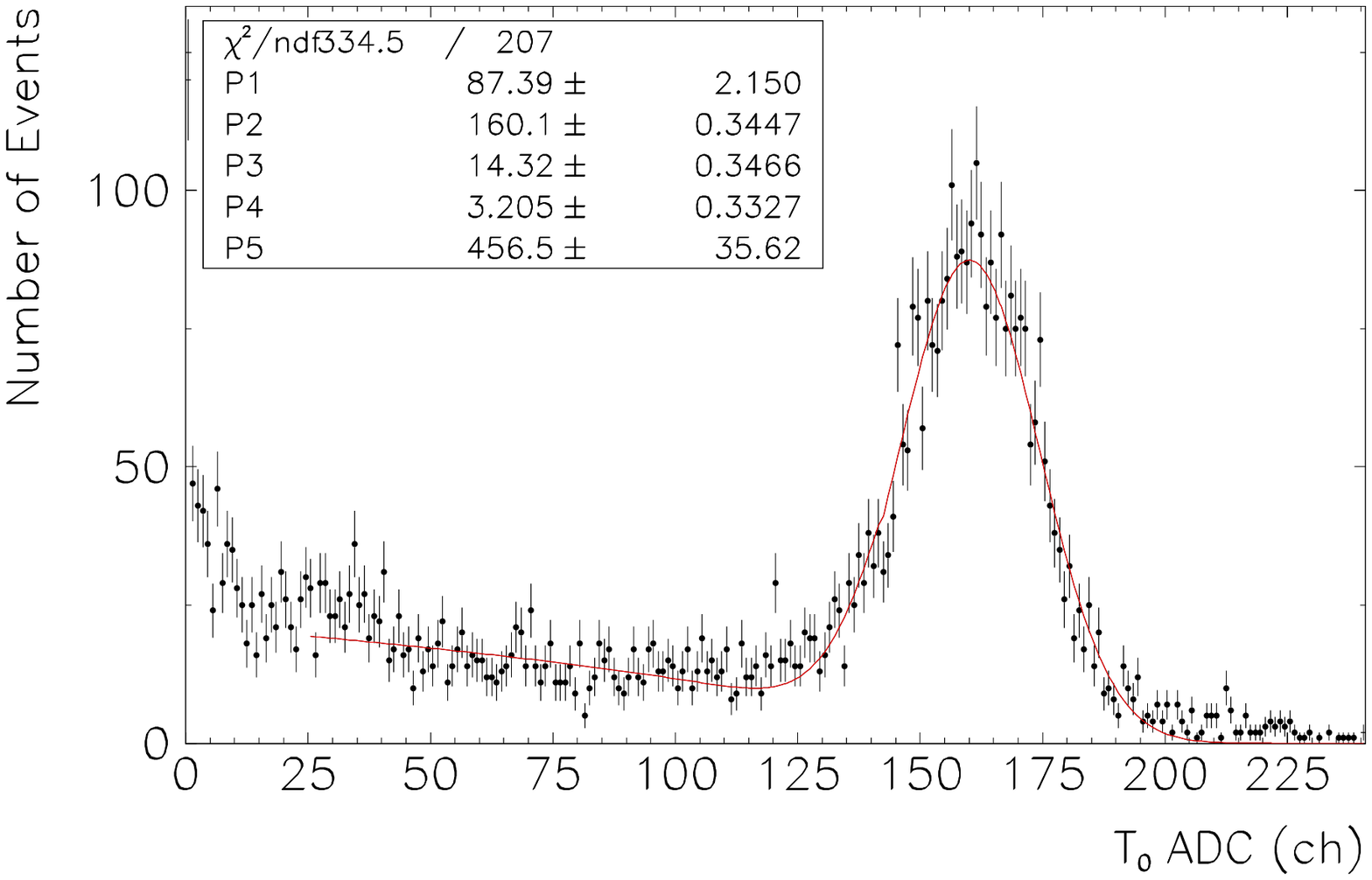,height=11cm}}
\vglue 1.0cm
\centerline{FIGURE~\ref{fig:tgt_adc}}
\vspace*{\stretch{2}}
\clearpage

\vspace*{\stretch{1}}
\centerline{\psfig{figure=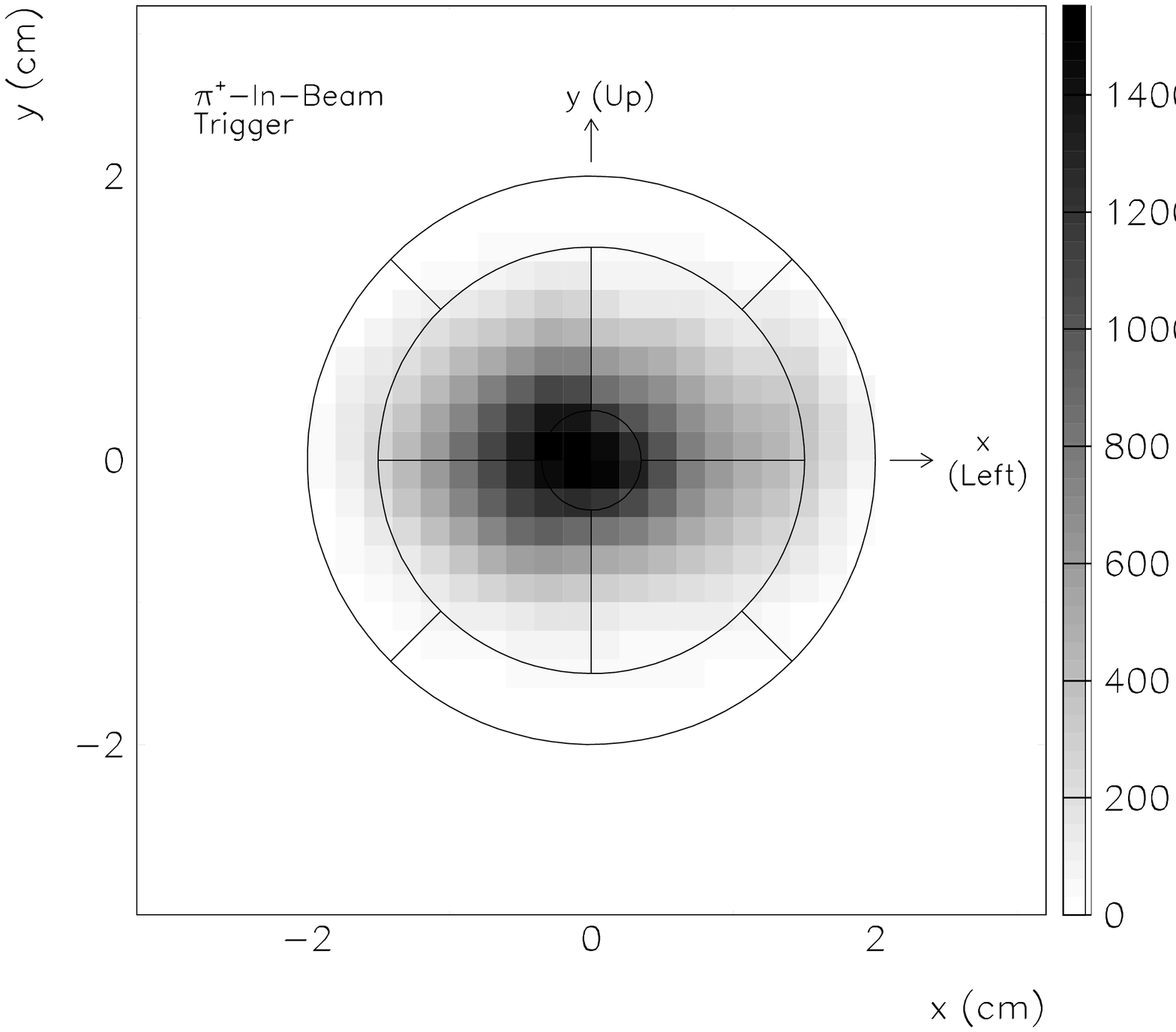,height=14cm}}
\centerline{FIGURE~\ref{fig:extr_2d}}
\vspace*{\stretch{2}}
\clearpage

\vspace*{\stretch{1}}
\centerline{\psfig{figure=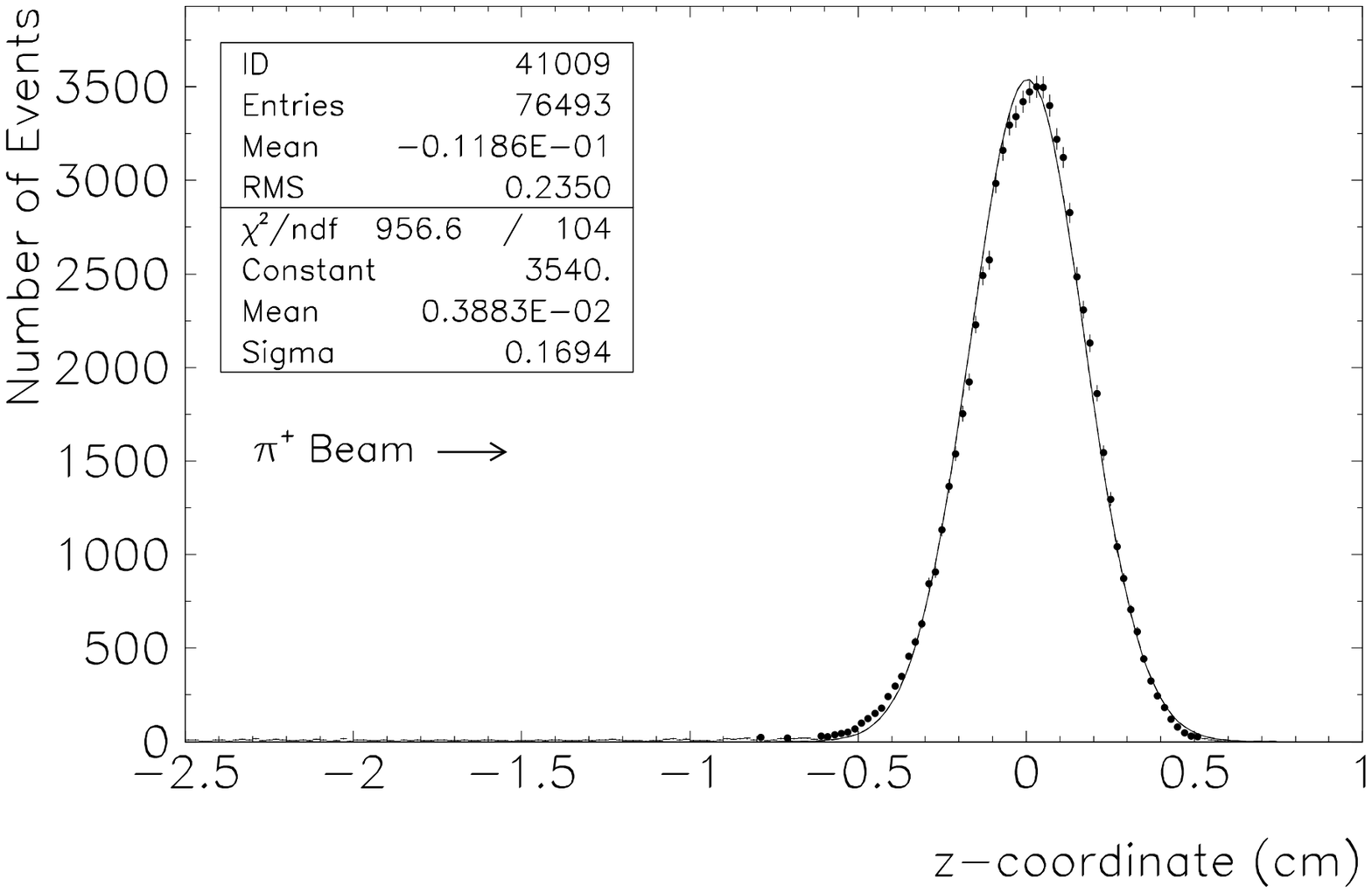,height=22cm}}
\vglue -9.0cm
\centerline{FIGURE~\ref{fig:pistop_z}}
\vspace*{\stretch{2}}
\clearpage

\vspace*{\stretch{1}}
\centerline{\psfig{figure=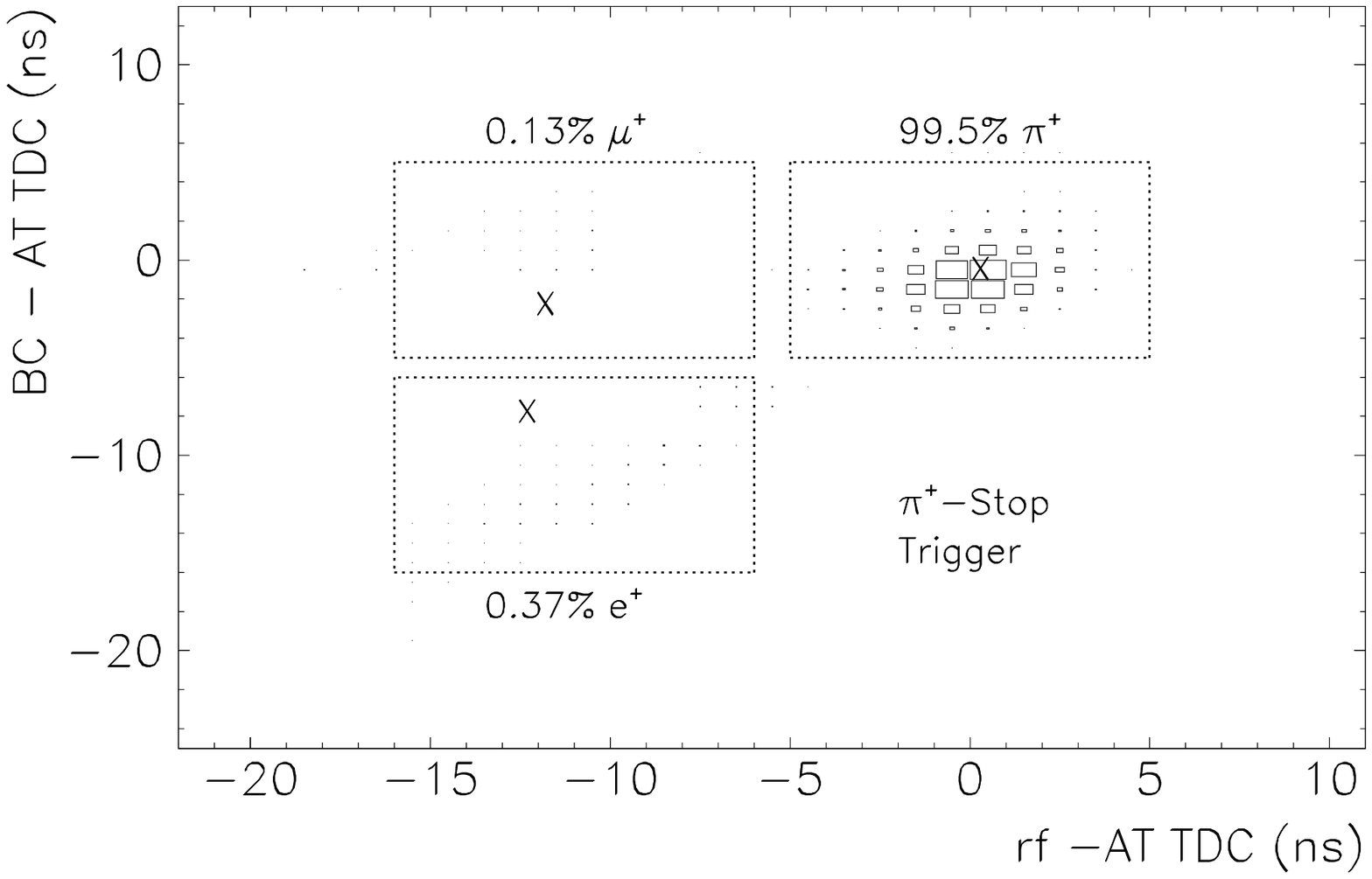,height=22cm}}
\vglue -10.0cm
\bigskip
\centerline{FIGURE~\ref{fig:contam}}
\vspace*{\stretch{2}}
\clearpage

\vspace*{\stretch{1}}
\centerline{\psfig{figure=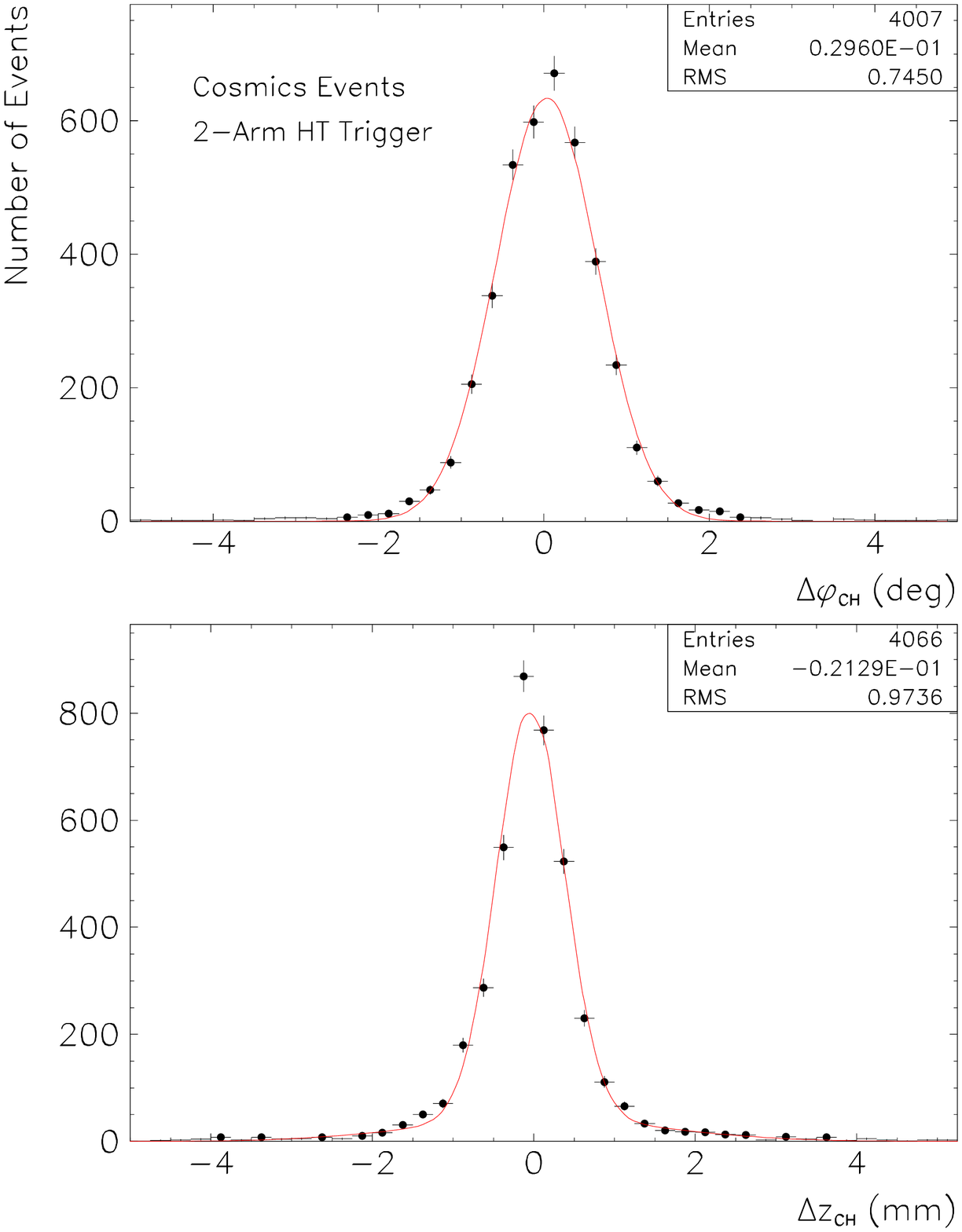,height=21.5cm}}
\centerline{FIGURE~\ref{fig:mwpc_r}}
\vspace*{\stretch{2}}
\clearpage

\vspace*{\stretch{1}}
\centerline{\psfig{figure=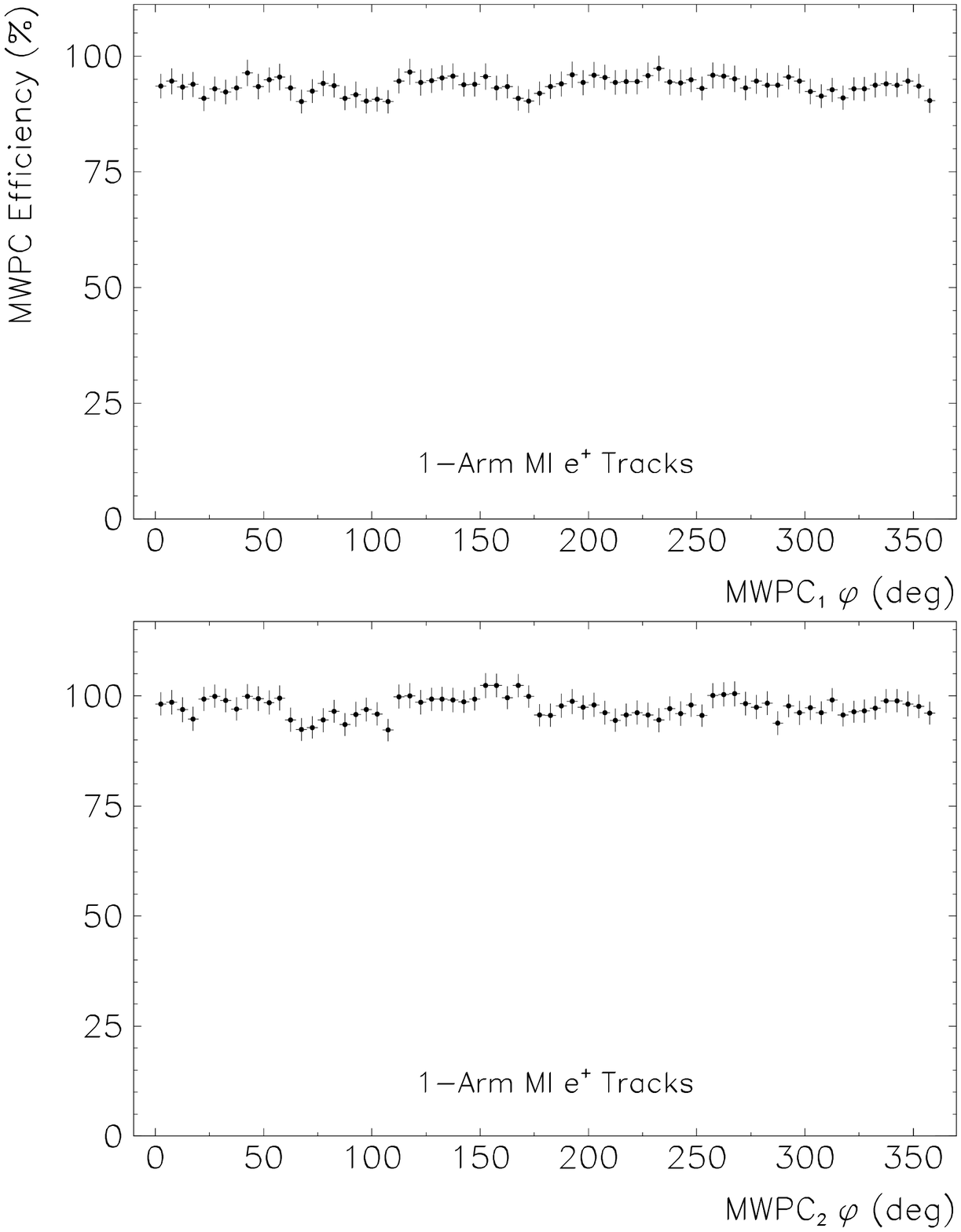,height=21.5cm}}
\bigskip
\centerline{FIGURE~\ref{fig:mwpc_phi}}
\vspace*{\stretch{2}}
\clearpage

\vspace*{\stretch{1}}
\begin{figure}
\hbox{\ }\vglue 2cm
\noindent \hglue 2cm (i)
\hbox{\ }\vglue -1.0cm
\centerline{\psfig{figure=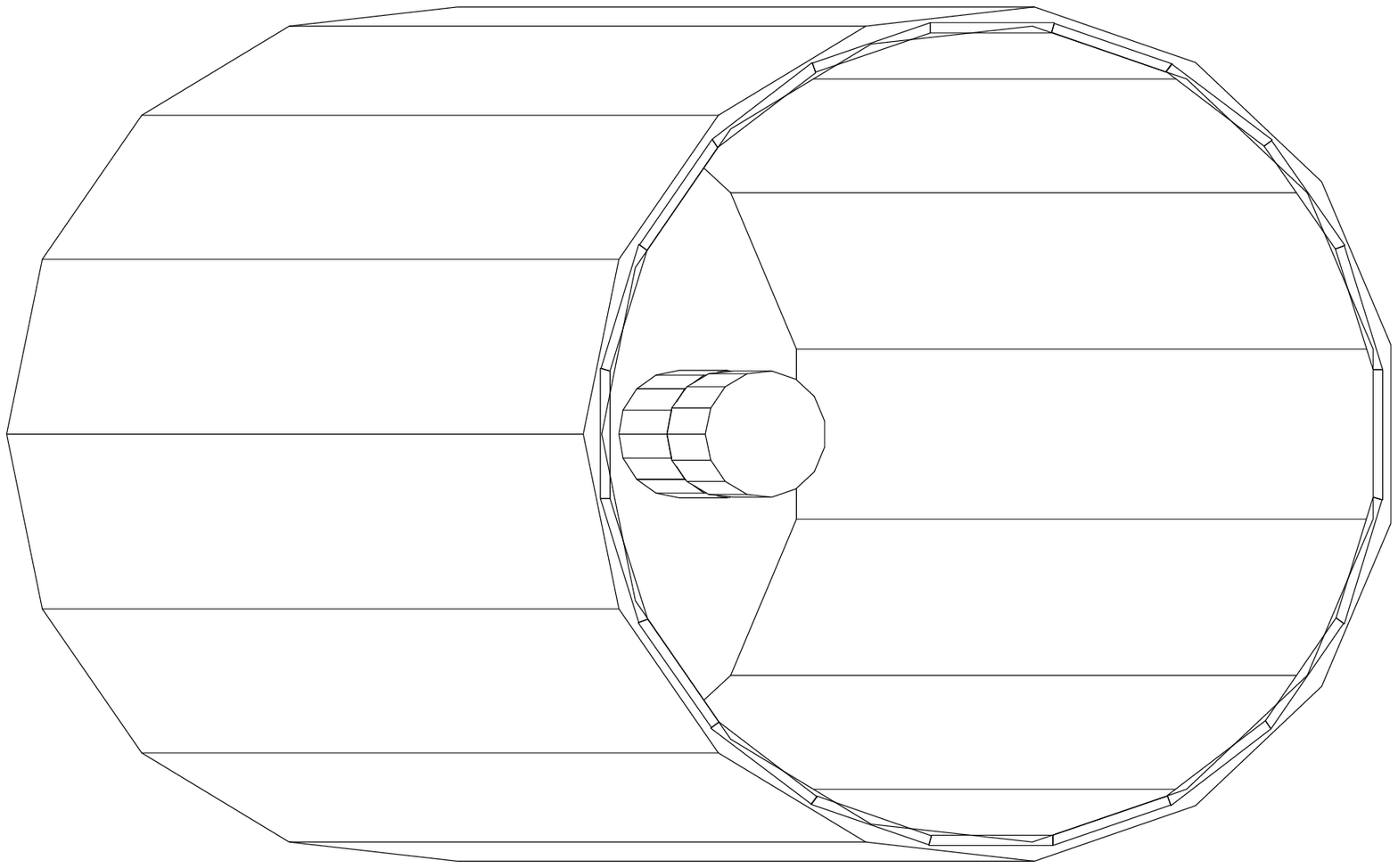,width=8cm}}
\hbox{\ }\vglue -1.0cm
\noindent \hglue 2cm (ii)
\hbox{\ }\vglue 1.5cm
\centerline{\psfig{figure=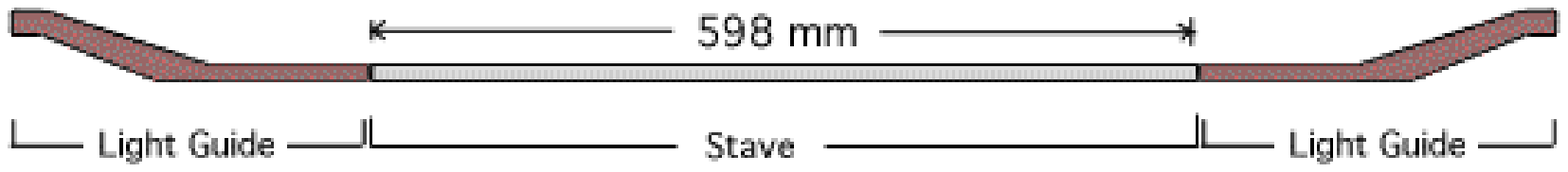,width=3.9in}}
\hbox{\ }\vglue 1.5cm
\noindent \hglue 2cm (iii)
\vglue 0.5cm
\centerline{\psfig{figure=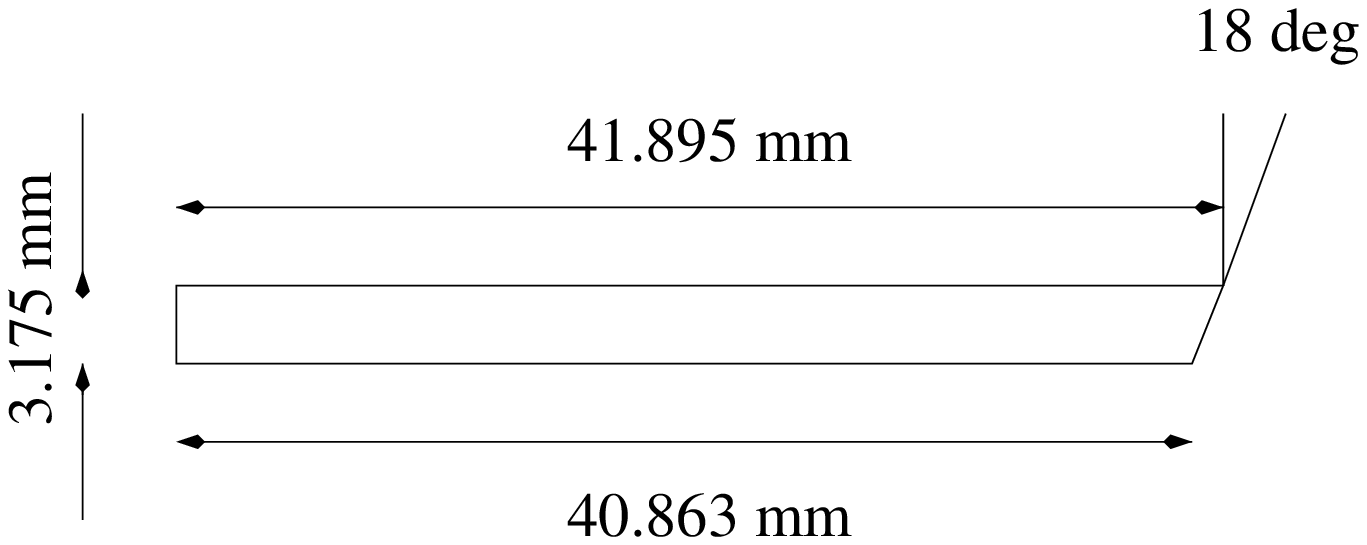,width=2.6in}}
\label{fig:pv_stave}
\vglue 2.0cm
\centerline{FIGURE~\ref{fig:pv_geom}}
\end{figure}
\vspace*{\stretch{2}}
\clearpage


\vspace*{\stretch{1}}
\centerline{\psfig{figure=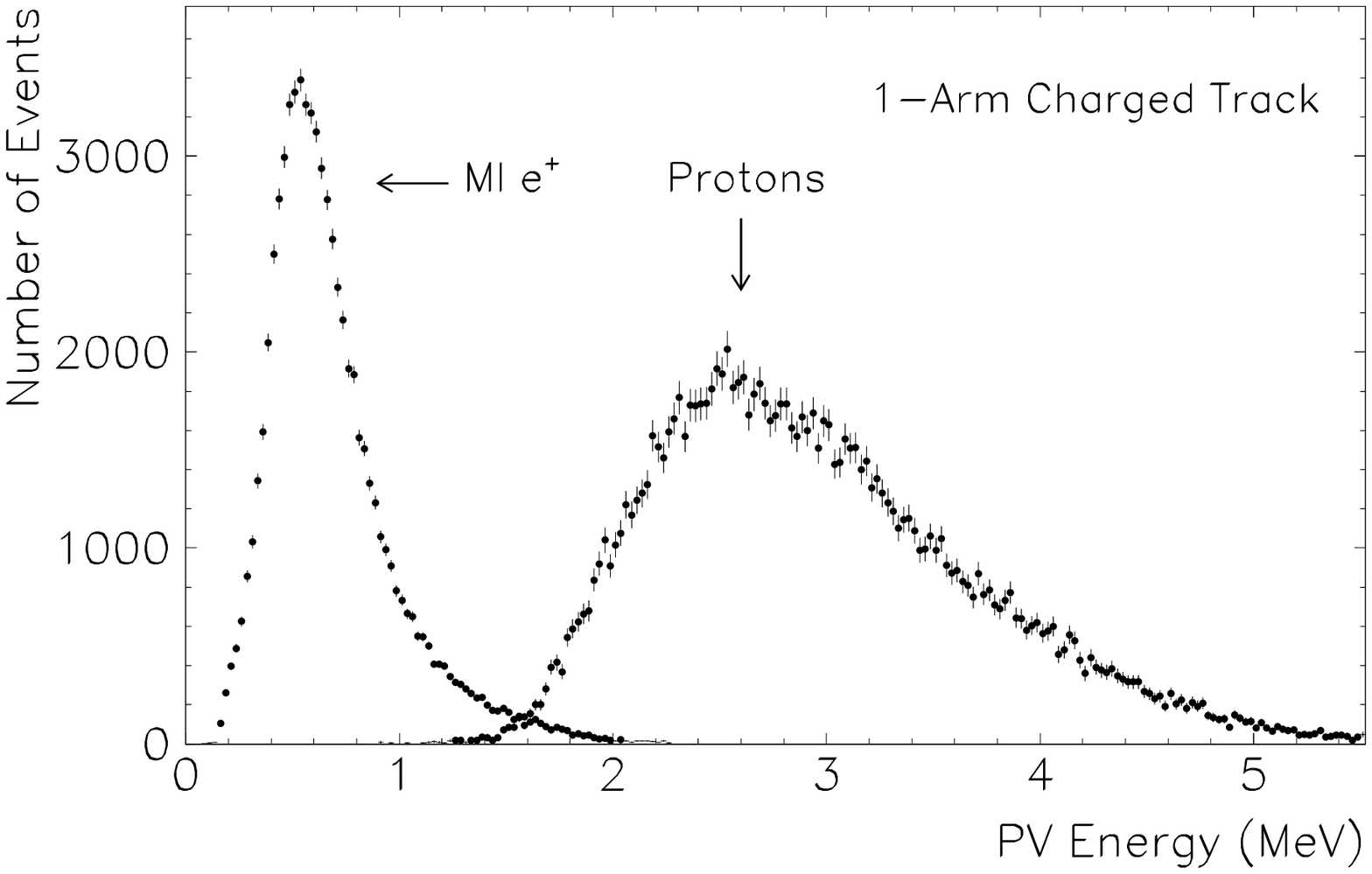,height=22cm}}
\vglue -9.cm
\centerline{FIGURE~\ref{fig:pv_adc}}
\vspace*{\stretch{2}}
\clearpage

\vspace*{\stretch{1}}
\centerline{\psfig{figure=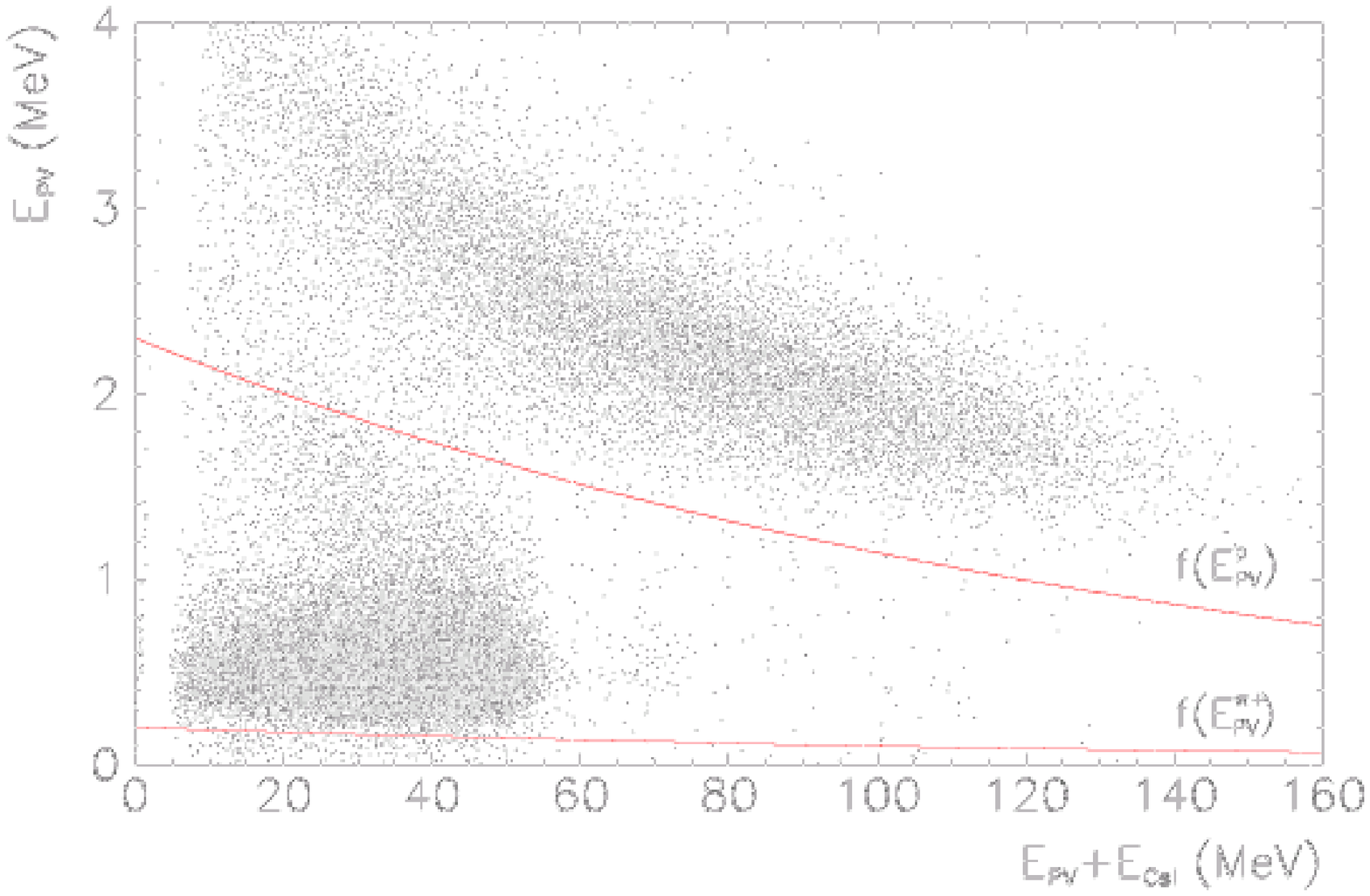,width=15.8cm}}
\vglue 1.0cm
\centerline{FIGURE~\ref{fig:part_id}}
\vspace*{\stretch{2}}
\clearpage

\vspace*{\stretch{1}}
\hbox{\ }\vglue -3.5cm
\centerline{\psfig{figure=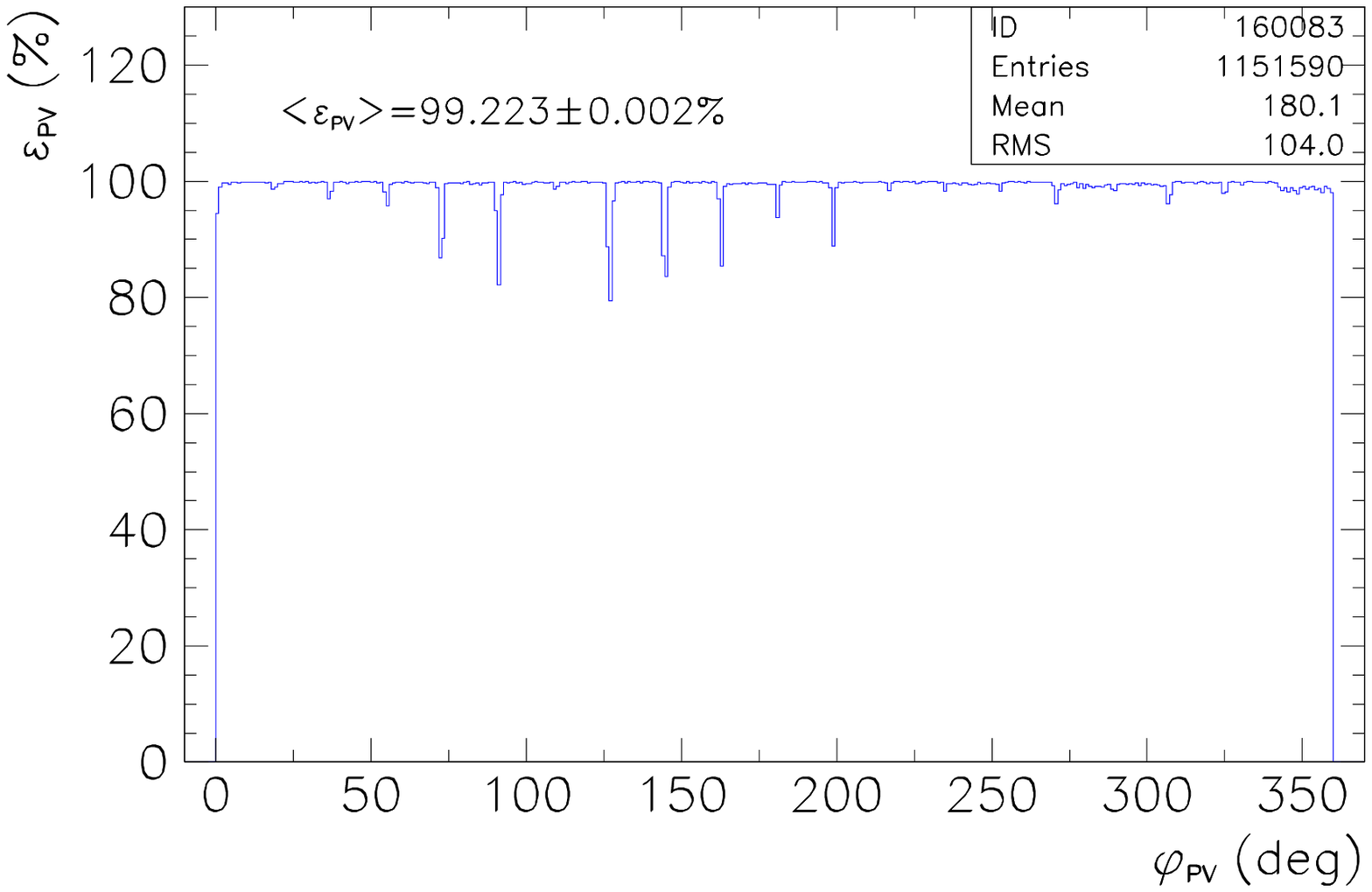,height=22cm}}
\vglue -8.0cm
\bigskip
\centerline{FIGURE~\ref{fig:pv_eff}}
\vspace*{\stretch{2}}
\clearpage

\vspace*{\stretch{1}}
\hbox{\ }
\vglue -2.0cm
\centerline{\psfig{figure=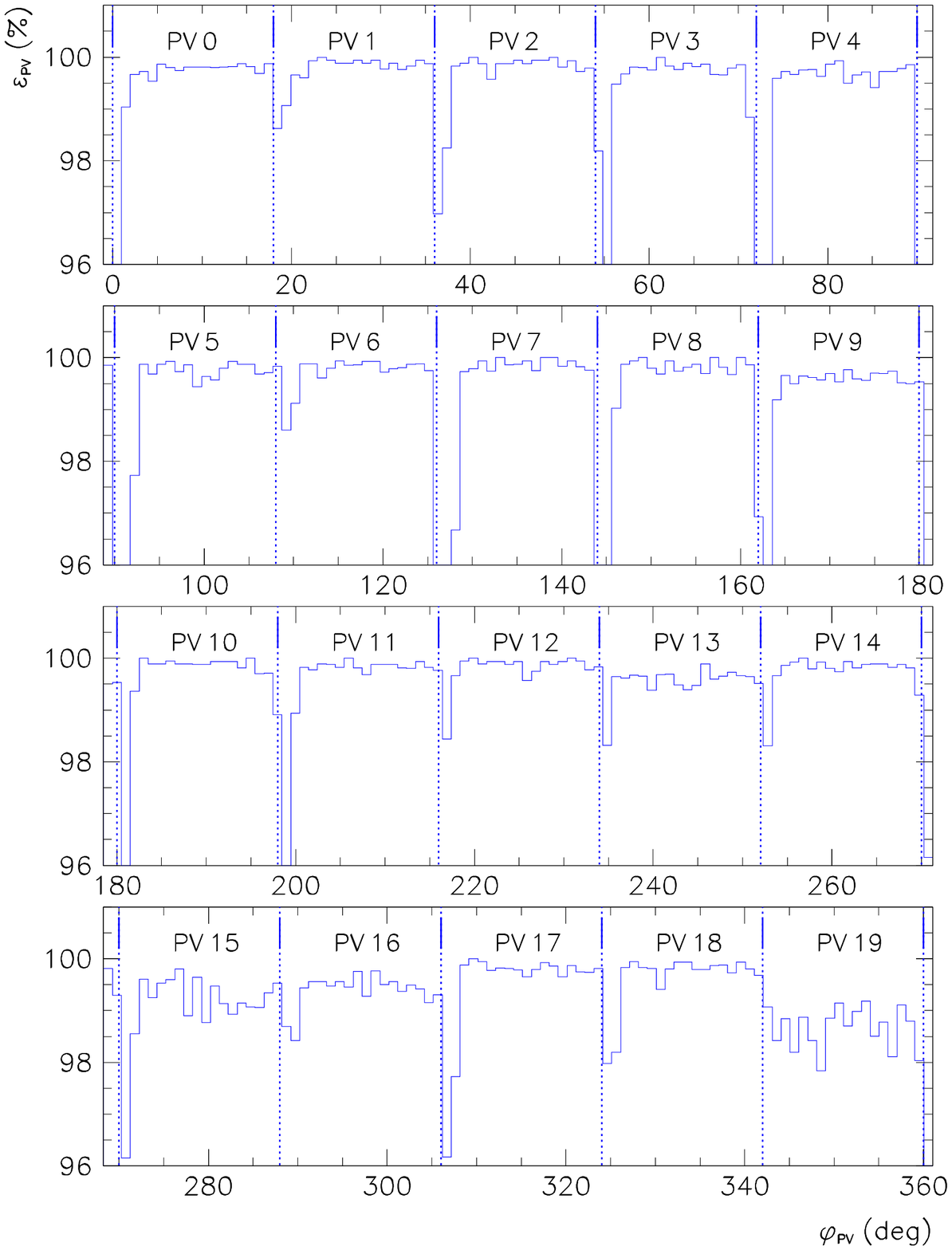,height=22cm}}
\bigskip\medskip
\centerline{FIGURE~\ref{fig:pv_eff_zoom}}
\vspace*{\stretch{2}}
\clearpage

\vspace*{\stretch{1}}
\vglue -2.5cm
\noindent \hglue 3cm (i)\hglue 7cm (ii)
\vglue 0.5cm
\hbox to \textwidth
 {\vbox{\hsize=3.1in
  \centerline{\psfig{figure=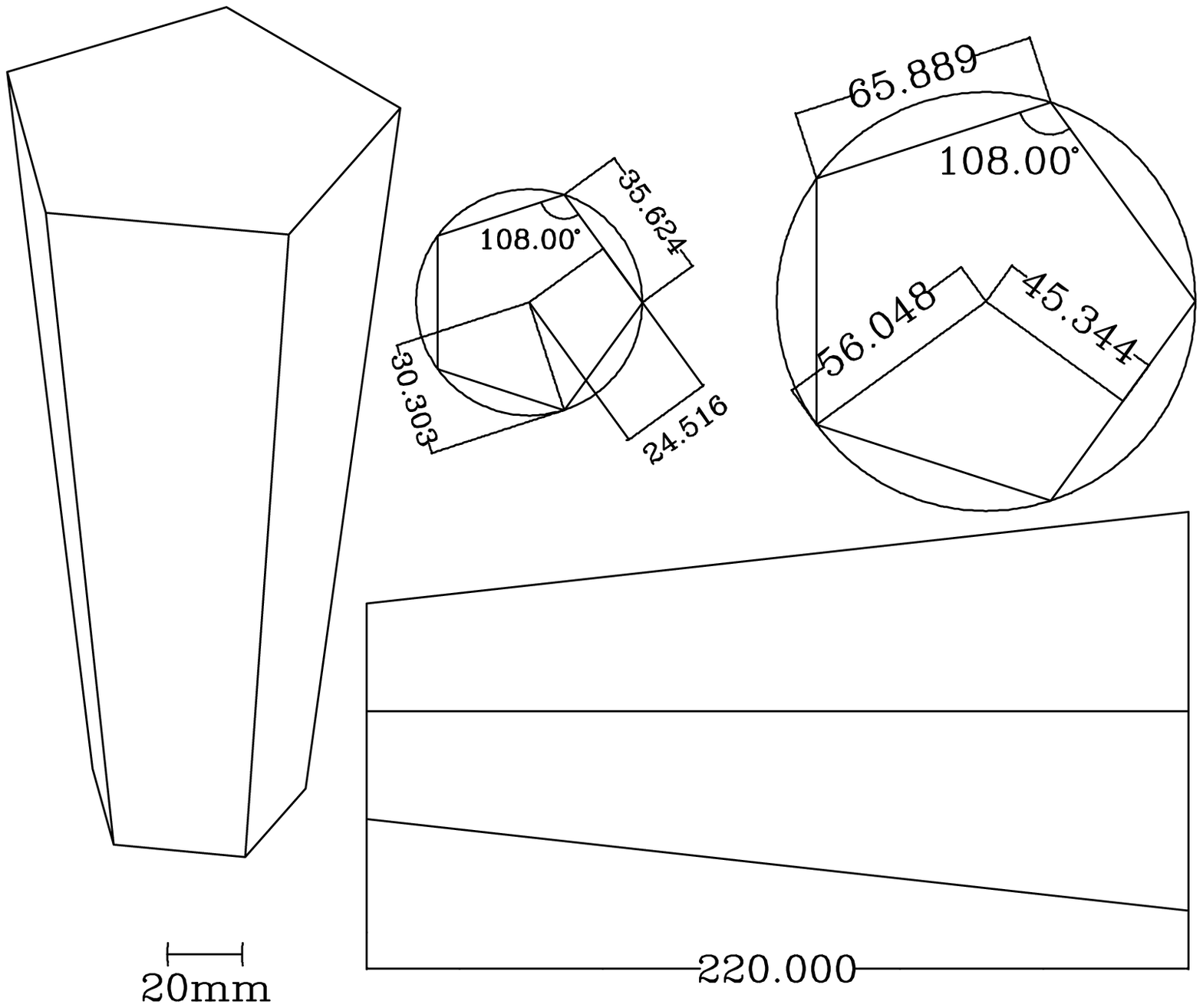,width=2.7in}}
  \vglue -2.5cm
       }\hfill
  \vbox{\hsize=3.1in
  \centerline{\psfig{figure=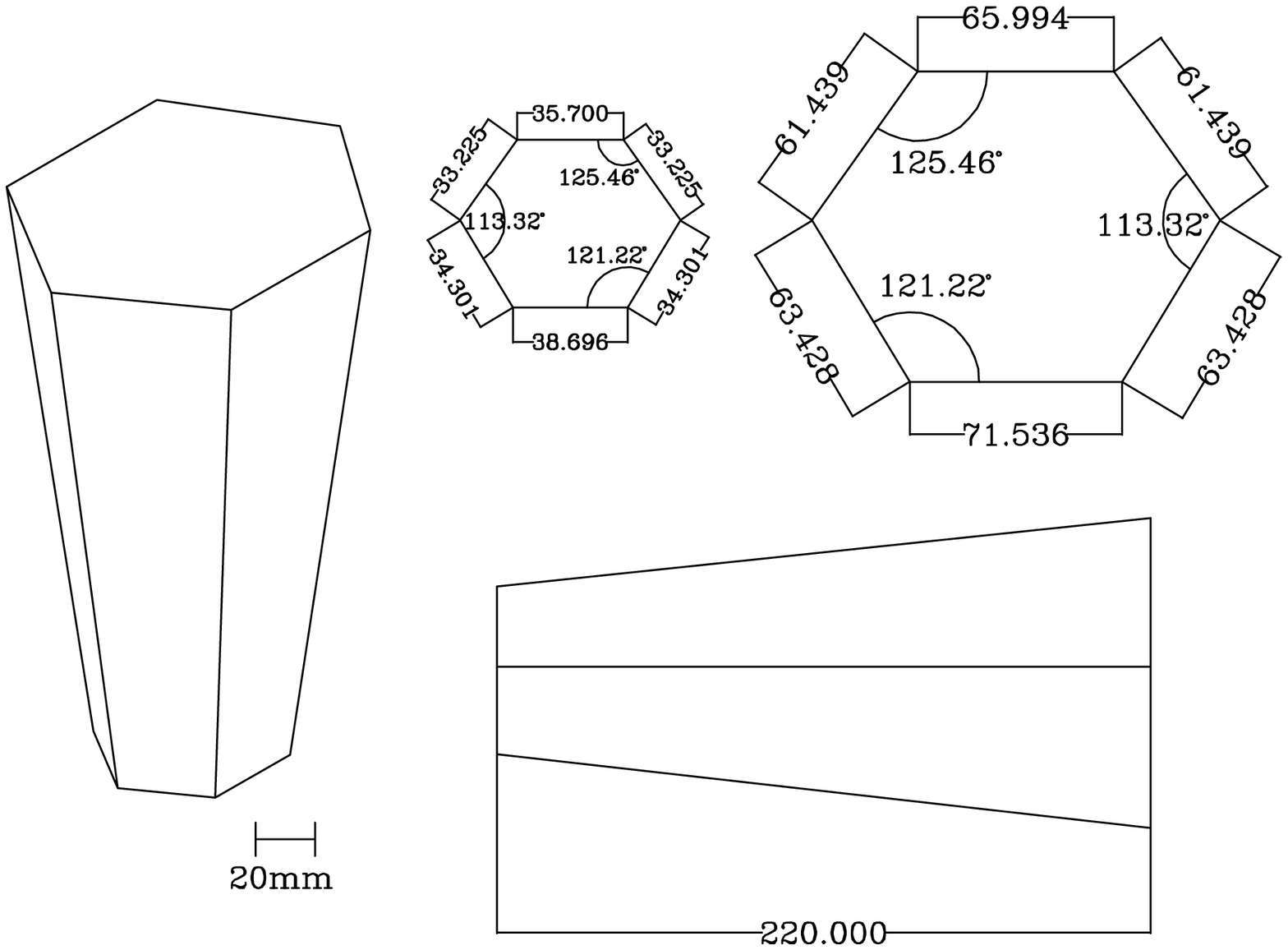,width=2.7in}}
  \vglue -2.5cm
       }
 }
\vglue 2.8cm

\vglue 0.5cm
\noindent \hglue 3cm (iii)\hglue 7cm (iv)
\vglue 0.5cm
\hbox to \textwidth
 {\vbox{\hsize=3.1in
  \centerline{\psfig{figure=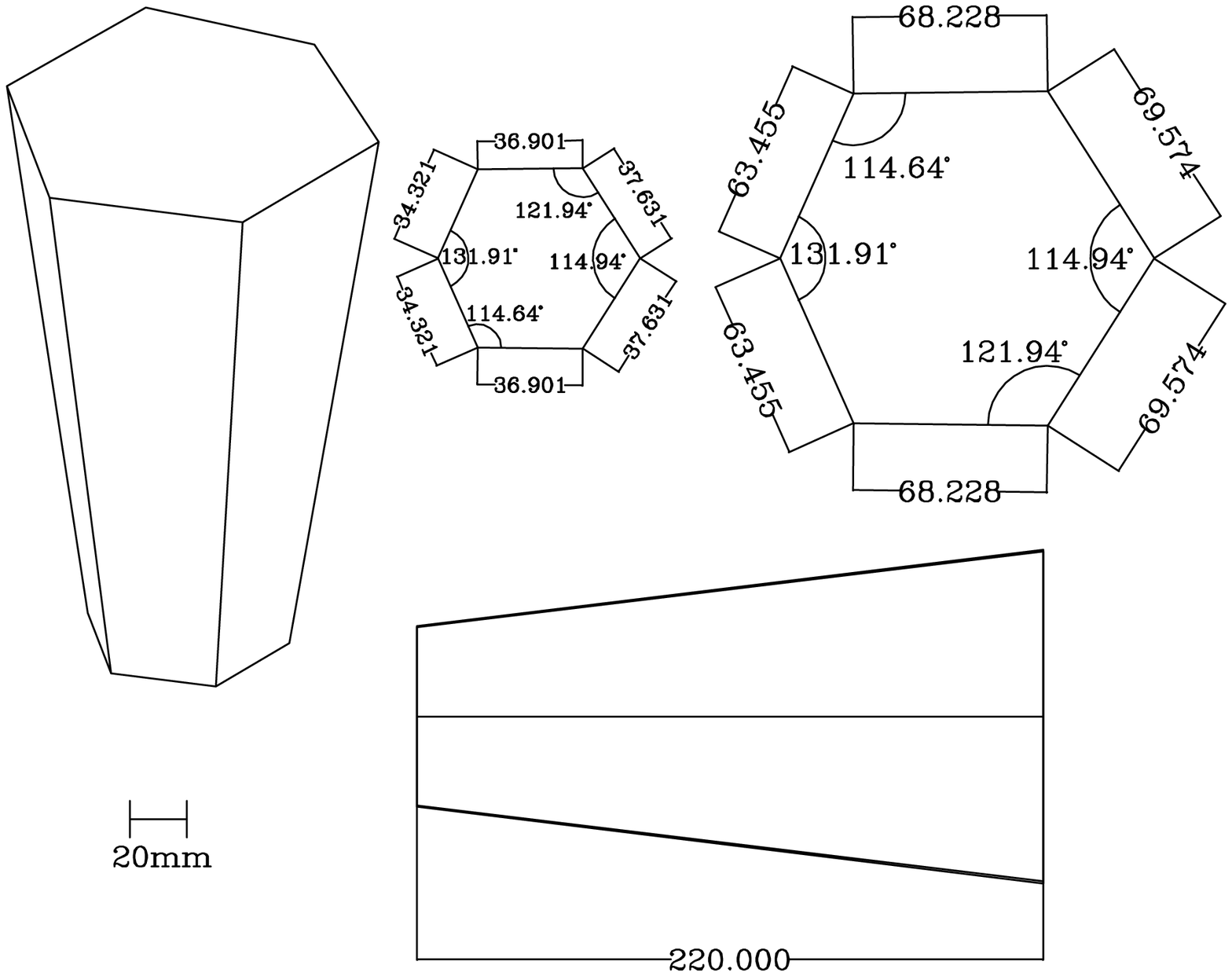,width=2.7in}}
  \vglue -2.5cm
       }\hfill
  \vbox{\hsize=3.1in
  \centerline{\psfig{figure=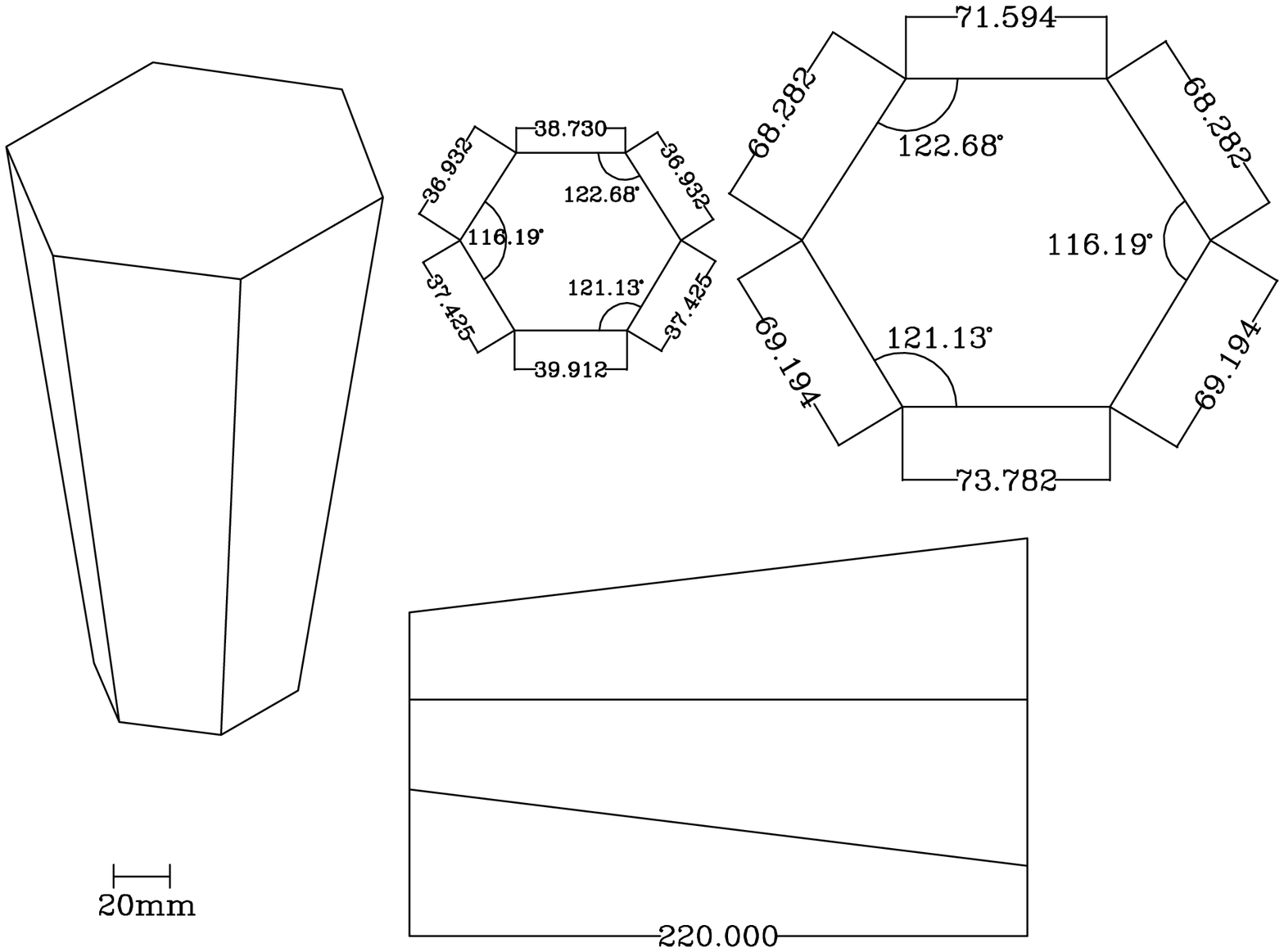,width=2.7in}}
  \vglue -2.5cm
       }
 }
\vglue 2.8cm

\vglue 0.5cm
\noindent \hglue 3cm (v)\hglue 7cm (vi)
\vglue 0.5cm
\hbox to \textwidth
 {\vbox{\hsize=3.1in
  \centerline{\psfig{figure=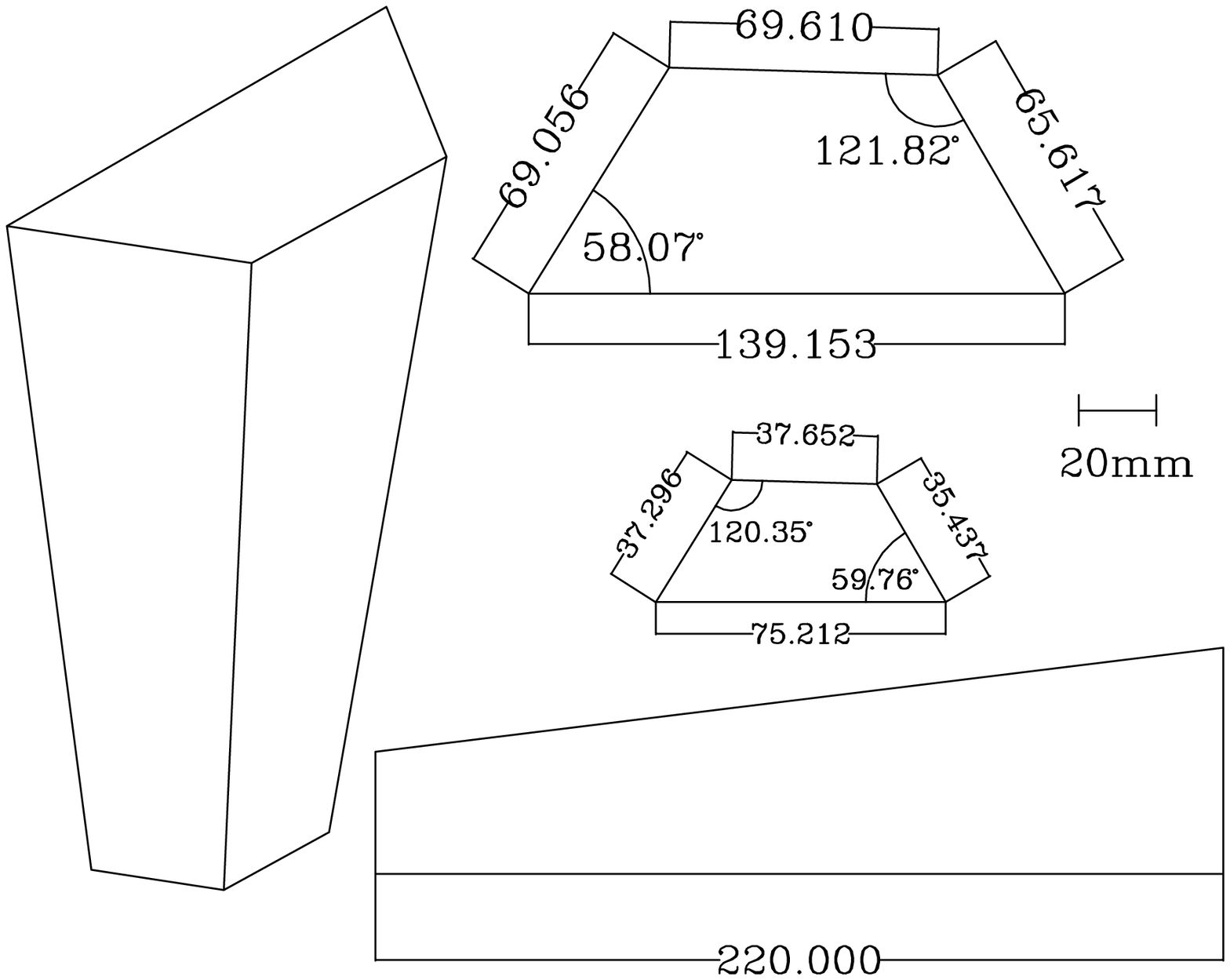,width=2.7in}}
  \vglue -2.5cm
       }\hfill
  \vbox{\hsize=3.1in
  \centerline{\psfig{figure=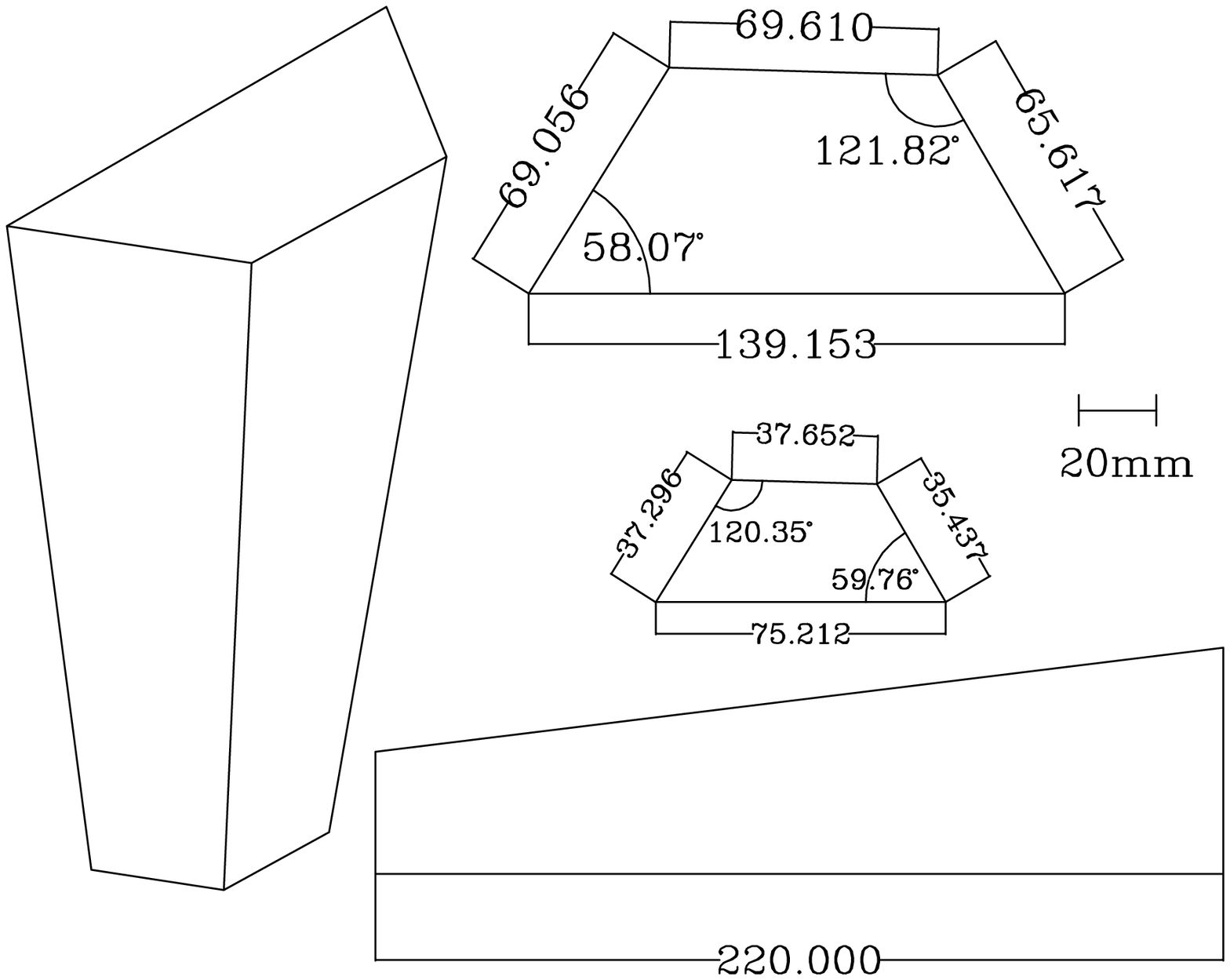,width=2.7in}}
  \vglue -2.5cm
       }
 }
\vglue 4.0cm
\centerline{FIGURE~\ref{fig:csi_shapes}a}
\vspace*{\stretch{2}}
\clearpage

\vspace*{\stretch{1}}
\vglue 0.5cm
\noindent \hglue 3cm (vii)\hglue 7cm (viii)
\vglue 0.5cm
\hbox to \textwidth
 {\vbox{\hsize=3.1in
  \centerline{\psfig{figure=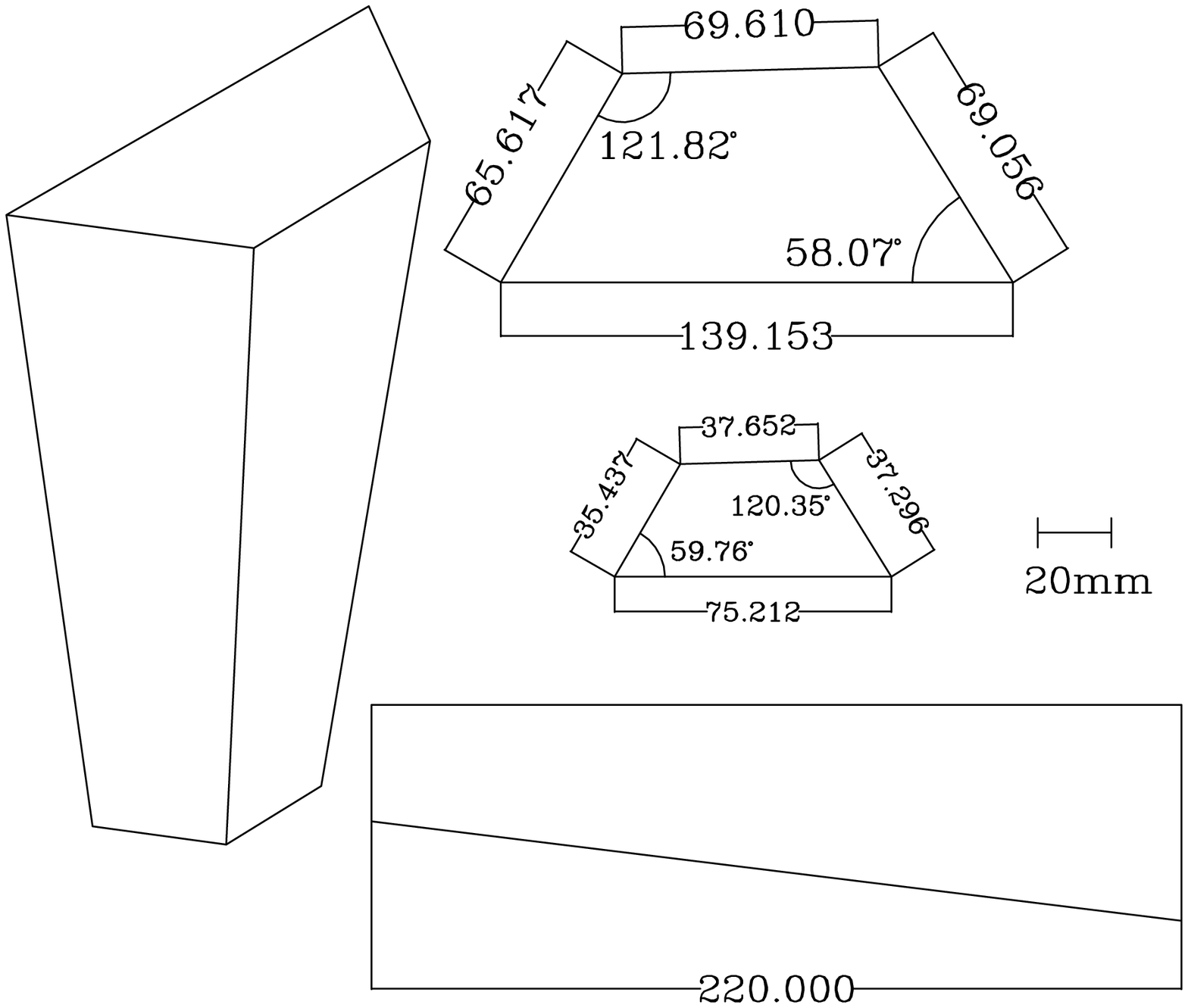,width=2.7in}}
  \vglue -2.5cm
       }\hfill
  \vbox{\hsize=3.1in
  \centerline{\psfig{figure=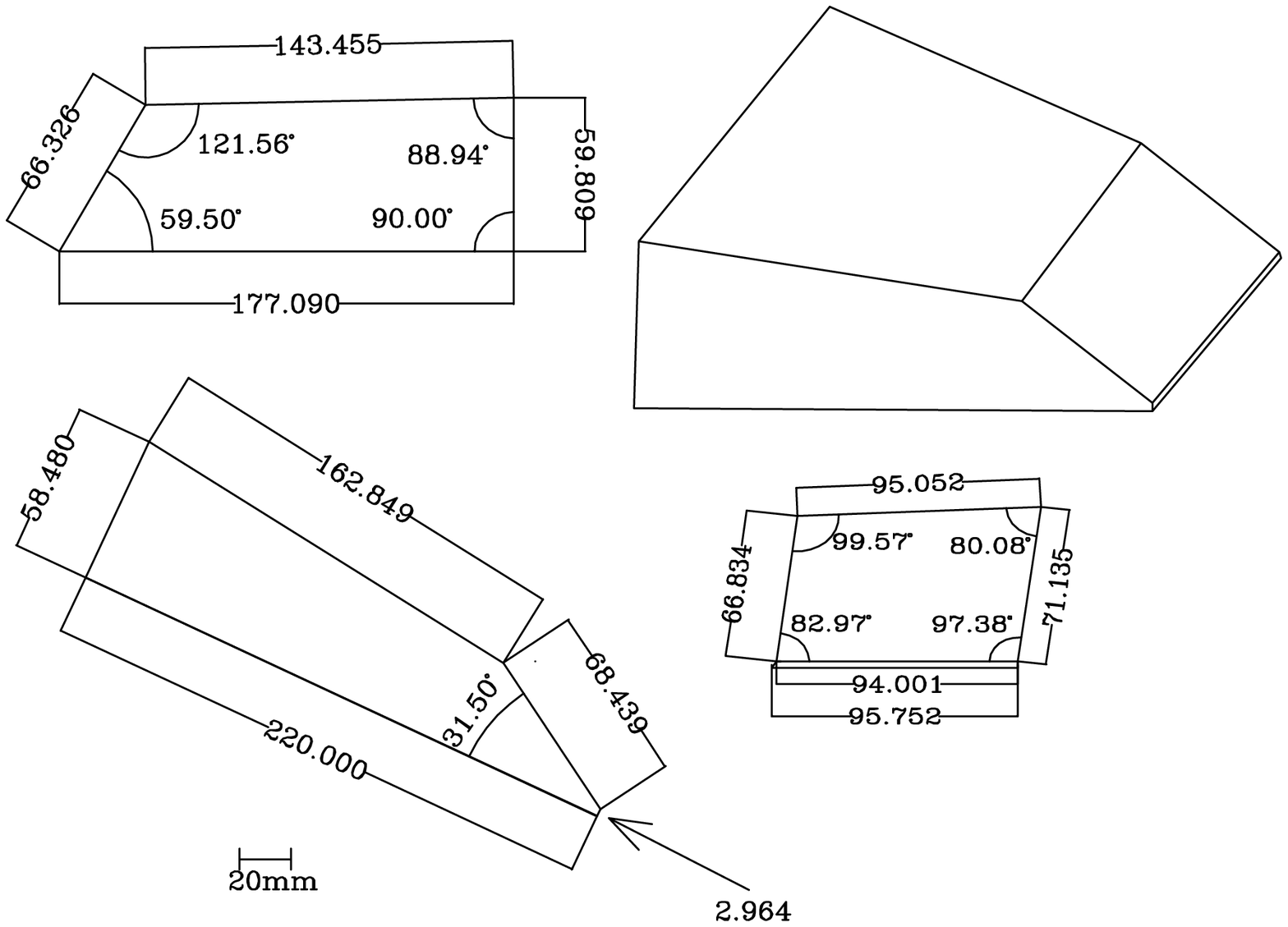,width=3.2in}}
  \vglue -2.5cm
       }
 }
\vglue 2.8cm

\vglue 0.5cm
\noindent \hglue 3cm (ix)\hglue 7cm (x)
\vglue 0.5cm
\hbox to \textwidth
 {\vbox{\hsize=3.1in
  \centerline{\psfig{figure=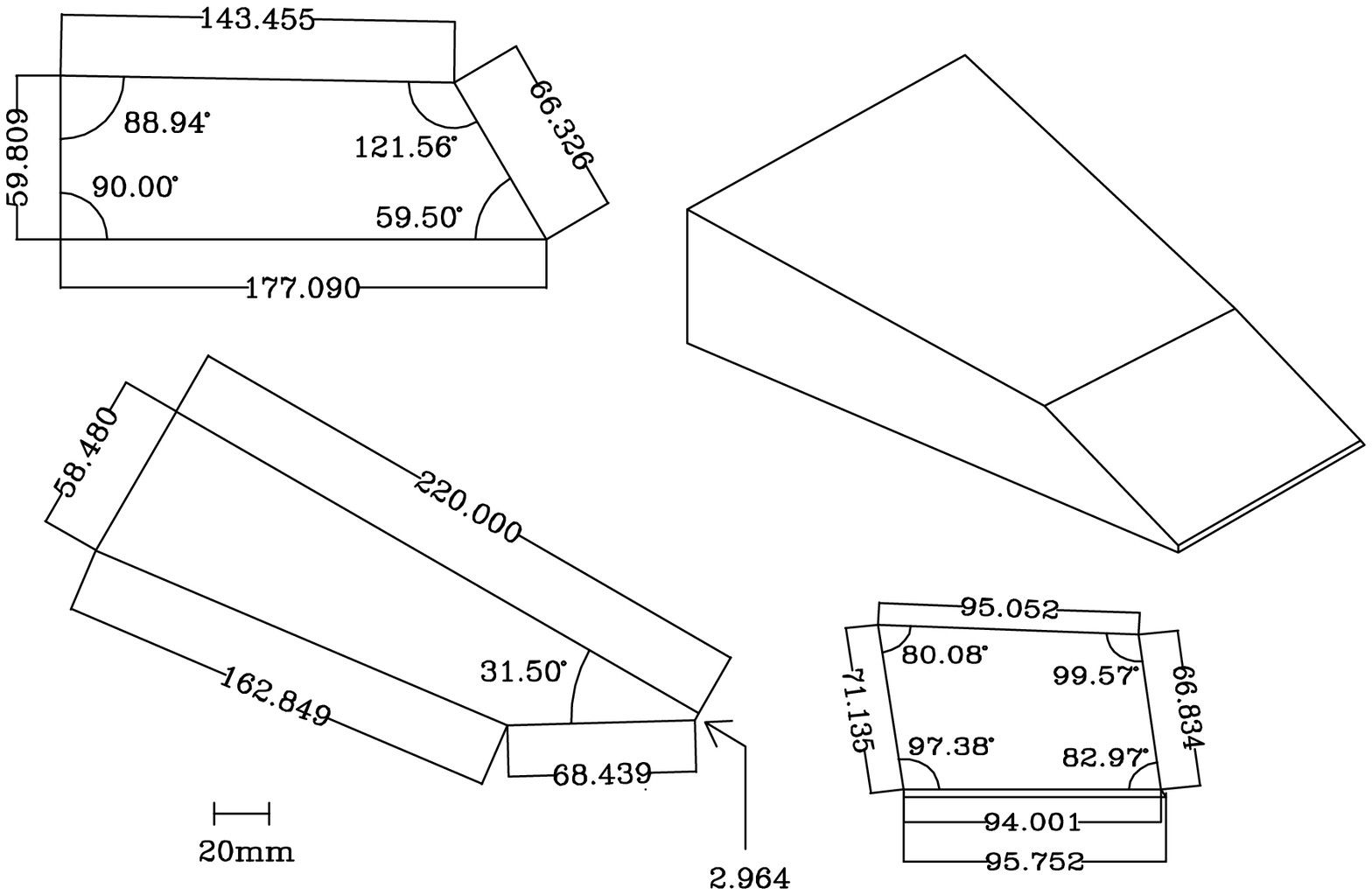,width=3.2in}}
  \vglue -2.5cm
       }\hfill
  \vbox{\hsize=3.1in
  \centerline{\psfig{figure=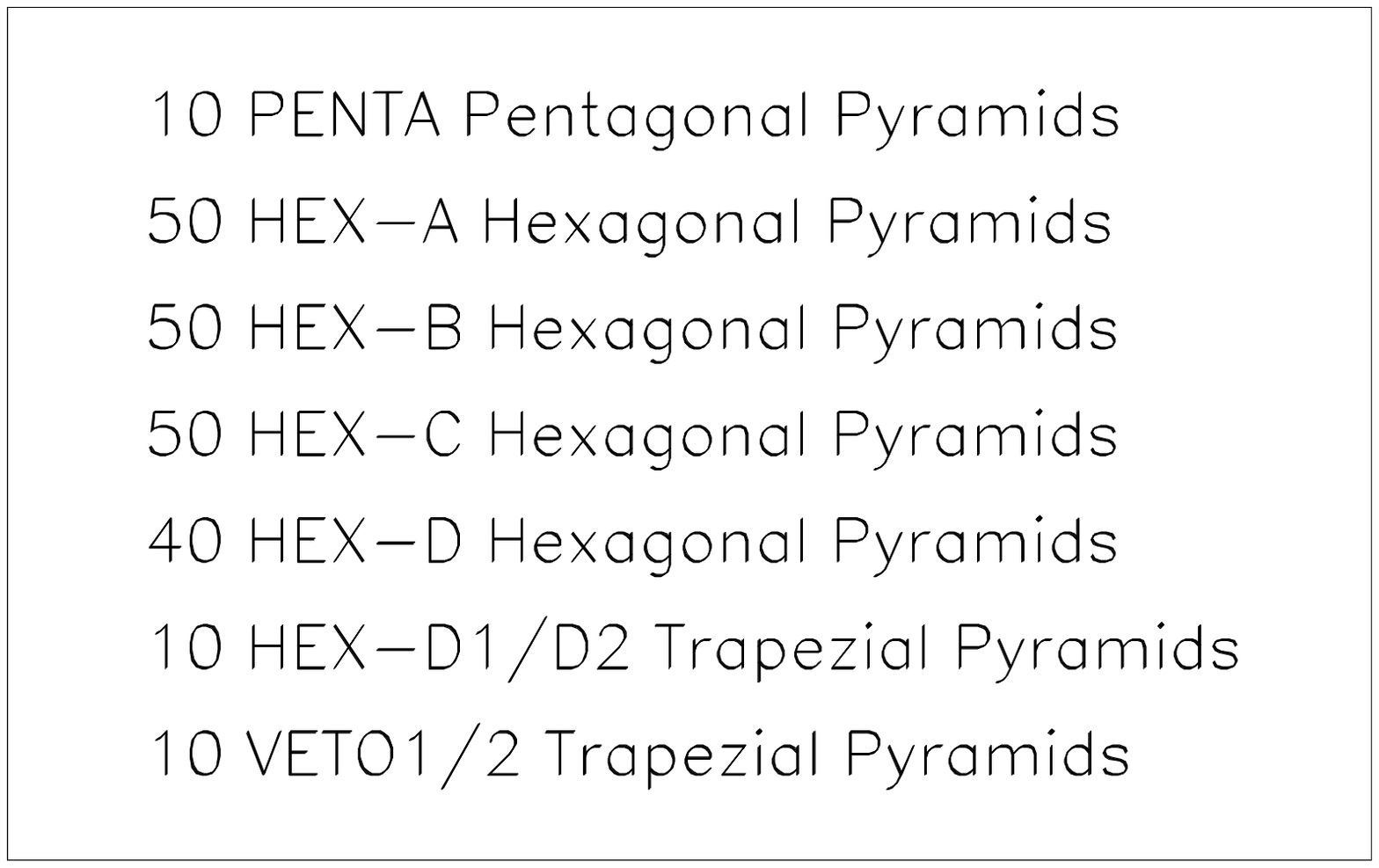,width=2.8in}}
  \vglue -7.8cm
       }
 }
\label{fig:csishapes}
\vglue 4.0cm
\centerline{FIGURE~\ref{fig:csi_shapes}b}
\vspace*{\stretch{2}}
\clearpage


\vspace*{\stretch{1}}
\hbox{\ }\noindent \hglue 3cm (i)
\centerline{\psfig{figure=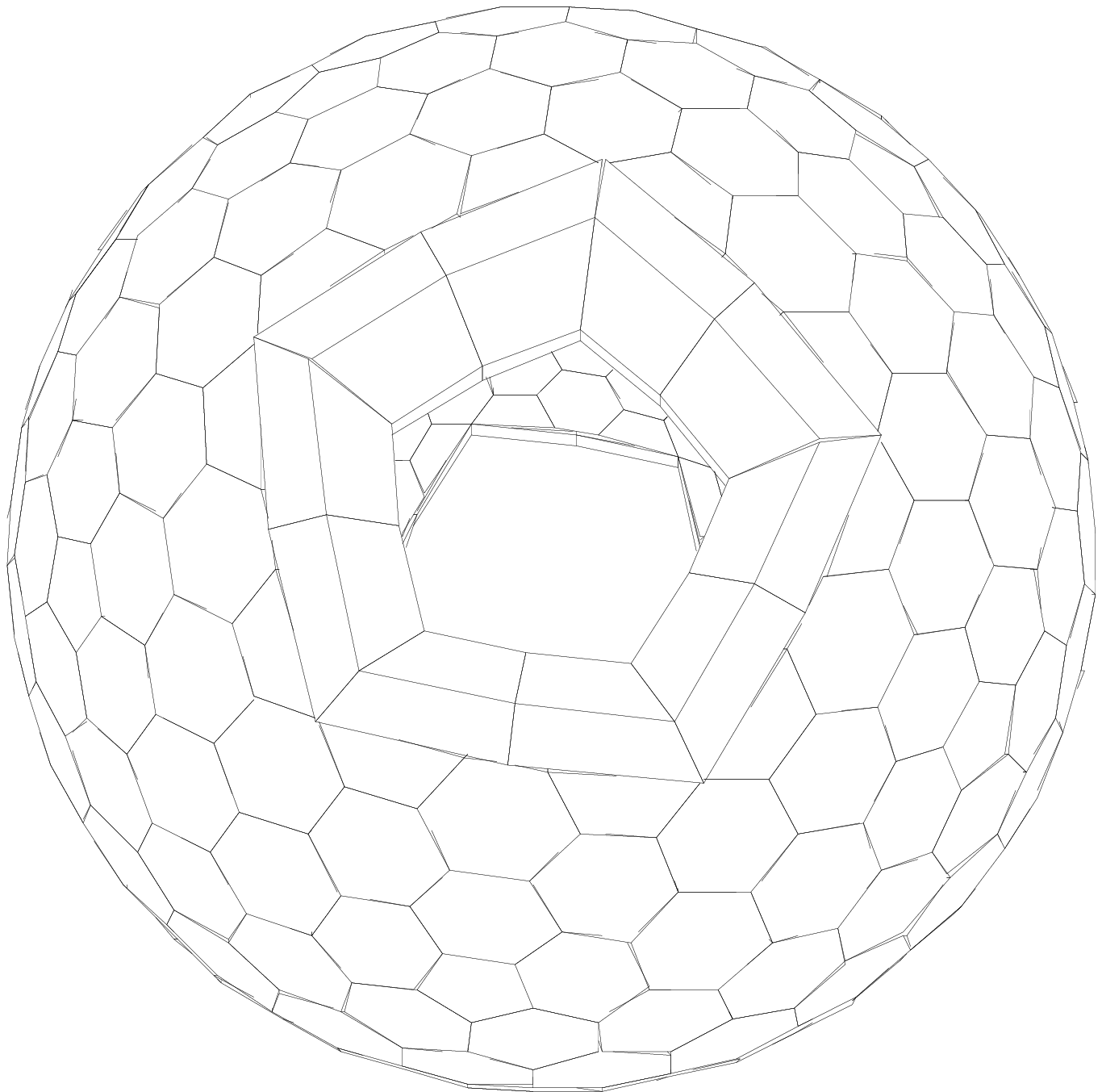,height=8cm}\hglue 3.5cm}
\bigskip\bigskip\bigskip

\noindent \hglue 3cm (ii)
\centerline{\psfig{figure=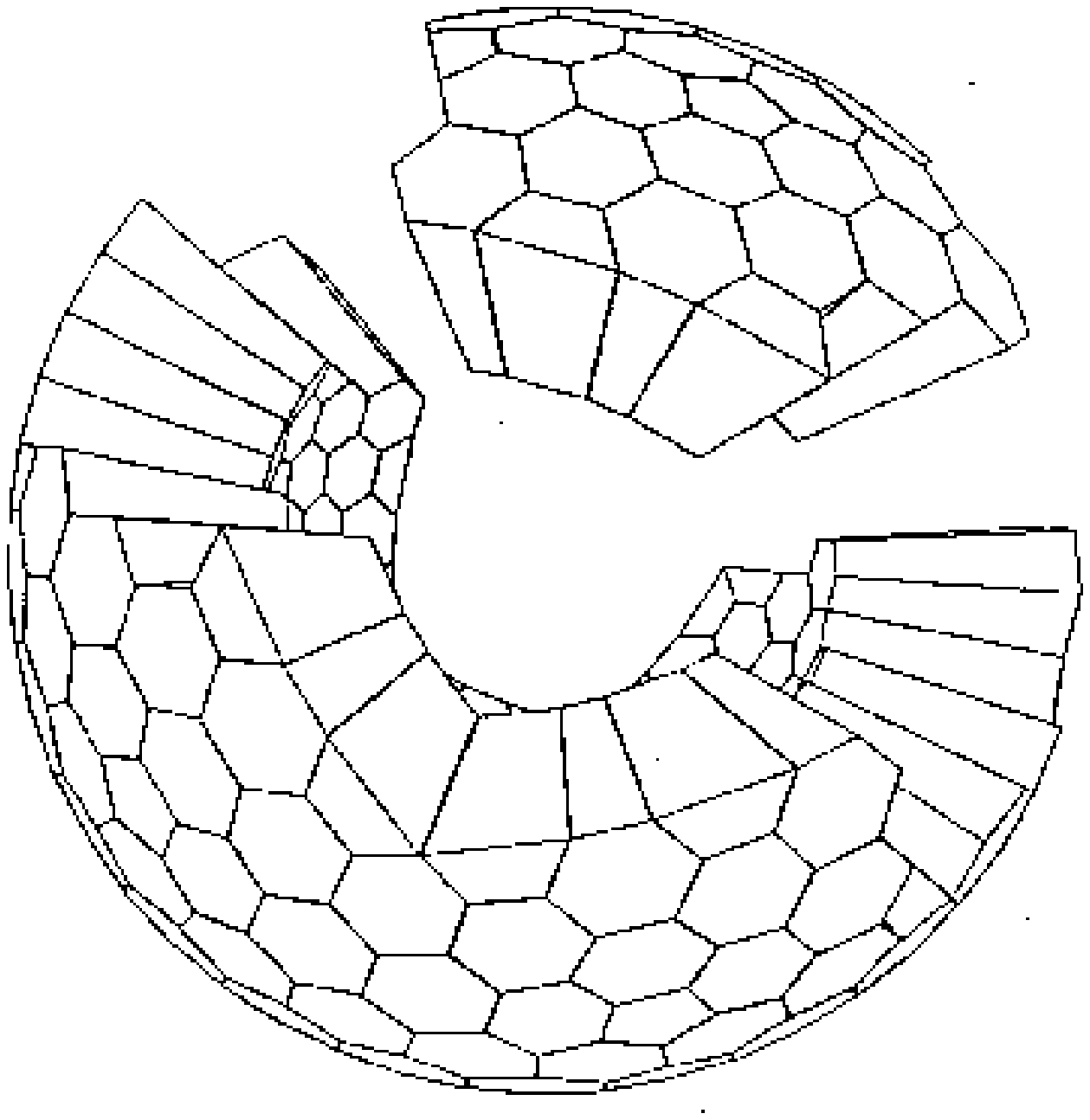,width=8.9cm}\hglue 3.5cm}
\vglue 2.0cm
\centerline{FIGURE~\ref{fig:balls}}
\vspace*{\stretch{2}}
\clearpage

\vspace*{\stretch{1}}
\hbox{\ }\vglue -3cm
\centerline{\psfig{figure=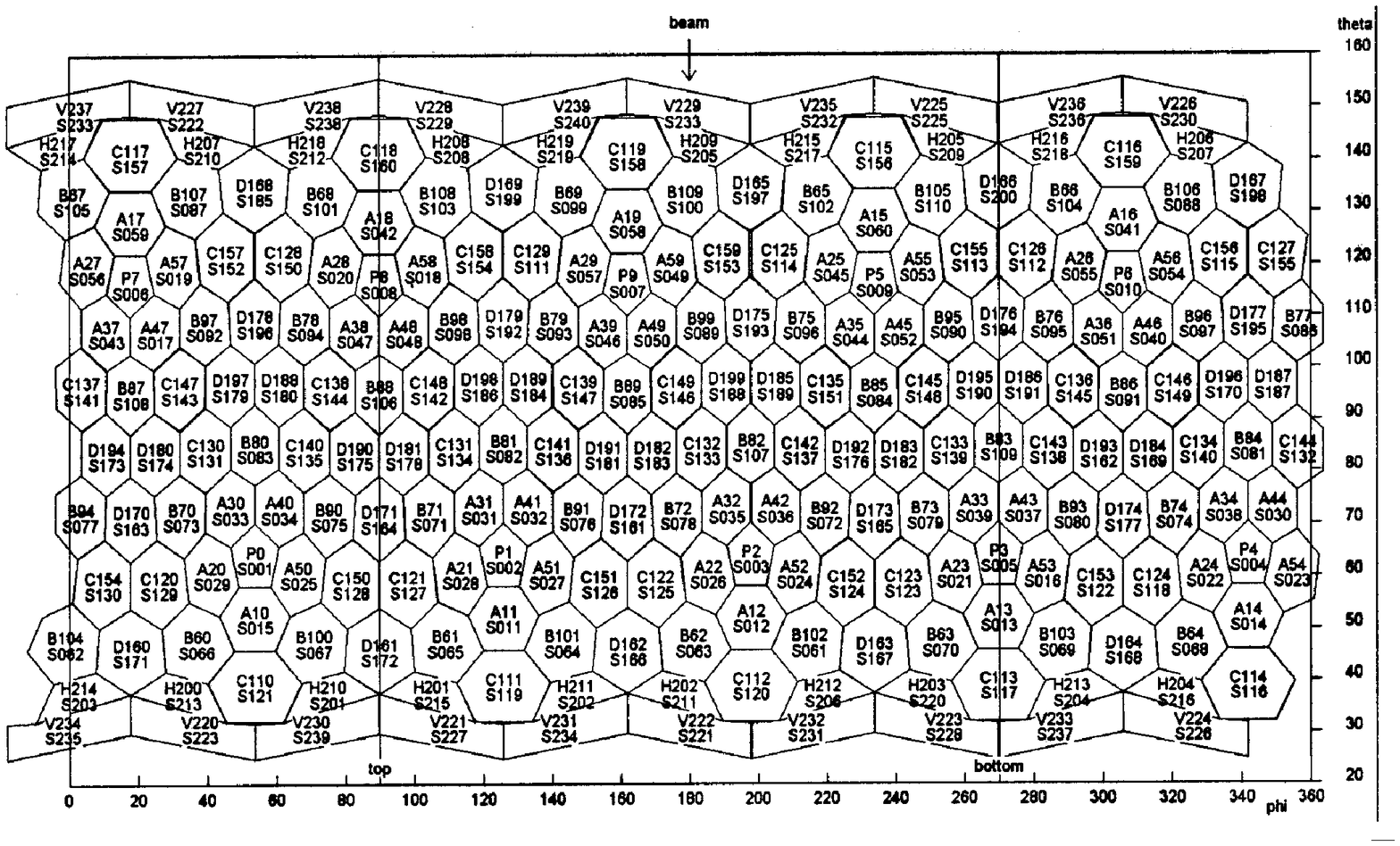,width=12cm}}
\vglue 12.0cm
\centerline{FIGURE~\ref{fig:calo_map}}
\vspace*{\stretch{2}}
\clearpage

\vspace*{\stretch{1}}
\vglue -1cm
\hbox{(i)}
\centerline{\psfig{figure=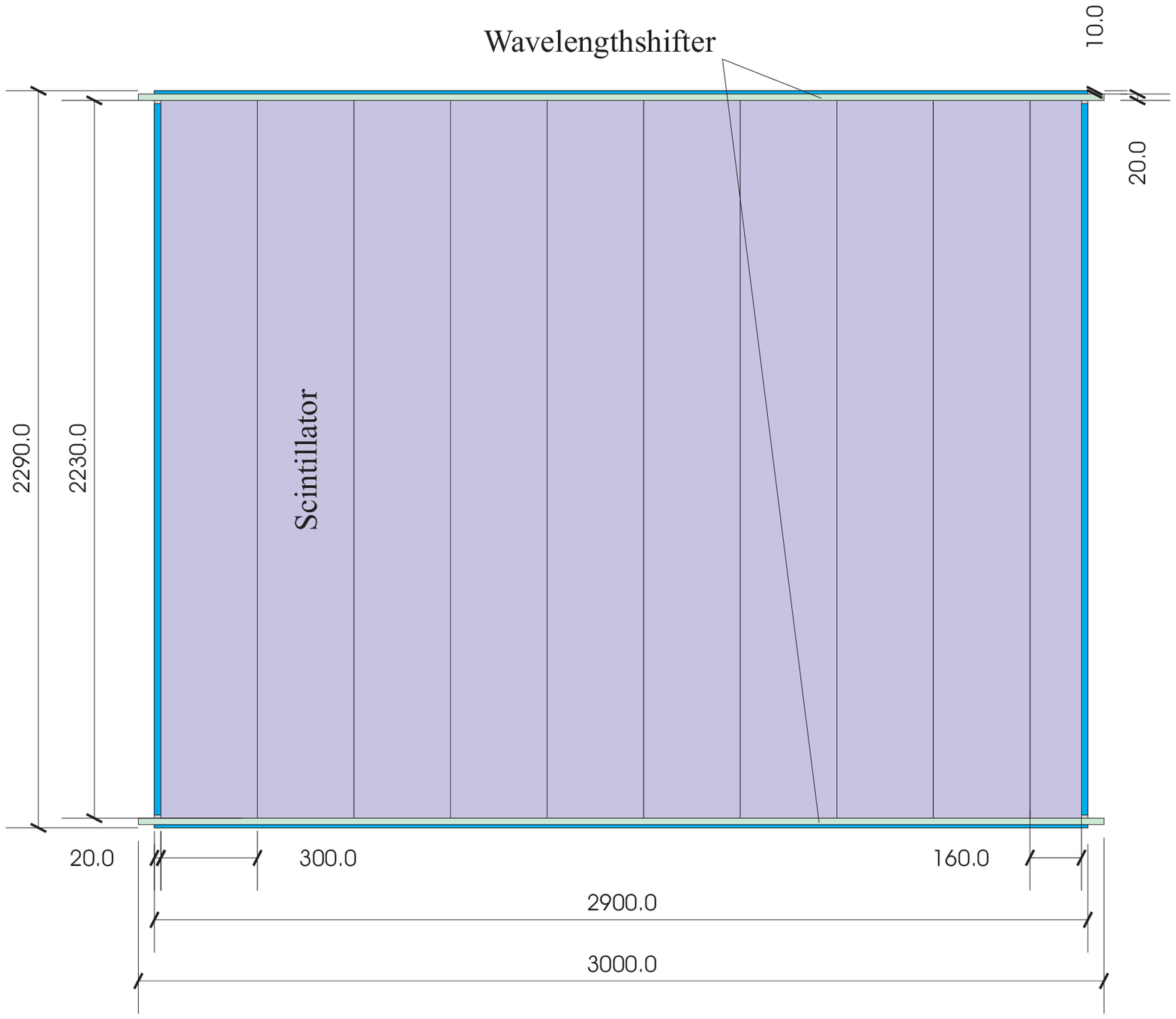,height=10cm}}
\vglue 1.cm
\hbox{(ii)}
\centerline{\psfig{figure=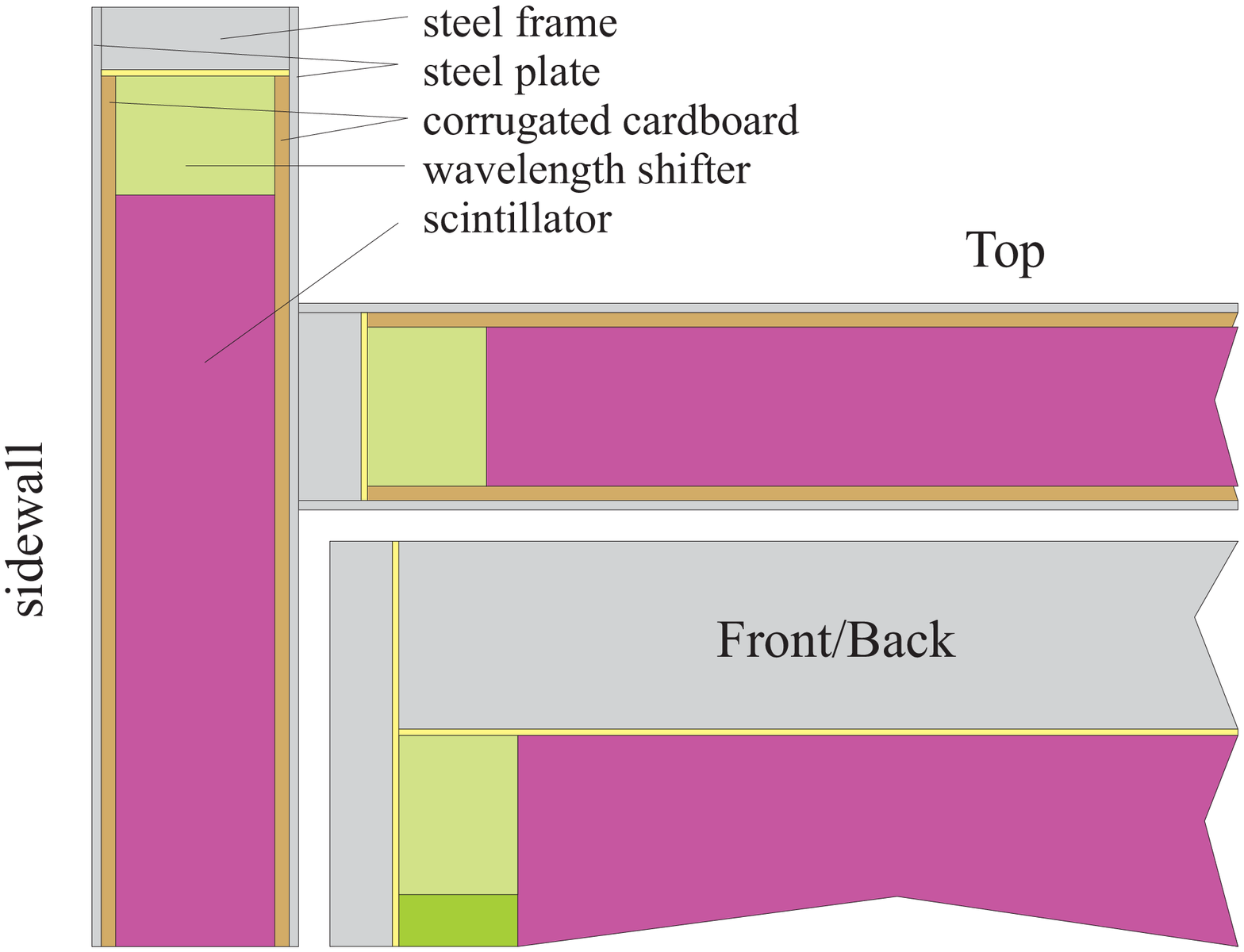,height=8cm}}
\vglue 1.5cm
\centerline{FIGURE~\ref{fig:cv_princ}}
\vspace*{\stretch{2}}
\clearpage

\vspace*{\stretch{1}}
\centerline{\psfig{figure=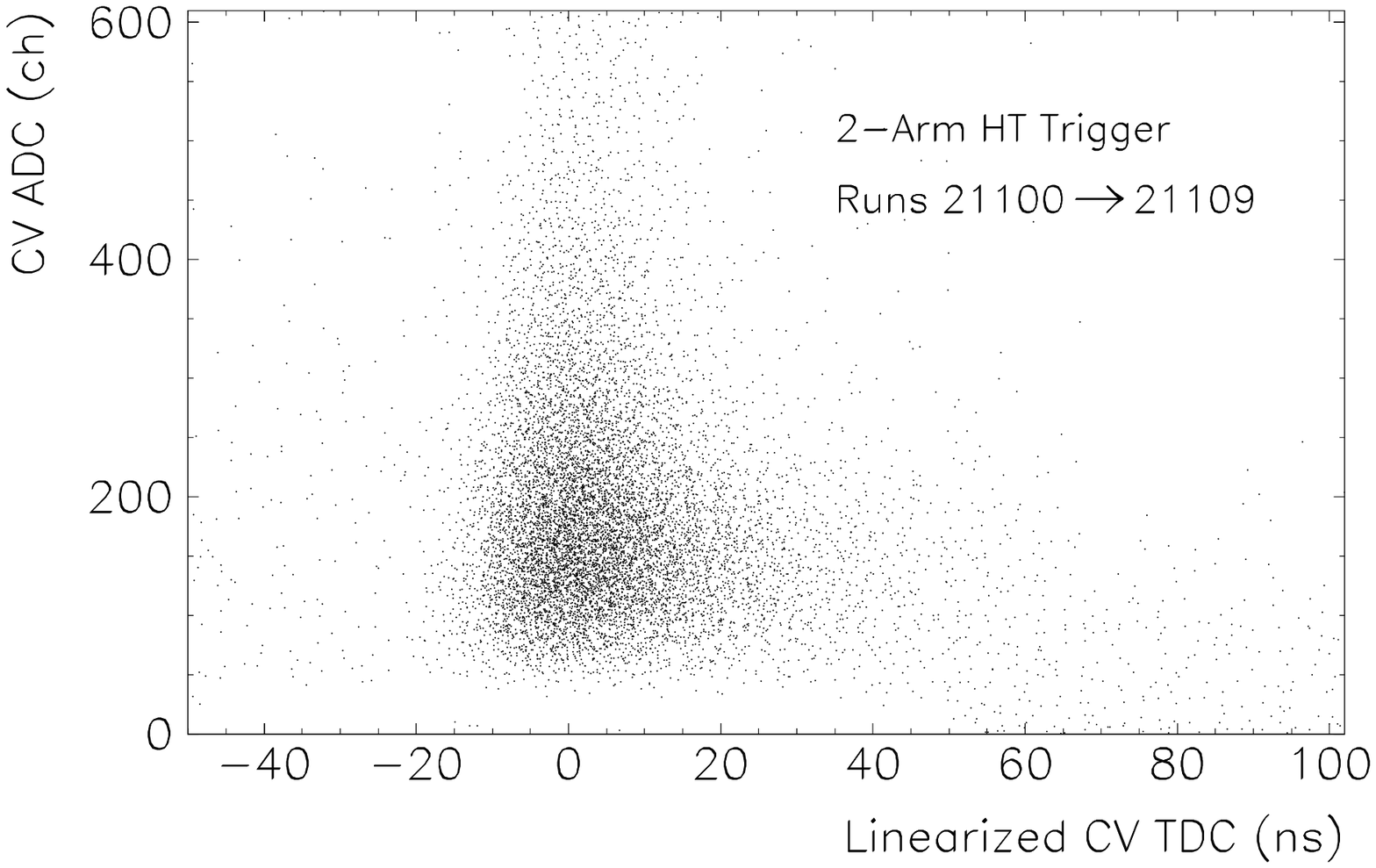,height=22cm}}
\vglue -9cm
\centerline{FIGURE~\ref{fig:cosm_e_t}}
\vspace*{\stretch{2}}
\clearpage

\vspace*{\stretch{1}}
\hbox{\ }\vglue -1.5cm
\centerline{\psfig{figure=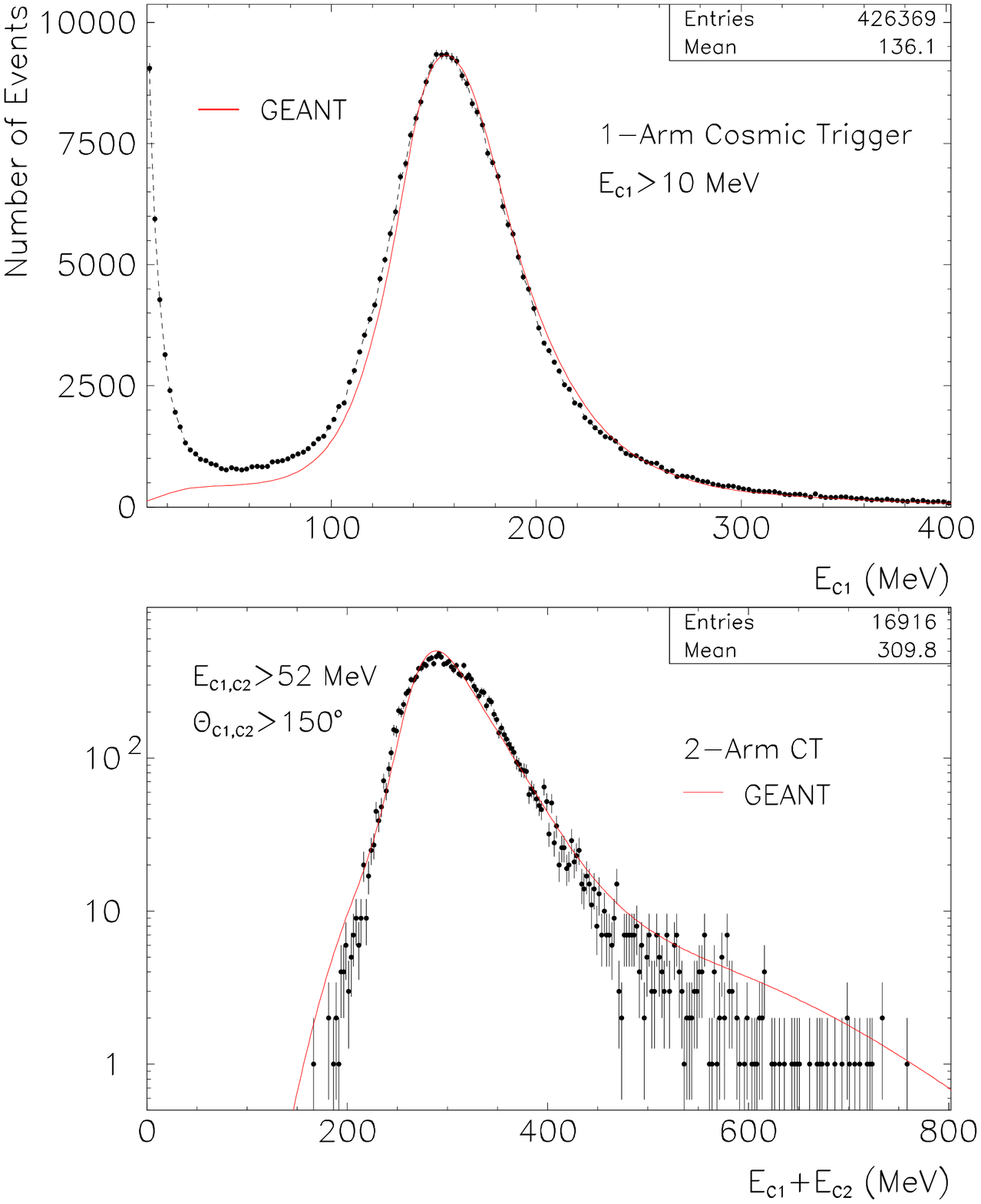,height=22cm}}
\bigskip
\centerline{FIGURE~\ref{fig:cosm_en}}
\vspace*{\stretch{2}}
\clearpage

\vspace*{\stretch{1}}
\centerline{\psfig{figure=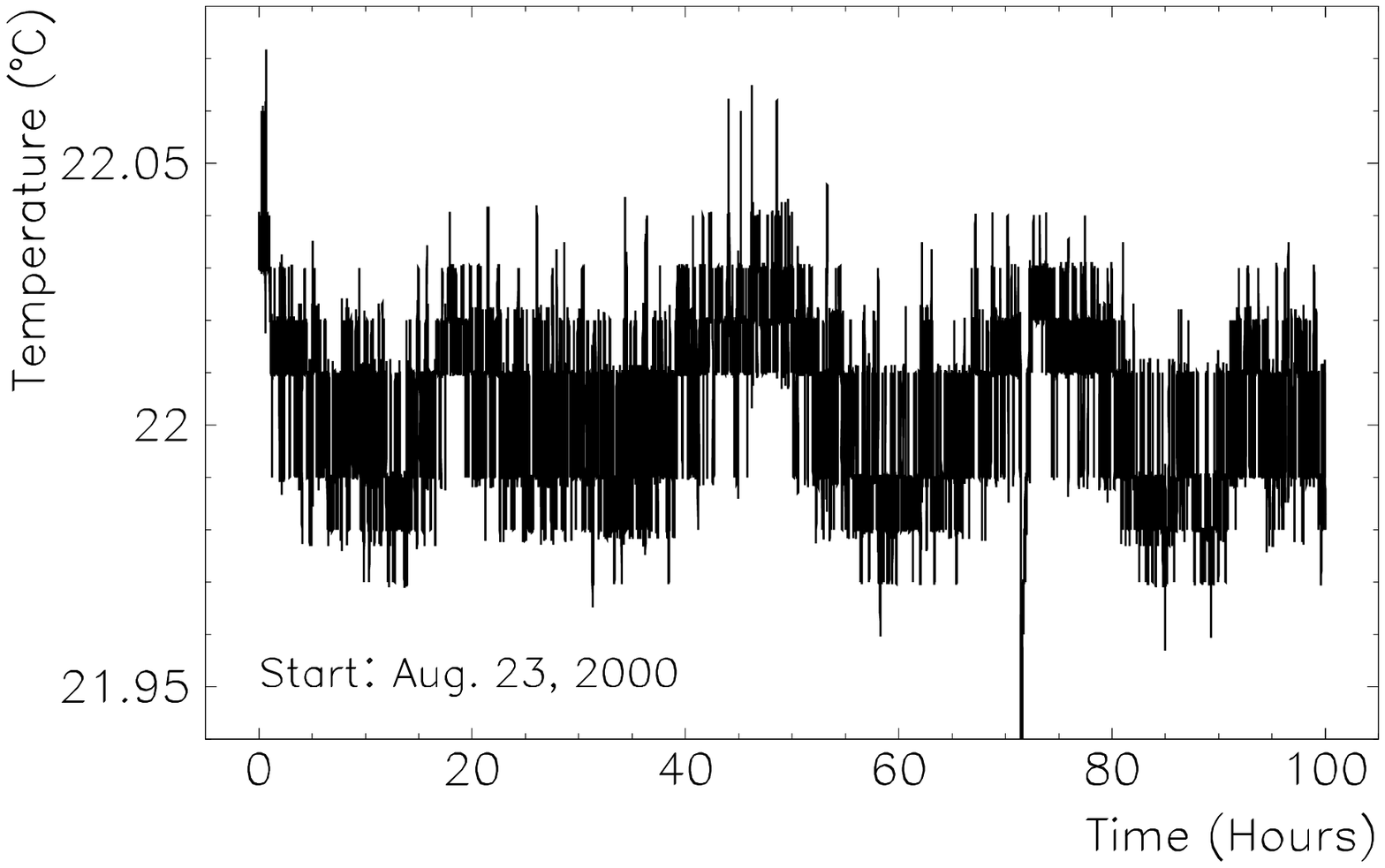,height=22cm}}
\vglue -9cm
\centerline{FIGURE~\ref{fig:temp}}
\vspace*{\stretch{2}}
\clearpage

\vspace*{\stretch{1}}
\centerline{\psfig{figure=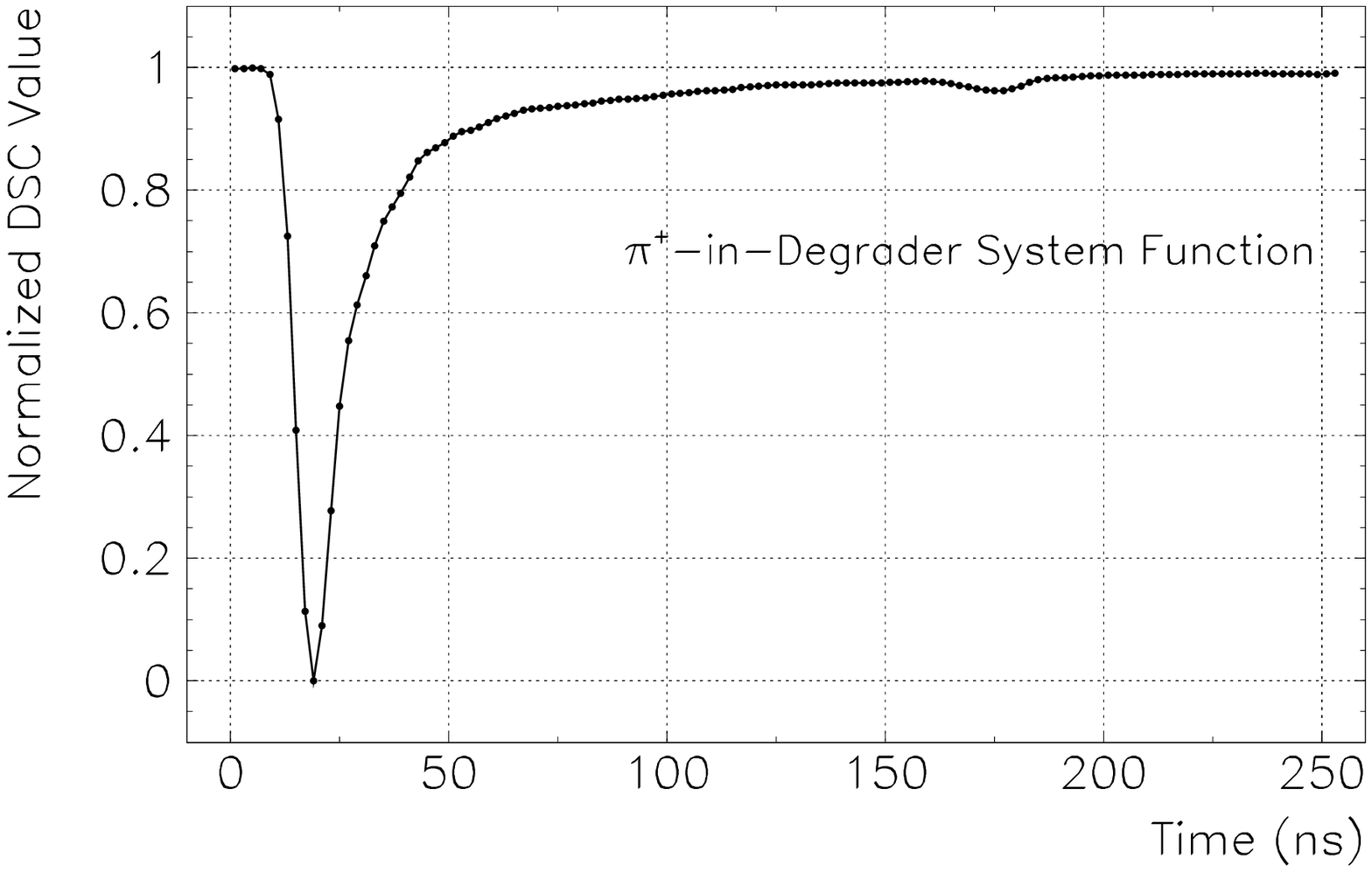,height=22cm}}  
\vglue -9.5cm
\centerline{FIGURE~\ref{fig:dsc_fun}}
\vspace*{\stretch{2}}
\clearpage

\vspace*{\stretch{1}}
\centerline{\psfig{figure=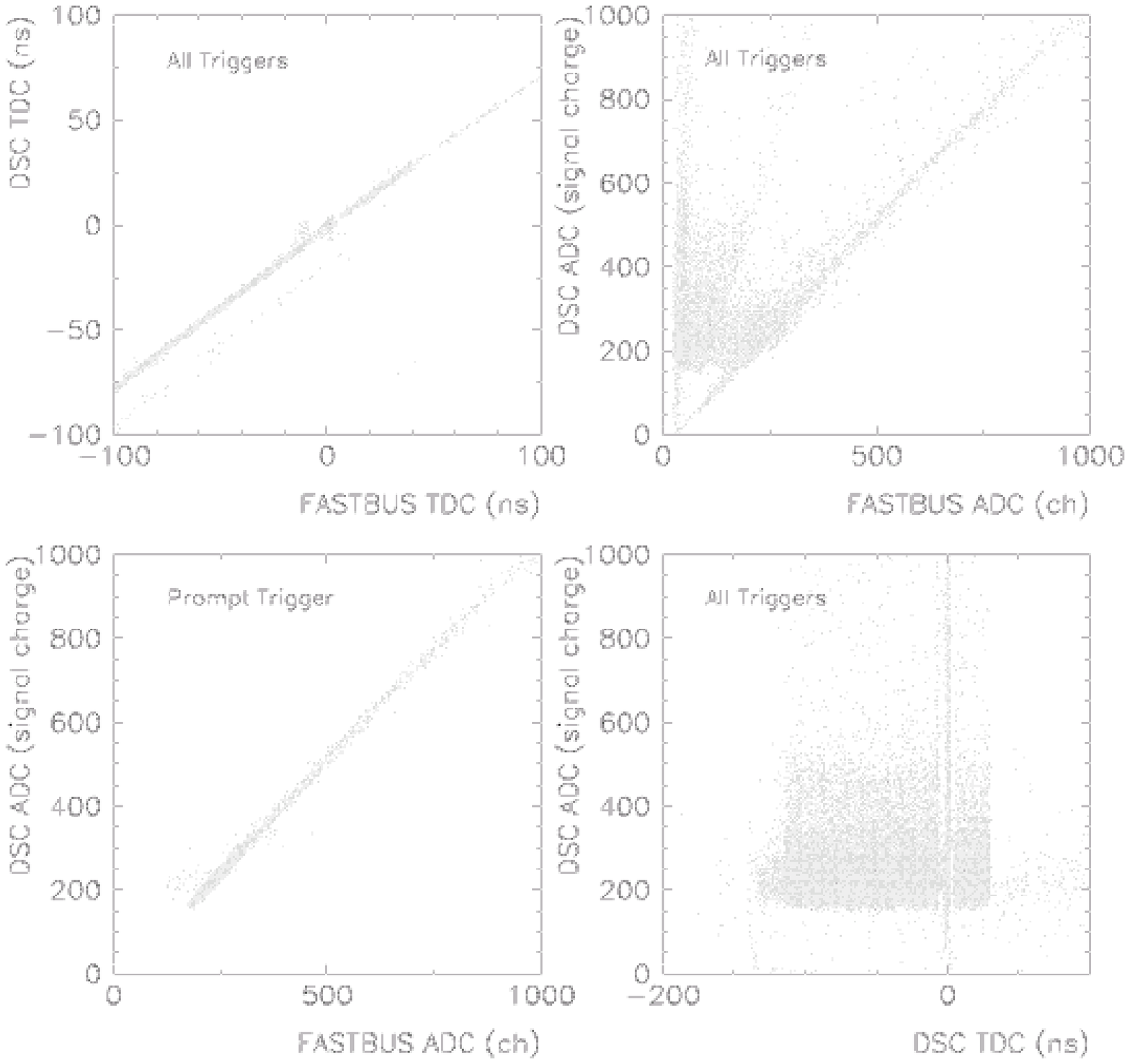,width=15.8cm}}  
\vglue 1.5cm
\centerline{FIGURE~\ref{fig:dsc_comp}} 
\vspace*{\stretch{2}}
\clearpage

\vspace*{\stretch{1}}
\centerline{\psfig{figure=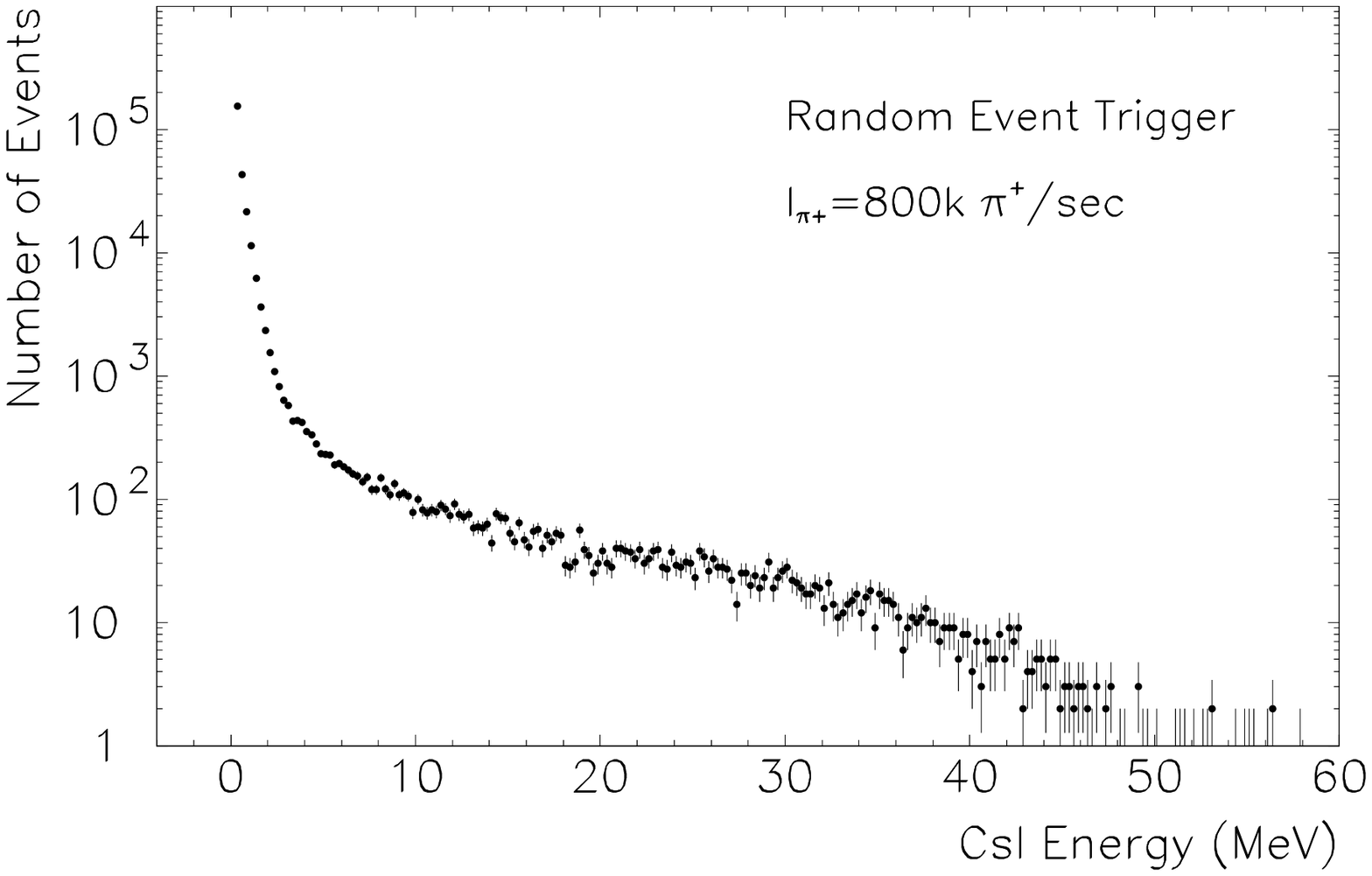,height=22cm}}  
\vglue -9cm
\centerline{FIGURE~\ref{fig:rcsi}}
\vspace*{\stretch{2}}
\clearpage

\vspace*{\stretch{1}}
\centerline{\psfig{figure=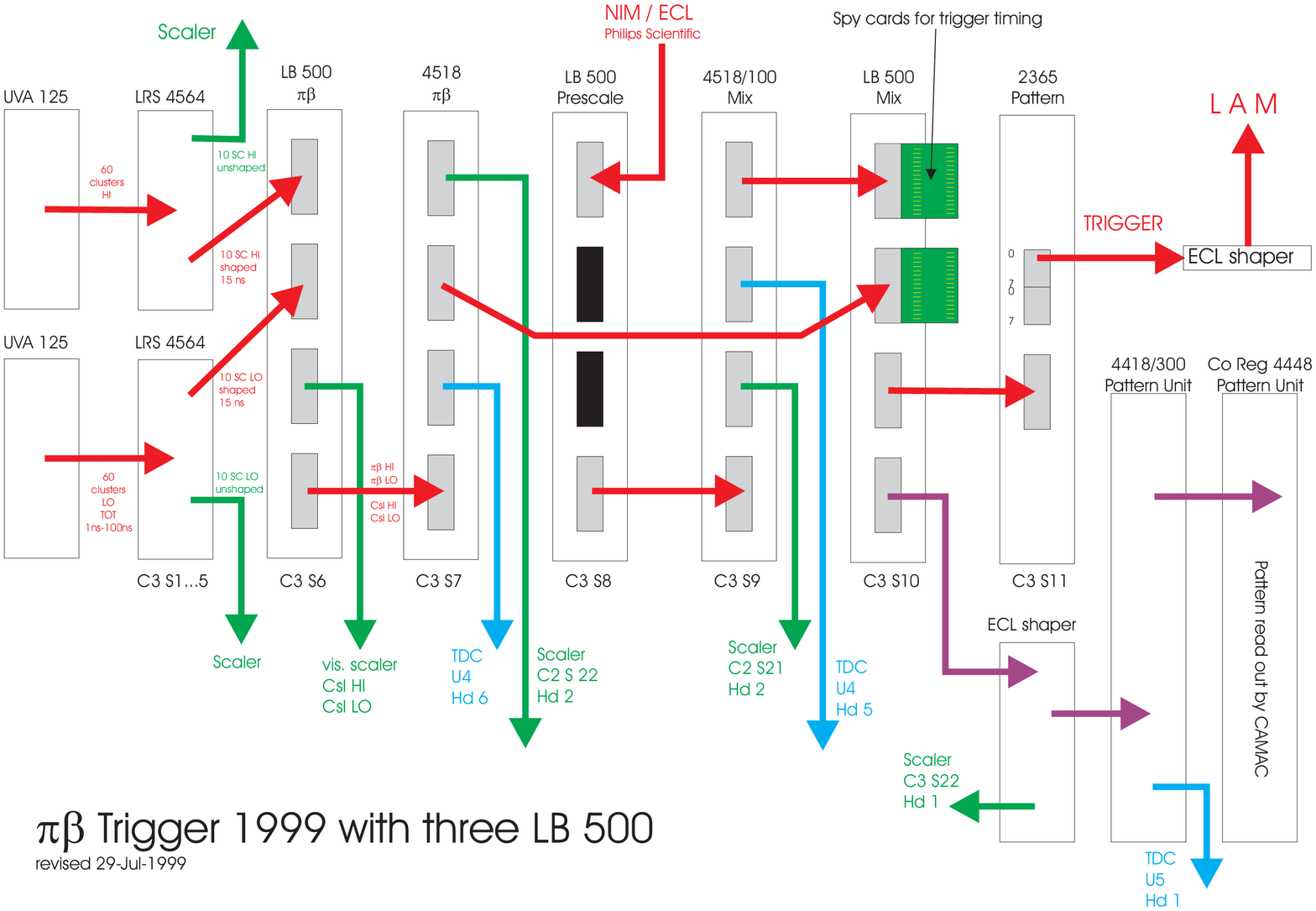,width=16cm}}  
\vglue 2cm
\centerline{FIGURE~\ref{fig:trig_ele}}
\vspace*{\stretch{2}}
\clearpage

\vspace*{\stretch{1}}
\centerline{\psfig{figure=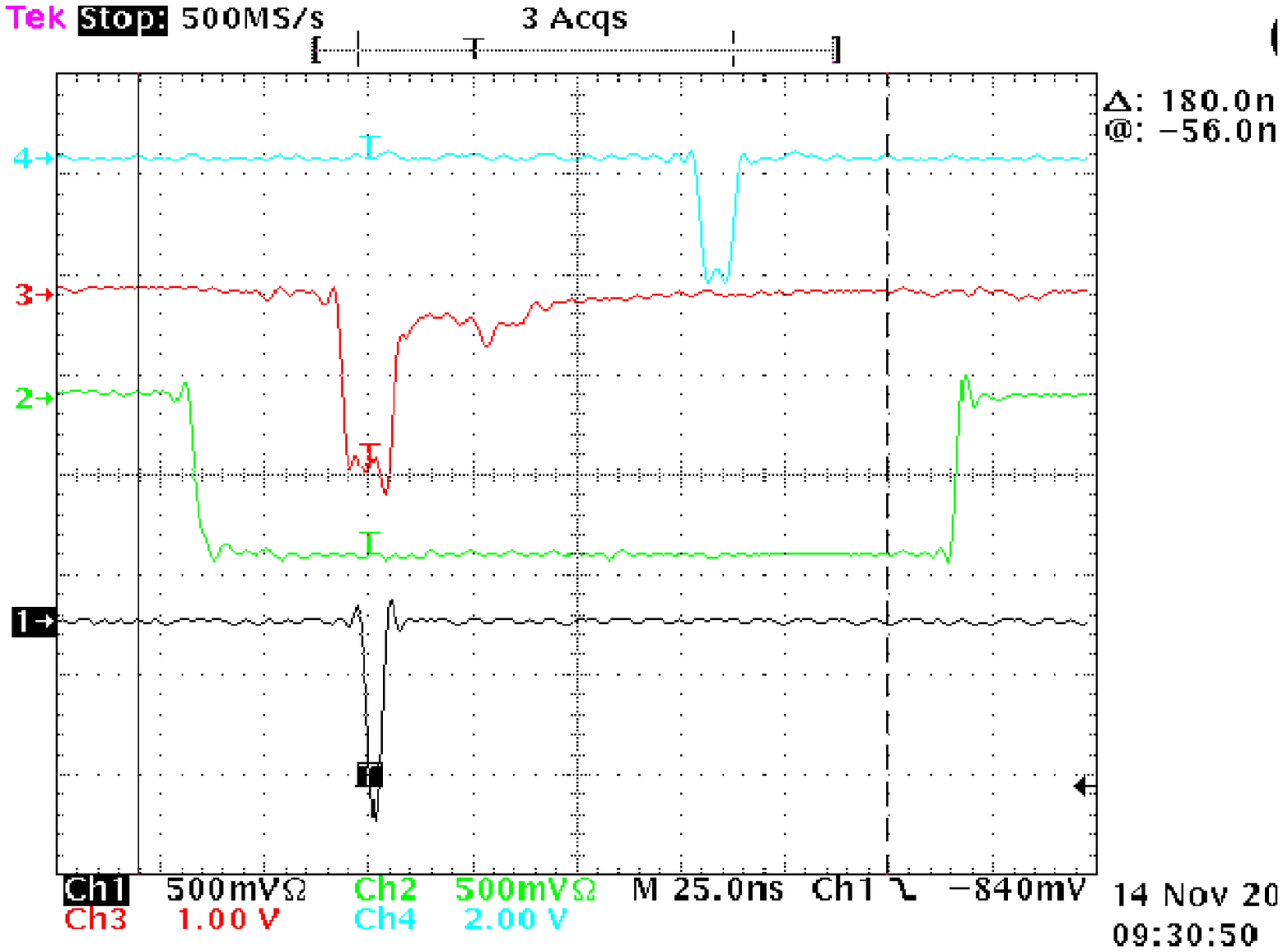,height=12cm}}
\bigskip\bigskip
\centerline{FIGURE~\ref{fig:trig_log}}
\vspace*{\stretch{2}}
\clearpage

\vspace*{\stretch{1}}
\hbox{\ }\vglue -1.5cm
\centerline{\psfig{figure=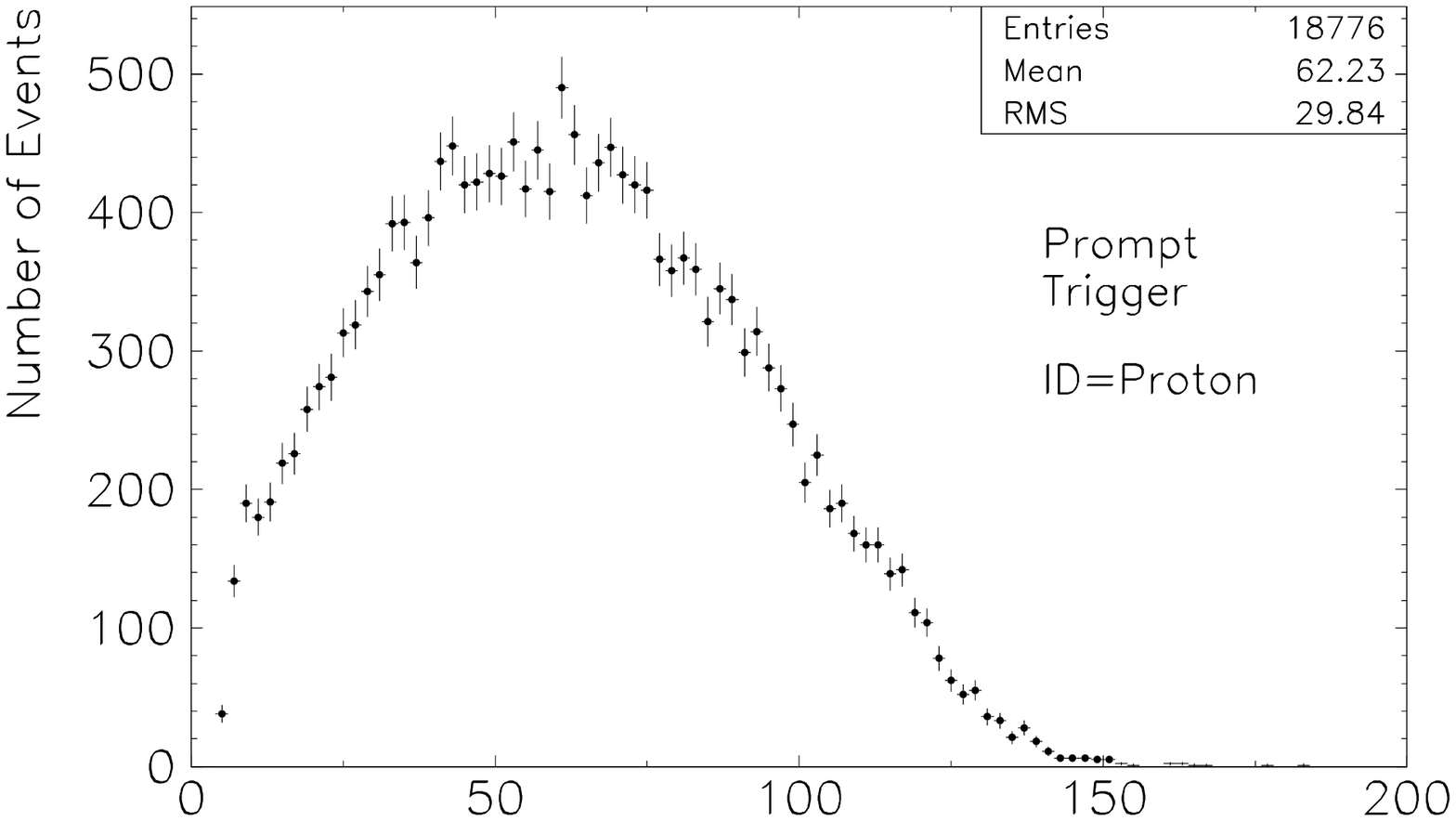,width=10cm}}
\vglue -6.0cm
\centerline{\psfig{figure=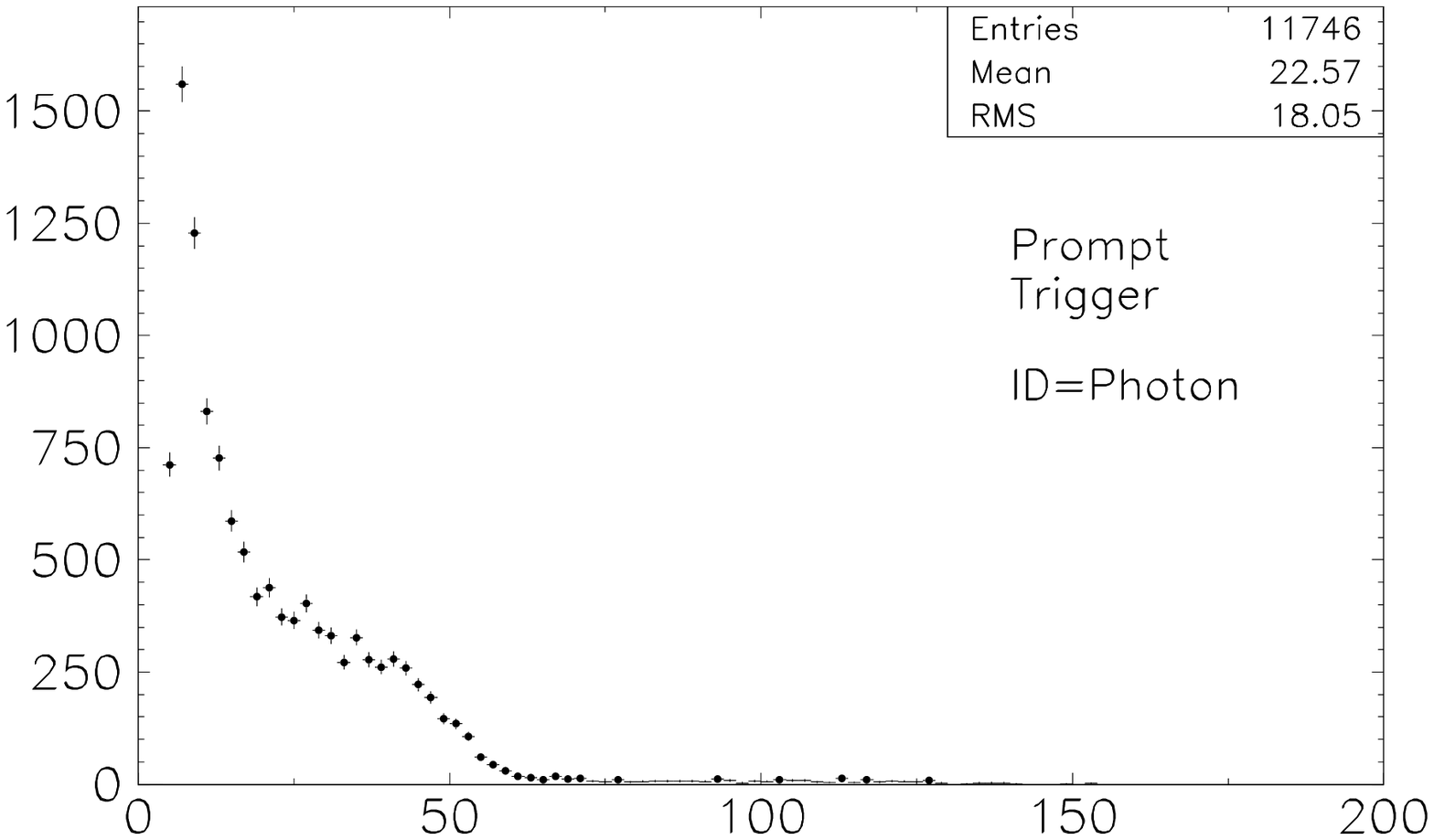,width=10cm}}
\vglue -6.0cm
\centerline{\psfig{figure=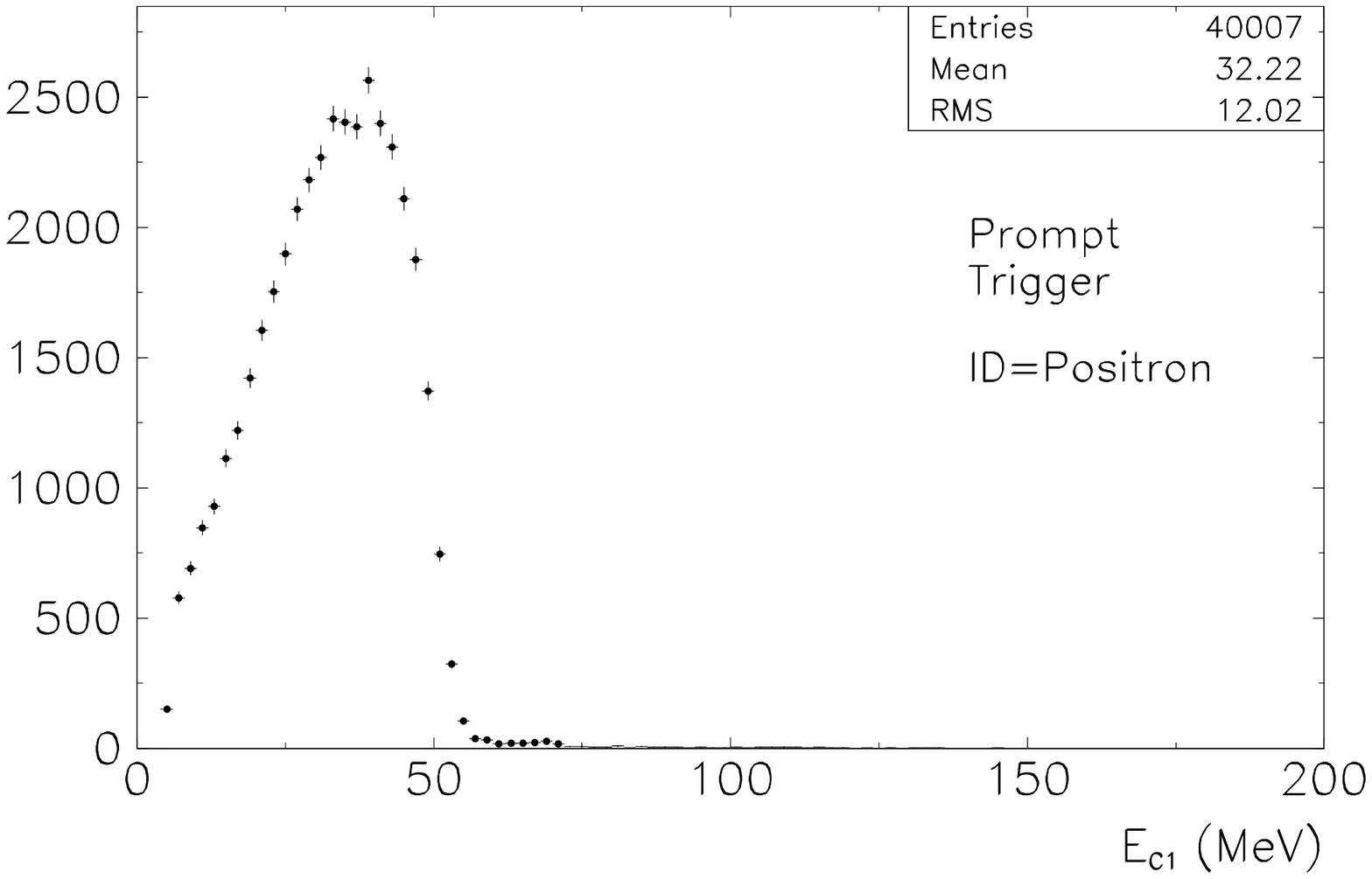,width=10cm}}
\vglue -6.0cm
\bigskip
\centerline{FIGURE~\ref{fig:prot_en}}
\vspace*{\stretch{2}}
\clearpage

\vspace*{\stretch{1}}
\centerline{\psfig{figure=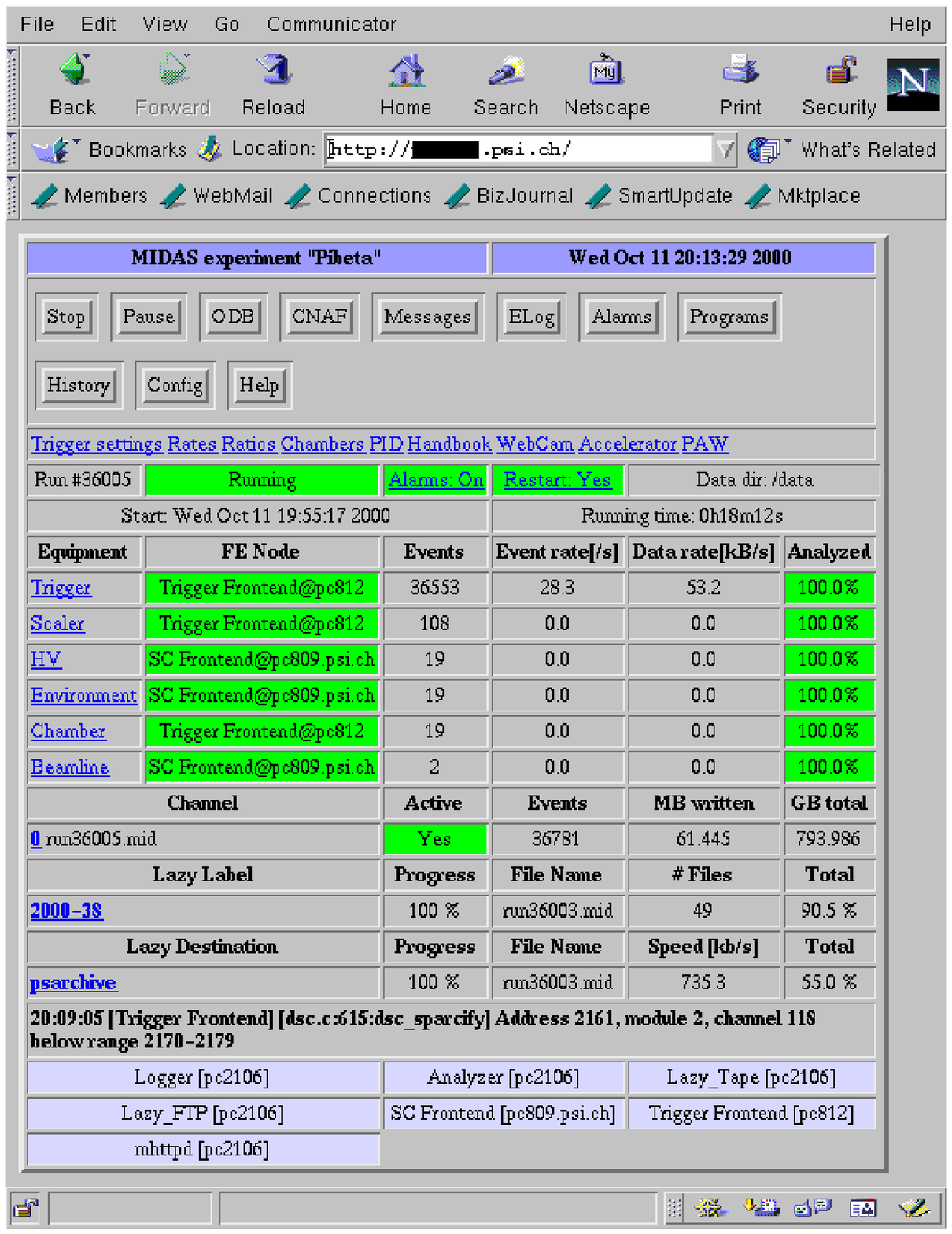,width=17cm}}   
\vglue -1.5cm
\centerline{FIGURE~\ref{fig:control}}
\vspace*{\stretch{2}}
\clearpage

\vspace*{\stretch{1}}
\vglue -1cm
\hbox{(i)}
\centerline{\psfig{figure=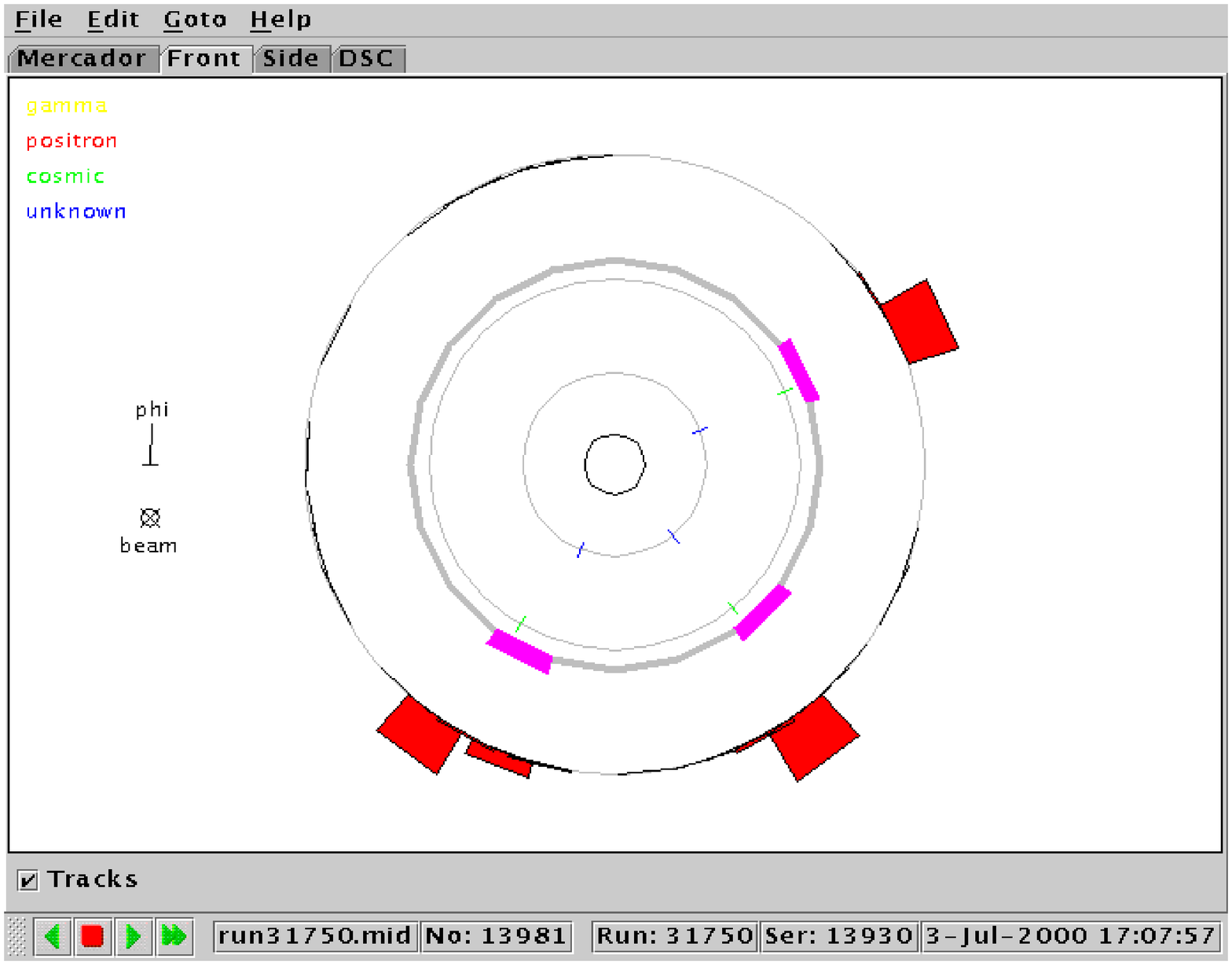,width=12cm}}  
\vglue 0.5cm
\hbox{(ii)}
\vglue 0.5cm
\centerline{\psfig{figure=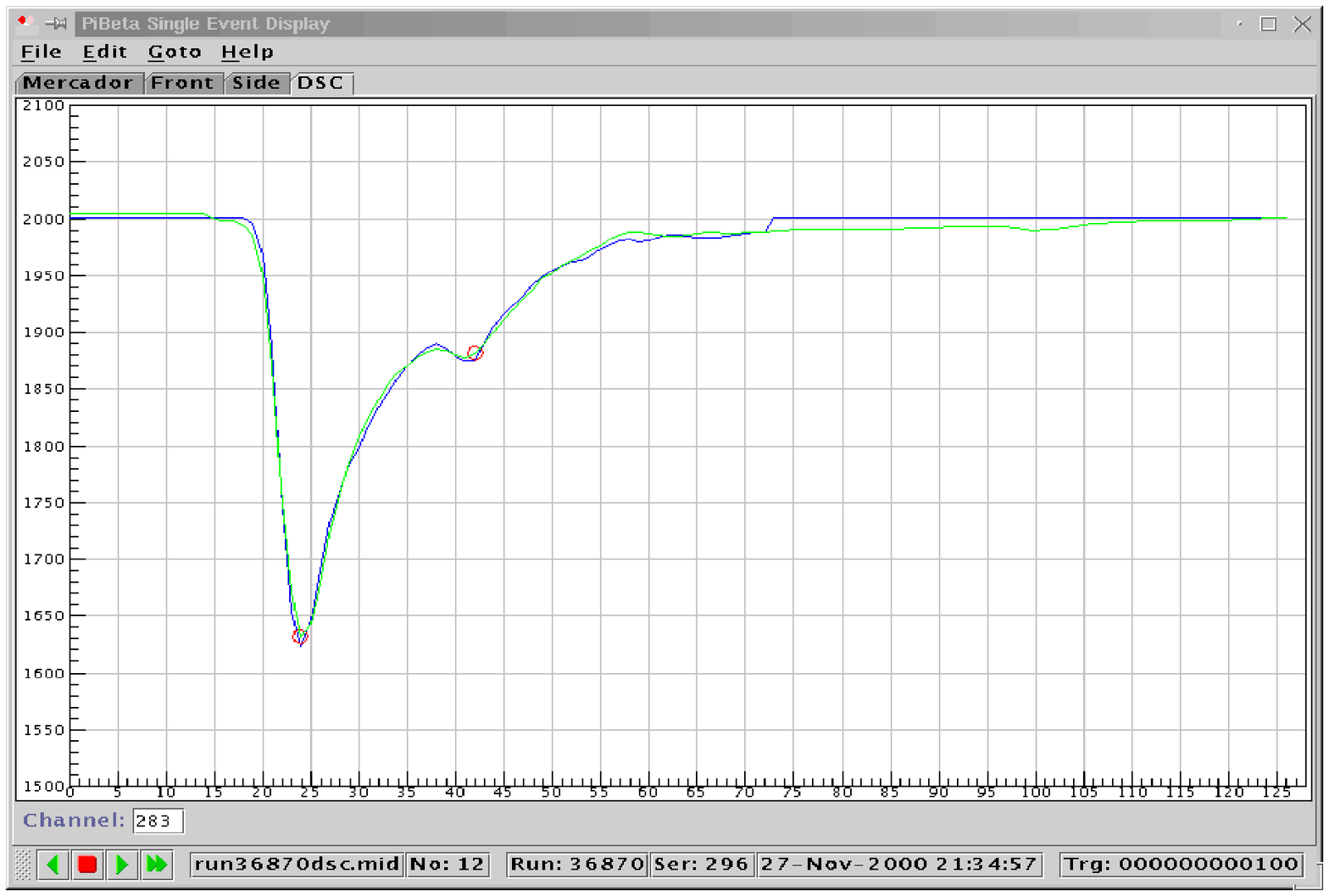,width=12cm}} 
\vglue 1.5cm
\centerline{FIGURE~\ref{fig:3e}}
\vspace*{\stretch{2}}
\clearpage

\vspace*{\stretch{1}}
\centerline{\psfig{figure=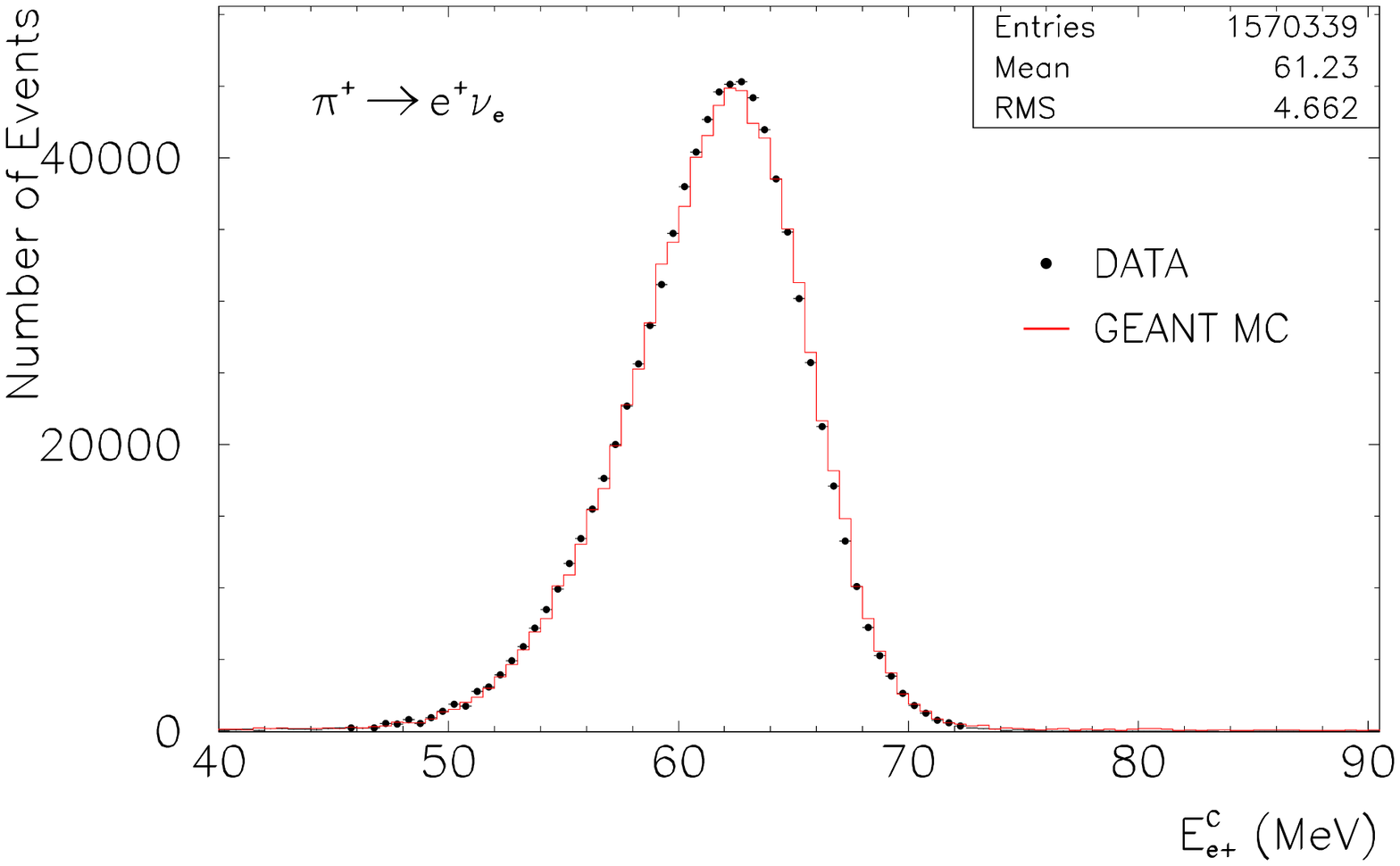,width=16cm}}
\vglue -9.5cm
\centerline{FIGURE~\ref{fig:pienu_en}}
\vspace*{\stretch{2}}
\clearpage

\vspace*{\stretch{1}}
\centerline{\psfig{figure=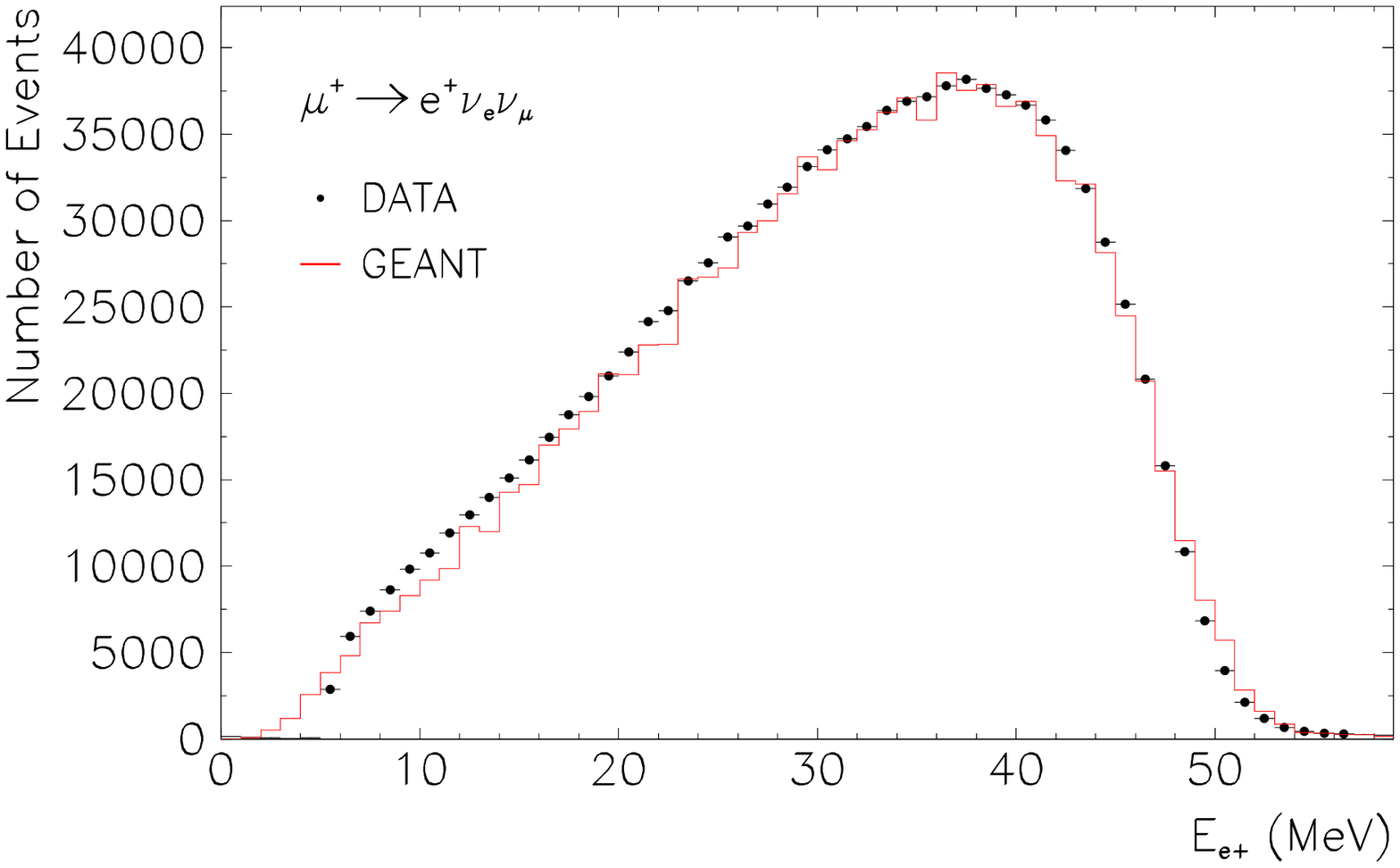,width=16cm}}
\vglue -9.5cm
\centerline{FIGURE~\ref{fig:michel_en}}
\vspace*{\stretch{2}}
\clearpage

\vspace*{\stretch{1}}
\centerline{\psfig{figure=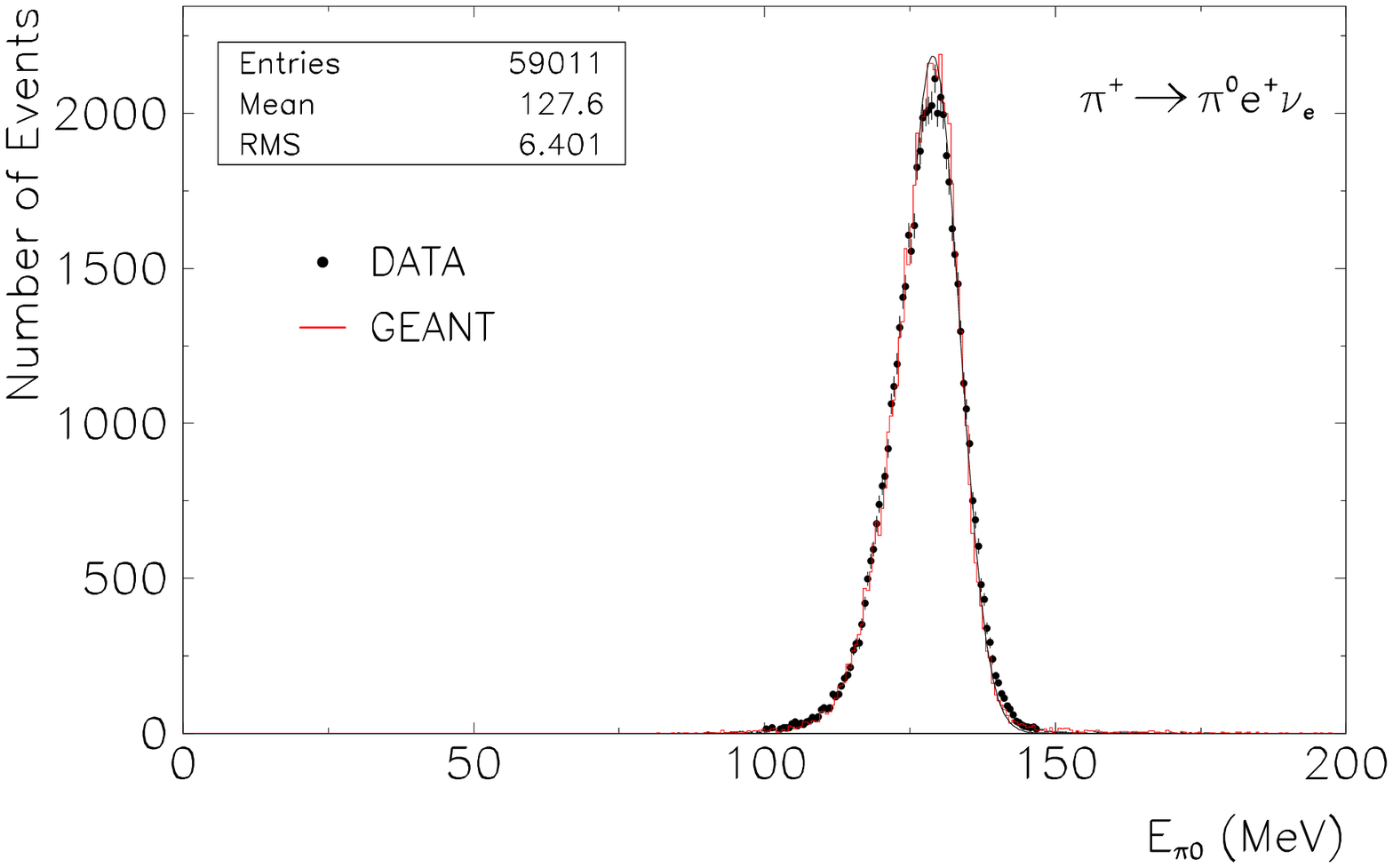,width=16cm}}
\vglue -9.8cm
\centerline{\psfig{figure=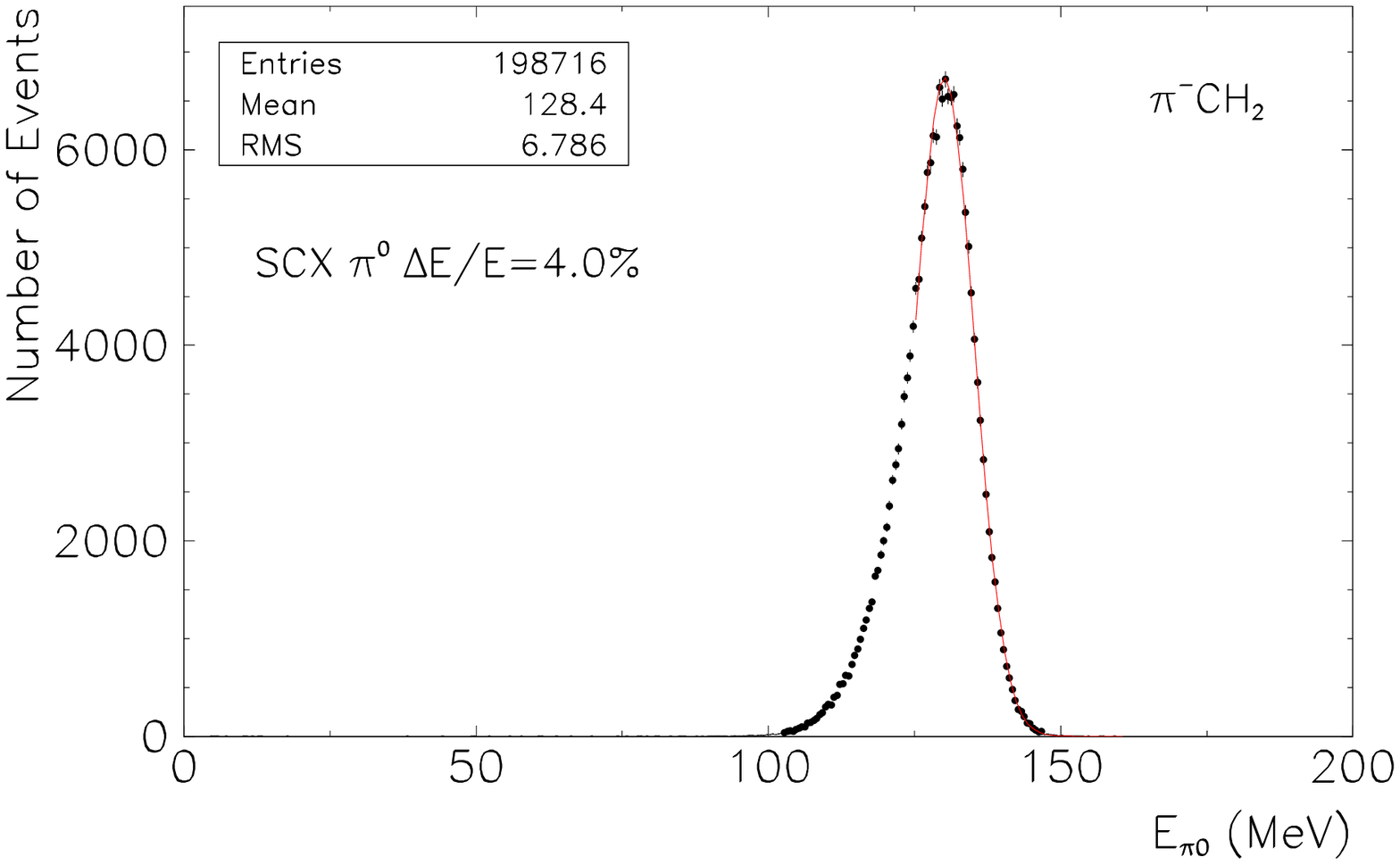,width=16cm}}
\vglue -9.8cm
\centerline{FIGURE~\ref{fig:pb_en}}
\vspace*{\stretch{2}}
\clearpage

\vspace*{\stretch{1}}
\centerline{\psfig{figure=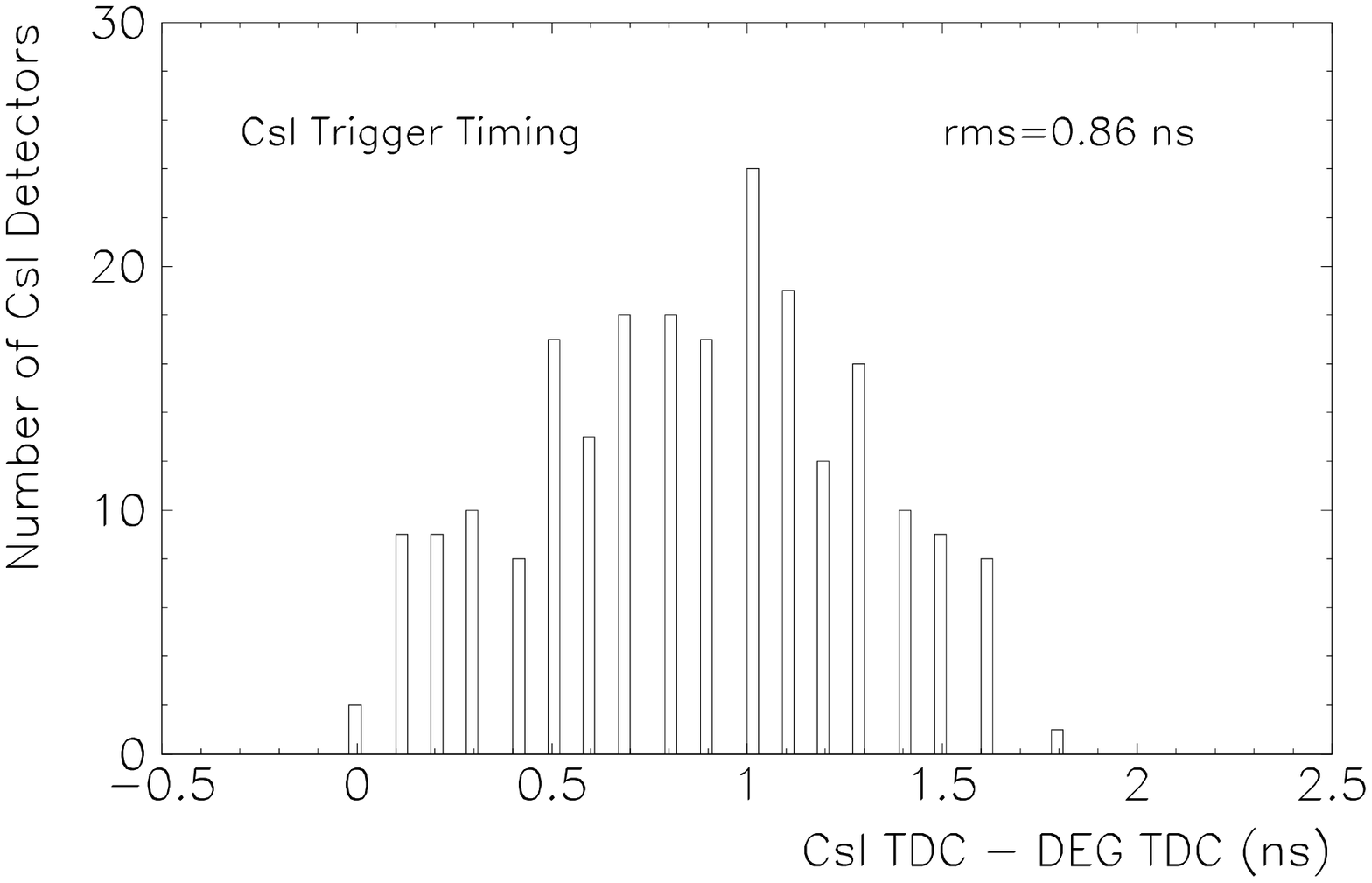,height=22cm}}
\vglue -9cm
\centerline{FIGURE~\ref{fig:csi_tr_tim}}
\vspace*{\stretch{2}}
\clearpage

\vspace*{\stretch{1}}
\centerline{\psfig{figure=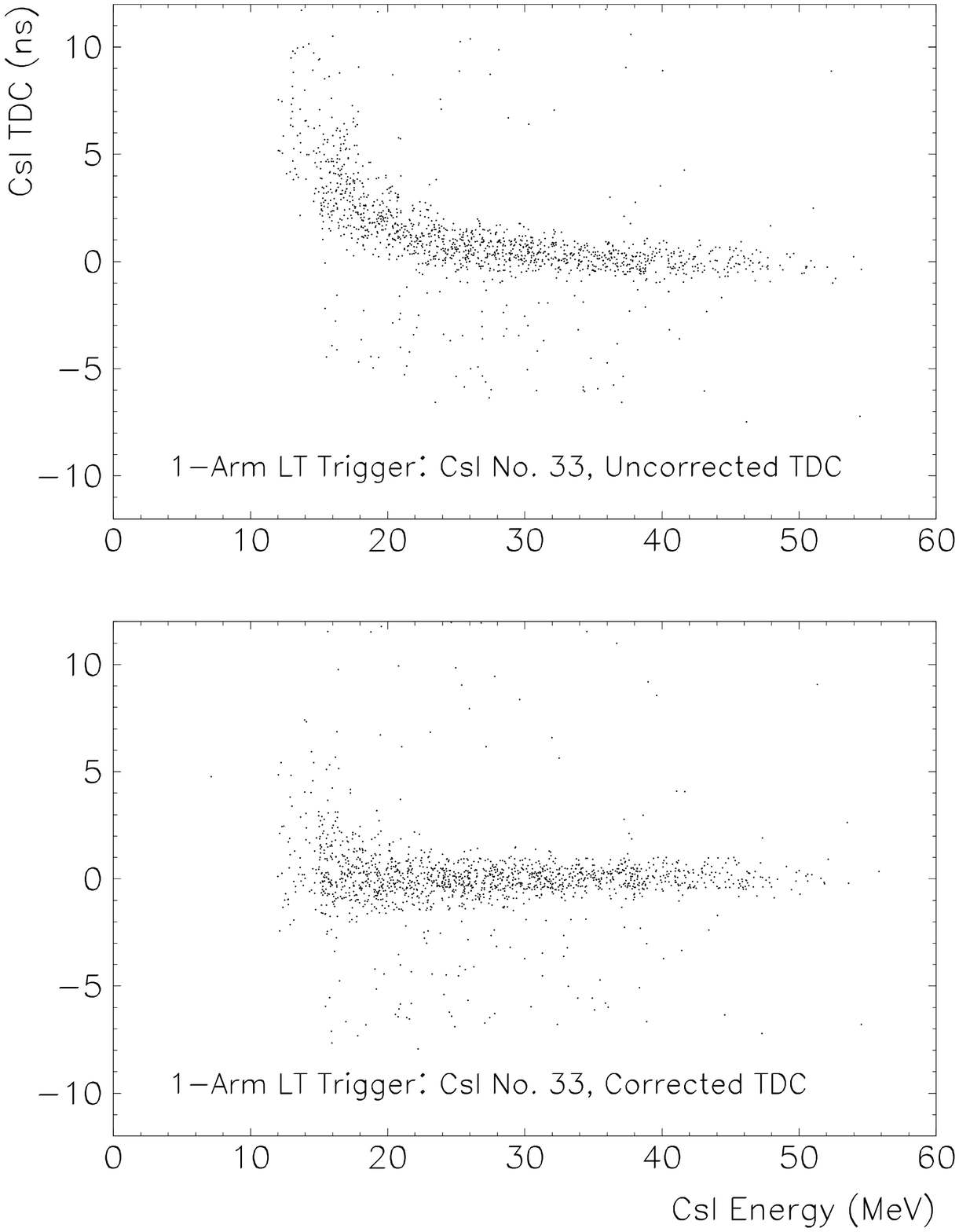,width=16cm}}
\bigskip
\centerline{FIGURE~\ref{fig:slew}}
\vspace*{\stretch{2}}
\clearpage

\vspace*{\stretch{1}}
\hbox{\ }\vglue -1.0cm
\centerline{\psfig{figure=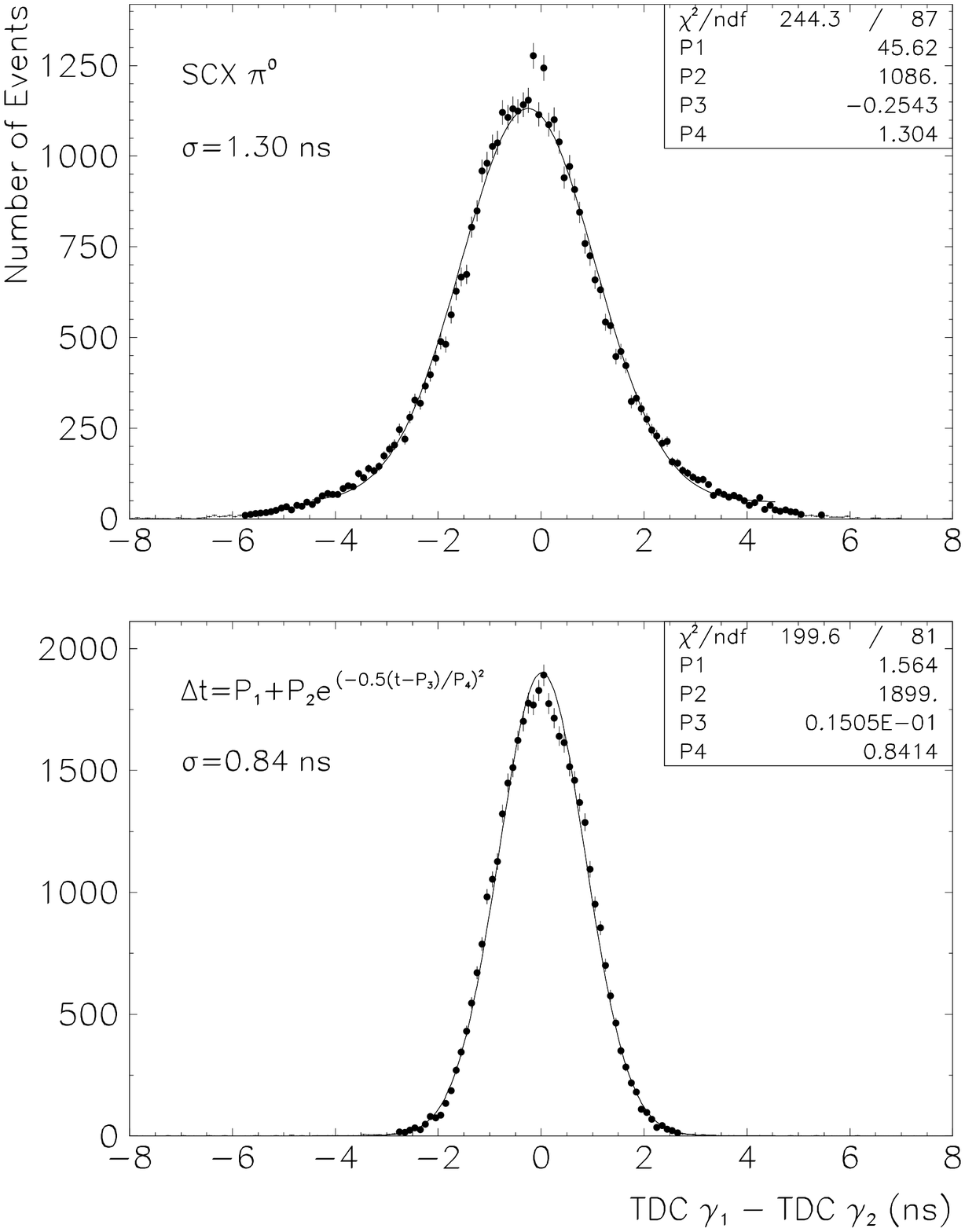,height=22cm}}
\centerline{FIGURE~\ref{fig:cal_t}}
\vspace*{\stretch{2}}
\clearpage

\vspace*{\stretch{1}}
\centerline{\psfig{figure=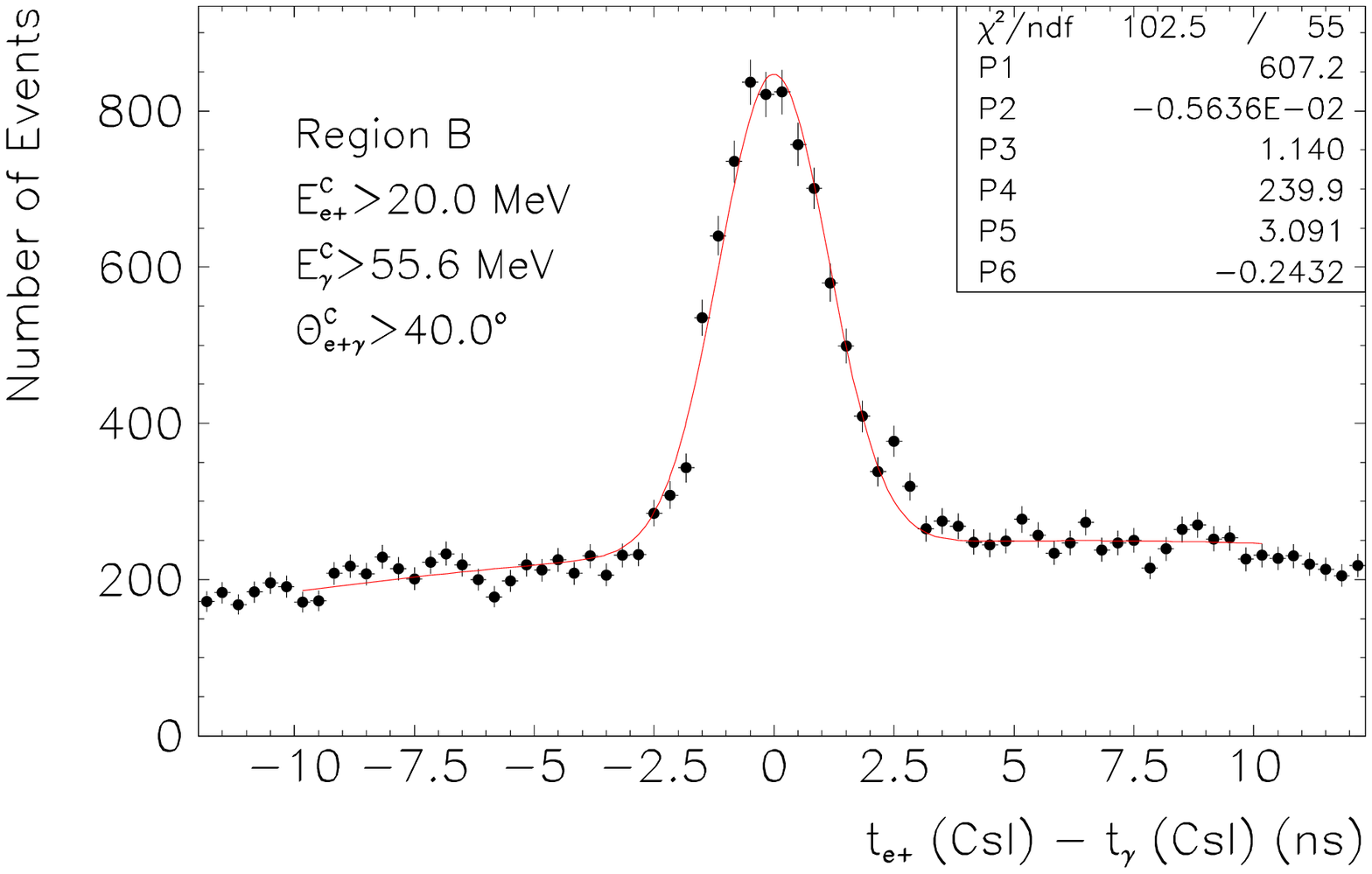,width=16cm}}
\vglue -9.8cm
\centerline{\psfig{figure=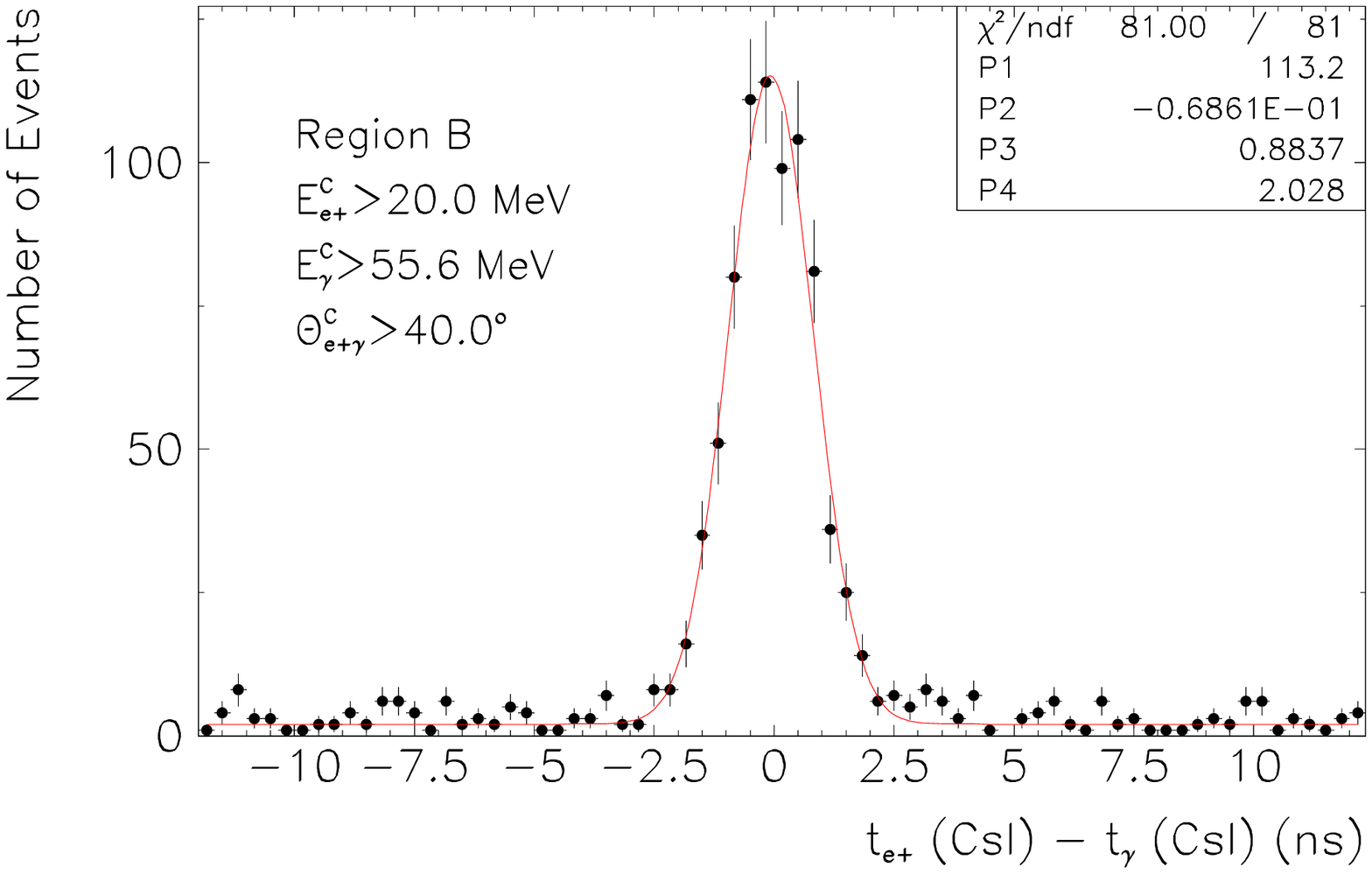,width=16cm}}
\vglue -9.8cm
\centerline{FIGURE~\ref{fig:pienug_coin}}
\vspace*{\stretch{2}}
\clearpage

\vspace*{\stretch{1}}
\centerline{\psfig{figure=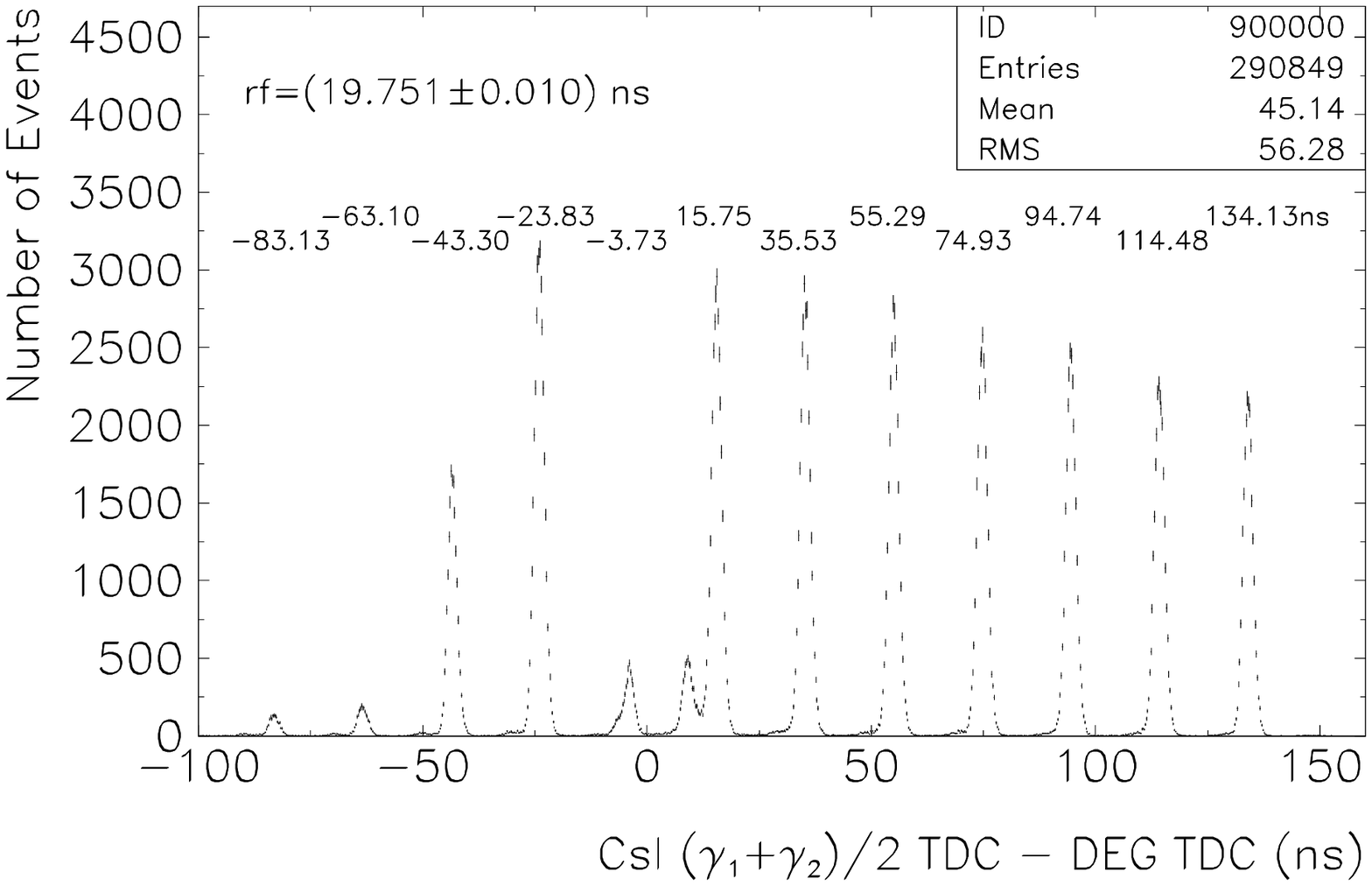,height=22cm}}  
\vglue -9cm
\centerline{FIGURE~\ref{fig:beam_time}}
\vspace*{\stretch{2}}
\clearpage

\vspace*{\stretch{1}}
\centerline{\psfig{figure=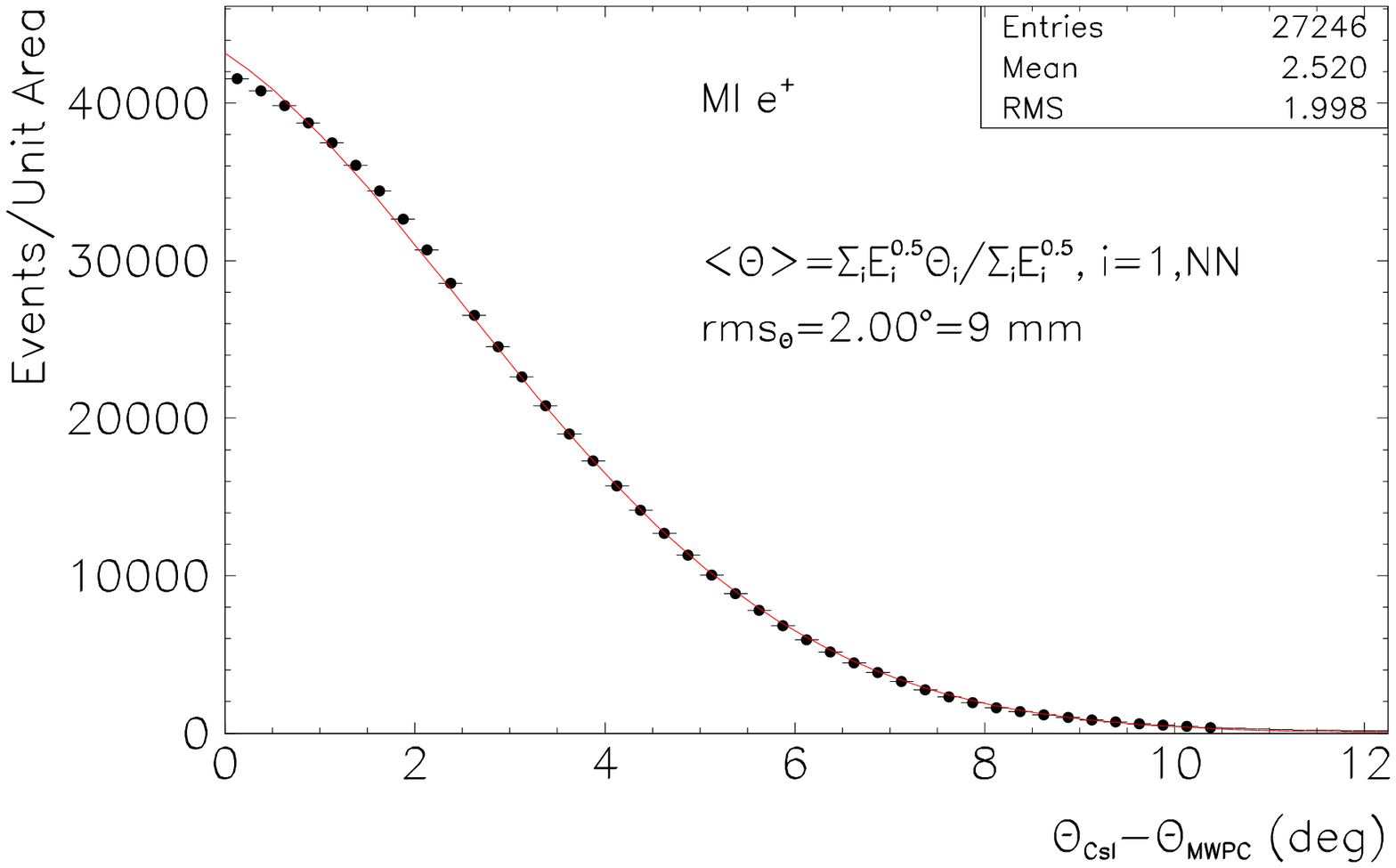,height=22cm}}  
\vglue -9cm
\centerline{FIGURE~\ref{fig:pos_resol}}
\vspace*{\stretch{2}}
\clearpage

\end{document}